\documentclass[structabstract]{aa}  
\usepackage{graphicx}
\usepackage{txfonts}
\usepackage{t1enc}
\usepackage{epsfig}
\usepackage{rotating}
\usepackage{verbatim}
\usepackage{subfig}
\usepackage{pifont}
\usepackage{natbib}
\usepackage{longtable}
\usepackage{supertabular}
\usepackage{lscape}
\usepackage{multirow}
\begin{document}
%%%%%%%%%% Local definitions %%%%%%%%%%%%%%%%%%%%%%%%%%%%%%%%%%%%%%%%%%
\newcommand{\Td}    {T_\mathrm{d}}
\newcommand{\Tex}   {T_\mathrm{ex}}
\newcommand{\Trot}  {T_\mathrm{rot}}
\newcommand{\mum}   {$\mu$m}
\newcommand{\kms}   {km~s$^{-1}$}
\newcommand{\cmg}   {cm$^{2}$~g$^{-1}$}
\newcommand{\cmt}   {cm$^{-3}$}
\newcommand{\jpb}   {$\rm Jy~beam^{-1}$}    
\newcommand{\lo}    {$L_{\sun}$}
\newcommand{\mo}    {$M_{\sun}$}
\newcommand{\nh}    {NH$_3$}
\newcommand{\nth}   {N$_2$H$^+$}
\newcommand{\chtoh} {CH$_3$OH}
\newcommand{\water} {H$_2$O}
\newcommand{\et}    {et al.}
\newcommand{\eg}    {e.\,g.,}
\newcommand{\ie}    {i.\,e.,}
\newcommand{\hi}    {\ion{H}{i}}
\newcommand{\hii}   {\ion{H}{ii}}
\newcommand{\uchii} {UC~\ion{H}{ii}}
\newcommand{\hchii} {HC~\ion{H}{ii}}
\newcommand{\raun}  {$^\mathrm{h~m~s}$}
\newcommand{\deun}  {$\mathrm{\degr~\arcmin~\arcsec}$}
\newcommand{\taba}  {\tablefootmark{a}}
\newcommand{\tabb}  {\tablefootmark{b}}
\newcommand{\tabc}  {\tablefootmark{c}}
\newcommand{\tabd}  {\tablefootmark{d}}
\newcommand{\tabe}  {\tablefootmark{e}}
\newcommand{\tabf}  {\tablefootmark{f}}
\newcommand{\tabg}  {\tablefootmark{g}}
\newcommand{\tabh}  {\tablefootmark{h}}
\newcommand{\tabi}  {\tablefootmark{i}}
\newcommand{\tabj}  {\tablefootmark{j}}
\newcommand{\tabk}  {\tablefootmark{k}}
\newcommand{\tabz}  {\phantom{\tablefootmark{z}}}
\newcommand{\supa}  {$^\mathrm{a}$}
\newcommand{\supb}  {$^\mathrm{b}$}
\newcommand{\supc}  {$^\mathrm{c}$}
\newcommand{\supd}  {$^\mathrm{d}$}
\newcommand{\supe}  {$^\mathrm{e}$}
\newcommand{\supf}  {$^\mathrm{f}$}
\newcommand{\supg}  {$^\mathrm{g}$}
\newcommand{\suph}  {$^\mathrm{h}$}
\newcommand{\phn}   {\phantom{0}}
\newcommand{\phnn}  {\phantom{0}\phantom{0}}
\newcommand{\phnnn} {\phantom{0}\phantom{0}\phantom{0}}
\newcommand{\phnnnn}{\phantom{0}\phantom{0}\phantom{0}\phantom{0}}
\newcommand{\phe}   {\phantom{$^\mathrm{c}$}}
\newcommand{\phb}   {\phantom{$>$}}
\newcommand{\phl}   {$<$}
\newcommand{\phm}   {$>$}
\newcommand{\phnm}  {\phantom{0}\phantom{$.$}}
\newcommand{\phbn}  {\phantom{$>$}\phantom{0}}
\newcommand{\phbnn} {\phantom{$>00$}}
\newcommand{\phnb}  {\phantom{0}\phantom{$>$}}
\newcommand{\phmm}  {\phantom{\pm0.0}}
\newcommand{\phmn}  {\phantom{0\pm0.00}}
\newcommand{\swp}   {~;}
\newcommand{\sep}   {\,/\,}
\newcommand{\cmark} {\ding{51}}
\def\HII{H{\sc ii}}
\def\Nly{\mbox{$N_{\rm Ly}$}}
\def\Lbol{\mbox{$L_{\rm bol}$}}
\def\ne{\mbox{$n_{\rm e}$}}
\def\Ro{R_{\rm o}}
\def\Rd{R_{\rm d}}
\def\Rh{R_{\rm h}}
\def\zd{z_{\rm d}}
\def\NHII{\mbox{$N_{\rm HII}$}}
\def\d{{\rm d}}
\def\e{{\rm e}}
\def\Log{{\rm Log}}
\def\fvol{\frac{\d N}{\d V}}
\def\flum{\frac{\d N}{\d\Log L}}
\def\fne{\frac{\d N}{\d\Log n_{\rm e}}}
\def\fD{\mbox{$\frac{\d N}{\d\Log D_{\rm HII}}$}}
%%%%%%%%%%%%%%%%%%%%%%%%%%%%%%%%%%%%%%%%%%%%%%%%%%%%%%%%%%%%%%%%%%%%%%%
%
%
   \title{Evolution and excitation conditions of outflows \\ in high-mass star-forming regions}

%   \subtitle{First results of IRAM~30\,m observations}

   \author{\'A.S\'anchez-Monge\inst{1} 
           \and A.~L\'opez-Sepulcre\inst{2}
           \and R.~Cesaroni\inst{1}
           \and C.~M.~Walmsley\inst{1,3}
           \and C.~Codella\inst{1}
           \and M.~T. Beltr\'an\inst{1}
           \and M.~Pestalozzi\inst{4}
           \and S.~Molinari\inst{4}
          }

   \offprints{\'Alvaro S\'anchez-Monge \email{asanchez@arcetri.astro.it}}

   \institute{Osservatorio Astrofisico di Arcetri, INAF, Largo Enrico Fermi 5, I-50125, Firenze, Italy
              \and UJF-Grenoble 1 / CNRS-INSU, Institut de Plan\'etologie et d'Astrophysique de Grenoble (IPAG) UMR 5274, Grenoble, F-38041, France
              \and Dublin Insitute for Advanced Studies (DIAS), 31 Fitzwilliam Place, Dublin 2, Ireland
              \and Istituto di Astrofisica e Planetologia Spaziali (IAPS-INAF), Via Fosso del Cavaliere 100, I-00133, Roma, Italy
             }

   \date{Received; accepted }

  \abstract
  % context heading (optional)
  % {} leave it empty if necessary  
{Theoretical models suggest that massive stars form via disk-mediated accretion, in a similar fashion to low-mass stars. In this scenario, bipolar outflows ejected along the disk axis play a fundamental role, and their study can help to characterize the different evolutionary stages involved in the formation of a high-mass star. A recent study toward massive molecular outflows has revealed a decrease of the SiO line intensity as the object evolves.}
  % aims heading (mandatory)
{The present study aims at characterizing the variation of the molecular outflow properties with time, and at studying the SiO excitation conditions in outflows associated with high-mass young stellar objects (YSOs).}
  % methods heading (mandatory)
{We used the IRAM~30-m telescope on Pico Veleta (Spain) to map 14 high-mass star-forming regions in the SiO\,(2--1), SiO\,(5--4) and HCO$^+$\,(1--0) lines, which trace the molecular outflow emission. The FTS backend, covering a total frequency range of $\sim$15~GHz, allowed us to simultaneously map several dense gas (\eg\ N$_2$H$^+$, C$_2$H, NH$_2$D, H$^{13}$CN) and hot core (CH$_3$CN) tracers. We used the Hi-GAL data to improve the previous spectral energy distributions, and obtain a more accurate dust envelope mass and bolometric luminosity for each source. We calculated the luminosity-to-mass ratio, which is believed to be a good indicator of the evolutionary stage of the YSO.}
  % results heading (mandatory)
{We detect SiO and HCO$^+$ outflow emission in all the fourteen sources, and bipolar structures in six of them. The outflow parameters are similar to those found toward other massive YSOs with luminosities $10^{3}$--$10^{4}$~\lo. We find an increase of the HCO$^+$ outflow energetics as the object evolve, and a decrease of the SiO abundance with time, from $10^{-8}$ to $10^{-9}$. The SiO\,(5--4) to (2--1) line ratio is found to be low at the ambient gas velocity, and increases as we move to red/blue-shifted velocities, indicating that the excitation conditions of the SiO change with the velocity of the gas. In particular, the high-velocity SiO gas component seems to arise from regions with larger densities and/or temperatures, than the SiO emission at the ambient gas velocity.}
  % conclusions heading (optional), leave it empty if necessary 
{The properties of the SiO and HCO$^+$ outflow emission suggest a scenario in which SiO is largely enhanced in the first evolutionary stages, probably due to strong shocks produced by the protostellar jet. As the object evolves, the power of the jet would decrease and so does the SiO abundance. During this process, however, the material surrounding the protostar would have been been swept up by the jet, and the outflow activity, traced by entrained molecular material (HCO$^+$), would increase with time.}


   \maketitle
%
%________________________________________________________________

%----------------------------------------------------------------------------
\begin{table*}[ht!]
\caption{\label{t:sample}List of observed sources, and spectral energy distribution parameters}
\centering
\begin{tabular}{c l c c c c c c c c c c c}
\hline\hline
&
&$\alpha$
&$\delta$
&$V_\mathrm{lsr}$\supb
&$D$
&
&\multicolumn{3}{c}{SED fit\supd}
&$L_\mathrm{bol}$\supe
&$L/M$
\\
\cline{8-10}
\#
&Source\supa
&(J2000)
&(J2000)
&(km~s$^{-1}$)
&(kpc)
&IR\supc
&$\beta$
&$T_\mathrm{dust}$ (K)
&$M$ (\mo)
&(\lo)
&(\lo~\mo$^{-1}$)
\\
\hline
%																																	   Vlsr    ANA REID BRAND
01	&18151$-$1208\_1		&18:17:58.0 &$-$12:07:27.0 &$+$33.0 		&3.0		&L		&1.7		&32		&\phn478		&26090		&54.6		\\ %+33.3  3.0 2.84 3.07 -
02	&G19.27$+$0.1M2		&18:25:52.6 &$-$12:04:48.0 &$+$26.9 		&2.4		&D		&2.5		&35		&\phnn77		&\phnn157	&\phn2.0		\\ %+26.9  2.4 2.40 2.58 -
03	&G19.27$+$0.1M1		&18:25:58.5 &$-$12:03:59.0 &$+$26.5 		&2.4		&D		&1.3		&22		&\phn193		&\phnn374	&\phn1.9		\\ %+26.5  2.4 2.37 2.55 -
04	&18236$-$1205		&18:26:25.4 &$-$12:03:51.4 &$+$26.5 		&2.5		&L		&1.0		&29		&\phn935		&\phn4507	&\phn4.8		\\ %+26.5  2.5 2.37 2.54 -
05	&18264$-$1152		&18:29:14.4 &$-$11:50:21.3 &$+$43.7 		&3.5		&L		&1.3		&27		&1634		&11910		&\phn7.3		\\ %+43.9  3.5 3.32 3.59 -
06	&G23.60$+$0.0M1		&18:34:11.6 &$-$08:19:06.0 &$+$106.5\phn &6.2		&D		&2.0		&18		&1820		&\phn5207	&\phn2.9		\\ %+106.5 6.2 5.64 6.26 * 3.9
07	&18316$-$0602		&18:34:20.5 &$-$05:59:30.4 &$+$42.5 		&3.1		&L		&1.2		&33		&1613		&31820		&19.7		\\ %+42.5  3.1 2.95 3.19 -
08	&G23.60$+$0.0M2		&18:34:21.1 &$-$08:18:07.0 &$+$53.5 		&3.9		&D		&1.9		&23		&\phn287		&\phn3049	&10.6		\\ %+53.6  3.9 3.57 3.87 -
09	&G24.33$+$0.1M1		&18:35:07.9 &$-$07:35:04.0 &$+$113.6\phn &6.7		&D		&2.2		&23		&2674		&48050		&18.0		\\ %+113.6 6.7 5.93 6.71 * 3.8
10	&G34.43$+$0.2M1\supf	&18:53:18.0 &$+$01:25:23.0 &$+$57.9 		&3.7		&D		&1.8		&26		&1369		&24050		&17.6		\\ %+58.1  3.7 3.59 3.92 -
11	&18507$+$0121		&18:53:19.6 &$+$01:24:37.1 &$+$57.6 		&3.7		&L		&1.0		&29		&3065		&14530		&\phn4.7		\\ %+58.1  3.7 3.58 3.90 -
12	&G34.43$+$0.2M3		&18:53:20.4 &$+$01:28:23.0 &$+$59.2 		&3.7		&D		&1.5		&21		&\phn612		&\phn1357	&\phn2.2		\\ %+59.4  3.7 3.66 4.00 -
13	&19095$+$0930		&19:11:54.0 &$+$09:35:52.0 &$+$43.9 		&3.3		&L		&1.6		&33		&\phn971		&50680		&52.2		\\ %+43.9  3.3 3.07 3.30 -
14	&23139$+$5939		&23:16:11.1 &$+$59:55:30.8 &$-$44.5 		&4.8		&L		&1.6		&30		&\phn972		&26040		&26.8		\\ %-44.1  4.8 3.54 4.50 *
\hline
\end{tabular}
\begin{list}{}{}
\item[\supa] Name of sources starting with numbers (\eg\ 18151$-$1208) refer to the IRAS name \citep{neugebauer1984}.
\item[\supb] Systemic velocity ($V_\mathrm{LSR}$) derived from the hyperfine fit to the N$_2$H$^+$\,(1--0) and C$_2$H\,(1--0) lines (see Sect.~\ref{s:resmolecules}).
\item[\supc] Sources classified as IR-luminous (IRL) or IR-dark (IRD) according to its emission or not at mid-IR wavelengths according to \citet{lopezsepulcre2010}.
\item[\supd] Parameters obtained from the single-temperature, modified black body fit to the spectral energy distribution (see Sect.~\ref{s:resSEDs}).
\item[\supe] Bolometric luminosity derived by integrating over the full observed spectral distribution (see Sect.~\ref{s:resSEDs}).
\item[\supf] Observed simultaneously with (in the same map as) 18507$+$0121.
\end{list}
\end{table*}
%----------------------------------------------------------------------------

%%%%%%%%%%%%%%%%%%%%%%%%%%%%%%%%%%%%%%%%%%%%%%%%%%%%%%%%%%%%%%%%%%%%%%%
\section{Introduction\label{s:intro}}

Establishing an evolutionary sequence for high-mass young stellar objects (YSOs) is one of the hot topics of current star formation research. It has been proposed that high accretion rates \citep[\eg][]{mckeetan2003} and/or accretion through massive disks \citep[\eg][]{krumholz2005} can explain the formation of massive stars. Recently, \citet{kuiper2010, kuiper2011} have demonstrated that stars with masses up to 140~\mo, can be formed via disk-mediated accretion. In this context a fundamental role is played by a bipolar outflow ejected along the disk axis. So far, a number of deeply embedded massive disk/outflow systems \citep[see][]{cesaroni2007} have been found, lending support to such models.

Studying the properties of molecular outflows can provide information on the different evolutionary stages during the formation process of a massive star. \citet{beuthershepherd2005} propose an evolutionary sequence for outflows driven by high-mass protostars in which a well-collimated outflow/jet gradually evolves into a wide-angle outflow/wind as the ionizing radiation powered by the central massive stellar object becomes more dominant. However, understanding the outflow population in high-mass star-forming regions is not always easy, due to the high level of clustering in these regions. Many high-mass star-forming regions show evidence for several outflows (\eg\ 05358$+$3543: \citealt{beuther2002c}, AFGL\,5142: \citealt{zhang2007}). In such a situation, lack of outflow bipolar signatures is to be expected in clustered star-forming regions, being thus interesting the identification of these bipolar structures.

Most of the observations of massive molecular outflows carried up to date, have focused on the CO and HCO$^+$ species, and their isotopologues. These molecules typically trace the low-velocity, extended, entrained gas component of the outflow and, in some cases, appear contaminated by the surrounding (infalling) envelope. Observations of a more reliable jet tracer are thus necessary to better characterize the jet/outflow properties. SiO emission is ideal for this purpose, because its formation is attributed to sputtering or vaporization of Si atoms from grains due to fast shocks \citep{gusdorf2008a, gusdorf2008b, guillet2009}, and thus suffers minimal contamination from quiescent or infalling envelopes. Recently, \citet{lopezsepulcre2011} studied the SiO\,(2--1) and (3--2) line emission toward a sample of 57 high-mass YSOs. These authors found that the intensity of the SiO line becomes fainter for increasing luminosity-to-mass ratio ($L/M$), considered an indicator of the evolutionary stage of YSOs. The variation of the SiO line intensity was interpreted as a decrease in the SiO abundance with time (as proposed by \citealt{sakai2010}) and/or a decrease in the jet/outflow mass with time. 

With this in mind, we performed SiO map observations toward a sub-sample of the \citet{lopezsepulcre2011} sample, to derive the properties of the SiO molecular outflows and better characterize the variation of SiO with time. We also map the outflow emission in HCO$^+$, \ie\ a tracer of the most extended and entrained gas, and the dense gas emission in different species such as N$_2$H$^+$ or C$_2$H. In Sect.~\ref{s:obs}, we describe the observations and the sample of sources. In Sect.~\ref{s:res} we present the main results derived from the IRAM\,30~m observations as well as from the fit of the spectral energy distributions. Finally, in Sect.~\ref{s:dis} we discuss our results focusing our analysis on the properties of molecular outflows, and in Sect.~\ref{s:con} we summarize our main results.

%%%%%%%%%%%%%%%%%%%%%%%%%%%%%%%%%%%%%%%%%%%%%%%%%%%%%%%%%%%%%%%%%%%%%%%
\section{Observations\label{s:obs}}

The IRAM~30\,m telescope (Granada, Spain) was used on March 21--25, 2012 to observe in the On-The-Fly (OTF) mapping mode (project 181-11) a sample (see Table~\ref{t:sample}) of 14 high-mass YSOs selected from the work of \citet{lopezsepulcre2011}. The OTF maps have sizes of $1\arcmin\times1\arcmin$ and were obtained with the EMIR heterodyne receiver tuned simultaneously at two different frequencies: 91.18~GHz (E090 band) and 222.23~GHz (E230 band). At these frequencies, the telescope delivers an angular resolution, beam efficiency and forward efficiency of $\theta_\mathrm{HPBW}$=28\arcsec, $B_\mathrm{eff}$=0.81 and $F_\mathrm{eff}$=0.95 in the E090 band, and $\theta_\mathrm{HPBW}$=10\arcsec, $B_\mathrm{eff}$=0.63 and $F_\mathrm{eff}$=0.94 in the E230 band. The FTS spectrometer was set as the spectral backend, resulting in a channel resolution of 195~kHz ($\sim$0.6~\kms\ at 3~mm, and $\sim$0.2~\kms\ at 1~mm), and a total bandwidth of $\sim$15~GHz: from 85.78~GHz to 93.56~GHz at 3~mm (band E090) and from 216.79~GHz to 224.57~GHz at 1~mm (band E230). We used the position-switching mode, with an OFF position located at ($-$600\arcsec,$+$600\arcsec) from the center coordinates listed in Table~\ref{t:sample}. The maps were scanned both along the right ascension and the declination axes to smear scanning effects on the resulting images. System temperatures typically ranged from 100~K to 250~K, reaching in some cases values as high as $\sim$900~K. The weather conditions during the observations, with zenith opacities of 0.15--0.60 at 230~GHz, were not always appropriate for 1~mm observations, and thus, for several sources the noise at 1~mm is too high to detect molecular species other than the strong $^{13}$CO and C$^{18}$O lines. In Table~\ref{t:outflowlevels}, we list the rms noise levels. The accuracy of the pointing was checked every 1.5 or 2 hours. We reduced the data using the CLASS and GREG programs of the GILDAS software package developed by the IRAM and Observatoire de Grenoble. All the resulting spectra have been smoothed to a resolution of 0.8~\kms.

In Table~\ref{t:sample}, we list the 14 objects mapped with the IRAM~30-m telescope (number identifier and name in Cols.~1 and 2), the equatorial coordinates of the center of the maps (Cols.~3 and 4), the systemic velocity (Col.~5) and the heliocentric distance (Col.~6). In Col.~7, we classify each source as IR-luminous (hereafter IRL) or IR-dark (hereafter IRD) according to its emission or not, respectively, at mid-IR wavelengths \citep[see][]{lopezsepulcre2010}.

%%%%%%%%%%%%%%%%%%%%%%%%%%%%%%%%%%%%%%%%%%%%%%%%%%%%%%%%%%%%%%%%%%%%%%%
\section{Results\label{s:res}}

%----------------------------------------------------------------------
\begin{figure*}[htp!]
\begin{center}
\begin{tabular}{c c}
 \epsfig{file=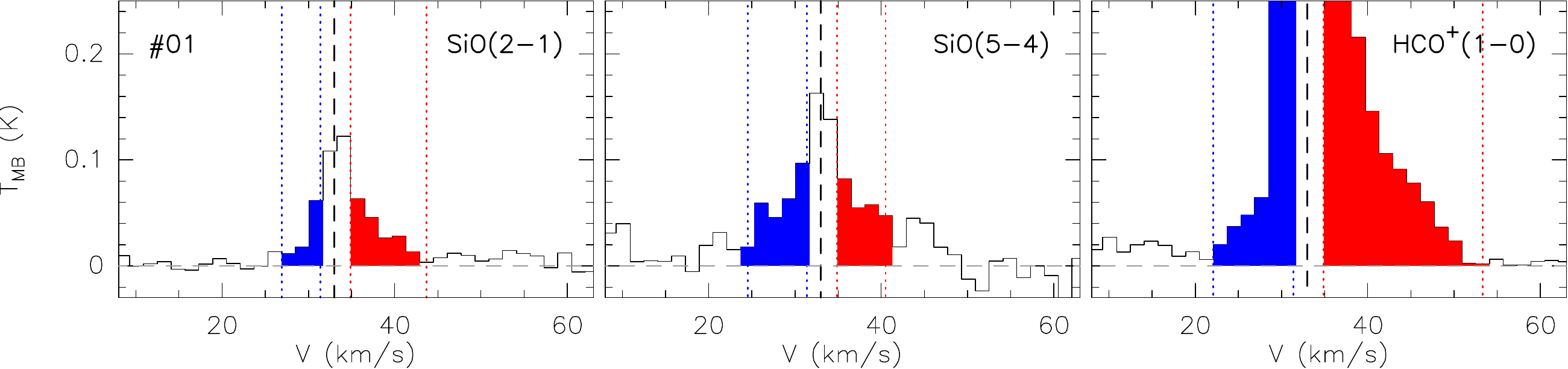, width=0.92\columnwidth, angle=0} &
 \epsfig{file=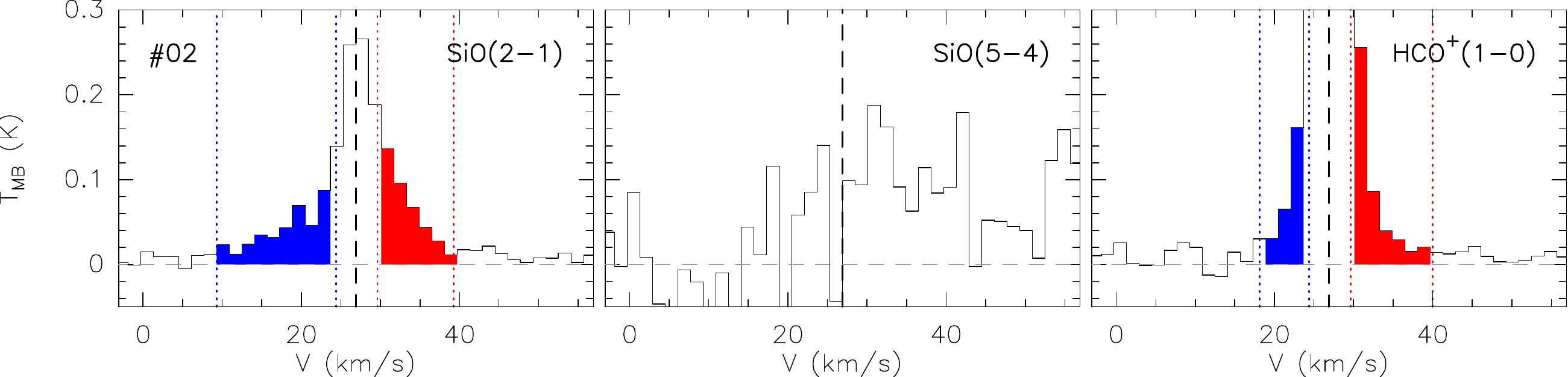, width=0.92\columnwidth, angle=0} \\
 \epsfig{file=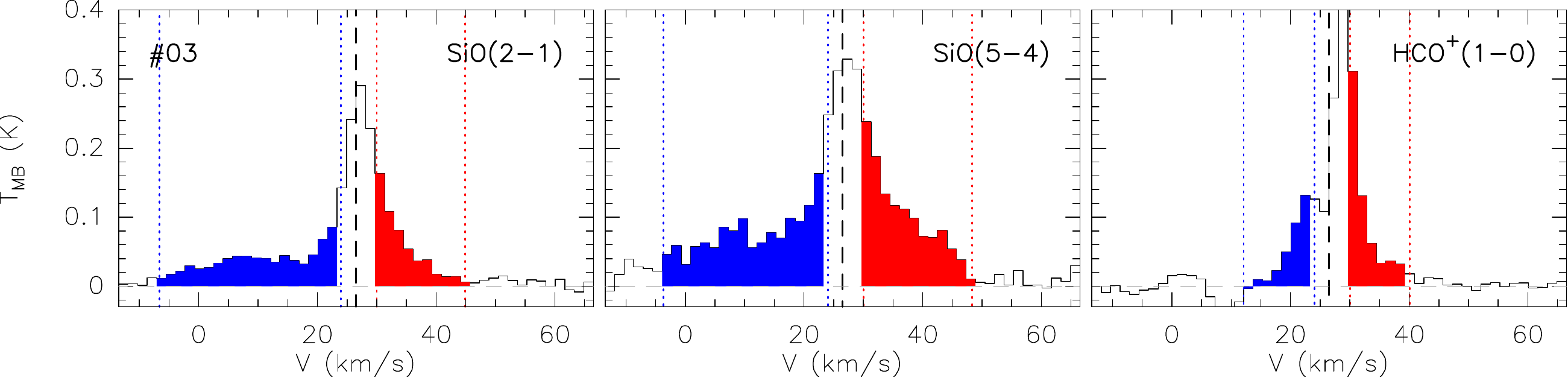, width=0.92\columnwidth, angle=0} &
 \epsfig{file=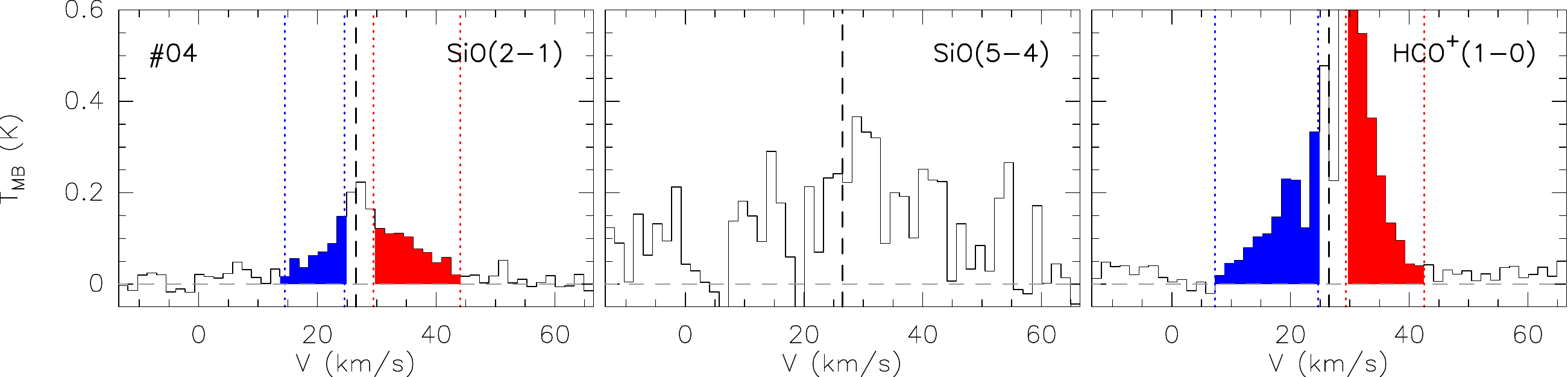, width=0.92\columnwidth, angle=0} \\
 \epsfig{file=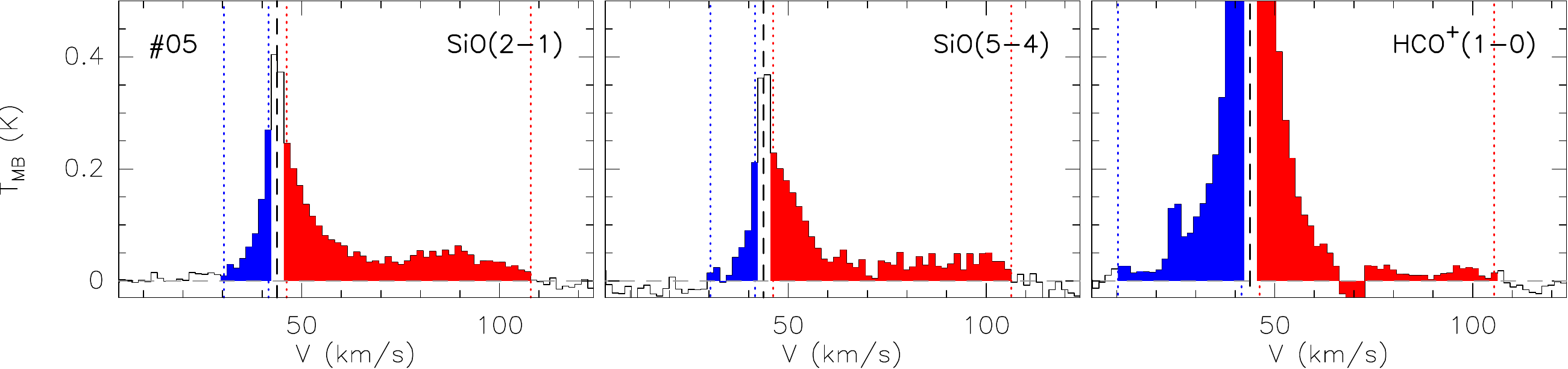, width=0.92\columnwidth, angle=0} &
 \epsfig{file=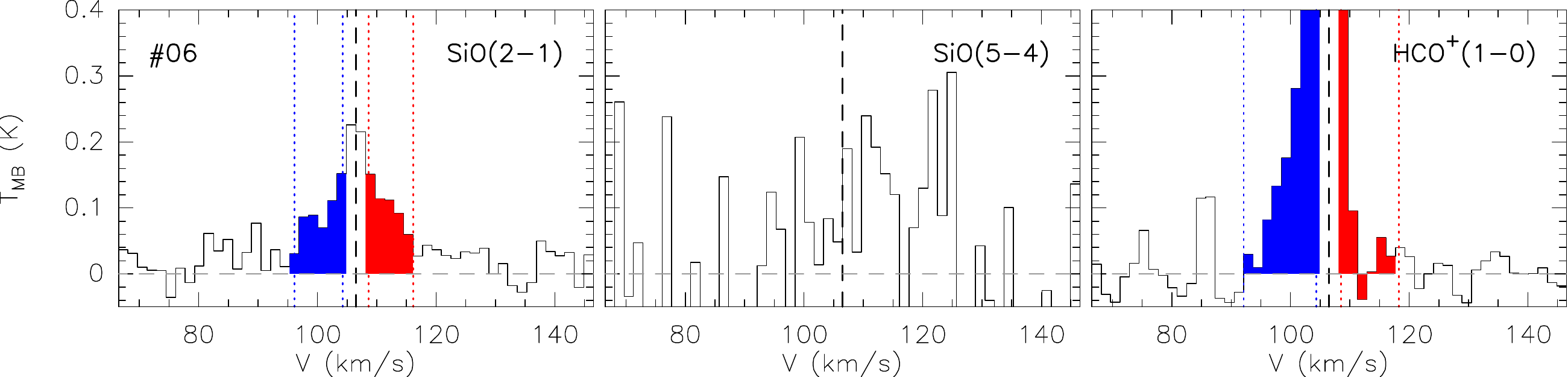, width=0.92\columnwidth, angle=0} \\
 \epsfig{file=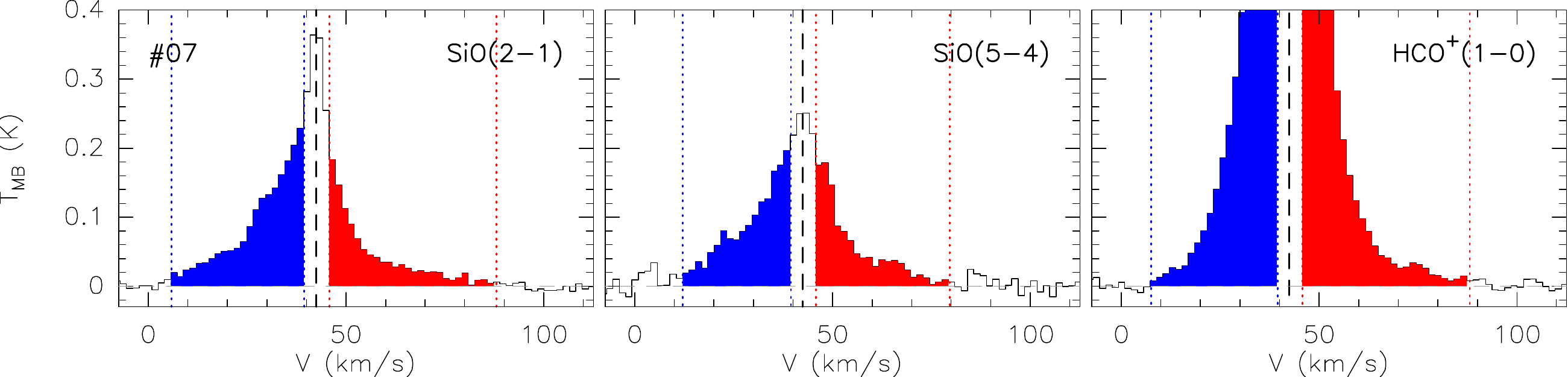, width=0.92\columnwidth, angle=0} &
 \epsfig{file=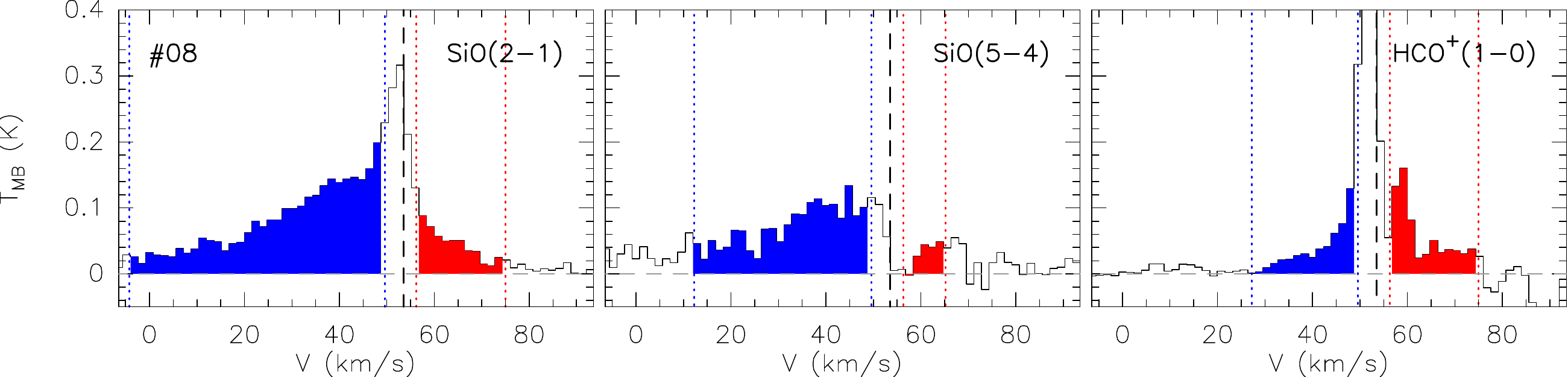, width=0.92\columnwidth, angle=0} \\
 \epsfig{file=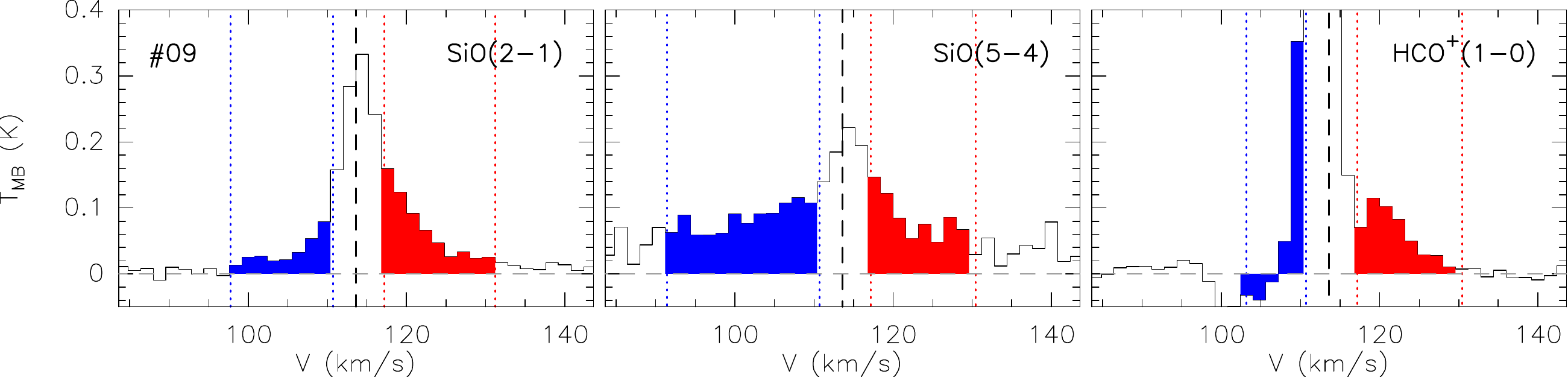, width=0.92\columnwidth, angle=0} &
 \epsfig{file=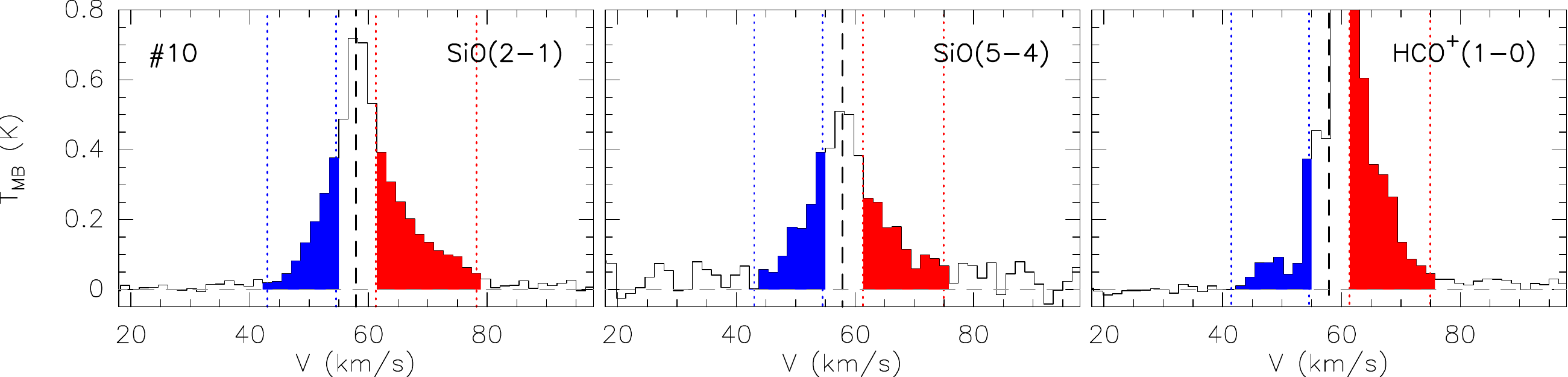, width=0.92\columnwidth, angle=0} \\
 \epsfig{file=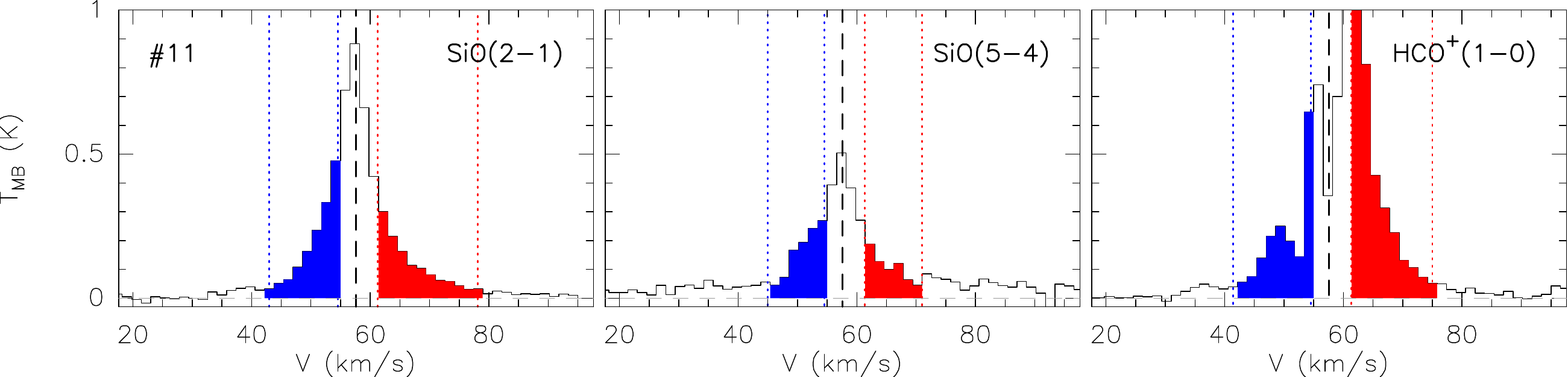, width=0.92\columnwidth, angle=0} &
 \epsfig{file=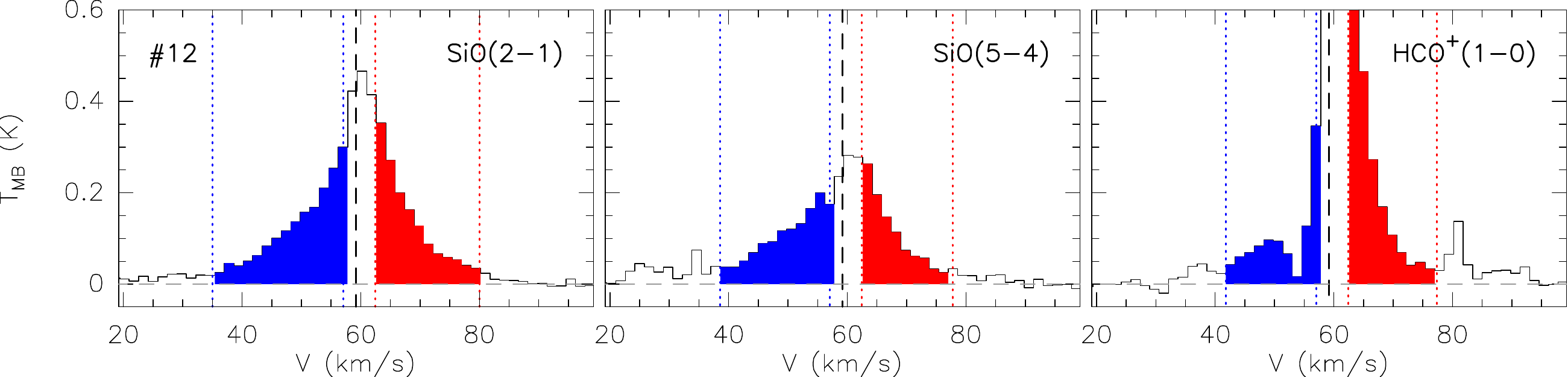, width=0.92\columnwidth, angle=0} \\
 \epsfig{file=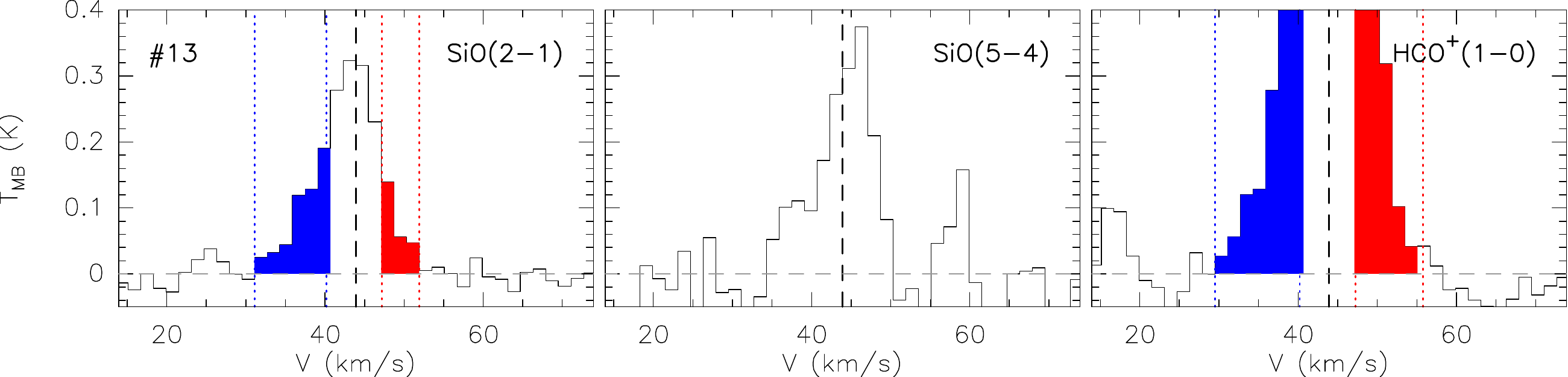, width=0.92\columnwidth, angle=0} &
 \epsfig{file=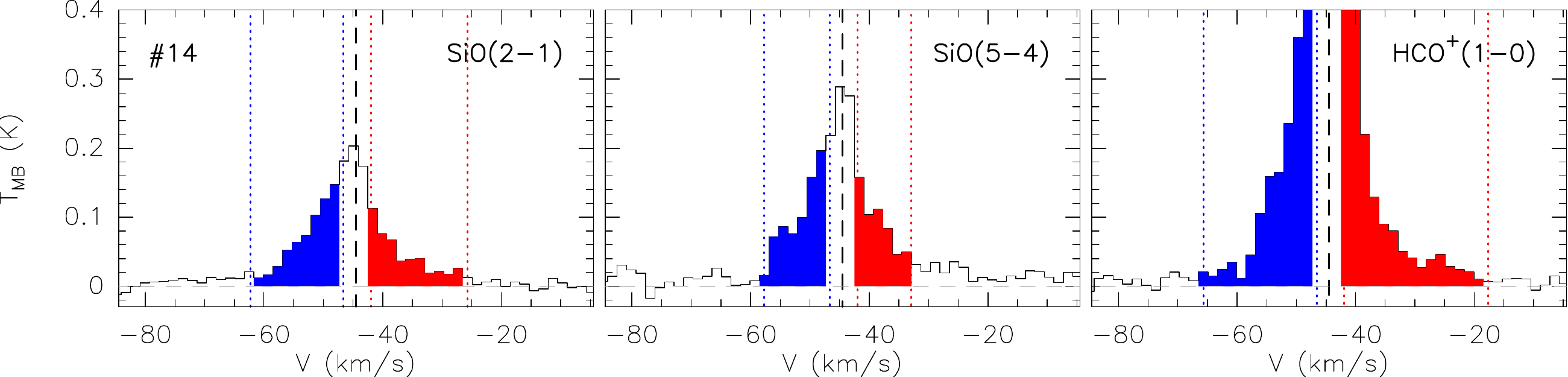, width=0.92\columnwidth, angle=0} \\
\end{tabular}
\end{center}
\caption{\label{f:outspec} SiO\,(2--1), SiO\,(5--4) and HCO$^+$\,(1--0) spectra (smoothed to a velocity resolution of 1.6~\kms). The black dashed vertical line marks the systemic velocity $V_\mathrm{LSR}$. The blue and red dotted vertical lines mark the high-velocity outflow wings (see Sect.~\ref{s:resoutflows}), which are also highlighted with colored blue and red histograms. The number of each source (as listed in Table~\ref{t:sample}) is shown in the top-left corner of the SiO\,(2--1) panels.}
\end{figure*}
%----------------------------------------------------------------------

%----------------------------------------------------------------------
\begin{figure*}[htp!]
\begin{center}
\begin{tabular}{l l}
 \epsfig{file=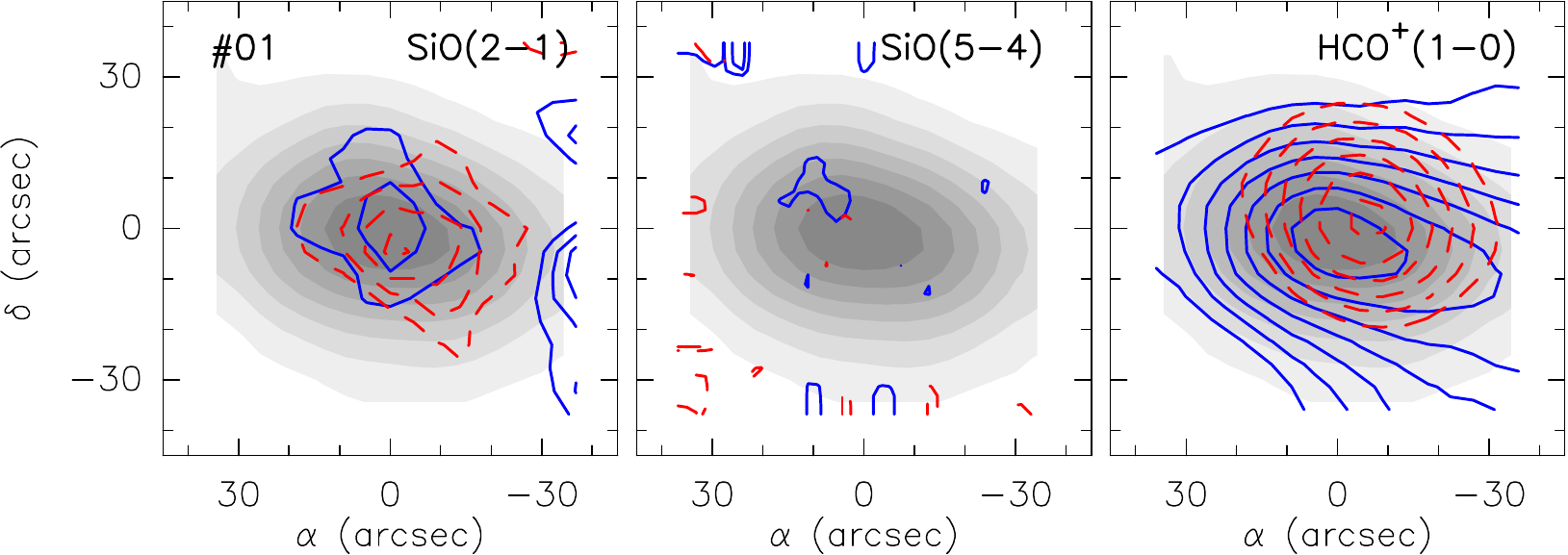, height=2.9cm, angle=0} &
 \epsfig{file=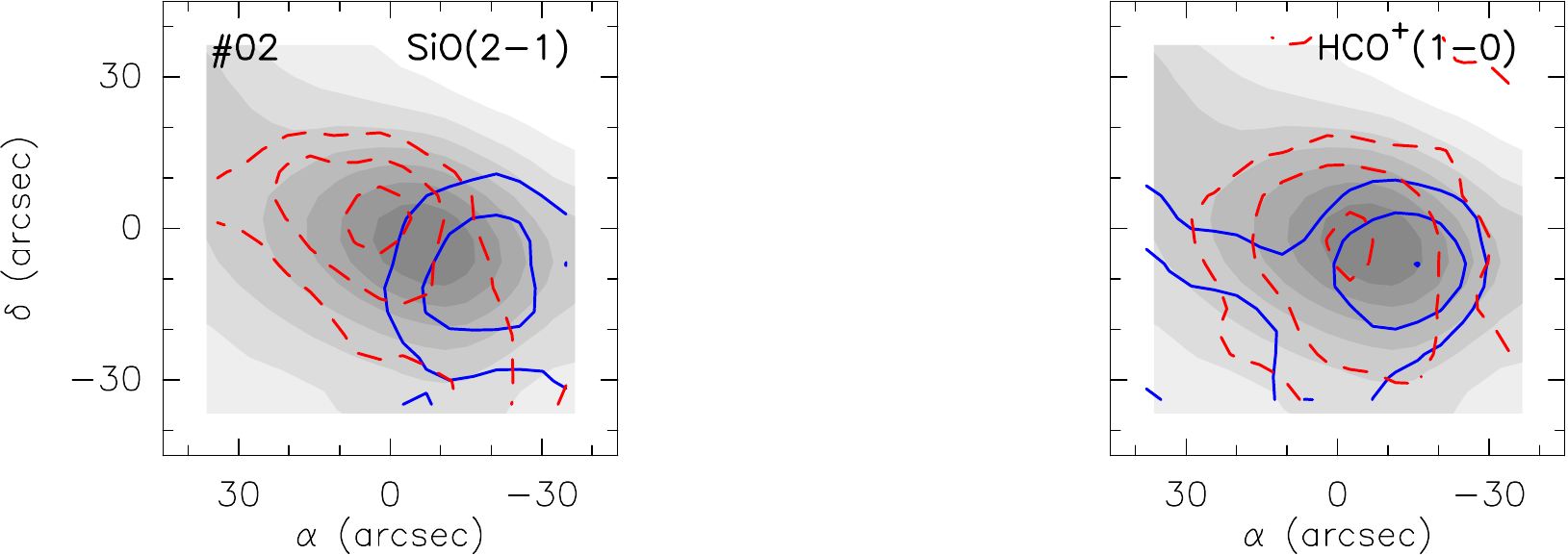, height=2.9cm, angle=0} \\
 \epsfig{file=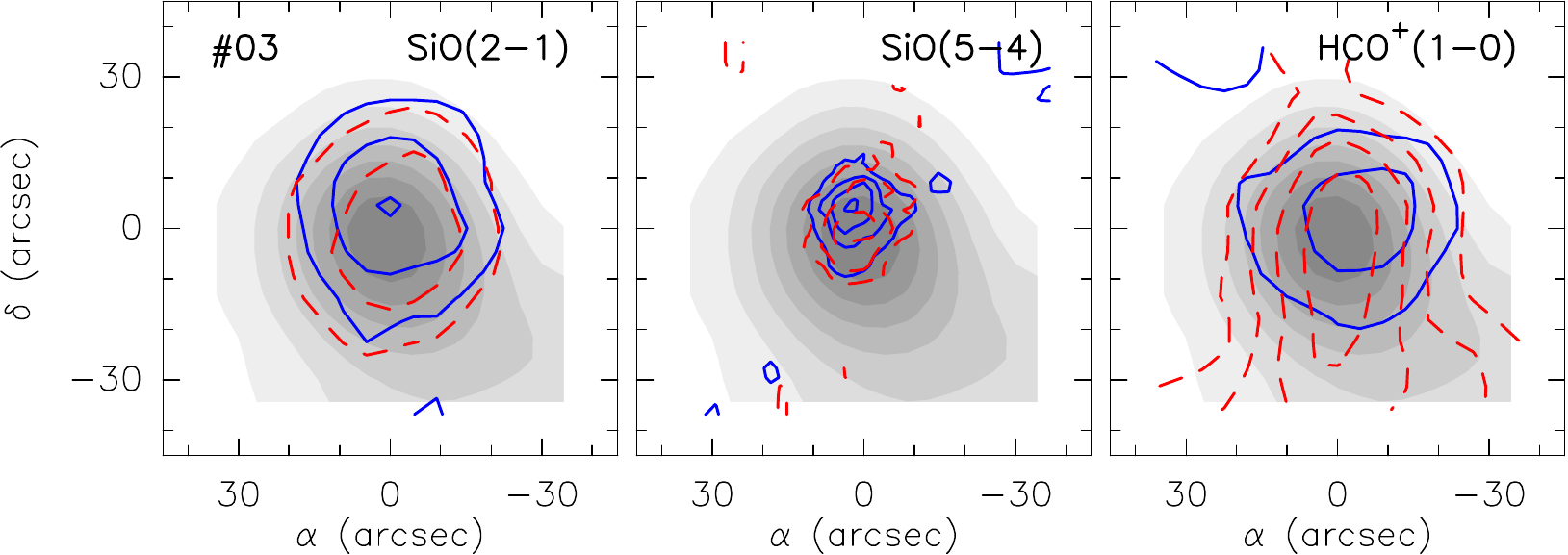, height=2.9cm, angle=0} &
 \epsfig{file=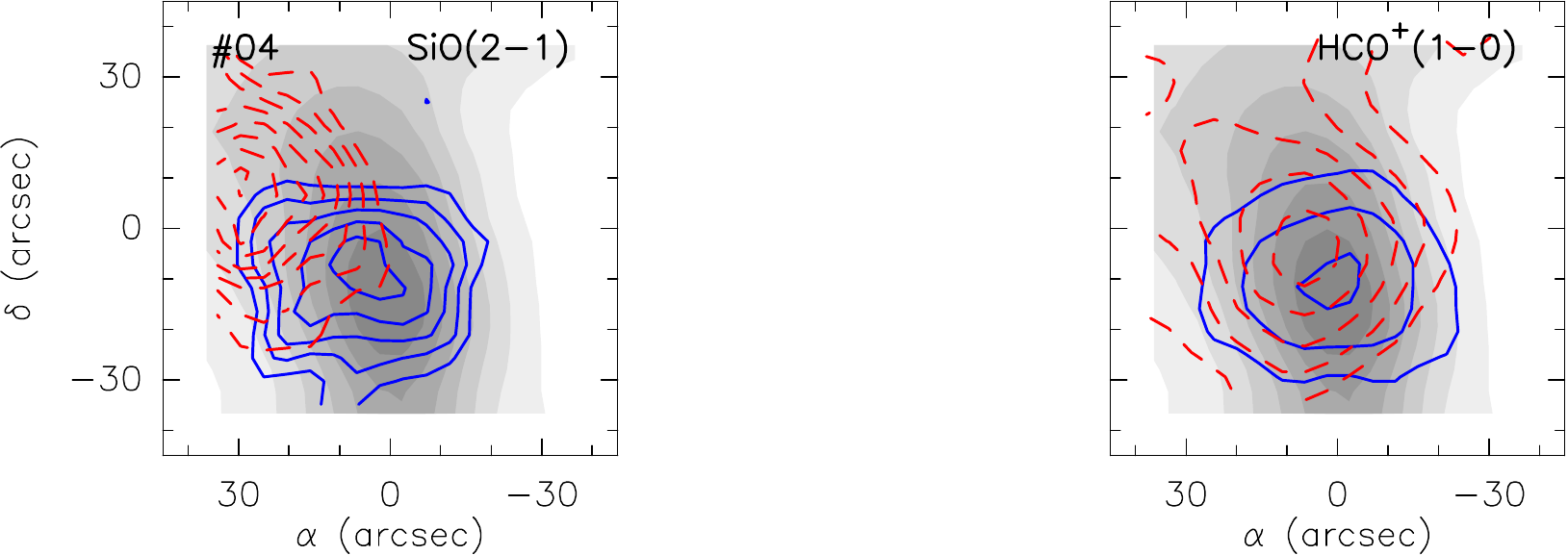, height=2.9cm, angle=0} \\
 \epsfig{file=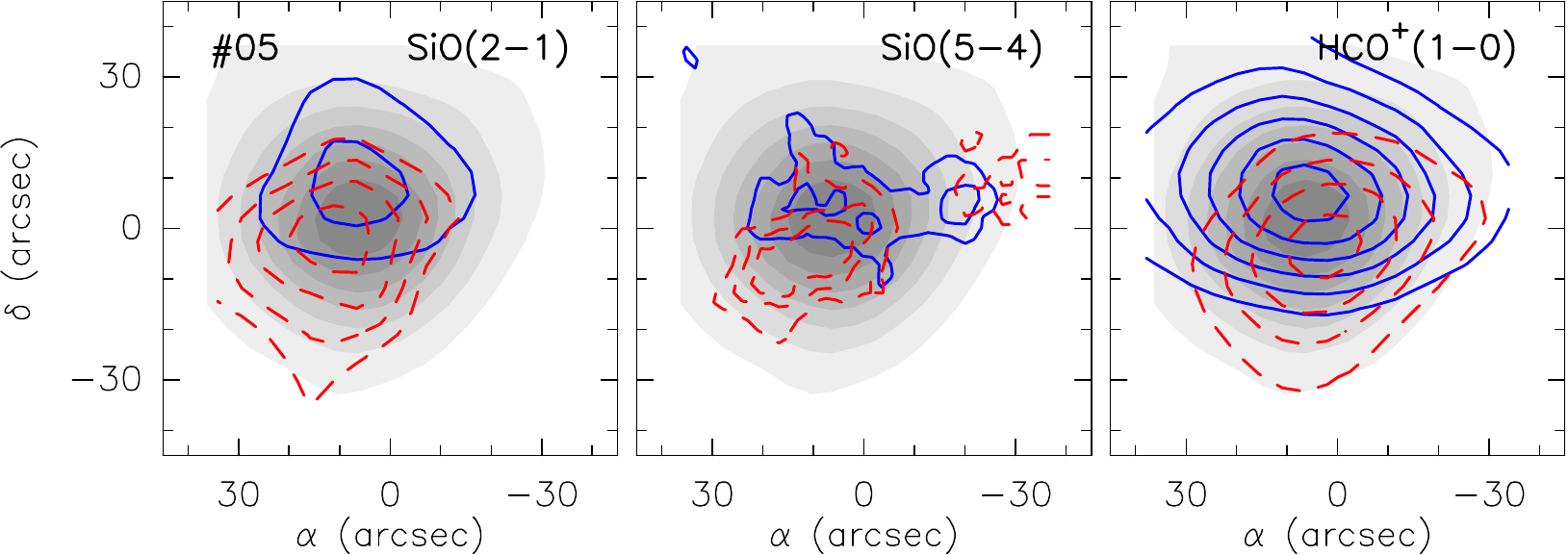, height=2.9cm, angle=0} &
 \epsfig{file=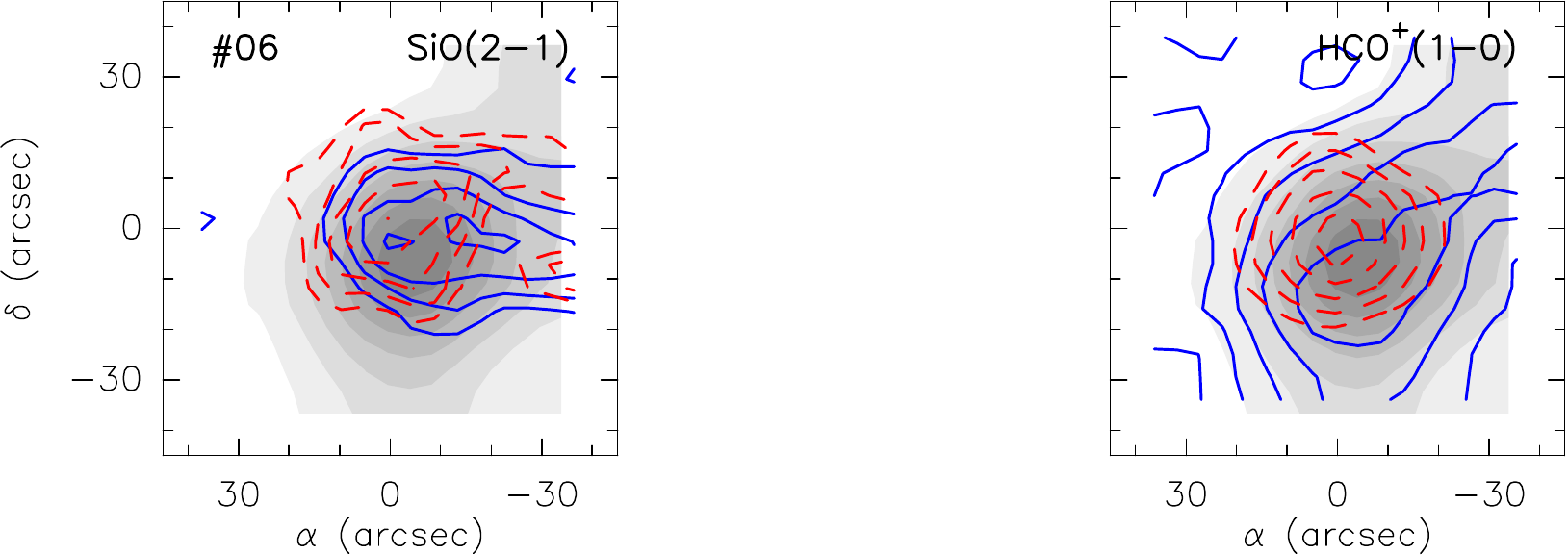, height=2.9cm, angle=0} \\
 \epsfig{file=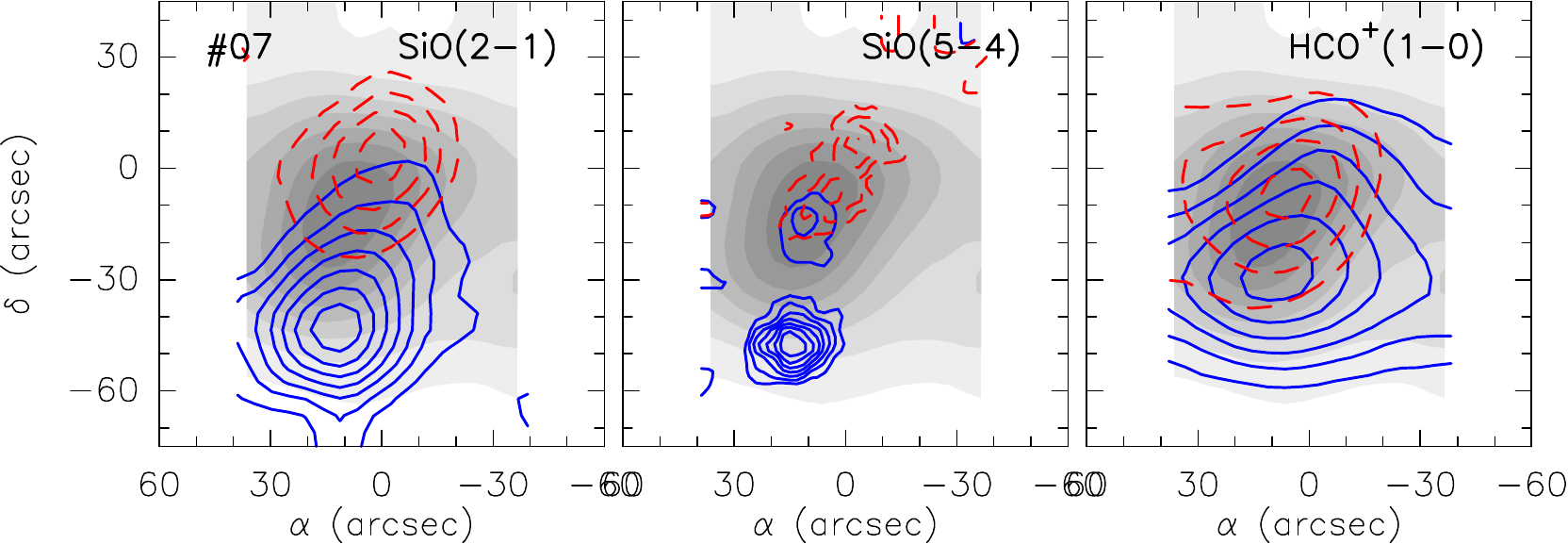, height=2.9cm, angle=0} &
 \epsfig{file=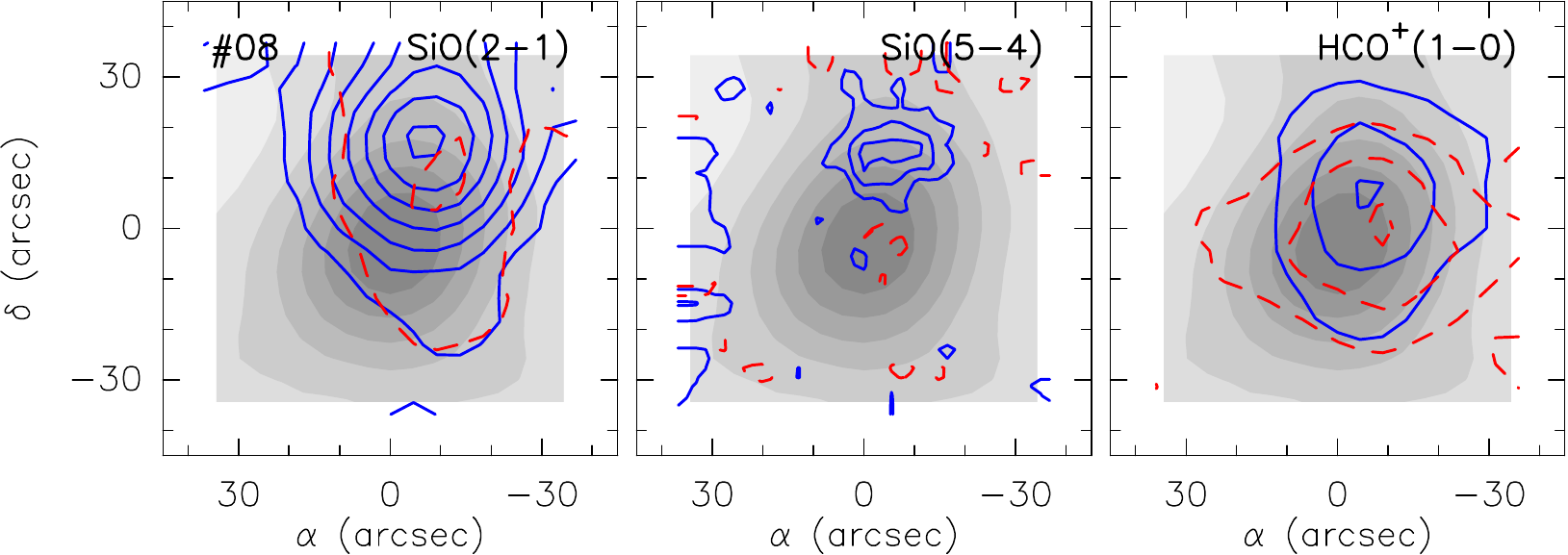, height=2.9cm, angle=0} \\
 \epsfig{file=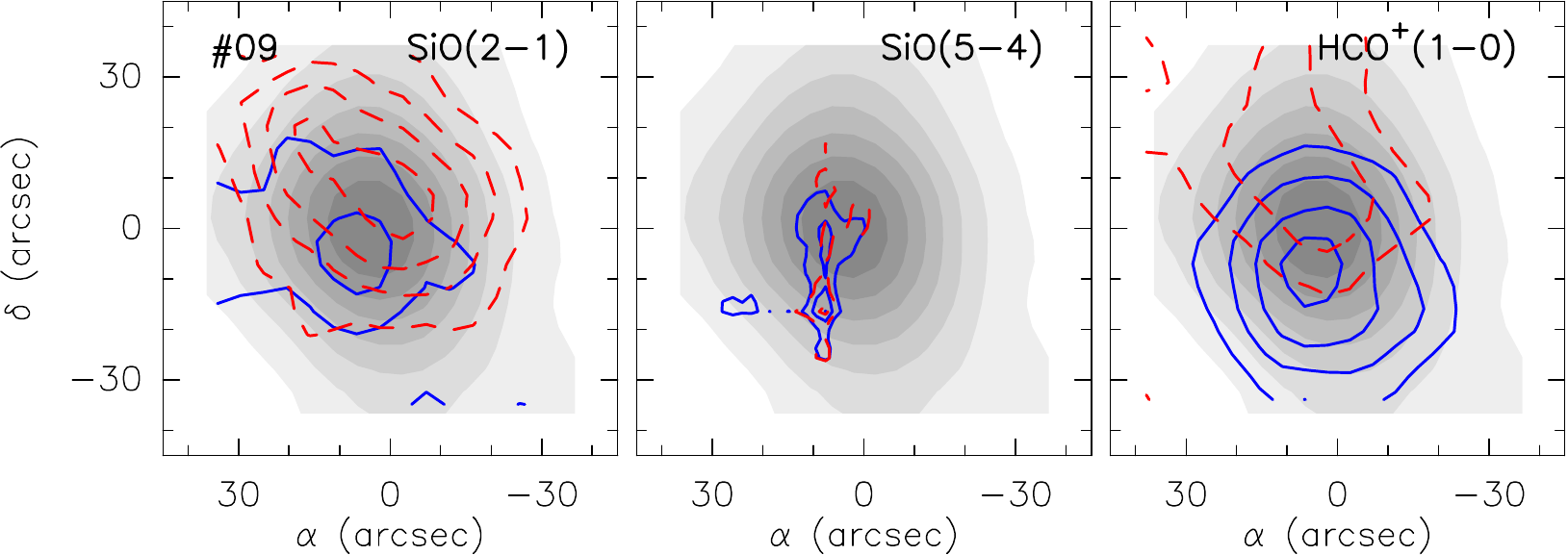, height=2.9cm, angle=0} &
 \epsfig{file=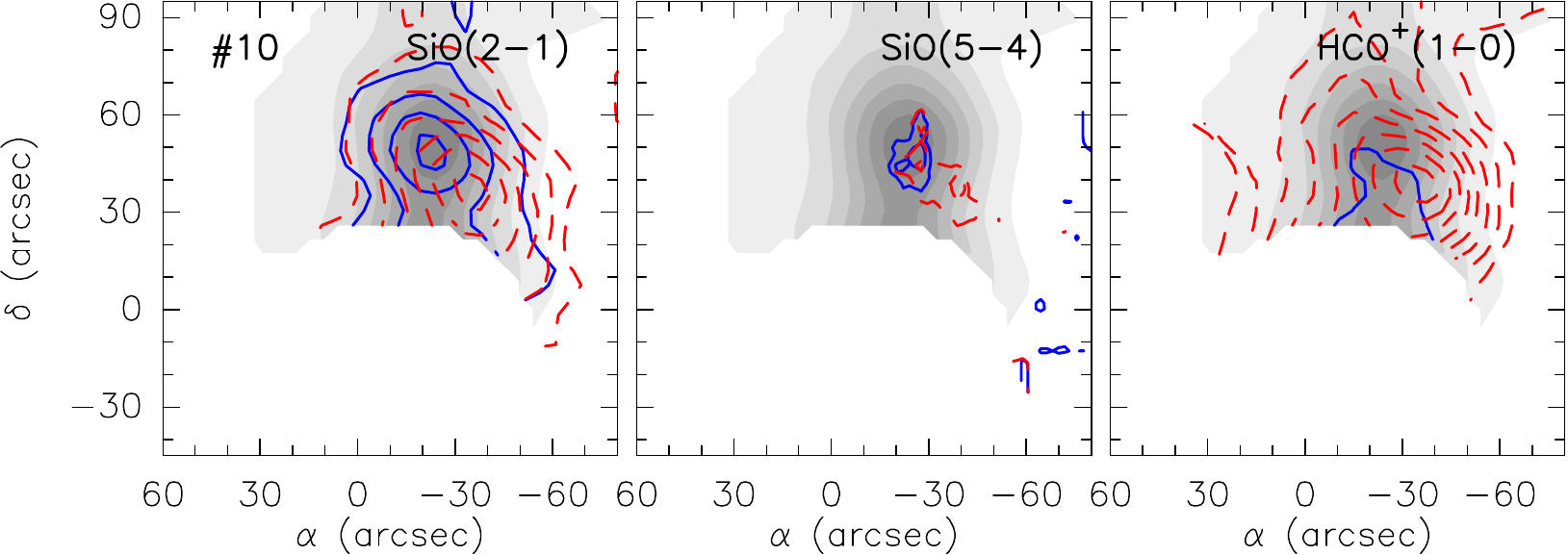, height=2.9cm, angle=0} \\
 \epsfig{file=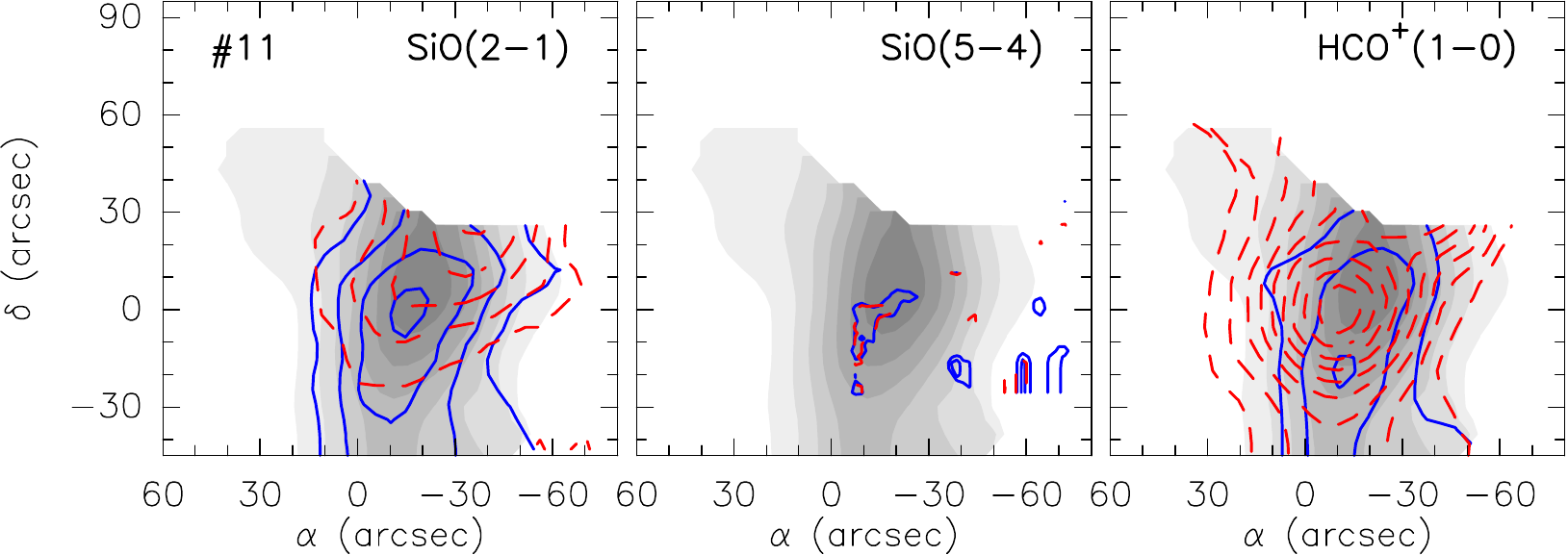, height=2.9cm, angle=0} &
 \epsfig{file=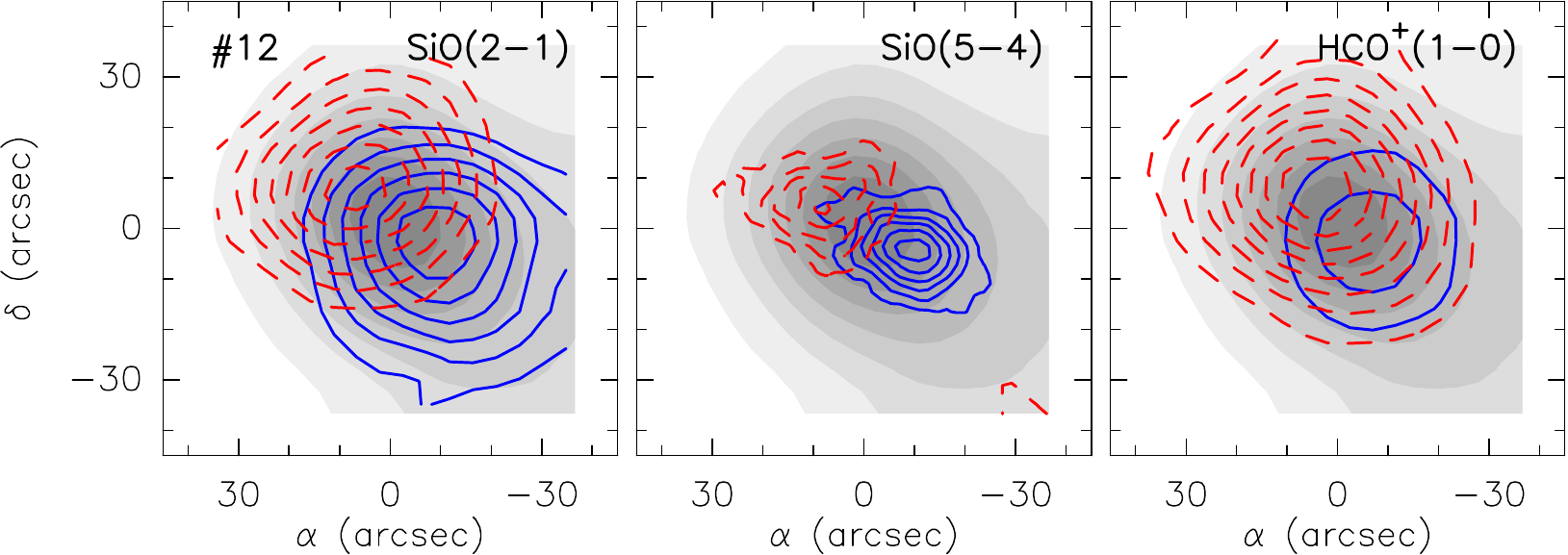, height=2.9cm, angle=0} \\
 \epsfig{file=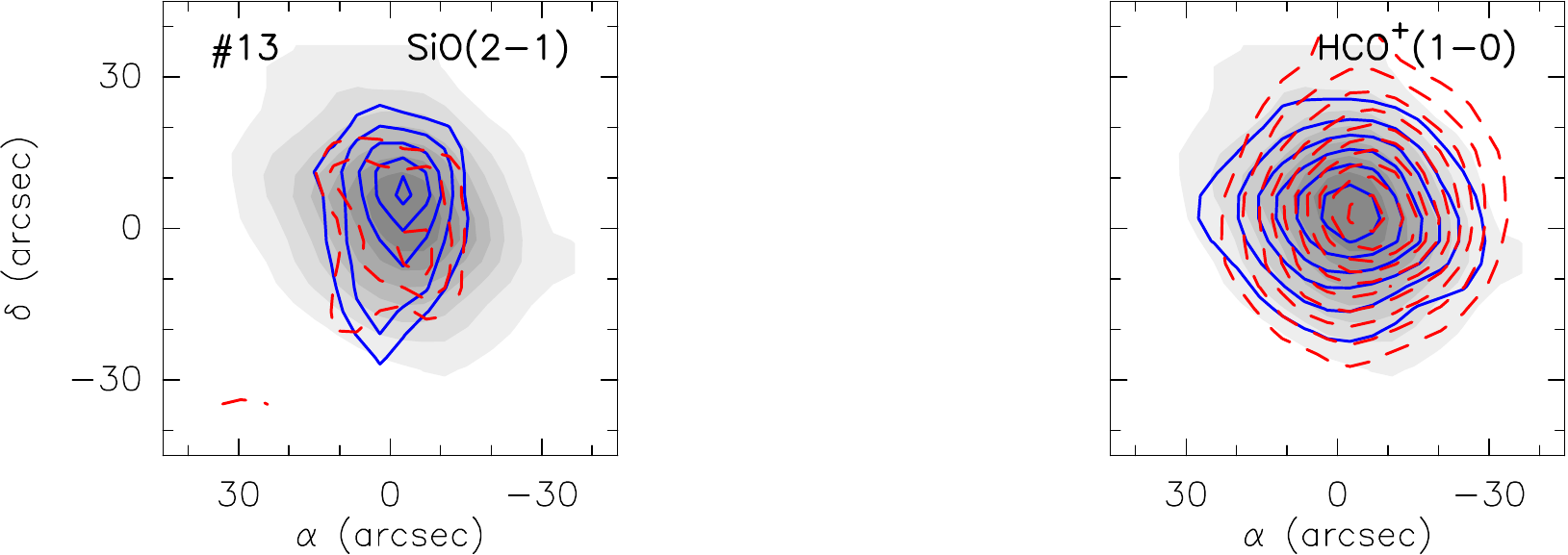, height=2.9cm, angle=0} &
 \epsfig{file=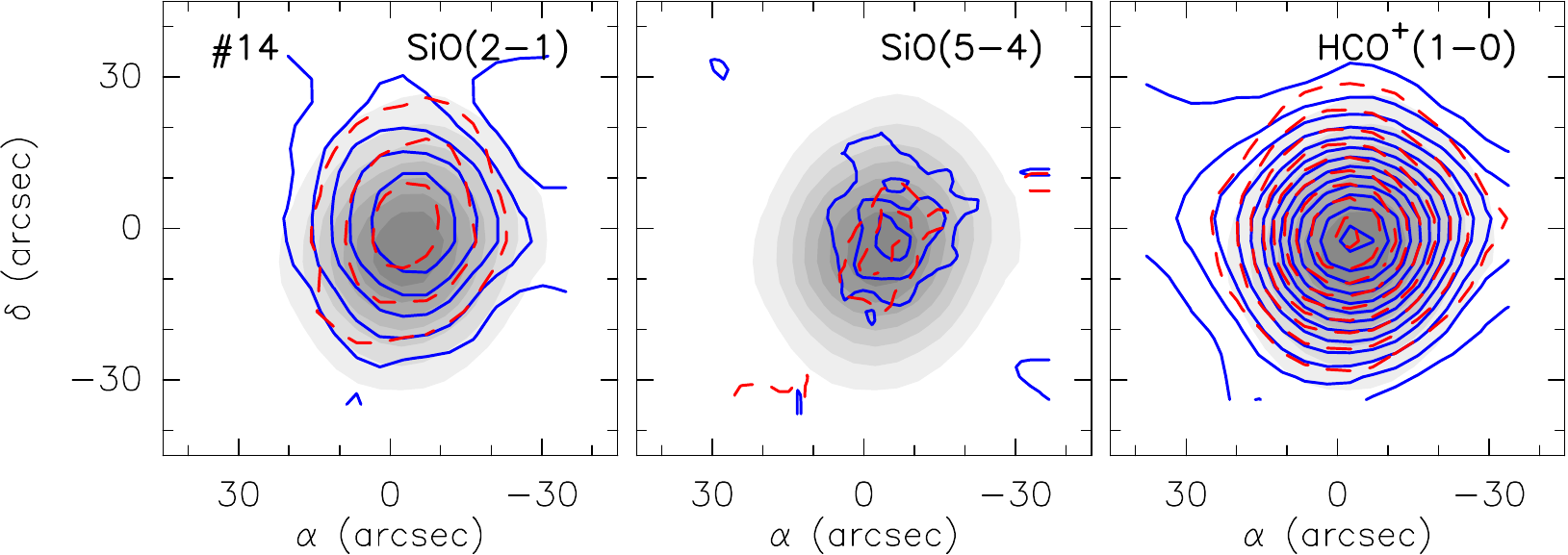, height=2.9cm, angle=0} \\
\end{tabular}
\end{center}
\caption{\label{f:outmap} SiO\,(2--1), SiO\,(5--4) and HCO$^+$\,(1--0) outflow maps. Blue- and red-shifted integrated wing emission is shown as blue-solid and red-dashed contours, respectively. The starting and step levels are listed in Table~\ref{t:outflowlevels}. The gray scale corresponds to the N$_2$H$^+$\,(1--0) integrated emission as shown in Fig.~\ref{f:summary}. The number of the region (as listed in Table~\ref{t:sample}) is indicated in the top-left corner of the SiO\,(2--1) panel. The x and y axis are the right ascension and declination offsets in arcsec, with the center (0,0) corresponding to the coordinates listed in Table~\ref{t:sample}. For four sources we do not show the SiO\,(5--4) panel because we do not detect outflow emission in this line. Note that sources \#10 and \#11 were observed simultaneously, in the same map.}
\end{figure*}
%----------------------------------------------------------------------

\subsection{Molecular line emission\label{s:resmolecules}}

The frequency range surveyed at 3 and 1~mm ($\sim$15~GHz in total) allowed us to simultaneously map several molecular transitions, typically found associated with both dense gas and outflow emission. In the bottom panels of Fig.~\ref{f:summary}, we show the spectra in the 86.48--94.48~GHz frequency range for each source. At 1~mm, the spectra are mainly dominated by noise (due to bad weather conditions, see Sect.~\ref{s:obs}). We have identified the lines detected with signal-to-noise ratio $\ga$5, which correspond to typical main beam temperatures $\sim$0.1~K. Most of the lines detected toward all the sources of the sample correspond to simple molecules such as N$_2$H$^+$, C$_2$H, NH$_2$D or H$^{13}$CN, which are typically found tracing the dense gas in YSOs \citep[\eg][]{tafalla2004, padovani2011, busquet2011, fontani2012}, and molecules such as SiO and HCO$^+$ typically used to trace the outflow emission \citep[\eg][]{tafalla2010, lopezsepulcre2010, codella2013}. For twelve of the fourteen sources, we also detected CH$_3$CN emission which is a tracer typically found in association with hot cores \citep[\eg][]{olmi1993, olmi1996, sanchezmonge2010, sanchezmonge2013}. Eight of them show emission of high excitation ($K$>2) transitions, which suggests the presence of a dense core with a high temperature. In Table~\ref{t:molecules}, we list the molecular lines detected toward each source.

In the top panels of Fig.~\ref{f:summary}, we present the zero-order moment (velocity-integrated intensity) maps of outflow tracers (SiO\,(2--1) and HCO$^+$\,(1--0)) and ambient velocity dense clump tracers (HCN\,(1--0), H$^{13}$CN\,(1--0), H$^{13}$CO$^+$\,(1--0), C$^{18}$O\,(2--1), C$_2$H\,(1--0), N$_2$H$^+$\,(1--0) and NH$_2$D\,(1,1)), complemented with continuum maps at 8.0~$\mu$m (from \emph{Spitzer}/IRAC: Infrared Array Camera, \citealt{fazio2004}) and 500~$\mu$m (from \emph{Herschel}/Hi-GAL, \citealt{molinari2010a, molinari2010b}), or alternatively at 8.3~$\mu$m (from MSX: Midcourse Space Experiment, \citealt{price1999}) and 850~$\mu$m (from JCMT/SCUBA: Submillimetre Common User Bolometer Array, \citealt{difrancesco2008}) when the former images are not available. In the following sections, we will analyze in more detail the molecular lines that can be used to study the properties of the molecular outflows (\ie\ SiO and HCO$^+$), while a more detailed analysis of the dense gas tracers will be presented in a forthcoming paper. Out of all the molecular lines tracing the dense gas, N$_2$H$^+$\,(1--0) and C$_2$H\,(1--0) lines are clearly detected in all the regions and have hyperfine structure. We used the HFS method of the CLASS program to take into account the hyperfine structure of the lines. The simultaneous fit to all the hyperfine components provides us with the systemic velocity of the dense gas. The adopted systemic velocity is listed in Col.~5 of Table~\ref{t:sample}. In addition, from the hyperfine structure fit we obtained a measurement of the line opacity, which allowed us to derive the column density. In Table~\ref{t:densegas}, we list the results of the fit, as well as the excitation temperature, the molecular column density, and the gas mass\footnote{The molecular gas mass of the core has been determined from the N$_2$H$^+$ and C$_2$H column densities (see Table~\ref{t:densegas}), and using the relation $M_\mathrm{gas}=(N_\mathrm{mol}/X)\,\mu\,m_\mathrm{H}\,\mathrm{Area}$, with $\mu$ the molecular weight ($=16$ for N$_2$H$^+$ and $=17$ for C$_2$H), $m_\mathrm{H}$ the mass of the hydrogen atom, $X$ the abundance of the molecule with respect to H$_2$ (assumed to be $3\times10^{-10}$ for N$_2$H$^+$ and $3\times10^{-9}$ for C$_2$H; \eg\ \citealt{huggins1984, caselli2002b, beuther2008, padovani2009, busquet2011, frau2012}), and Area$=\pi\mathrm{r}^2$. The deconvolved radius of the clump, $r$, has been calculated from the 50\% contour level of the N$_2$H$^+$ and C$_2$H integrated maps shown in Fig.~\ref{f:summary}.}. The average N$_2$H$^+$ and C$_2$H column densities are $8.5\times10^{12}$~cm$^{-2}$ and $1.0\times10^{14}$~cm$^{-2}$, respectively.

%Note that all the molecular lines presented in these panels are transitions at 3~mm, except for C$^{18}$O\,(2--1). At these frequencies most of the molecular line emission comes from a compact condensation barely resolved, and typically, only the SiO\,(2--1) and HCO$^+$\,(1--0) maps show some structure, probably due to bipolar structures associated with molecular outflows (see Sect.~\ref{s:resoutflows}).
%For the molecular lines tracing dense gas, we extracted the spectra averaged over the area enclosed by the 50\% contour level in the velocity-integrated intensity map, and calculated the integrated line intensities. In Table~\ref{t:densearea}, we list the integrated area and the full widths at zero power (FWHP).

%----------------------------------------------------------------------------
\begin{table}
\caption{\label{t:outflowlevels}Rms noises and contour levels used in Fig.~\ref{f:outmap}}
\centering
\begin{tabular}{c c c c c c c c c c c c}
\hline\hline

&\multicolumn{2}{c}{SiO\,(2--1)}
&
&\multicolumn{2}{c}{SiO\,(5--4)}
&
&\multicolumn{2}{c}{HCO$^+$\,(1--0)}
\\
\cline{2-3}\cline{5-6}\cline{8-9}
\texttt{ID}
&rms\supa
&levels\supb
&
&rms\supa
&levels\supb
&
&rms\supa
&levels\supb
\\
\hline
01	&\phn30.7	&3(1)	&&105.4		&3(1)	&&\phn50.8	&25(5)	\\	%1
02	&\phn43.9	&5(3)	&&320.1		&\ldots	&&\phn53.5	&5(3)	\\	%10
03	&\phn28.7	&5(5)	&&\phn78.4	&3(3)	&&\phn37.9	&5(3)	\\	%9
04	&\phn72.7	&3(1)	&&509.6		&\ldots	&&105.8		&5(3)	\\	%2
05	&\phn31.3	&10(5)	&&\phn78.8	&3(3)	&&\phn85.8	&10(5)	\\	%3
06	&\phn78.0	&3(1)	&&603.1		&\ldots	&&\phn91.5	&5(3)	\\	%11
07	&\phn31.1	&10(5)	&&\phn82.1	&5(3)	&&108.0		&10(5)	\\	%4
08	&\phn33.6	&5(5)	&&103.2		&3(3)	&&\phn49.8	&5(3)	\\	%12
09	&\phn31.6	&5(3)	&&100.1		&5(3)	&&\phn43.3	&5(3)	\\	%13
10	&\phn46.5	&5(5)	&&180.5		&5(3)	&&\phn89.5	&5(3)	\\	%15
11	&\phn46.8	&5(5)	&&179.7		&5(3)	&&\phn89.8	&5(3)	\\	%5
12	&\phn30.9	&10(5)	&&\phn80.5	&5(3)	&&\phn42.2	&10(5)	\\	%14
13	&\phn64.7	&3(1)	&&275.4		&\ldots	&&\phn95.2	&5(3)	\\	%7
14	&\phn28.1	&3(3)	&&\phn68.9	&3(3)	&&\phn46.0	&5(3)	\\	%8
\hline
\end{tabular}
\begin{list}{}{}
\item[\supa] Rms noise level in mK per 0.8~\kms\ channel.
\item[\supb] Contour levels used in Fig.~\ref{f:outmap}. The first number corresponds to the first contour in terms of $\sigma$, and the number in parenthesis indicates the step level used. The value of $\sigma$ has been calculated from the rms noise per channel, taking into account the velocity range covered to obtain the integrated blue- and red-shifted emission (Tables~\ref{t:outflowsio21}--\ref{t:outflowhco10}).
\end{list}
\end{table}
%----------------------------------------------------------------------------

\subsection{Molecular outflow detections\label{s:resoutflows}}

We have assessed the presence of molecular outflows by searching for high-velocity wings in the SiO\,(2--1), SiO\,(5--4), and HCO$^+$\,(1--0) spectra averaged over the area of  the 50\% contour level in the integrated intensity maps shown in Fig.~\ref{f:summary}. In Fig.~\ref{f:outspec}, we present the spectra of these three outflow tracers for all the sources in the sample. We detect high-velocity wings in SiO\,(2--1) and HCO$^+$\,(1--0) in all the sources, confirming the presence of outflows in these regions as previously reported by \citet{lopezsepulcre2010, lopezsepulcre2011}. Note that for a few regions (\eg\ G19.27$+$0.1\,M2, G23.60$+$0.0\,M1), the bad weather conditions during observations resulted in noisy SiO\,(5--4) spectra (cf.\ noises listed in Table~\ref{t:outflowlevels}) that prevented detection of high-velocity wings in SiO\,(5--4). The new HCO$^+$\,(1--0) spectra have 1$\sigma$ rms noises ($\sim$70~mK) a few times better than the previous observations of \citet[][$\sim$480~mK]{lopezsepulcre2010}, improving the detection of faint emission at high velocities.

%----------------------------------------------------------------------
\begin{figure}[t]
\begin{center}
\begin{tabular}{c}
 \epsfig{file=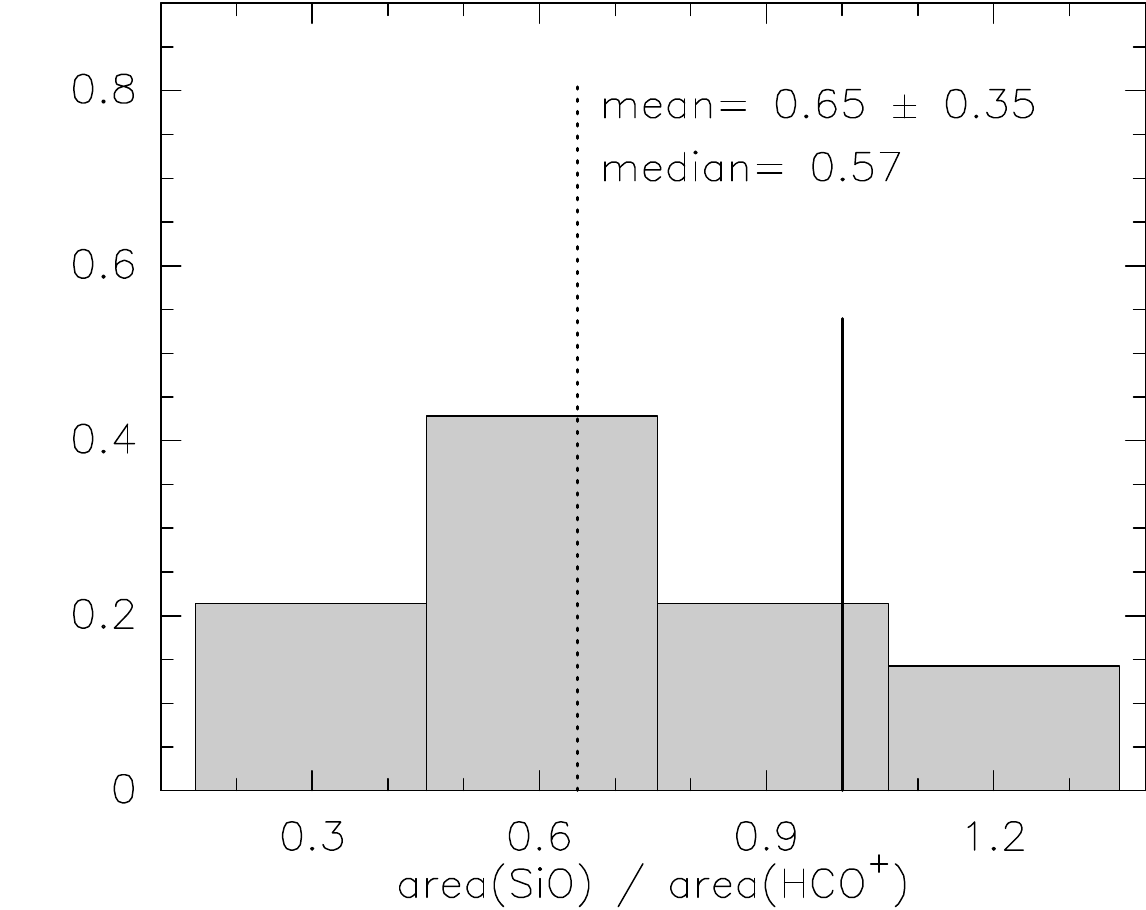, width=0.8\columnwidth, angle=0} \\
\end{tabular}
\end{center}
\caption{\label{f:areaSiOHCO} Distribution of the SiO\,(2--1) to HCO$^+$\,(1--0) area ratio for the outflow emission. The solid vertical line corresponds to the equality (\ie\ equal sizes for the SiO and HCO$^+$ outflow lobes). The dotted vertical line marks the mean value of the distribution. The mean ($\pm$ standard deviation) and median values are shown in the panel. The total number of sources is 14.}
\end{figure}
%----------------------------------------------------------------------

The red and blue high-velocity wings are shown in each spectra of Fig.~\ref{f:outspec}. For each source with line wings, we compared the spectra of the outflow tracers with those from dense gas tracers such as C$_2$H, N$_2$H$^+$ and H$^{13}$CO$^+$ and, following the approach of \citet{lopezsepulcre2009}, we defined the low-velocity limits where the line intensities of the C$_2$H\,($1_{2,2}$--$0_{1,1}$), N$_2$H$^+$\,($1_{0}$--$0_{1}$) and H$^{13}$CO$^+$\,(1--0) transitions fall below 2$\sigma$. In a few cases where also these lines display non-Gaussian extended wings, the beginning of the departure from a Gaussian fit on the observed spectra has been chosen as low-velocity limits for the outflow. The high-velocity limits have been chosen where the line intensity of the outflow tracers falls below 2$\sigma$. After inspection of the channel maps, these limits were in some cases slightly modified, based on the observed spatial distribution of the blue- and red-shifted emission. The resulting outflow velocity ranges ($\Delta V_\mathrm{blue}$ and $\Delta V_\mathrm{red}$) are listed in Cols.~4 and 5 of Tables~\ref{t:outflowsio21}, \ref{t:outflowsio54} and \ref{t:outflowhco10}, and have been used to obtain the outflow maps in Fig.~\ref{f:outmap}, as well as to compute the outflow parameters (see Sect.~\ref{s:disoutpar}).

The outflow maps are presented in Fig.~\ref{f:outmap}, where the blue and red wing emission of SiO\,(2--1), SiO\,(5--4) and HCO$^+$\,(1--0) lines (solid-blue and red-dashed contours, respectively) are superimposed on the N$_2$H$^+$\,(1--0) integrated emission (gray scale). We identify a bipolar structure in the outflow emission for six regions: G19.27$+$0.1M2, 18236$-$1205, 18264$-$1152, 18316$-$0602, G24.33$+$0.1M1, and G34.43$+$0.2M3. For all these sources, but one (18236$-$1205), the geometrical center of the two outflow lobes is close to the peak of the dense clump (see gray scale tracing the N$_2$H$^+$ emission in Fig.~\ref{f:outmap}). For a few sources, and despite the presence of extended high-velocity wings, we do not find bipolar morphology in the maps (\eg\ 23139$+$5939). This fact suggests that the outflow is directed close to the line of sight, or that a more complex structure (with multiple outflows) produces the high-velocity emission that we detect. Higher angular resolution observations are necessary to study the presence of multiple outflows in some of these regions. In general, the SiO and HCO$^+$ profiles and spatial distributions are similar though in detail different. The extent in velocity of the profiles is similar, and both species have the same orientation in the bipolar cases. However, the HCO$^+$ spatial distribution is more extended than the SiO one. This is shown in Fig.~\ref{f:areaSiOHCO}, where we plot the distribution of the SiO to HCO$^+$ area ratio. The area is calculated as the sum of the red and blue outflow lobe areas, determined by the deconvolved sizes listed in Col.~6 of Tables~\ref{t:outflowsio21} and \ref{t:outflowhco10}. The mean and median values of the distribution are 0.65 and 0.57. This supports the idea that SiO is associated with a more collimated agent (primary jet, although not resolved in our observations) and that HCO$^+$ is associated with extended entrained gas, as found in different star-forming regions with higher angular resolutions \citep[\eg][]{cesaroni1999}. 

%----------------------------------------------------------------------------
\begin{table*}
\caption{\label{t:outflowsio21}SiO\,(2--1) molecular outflows: velocity ranges of the wings, lobe sizes, outflow energetics and kinematics}
\centering
\begin{tabular}{c c c c c c c c c c}
\hline\hline

&$V_\mathrm{LSR}$
&\multicolumn{1}{c}{$\Delta V_\mathrm{blue}$}
&\multicolumn{1}{c}{$\Delta V_\mathrm{red}$}
&$V_\mathrm{max-b/r}$\supa
&size$_\mathrm{blue/red}$\supb
&$M_\mathrm{out}$
&$p_\mathrm{out}$
&$E_\mathrm{out}$
&$t_\mathrm{kin}$
\\
\texttt{ID}
&(\kms)
&\multicolumn{1}{c}{(\kms)}
&\multicolumn{1}{c}{(\kms)}
&(\kms)
&(pc)
&(\mo)
&(\mo~\kms)
&(10$^{46}$~erg)
&(10$^{4}$~yr)
\\
\hline
01	&\phn$+$33.0	&[$+$26.9\swp$+$31.4]		&[$+$34.9\swp$+$43.7]		&\phn6.1	\sep10.7	&0.31\sep0.20	&\phnn9.7	&\phn36		&\phn0.18	&3.0		\\
02	&\phn$+$26.9	&\phn[$+$9.3\swp$+$24.4]		&[$+$29.6\swp$+$39.3]		&17.6\sep12.4	&0.19\sep0.28	&\phnn3.8	&\phn26		&\phn0.23	&1.5		\\
03	&\phn$+$26.5	&\phn[$-$6.6\swp$+$24.0]		&[$+$30.0\swp$+$44.9]		&33.1\sep18.4	&0.23\sep0.18	&\phn13.\phn	&140			&\phn2.3\phn	&0.9		\\
04	&\phn$+$26.5	&[$+$14.5\swp$+$24.6]		&[$+$29.4\swp$+$44.1]		&12.0\sep17.6	&0.27\sep0.36	&\phn35.\phn	&280			&\phn2.8\phn	&2.2		\\
05	&\phn$+$43.7	&[$+$30.3\swp$+$41.6]		&\phn[$+$46.2\swp$+$107.9]	&13.4\sep64.2	&0.35\sep0.51	&110.\phn	&2130\phn	&75.\phnn	&2.1		\\
06	&$+$106.5	&\phn[$+$96.1\swp$+$104.3]	&[$+$108.6\swp$+$116.1]		&10.4\sep\phn9.6	&0.66\sep0.73	&\phn36.\phn	&210			&\phn1.4\phn	&6.8		\\
07	&\phn$+$42.5	&\phn[$+$5.9\swp$+$39.5]		&[$+$45.8\swp$+$88.1]		&36.6\sep45.6	&0.33\sep0.37	&103.\phn	&1400\phn	&27.\phnn	&0.8		\\
08	&\phn$+$53.5	&\phn[$-$4.2\swp$+$49.6]		&[$+$56.3\swp$+$75.0]		&57.7\sep21.5	&0.53\sep0.47	&\phn24.\phn	&470			&14.\phnn	&1.6		\\
09	&$+$113.6	&\phn[$+$97.8\swp$+$110.7]	&[$+$117.2\swp$+$131.2]		&15.8\sep17.6	&1.14\sep0.97	&\phn54.\phn	&420			&\phn4.1\phn	&6.2		\\
10	&\phn$+$57.9	&[$+$43.0\swp$+$54.5]		&[$+$61.3\swp$+$78.2]		&14.9\sep20.3	&0.48\sep0.35	&\phn76.\phn	&630			&\phn6.5\phn	&2.3		\\
11	&\phn$+$57.6	&[$+$43.0\swp$+$54.5]		&[$+$61.3\swp$+$78.2]		&14.6\sep20.6	&0.54\sep0.82	&\phn99.\phn	&750			&\phn7.1\phn	&4.0		\\
12	&\phn$+$59.2	&[$+$35.0\swp$+$57.1]		&[$+$62.4\swp$+$80.0]		&24.2\sep20.8	&0.38\sep0.42	&\phn50.\phn	&440			&\phn5.1\phn	&1.7		\\
13	&\phn$+$43.9	&[$+$31.1\swp$+$40.2]		&[$+$47.2\swp$+$51.9]		&12.8\sep\phn8.0	&0.32\sep0.19	&\phnn8.5	&\phn55		&\phn0.41	&2.7		\\
14	&\phn$-$44.5	&[$-$62.3\swp$-$46.6]		&[$-$42.0\swp$-$25.7]		&17.8\sep18.8	&0.17\sep0.38	&\phn63.\phn	&450			&\phn4.3\phn	&1.5		\\
\hline
\end{tabular}
\begin{list}{}{}
\item[\supa] Maximum outflow velocity for the blue and red lobes, respectively, calculated as the difference between the high-velocity limit of the outflow velocity range, $\Delta V$, and the systemic velocity, $V_\mathrm{LSR}$.
\item[\supb] Deconvolved size of the blue and red outflow lobes, respectively.
\end{list}
\end{table*}
\begin{table*}
\caption{\label{t:outflowsio54}SiO\,(5--4) molecular outflows: velocity ranges of the wings, lobe sizes, outflow energetics and kinematics}
\centering
\begin{tabular}{c c c c c c c c c c}
\hline\hline

&$V_\mathrm{LSR}$
&\multicolumn{1}{c}{$\Delta V_\mathrm{blue}$}
&\multicolumn{1}{c}{$\Delta V_\mathrm{red}$}
&$V_\mathrm{max-b/r}$\supa
&size$_\mathrm{blue/red}$\supb
&$M_\mathrm{out}$
&$p_\mathrm{out}$
&$E_\mathrm{out}$
&$t_\mathrm{kin}$
\\
\texttt{ID}
&(\kms)
&\multicolumn{1}{c}{(\kms)}
&\multicolumn{1}{c}{(\kms)}
&(\kms)
&(pc)
&(\mo)
&(\mo~\kms)
&(10$^{46}$~erg)
&(10$^{4}$~yr)
\\
\hline
01	&\phn$+$33.0	&[$+$24.5\swp$+$31.4]		&[$+$34.9\swp$+$40.5]		&\phn8.5\sep\phn7.5	&0.27\sep0.37	&\phnn2.4	&\phn11		&\phnn0.060	&3.6		\\
02	&\phn$+$26.9	&\ldots						&\ldots						&\ldots				&\ldots			&\ldots		&\ldots		&\ldots		&\ldots	\\
03	&\phn$+$26.5	&\phn[$-$3.7\swp$+$24.0]		&[$+$30.0\swp$+$48.3]		&30.2\sep21.8		&0.12\sep0.07	&\phnn4.9	&\phn54		&\phn0.83	&0.4		\\
04	&\phn$+$26.5	&\ldots						&\ldots						&\ldots				&\ldots			&\ldots		&\ldots		&\ldots		&\ldots	\\
05	&\phn$+$43.7	&[$+$30.3\swp$+$41.6]		&\phn[$+$46.2\swp$+$106.3]	&13.4\sep62.6		&0.28\sep0.30	&\phn21.\phn	&440			&16.\phnn	&1.3		\\
06	&$+$106.5	&\ldots						&\ldots						&\ldots				&\ldots			&\ldots		&\ldots		&\ldots		&\ldots	\\
07	&\phn$+$42.5	&[$+$12.1\swp$+$39.5]		&[$+$45.8\swp$+$79.7]		&30.4\sep37.2		&0.32\sep0.15	&\phn30.\phn	&370			&\phn6.3\phn	&0.7		\\
08	&\phn$+$53.5	&[$+$12.2\swp$+$49.6]		&[$+$56.3\swp$+$65.2]		&41.3\sep11.7		&0.12\sep0.23	&\phnn4.2	&\phn78		&\phn1.9\phn	&0.8		\\
09	&$+$113.6	&\phn[$+$91.4\swp$+$110.7]	&[$+$117.2\swp$+$130.4]		&22.2\sep16.8		&0.40\sep0.46	&\phn35.\phn	&370			&\phn4.8\phn	&2.2		\\
10	&\phn$+$57.9	&[$+$43.0\swp$+$54.5]		&[$+$61.3\swp$+$75.0]		&14.9\sep17.1		&0.25\sep0.18	&\phn15.\phn	&120			&\phn1.1\phn	&1.3		\\
11	&\phn$+$57.6	&[$+$45.0\swp$+$54.5]		&[$+$61.3\swp$+$71.0]		&12.6\sep13.4		&0.42\sep0.46	&\phn11.\phn	&\phn79		&\phn0.64	&3.3		\\
12	&\phn$+$59.2	&[$+$38.6\swp$+$57.1]		&[$+$62.4\swp$+$77.8]		&20.6\sep18.6		&0.28\sep0.21	&\phn16.\phn	&130			&\phn1.4\phn	&1.2		\\
13	&\phn$+$43.9	&\ldots						&\ldots						&\ldots				&\ldots			&\ldots		&\ldots		&\ldots		&\ldots	\\
14	&\phn$-$44.5	&[$-$57.7\swp$-$46.6]		&[$-$42.0\swp$-$32.9]		&13.2\sep11.6		&0.25\sep0.34	&\phn20.\phn	&120			&\phn0.89	&2.3		\\
\hline
\end{tabular}
\begin{list}{}{}
\item[\supa] Maximum outflow velocity for blue and red lobes (as in Table~\ref{t:outflowsio21}).
\item[\supb] Deconvolved size of the blue and red outflow lobes, respectively.
\end{list}
\end{table*}
\begin{table*}
\caption{\label{t:outflowhco10}HCO$^+$\,(1--0) molecular outflows: velocity ranges of the wings, lobe sizes, outflow energetics and kinematics}
\centering
\begin{tabular}{c c c c c c c c c c c}
\hline\hline

&$V_\mathrm{LSR}$
&\multicolumn{1}{c}{$\Delta V_\mathrm{blue}$}
&\multicolumn{1}{c}{$\Delta V_\mathrm{red}$}
&$V_\mathrm{max-b/r}$\supa
&size$_\mathrm{blue/red}$\supb
&$M_\mathrm{out}$
&$p_\mathrm{out}$
&$E_\mathrm{out}$
&$t_\mathrm{kin}$
\\
\texttt{ID}
&(\kms)
&\multicolumn{1}{c}{(\kms)}
&\multicolumn{1}{c}{(\kms)}
&(\kms)
&(pc)
&(\mo)
&(\mo~\kms)
&(10$^{46}$~erg)
&(10$^{4}$~yr)
\\
\hline
01	&$+$33.0			&[$+$22.1\swp$+$31.4]		&[$+$34.9\swp$+$53.3]		&10.9\sep20.3	&0.48\sep0.84	&\phn37.\phn	&110			&\phn0.57	&4.9		\\
02	&$+$26.9			&[$+$18.1\swp$+$24.4]		&[$+$29.6\swp$+$40.0]		&\phn8.8\sep13.1	&0.37\sep0.28	&\phnn0.7	&\phnn3		&\phnn0.016	&2.9		\\
03	&$+$26.5			&[$+$12.1\swp$+$24.0]		&[$+$30.0\swp$+$40.1]		&14.4\sep13.6	&0.45\sep0.32	&\phnn1.6	&\phnn9		&\phnn0.056	&2.7		\\
04	&$+$26.5			&\phn[$+$7.3\swp$+$24.6]		&[$+$29.4\swp$+$42.5]		&19.2\sep16.0	&0.46\sep0.36	&\phn10.\phn	&\phn65		&\phn0.53	&2.3		\\
05	&$+$43.7			&[$+$10.3\swp$+$41.6]		&\phn[$+$46.2\swp$+$105.5]	&33.4\sep61.8	&0.47\sep0.59	&\phn66.\phn	&440			&\phn6.8\phn	&1.2		\\
06	&$+$106.5\phn	&\phn[$+$92.1\swp$+$104.3]	&[$+$108.6\swp$+$118.3]		&14.4\sep11.8	&0.15\sep1.44	&\phn57.\phn	&230			&\phn1.2\phn	&5.5		\\
07	&$+$42.5			&\phn[$+$7.5\swp$+$39.5]		&[$+$45.8\swp$+$88.1]		&35.0\sep45.6	&0.58\sep0.75	&103.\phn	&930			&13.\phnn	&1.7		\\
08	&$+$53.5			&[$+$27.2\swp$+$49.6]		&[$+$56.3\swp$+$75.0]		&26.3\sep21.5	&0.60\sep0.45	&\phnn5.9	&\phn57		&\phn0.74	&2.2		\\
09	&$+$113.6\phn	&[$+$103.2\swp$+$110.7]		&[$+$117.2\swp$+$130.4]		&10.4\sep16.8	&0.91\sep0.90	&\phn16.\phn	&\phn89		&\phn0.64	&6.9		\\
10	&$+$57.9			&[$+$41.4\swp$+$54.5]		&[$+$61.3\swp$+$75.0]		&16.5\sep17.1	&0.64\sep0.56	&\phn22.\phn	&150			&\phn1.2\phn	&3.5		\\
11	&$+$57.6			&[$+$41.4\swp$+$54.5]		&[$+$61.3\swp$+$75.0]		&16.2\sep17.4	&0.64\sep0.81	&\phn32.\phn	&230			&\phn1.9\phn	&4.3		\\
12	&$+$59.2			&[$+$41.8\swp$+$57.1]		&[$+$62.4\swp$+$77.4]		&17.4\sep18.2	&0.41\sep0.35	&\phn20.\phn	&130			&\phn1.1\phn	&2.1		\\
13	&$+$43.9			&[$+$29.5\swp$+$40.2]		&[$+$47.2\swp$+$55.8]		&14.4\sep11.9	&0.29\sep0.26	&\phn18.\phn	&\phn89		&\phn0.51	&2.1		\\
14	&$-$44.5			&[$-$65.7\swp$-$46.6]		&[$-$42.0\swp$-$17.7]		&21.2\sep26.8	&0.39\sep0.47	&\phn37.\phn	&210			&\phn1.8\phn	&1.8		\\
\hline
\end{tabular}
\begin{list}{}{}
\item[\supa] Maximum outflow velocity for blue and red lobes (as in Table~\ref{t:outflowsio21}).
\item[\supb] Deconvolved size of the blue and red outflow lobes, respectively.
\end{list}
\end{table*}
%----------------------------------------------------------------------------

\subsection{Spectral energy distributions\label{s:resSEDs}}

The spectral energy distributions (SEDs) of the sources in this sample were built by \citet{lopezsepulcre2011}, with continuum data at different wavelengths (ranging from mid-IR to 1.2~mm) collected from different telescopes and instruments. In this work we improve the SED of each source by complementing the previous data with the Herschel infrared Galactic Plane Survey \citep[Hi-GAL,][]{molinari2010a, molinari2010b} data at 500, 350, 250, 160 and 70~$\mu$m. Compact source detection and extraction were performed with CuTEx \citep{elia2010, molinari2011} in each of the five Hi-GAL maps separately. In Table~\ref{t:IRfluxes}, we list the continuum fluxes at different wavelengths for each source. The new Hi-GAL data represents a dramatic improvement to the SED, in particular for four sources with no previous data at far-IR/sub-millimeter wavelengths.

We have fitted the SEDs with a single-temperature, modified black body function, given by $F_\nu(T)=\Omega_\mathrm{S}\,B_\nu(T)\,(1-e^{-\tau_\nu})$, where $F_\nu(T)$ is the flux at frequency $\nu$, $\Omega_\mathrm{S}$ is the solid angle subtended by the source, $B_\nu(T)$ is the black body brightness, and $\tau_\nu=\tau_0\,(\nu/\nu_0)^\beta$ is the frequency-dependent optical depth, with $\kappa_0=0.005$~cm$^2$~g$^{-1}$ at $\nu_0=230.6$~GHz \citep{kramer2003}, assuming a gas-to-dust mass ratio of 100. The best fit was obtained by varying $\beta$ in the range 1.0--2.5, the temperature in the range 10--100~K, and the mass in the range 0.1--10000~\mo, and minimizing the expression $\sum_\mathrm{i}(F_{\nu\,\mathrm{i}}^\mathrm{obs}-F_{\nu\,\mathrm{i}}^\mathrm{model})^2$. We aim at determining the properties (mass and temperature) of the dust envelope associated with the massive protostars, which dominates the emission from far-infrared to millimeter wavelengths. Thus, for our best fit calculations, we used only the fluxes at wavelengths $\ge60$~$\mu$m (\ie\ $\nu\le5000$~GHz). This permits us to avoid the contribution of a potential, warmer component not associated with the dust envelope, and dominant at shorter wavelengths (see the excess at $\lambda\le60$~$\mu$m for some sources in Fig.~\ref{f:seds}). The best-fit parameters are listed in Cols.~8--10 of Table~\ref{t:sample}. The bolometric luminosity, listed in Col.~11, has been derived by integrating over the full observed spectral distribution (considering also the emission from the warmer component), $L_\mathrm{bol}=4\pi D^2\int{F_\nu^\mathrm{obs}\,d\nu}$, where $D$ is the heliocentric distance to the source. The median dust envelope mass and bolometric luminosity of our sample are 970~\mo\ and 13000~\lo, respectively. The masses and luminosities of our sources are in the range 300--3000~\mo\ and $10^{3.0}$--$10^{4.5}$~\lo, but for G19.27$+$0.1M1 and G19.27$+$0.1M1. These two sources have luminosities $\sim$$10^{2}$~\lo\ and masses of $\sim$100~\mo, about one order of magnitude below the median values. We considered whether the real distance of these sources is the far (13~kpc) rather than the adopted near kinematic (2.4~kpc) distance. However, since these two sources belong to an infrared dark cloud \citep[\eg][]{rathborne2006}, it seems improbable that they are located at such a large distance. Probably, these two clumps will produce intermediate-mass stars rather than high-mass stars. We note that for a few sources we obtain low values of $\beta$ (in the range 1.0--1.3), which may be a consequence of assuming an isothermal fit. Finally, in the last column of Table~\ref{t:sample}, we indicate the luminosity-to-mass ratio, $L/M$, which is believed to be a good indicator of the evolutionary state of the YSO both for the low-mass \citep[\eg][]{saraceno1996} and high-mass \citep[\eg][]{molinari2008} regimes.

%----------------------------------------------------------------------------
\begin{table}
\caption{\label{t:abundances}SiO temperatures and abundances}
\centering
\begin{tabular}{c c c c c}
\hline\hline
\texttt{ID}
&$T$\,(K)\supa
&X(SiO)$_\mathrm{clump}$\supb
&X(SiO)$_\mathrm{HCO^+}$\supc
&ratio\supd
\\
\hline
01	&18.7		&$0.26\times10^{-9}$	&\phn$0.07\times10^{-9}$	&3.80	\\
02	&10.0		&$3.82\times10^{-9}$	&$20.22\times10^{-9}$	&0.19	\\
03	&13.0		&$1.59\times10^{-9}$	&$12.62\times10^{-9}$	&0.13	\\
04	&10.0		&$0.59\times10^{-9}$	&\phn$2.03\times10^{-9}$	&0.29	\\
05	&15.1		&$1.31\times10^{-9}$	&\phn$2.17\times10^{-9}$	&0.60	\\
06	&10.0		&$1.36\times10^{-9}$	&\phn$0.85\times10^{-9}$	&1.60	\\
07	&12.7		&$2.13\times10^{-9}$	&\phn$2.12\times10^{-9}$	&1.01	\\
08	&\phn9.6		&$7.32\times10^{-9}$	&$29.95\times10^{-9}$	&0.24	\\
09	&11.0		&$2.09\times10^{-9}$	&\phn$7.19\times10^{-9}$	&0.29	\\
10	&12.9		&$2.62\times10^{-9}$	&\phn$9.20\times10^{-9}$	&0.29	\\
11	&13.0		&$1.69\times10^{-9}$	&\phn$5.15\times10^{-9}$	&0.33	\\
12	&11.7		&$4.19\times10^{-9}$	&$10.64\times10^{-9}$	&0.39	\\
13	&10.0		&$0.47\times10^{-9}$	&\phn$0.22\times10^{-9}$	&2.08	\\
14	&13.6		&$1.07\times10^{-9}$	&\phn$1.79\times10^{-9}$	&0.60	\\
\hline
\end{tabular}
\begin{list}{}{}
\item[\supa] Temperature used to derive the outflow parameters (see Sect.~\ref{s:disoutpar}), which has been computed as the rotational temperature derived from the SiO\,(2--1) and (5--4) lines, except for those sources with no detection of the SiO\,(5--4) for which we assumed a value of 10~K.
\item[\supb] SiO abundance derived from the ratio between the number of SiO particles over the clump and the corresponding H$_2$ gas mass, which was obtained from the relation between the outflow mass and the mass of the clump $M_\mathrm{out}=0.3\,M_\mathrm{clump}^{0.8}$ \citep[\eg][]{beuther2002b, lopezsepulcre2010}.
\item[\supc] SiO abundance derived from the ratio of the SiO and HCO$^+$ column densities, and assuming a fixed HCO$^+$ abundance of $5\times10^{-9}$ \citep{irvine1987, tafalla2010, busquet2011, tafallahacar2013}.
\item[\supd] X(SiO)$_\mathrm{clump}$ to X(SiO)$_\mathrm{HCO^+}$ ratio.
\end{list}
\end{table}
%----------------------------------------------------------------------------

%%%%%%%%%%%%%%%%%%%%%%%%%%%%%%%%%%%%%%%%%%%%%%%%%%%%%%%%%%%%%%%%%%%%%%%
\section{Analysis and Discussion\label{s:dis}}

\subsection{Derivation of outflow parameters\label{s:disoutpar}}

\begin{figure}[t]
\begin{center}
\begin{tabular}{c}
 \epsfig{file=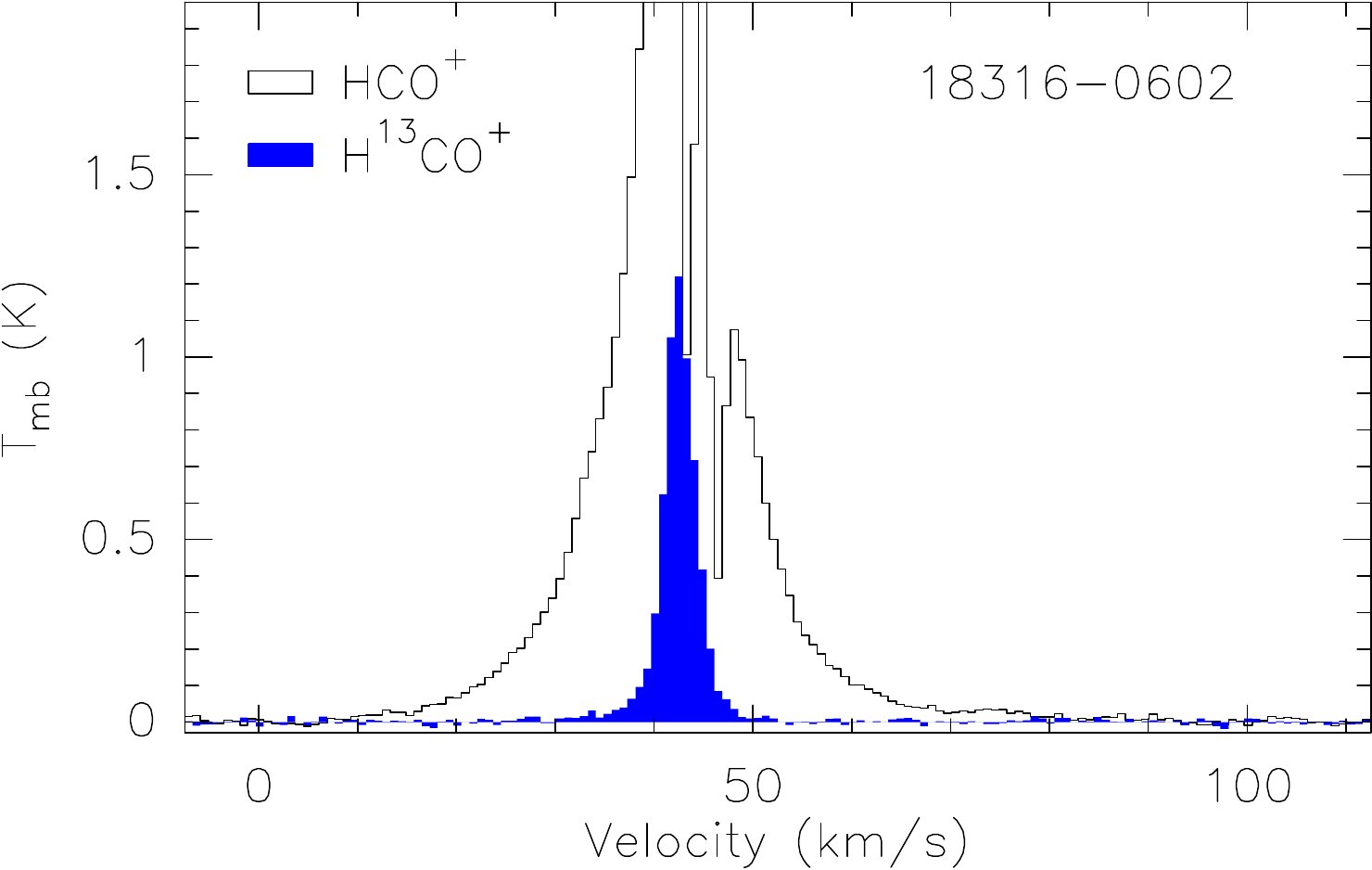, width=0.8\columnwidth, angle=0} \\
\end{tabular}
\end{center}
\caption{\label{f:hcospectra} Comparision of the HCO$^+$\,(1--0) and H$^{13}$CO$^+$\,(1--0) spectra for 18316$-$0602. No H$^{13}$CO$^+$\,(1--0) emission is detected at high velocities, which indicates that it is reasonable to assume optically thin emission for the HCO$^+$ line wings when deriving the outflow parameters (see Sect.~\ref{s:disoutpar}).}
\end{figure}

We used the procedure described in Eqs.~(4)--(6) of \citet{lopezsepulcre2009} to derive the outflow mass ($M_\mathrm{out}$), momentum ($p_\mathrm{out}$) and mechanical energy ($E_\mathrm{out}$) of the SiO and HCO$^+$ molecular outflows. These quantities have been calculated from the emission integrated inside the outflow velocity limits ($\Delta V_\mathrm{blue}$ and $\Delta V_\mathrm{red}$) and, spatially, over the area defined by the 5$\sigma$ contour level of the corresponding outflow lobe (Fig.~\ref{f:outmap}). In the calculations, the temperature for HCO$^+$ was fixed to 10~K (the outflow parameters increase by a factor 1.5 and 2 for temperature of 20~K and 30~K, respectively). For SiO, we calculated the rotation temperature from the intensity ratio of the SiO\,(2--1) and (5--4) lines for the ten sources in which both lines are detected, while we considered a fixed value of 10~K for the remaining four sources. The computed values are listed in Table~\ref{t:abundances}. The mean temperature for these ten sources is $\sim$13~K, which is similar to the assumed temperature of 10~K for the other sources. We assumed optically thin emission for both species. This is a common assumption for SiO outflows observed with single-dish telescopes (\ie\ angular resolutions $\ga$10\arcsec), and seems reasonable for the high-velocity component of our HCO$^+$ outflows, since no H$^{13}$CO$^+$\,(1--0) line emission is detected at high velocities (see \eg\ Fig.~\ref{f:hcospectra}).

For the abundances, we used a fixed value of $5\times10^{-9}$ for HCO$^+$. Several works \citep[\eg][]{irvine1987, tafalla2010, busquet2011, tafallahacar2013} report, from studies in different star-forming regions, an HCO$^+$ abundance in the range $1$-$9\times10^{-9}$, with no large enhancements when comparing the outflow gas with the dense core. In contrast, the abundance of SiO has been found to vary by several orders of magnitude, with values ranging from $10^{-12}$ to $10^{-7}$ \citep[\eg][]{ziurys1989, bachillerperezgutierrez1997, garay1998, codella2005, nisini2007}. For our sources, we refrained from assuming a given SiO abundance and estimated it using two approaches. First, we derived the SiO abundance from the ratio between the number of SiO particles over the clump and the corresponding H$_2$ gas mass. The former was derived from the SiO emission integrated in all the velocity range and, spatially, over the clump size, using the excitation temperatures listed in Table~\ref{t:abundances}. The latter was obtained from the relation between the outflow mass and the mass of the clump (listed in Table~\ref{t:sample}): $M_\mathrm{out}=0.3\,M_\mathrm{clump}^{0.8}$ \citep[\eg][]{beuther2002b, lopezsepulcre2009}. The SiO abundances derived with this method are listed in Col.~3 of Table~\ref{t:abundances}. In a second approach, we derived the SiO abundance by taking the ratio of the SiO and HCO$^+$ column densities in the high-velocity regime, and assuming a standard HCO$^+$ abundance of $5\times10^{-9}$. We did not consider the emission at systemic velocities to exclude the HCO$^+$ emission from the bulk gas. A caveat is that although HCO$^+$ seems to have quite a stable abundance, some studies have revealed some enhancement of HCO$^+$ in high velocity bullets \citep[\eg][]{tafalla2010, pacheco2012}. The abundances obtained with this method are listed in Col.~4 of Table~\ref{t:abundances}. The X(SiO)$_\mathrm{clump}$ to X(SiO)$_\mathrm{HCO^+}$ ratio varies from 4 to 0.1, with a median of 0.36 (cf.\ Col.~5 of Table~\ref{t:abundances}). The SiO abundance in our outflows is in the range $10^{-9}$--$10^{-8}$.

%18151-1208_1	18.7
%G19.27+0.1M2	10.0 -
%G19.27+0.1M1	13.0
%18236-1205		46.1 -
%18264-1152		15.1
%G23.60+0.0M1	10.0 -
%18316-0602		12.7
%G23.60+0.0M2	 9.6
%G24.33+0.1M1	11.0
%G34.43+0.2M1	12.9
%18507+0121		13.0
%G34.43+0.2M3	11.7
%19095+0930		13.1 -
%23139+5939		13.6

%----------------------------------------------------------------------
%----------------------------------------------------------------------

%----------------------------------------------------------------------
\begin{figure}[htp!]
\begin{center}
\begin{tabular}{c}
 \epsfig{file=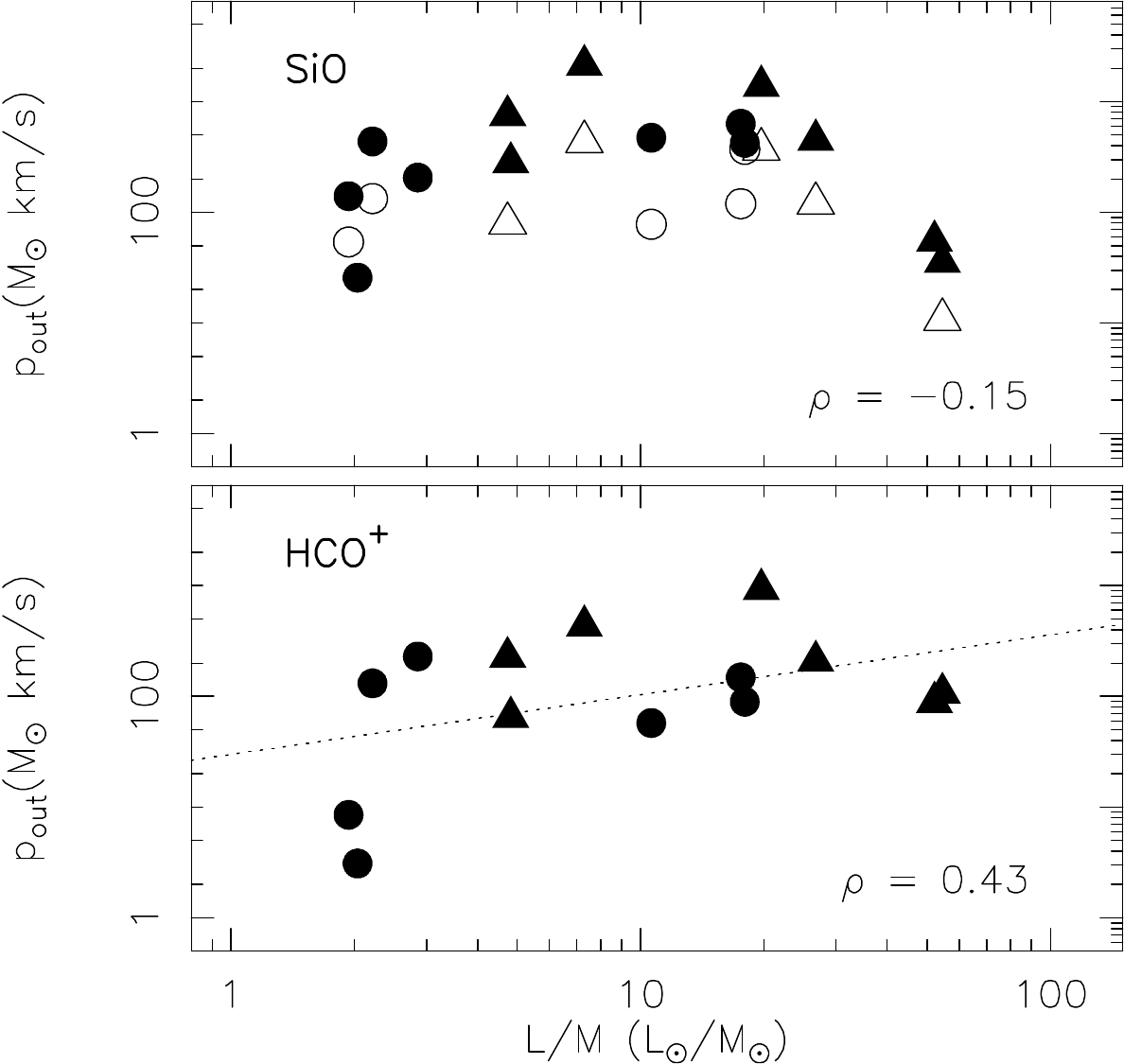, width=0.8\columnwidth, angle=0} \\
\end{tabular}
\end{center}
\caption{\label{f:outcorrelation} Outflow momentum against $L/M$, for the SiO\,(2--1) and HCO$^+$\,(1--0) outflows. Data for SiO\,(5--4) outflows are indicated as open symbols. Triangles and circles correspond to sources classified as IRL and IRD, respectively (see Table~\ref{t:sample}). The black dotted line shows the result of a least squares fit to the data, with the Pearson correlation coefficient ($\rho$) shown in the bottom right corner.}
\end{figure}
%----------------------------------------------------------------------

%----------------------------------------------------------------------
\begin{figure}[t]
\begin{center}
\begin{tabular}{c}
 \epsfig{file=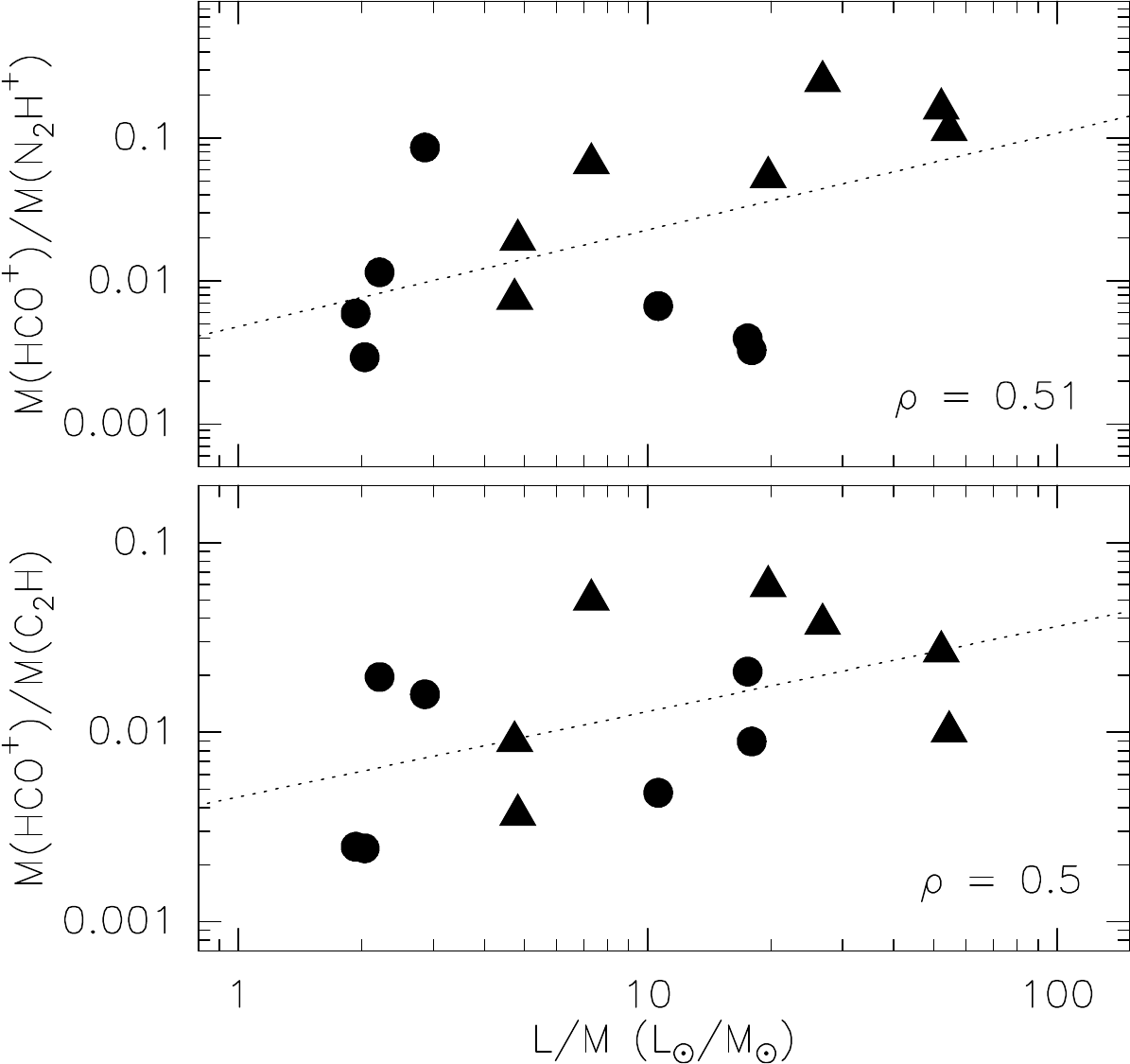, width=0.8\columnwidth, angle=0} \\
\end{tabular}
\end{center}
\caption{\label{f:outcorrelation_v2} Ratio of the outflow mass (derived from the HCO$^+$ outflow) to gas mass of the whole molecular clump (derived from N$_2$H$^+$ or C$_2$H; see Table~\ref{t:densegas}), against $L/M$. Symbols are the same as in Fig.~\ref{f:outcorrelation}. The black dotted lines show the result of least squares fits to the data, with the Pearson correlation coefficient ($\rho$) shown in the bottom right corner.}
\end{figure}
%----------------------------------------------------------------------

In Tables~\ref{t:outflowsio21}, \ref{t:outflowsio54} and \ref{t:outflowhco10}, we list the derived outflow parameters for the SiO\,(2--1), SiO\,(5--4) and HCO$^+$\,(1--0) lines, respectively. Assuming an uncertainty of 10\% on the flux and 20\% on the distance estimates, one finds that this implies uncertainties of 50\% for the derived outflow parameters. We have used the SiO abundances derived from the gas mass of the clump (Col.~3 of Table~\ref{t:abundances}). In Tables~\ref{t:outflowsio21}--\ref{t:outflowhco10}, we also list the maximum outflow velocities ($V_\mathrm{max-b}$ and $V_\mathrm{max-r}$) and the deconvolved lobe sizes (size$_\mathrm{blue}$ and size$_\mathrm{red}$). The maximum outflow velocities have been obtained from the difference between the high-velocity limit of the outflow velocity range, $\Delta V$, and the systemic velocity, $V_\mathrm{LSR}$. Typical values are in the range 10--20~\kms, with some extreme values as for example the $\sim$60~\kms\ of the red-shifted wing in 18264$-$1152. From the maximum outflow velocities and the lobe sizes we calculated the kinematic timescale as $t_\mathrm{kin}=\mathrm{size}/\mathrm{V}_\mathrm{max}$. The average of the numbers obtained for the blue and red lobes resulted in the kinematic timescale listed in Tables~\ref{t:outflowsio21}--\ref{t:outflowhco10}, which is typically a few $10^4$~yr. Finally, we note that the mass loss rate, mechanical force, and mechanical luminosity can be obtained as $\dot{M}_\mathrm{out}=M_\mathrm{out}/t_\mathrm{kin}$, $F_\mathrm{out}=p_\mathrm{out}/t_\mathrm{kin}$, and $L_\mathrm{mec}=E_\mathrm{out}/t_\mathrm{kin}$, respectively. In the derivation of the outflow parameters we have not corrected for the inclination of the outflow axis with respect to the line of sight. Some outflow parameters, such as the energy and mechanical luminosity, can be greatly affected if inclination $>60\degr$ (see Table~6 of \citealt{lopezsepulcre2010}, for the correction factor that must be applied for different inclinations).

The median outflow mass loss rate, mechanical force, and mechanical luminosity in our sample are $\sim$$9.1\times10^{-4}$~\mo~yr$^{-1}$, $\sim$$5.8\times10^{-3}$~\mo~\kms~yr$^{-1}$, and $\sim$4~\lo, respectively. These values are higher than typical values found in low-mass outflows\footnote{Molecular outflows driven by low-mass protostars have mass loss rates $\sim$$10^{-7}$--$10^{-6}$~\mo~yr$^{-1}$, outflow momentum rates $\sim$$10^{-5}$--$10^{-6}$~\mo~\kms~yr$^{-1}$, and mechanical luminosities $\sim$0.1--1~\lo\ \citep[\eg][]{bontemps1996, bjerkeli2013}.} and similar to those found toward objects with luminosities in the range $10^3$--$10^5$~\lo\ \citep[\eg][]{beuther2002b, wu2004, lopezsepulcre2009}, confirming that the molecular outflows reported in this work are associated with high-mass star formation.

% MEAN OUTFLOW PARAMETERS
% SiO(2-1) SiO(5-4) HCO+(1-0)
% 0.0016   0.0218   35.8
% 0.0011   0.0140   21.0
% 0.0013   0.0091    8.9
%mass loss rate, mechanical force and mechanical luminosity

%----------------------------------------------------------------------
\begin{figure}[t]
\begin{center}
\begin{tabular}{c}
 \epsfig{file=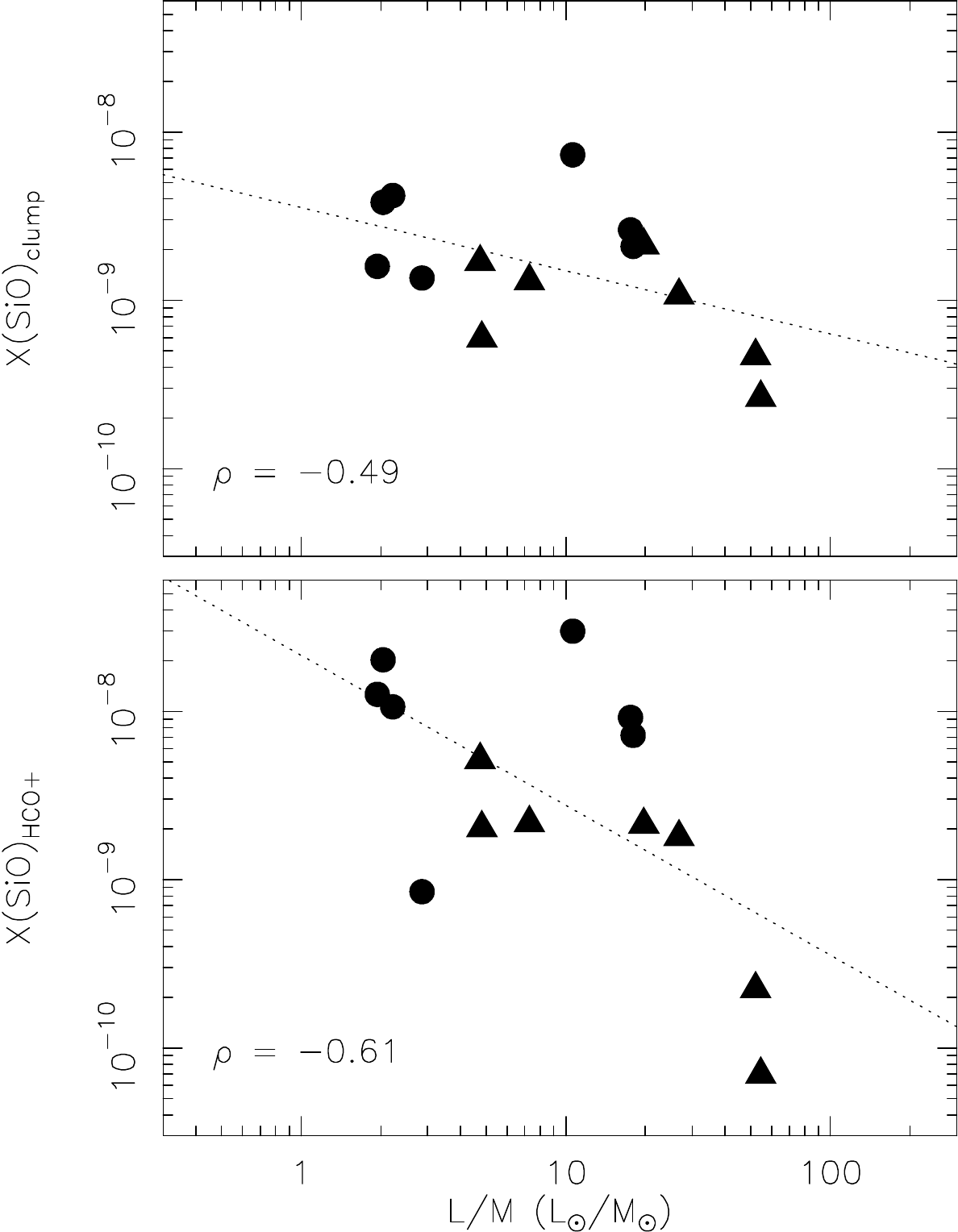, width=0.76\columnwidth, angle=0} \\
\end{tabular}
\end{center}
\caption{\label{f:outabundance} SiO abundance determined from the gas mass of the clump (top panel) and from the SiO to HCO$^+$ column density ratio (bottom panel) versus $L/M$. Symbols are the same as in Fig.~\ref{f:outcorrelation}. The black dotted lines show the result of least squares fits to the data, with the Pearson correlation coefficient ($\rho$) shown in the bottom left corner.}
\end{figure}
%----------------------------------------------------------------------

\subsection{Outflow activity with time\label{s:disouttime}}

Different studies \citep[\eg][]{saraceno1996, sridharan2002, molinari2008, giannetti2013} have proposed the luminosity-to-mass ratio, $L/M$, to be a good measure of the evolutionary phase  in the star formation process, with lower values corresponding to less evolved stages\footnote{We note that there are alternative interpretations to the $L/M$ ratio, \eg\ \citet{faundez2004} propose that this ratio could be an indicator of which type of star is the most conspicuous one within a star-forming region.}. With this in mind, \citet{lopezsepulcre2011} studied the variation of the SiO\,(2--1) luminosity with $L/M$ for a sample of 47 high-mass YSOs. These authors found a decrease of $L_\mathrm{SiO}/L_\mathrm{bol}$ with $L/M$ that can be interpreted as either a decrease of the SiO outflow energetics or a decrease of the SiO abundance with time. However, this result was based on single-pointing spectra, and did not make it possible to determine the outflow physical parameters, which are necessary to confirm whether the outflow properties vary with the evolution of the object. Such parameters can be now obtained from the outflow maps presented in this paper (see Sect.~\ref{s:disoutpar}).

In Fig.~\ref{f:outcorrelation}, we plot the outflow momentum with respect to $L/M$, for the SiO and HCO$^+$ outflows (similar plots are obtained for the outflow mass and energy). The low number of sources (only fourteen), and the restricted $L/M$ range covered in our observations (<$100$~\lo~\mo$^{-1}$), hinder a robust statistical analysis, but it is still possible to search for trends in the plots. No clear variations are found for the SiO outflow momentum as the object evolves (see Fig.~\ref{f:outcorrelation}-top panel), while the HCO$^+$ outflow momentum seems to slightly increase with time (with a Pearson correlation coefficient of 0.43, see dotted line in Fig.~\ref{f:outcorrelation}-bottom panel). We note that the trend is majorly dominated by the two sources with the lower $L/M$ values: G19.27$+$0.1M1 and G19.27$+$0.1M2, which could be associated with intermediate-mass rather than high-mass YSOs (see Sect.~\ref{s:resSEDs}). A similar trend is also seen in Fig.~\ref{f:outcorrelation_v2}, where we plot the ratio of the outflow mass (calculated from the HCO$^+$ outflow) to the gas mass of the clump, derived from the N$_2$H$^+$ and C$_2$H emission (see Table~\ref{t:densegas}), which is a distance independent parameter. However, observations of more sources, covering a larger $L/M$ range, are necessary to definitively confirm or rule out a possible increase of the HCO$^+$ outflow energetics with time. In Fig.~\ref{f:outabundance}-top, we show the SiO abundance as function of $L/M$, while in the bottom panel, we show the same plot using the SiO abundances derived from the SiO to HCO$^+$ column density ratio (see Sect.~\ref{s:disoutpar}). In both cases, we find a decrease of the SiO abundance with time (as suggested by \citealt{sakai2010}). A least squares fit to the data results in ${\rm X(SiO)_{clump}}=-(0.38\pm0.07)\,[L/M]-(8.45\pm0.07)$ and ${\rm X(SiO)_{HCO^+}}=-(0.89\pm0.21)\,[L/M]-(7.67\pm0.22)$, with correlation coefficients of $-0.5$ and $-0.6$, for the top and bottom panels respectively. These results are consistent with the variation reported by \citet{lopezsepulcre2011} and suggest a scenario in which SiO is largely enhanced in the first evolutionary stages, as expected if SiO is released from dust grains due to strong shocks produced by protostellar jets. As the object evolves, the power of the jet decreases, and so does the intensity of the shocks and the SiO abundance. Throughout this process, however, there is likely an increase of the amount of ambient material that is swept up by the jet, and thus the time-integrated outflow activity, which is traced by entrained molecular material (\eg\ HCO$^+$), likely increases with the age of the driving source.

%----------------------------------------------------------------------
\begin{figure*}[htp!]
\begin{center}
\begin{tabular}{c c c}
 \epsfig{file=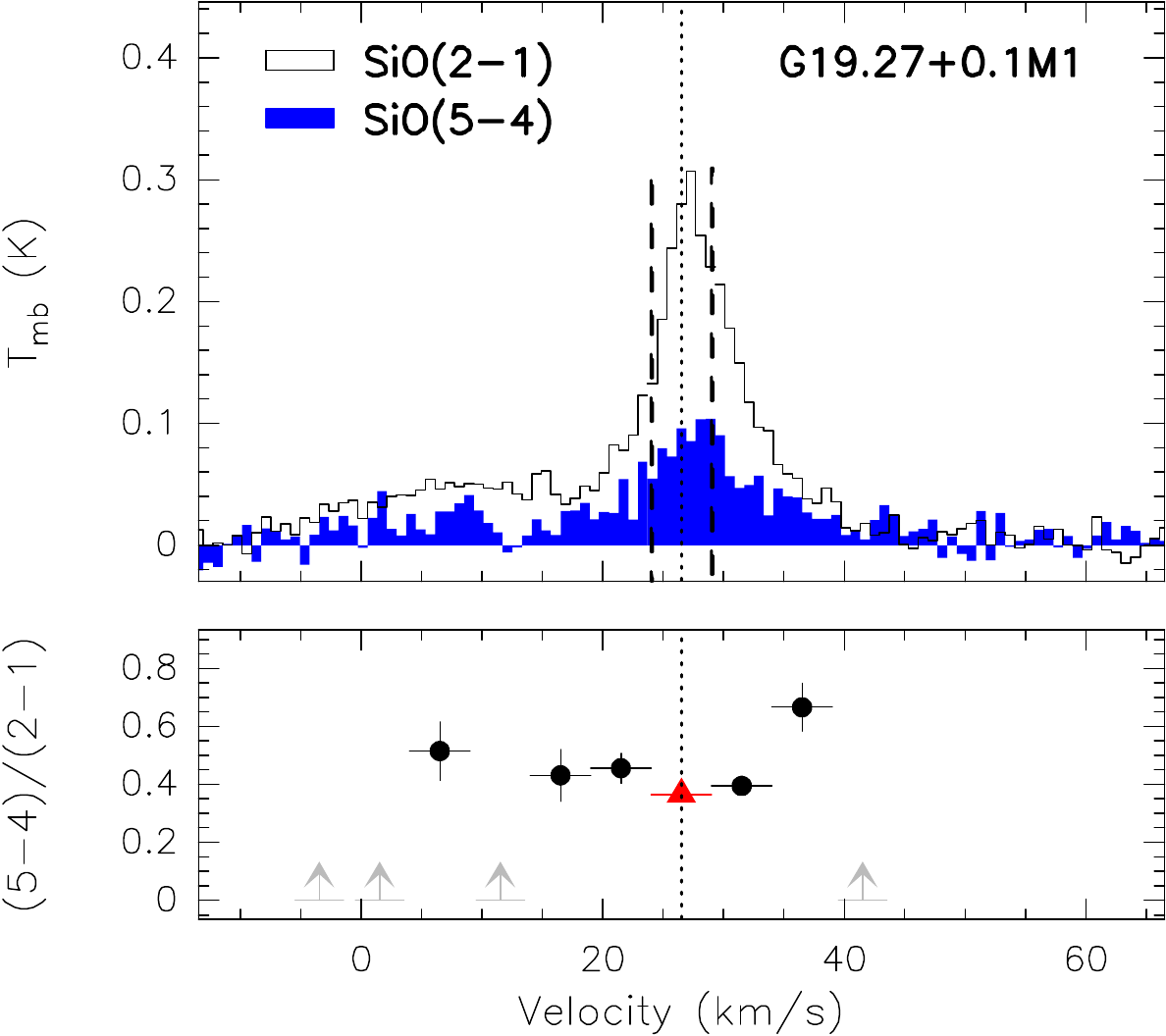, width=0.3\textwidth, angle=0} &
 \epsfig{file=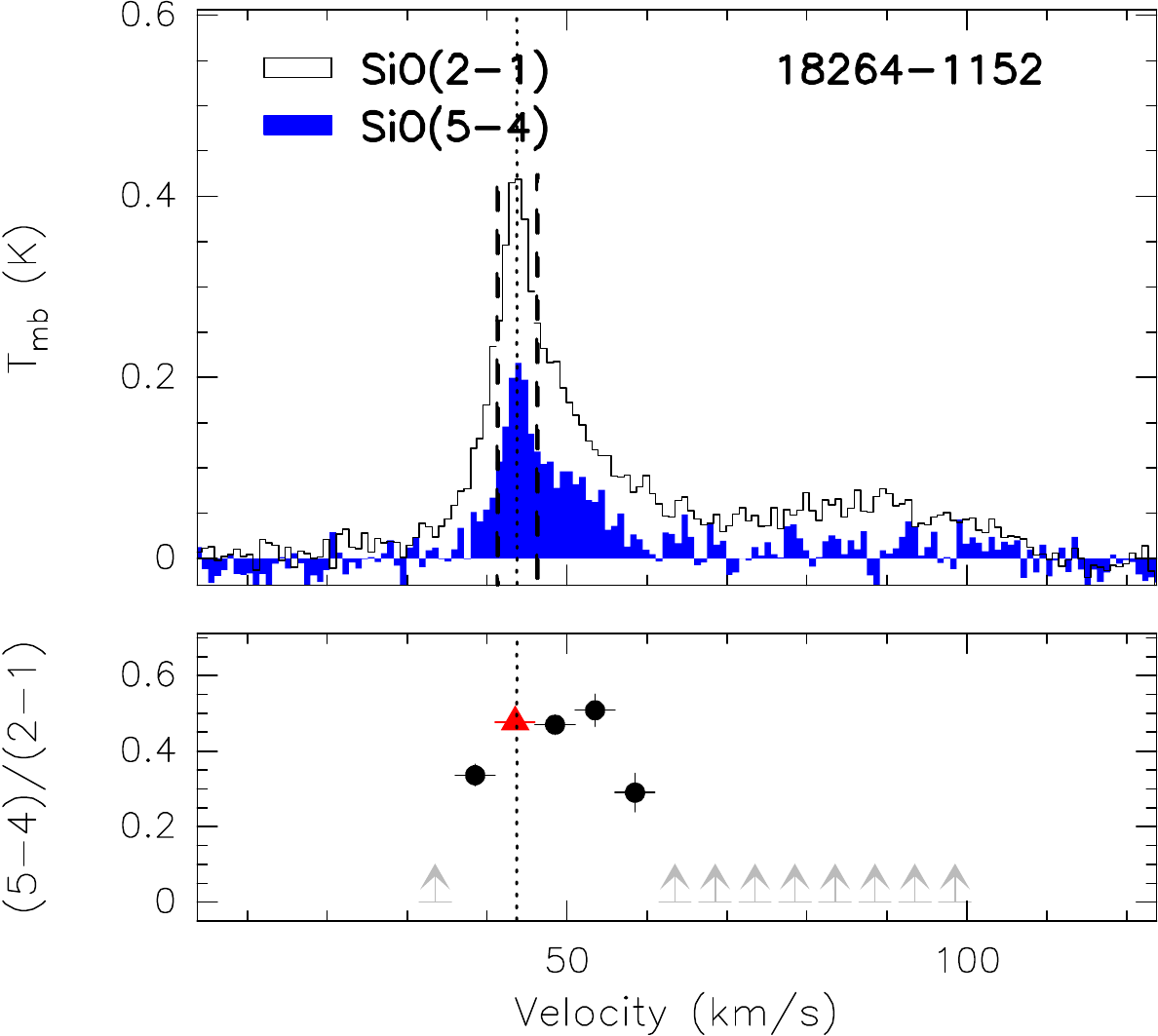, width=0.3\textwidth, angle=0} &
 \epsfig{file=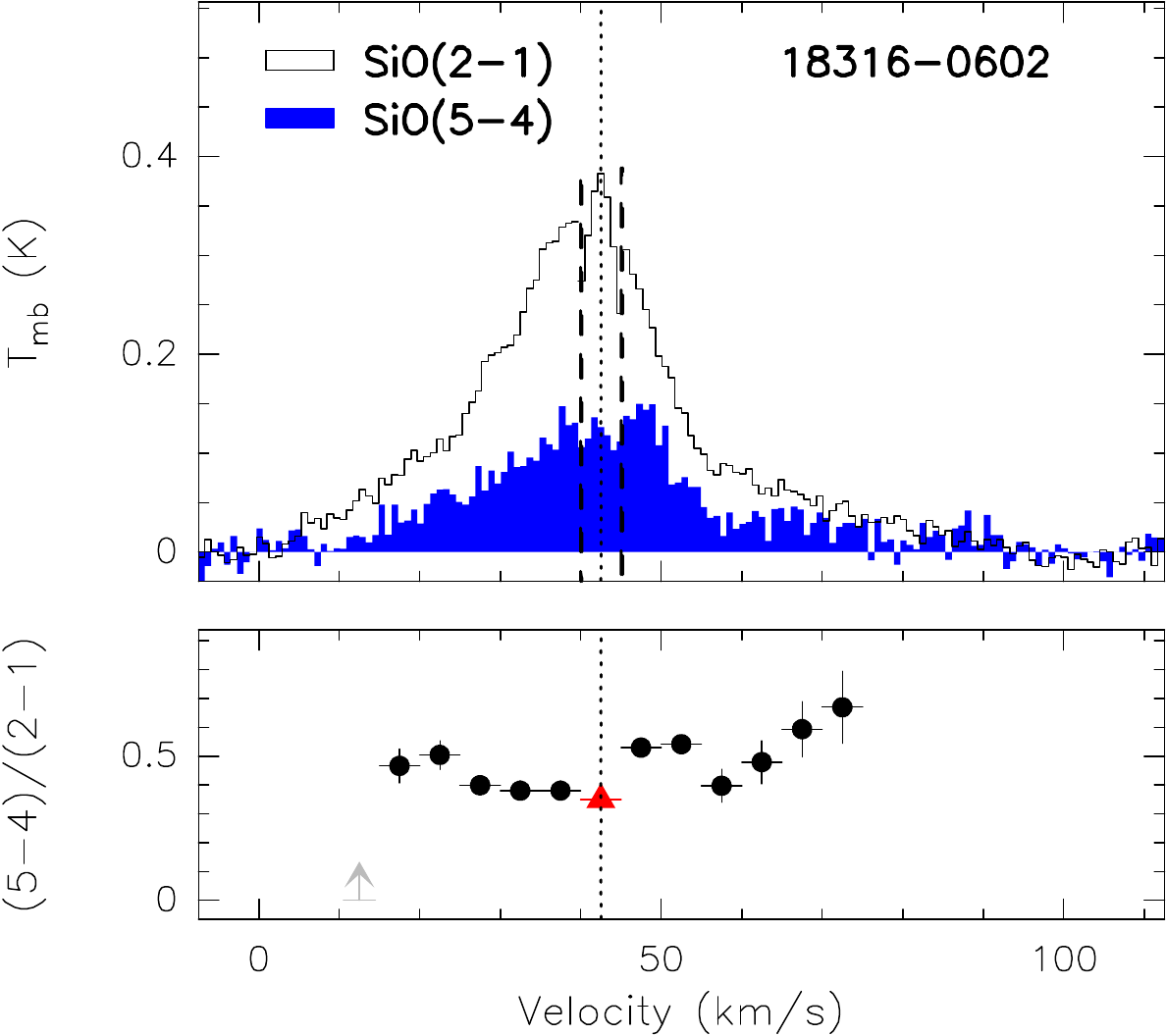, width=0.3\textwidth, angle=0} \\
 \epsfig{file=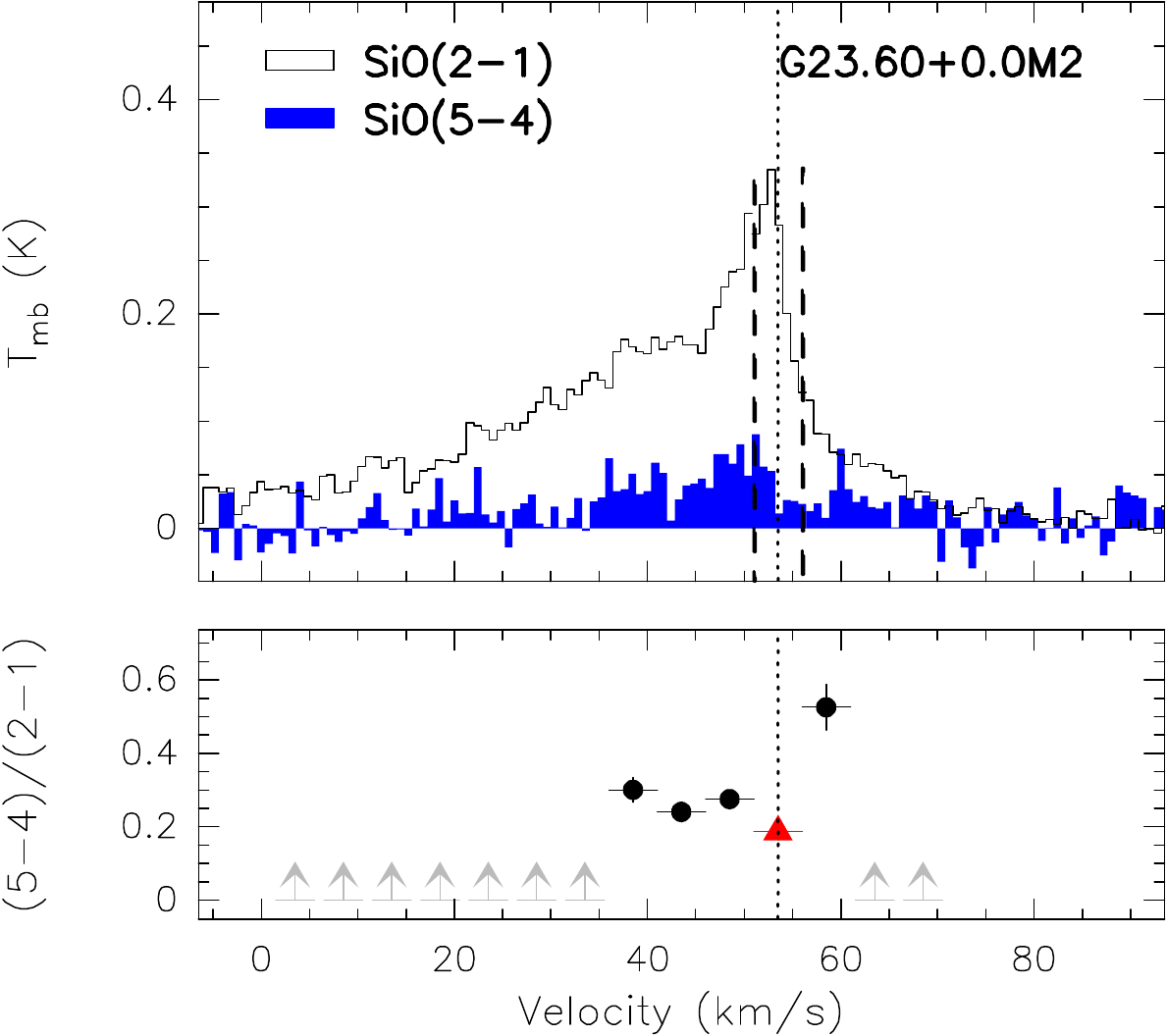, width=0.3\textwidth, angle=0} &
 \epsfig{file=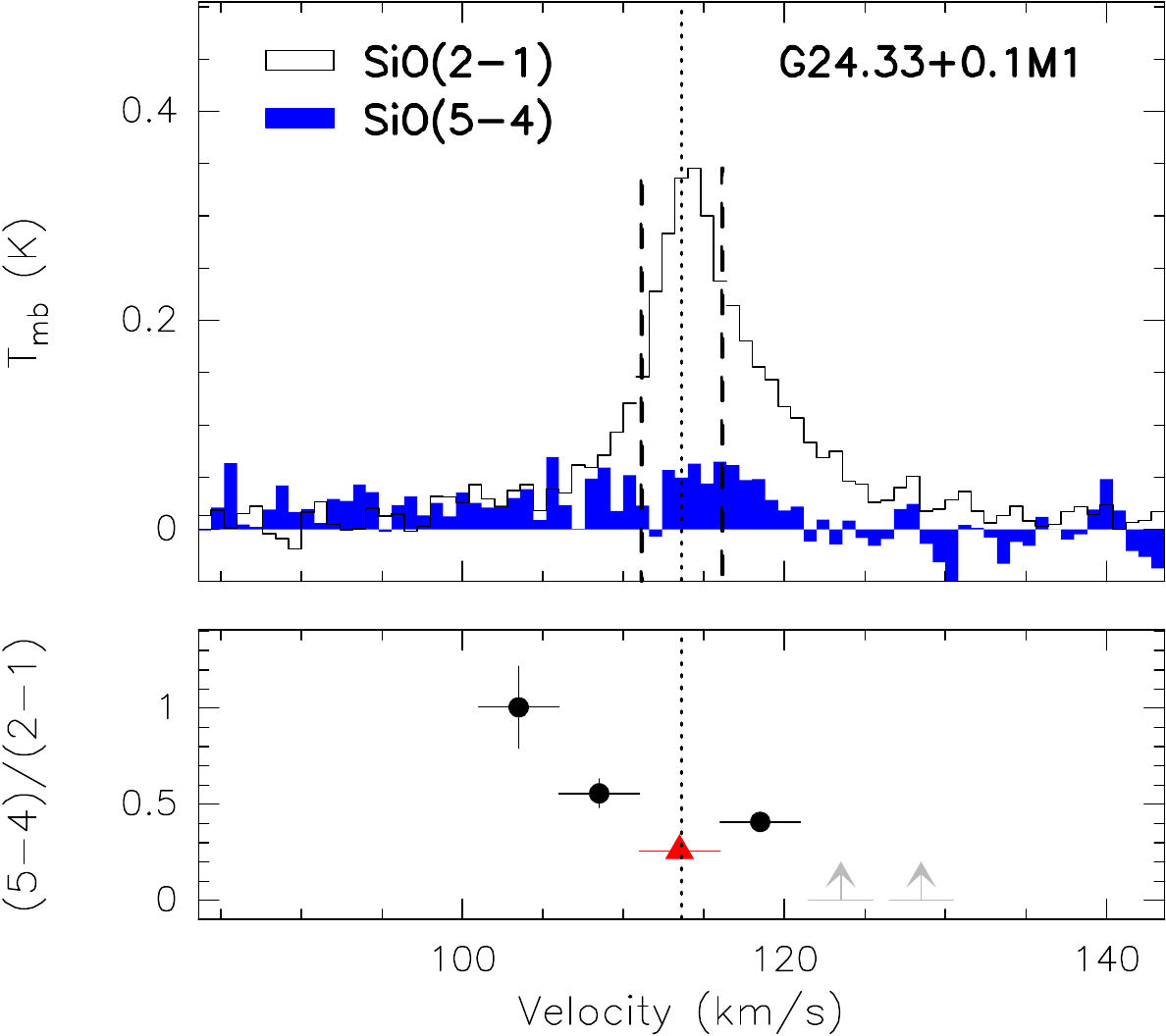, width=0.3\textwidth, angle=0} &
 \epsfig{file=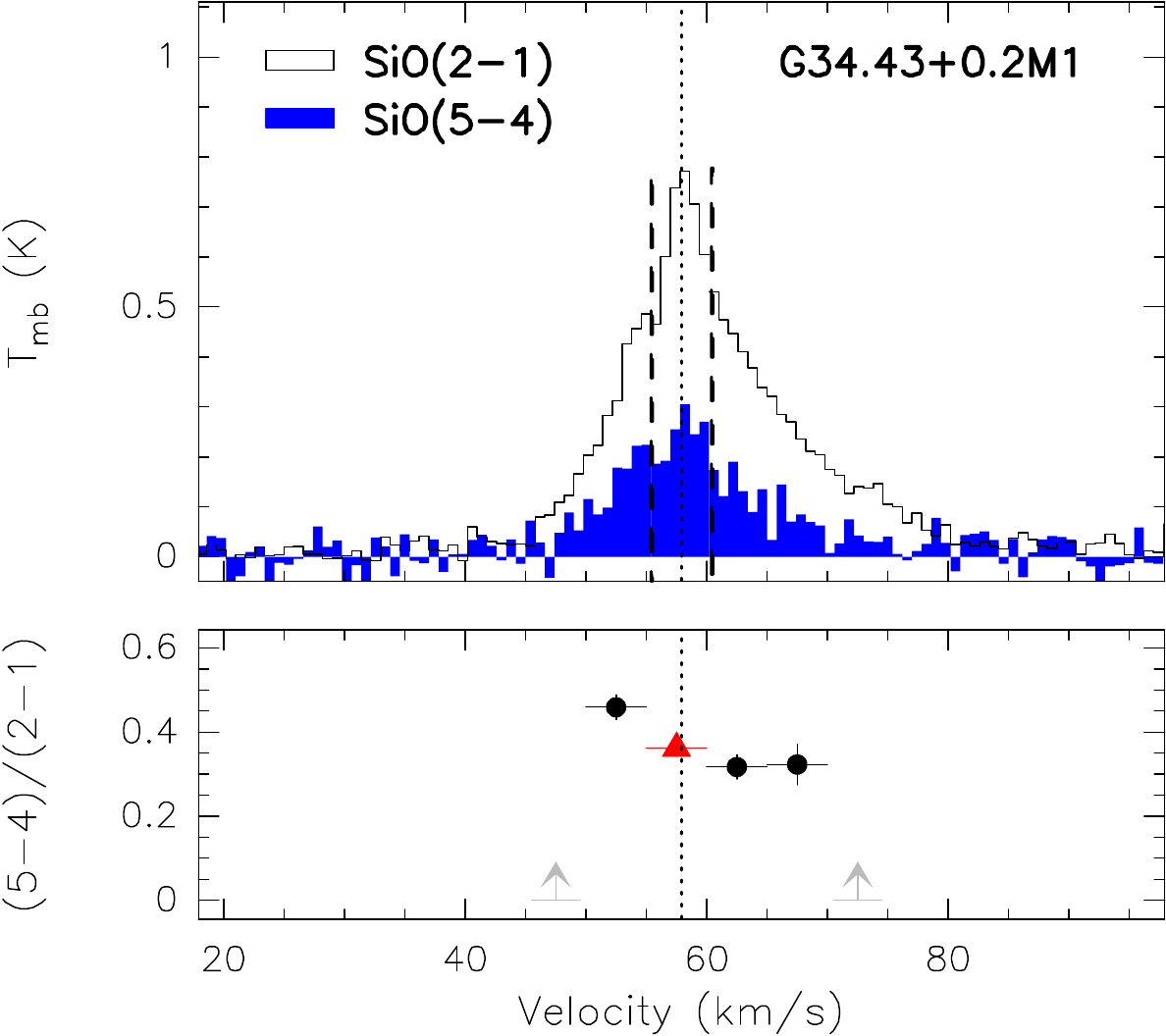, width=0.3\textwidth, angle=0} \\
 \epsfig{file=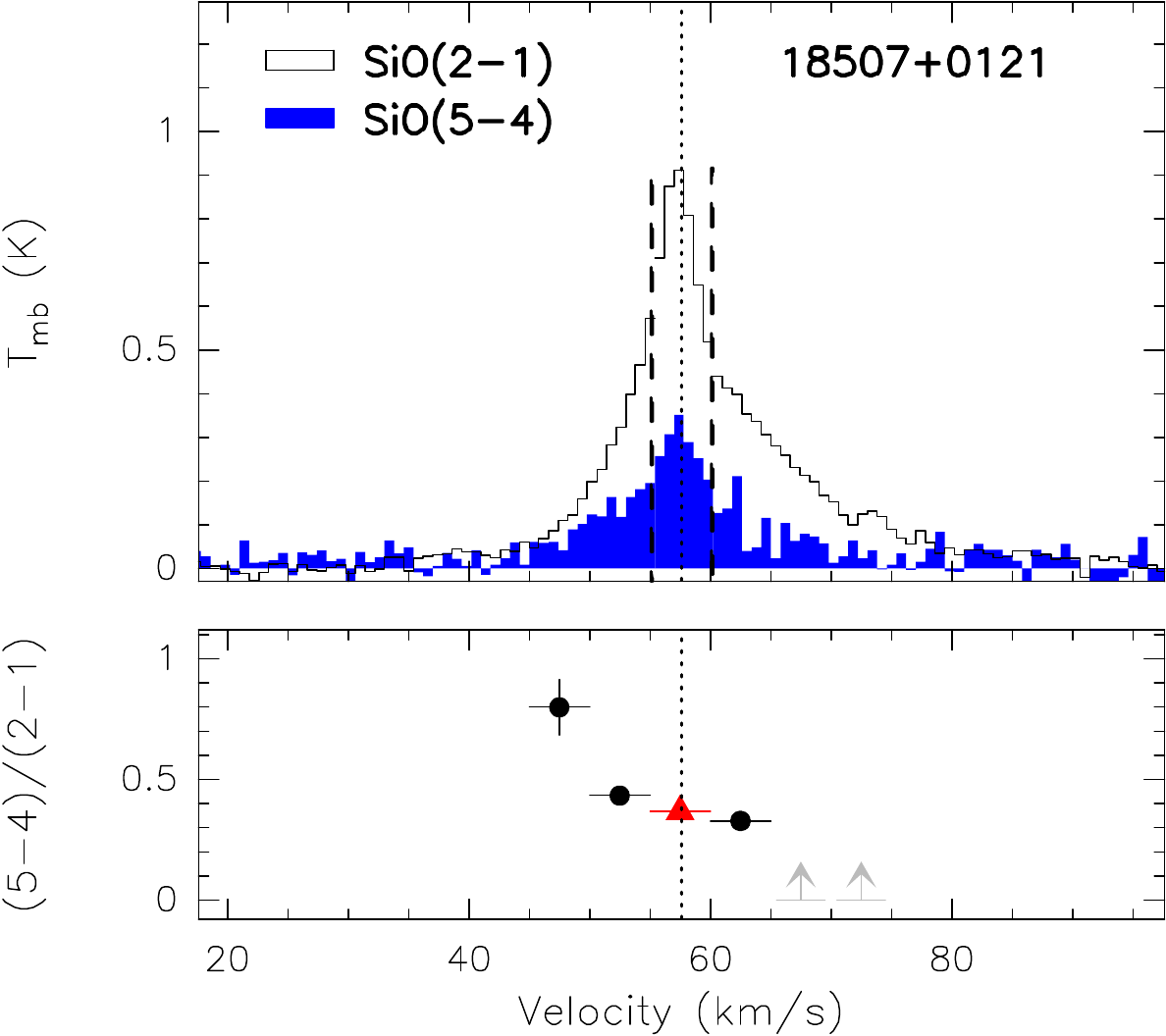, width=0.3\textwidth, angle=0} &
 \epsfig{file=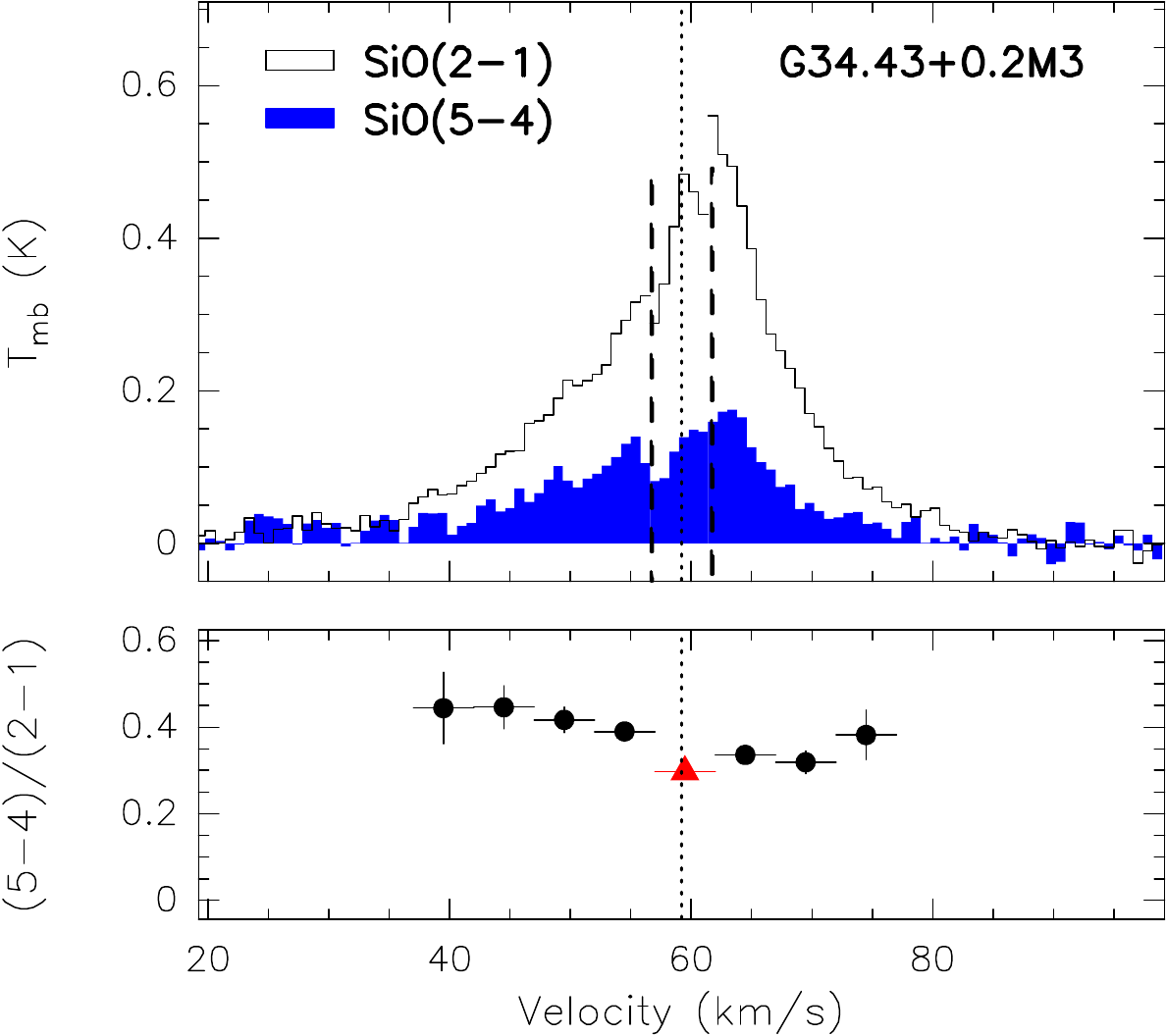, width=0.3\textwidth, angle=0} &
 \epsfig{file=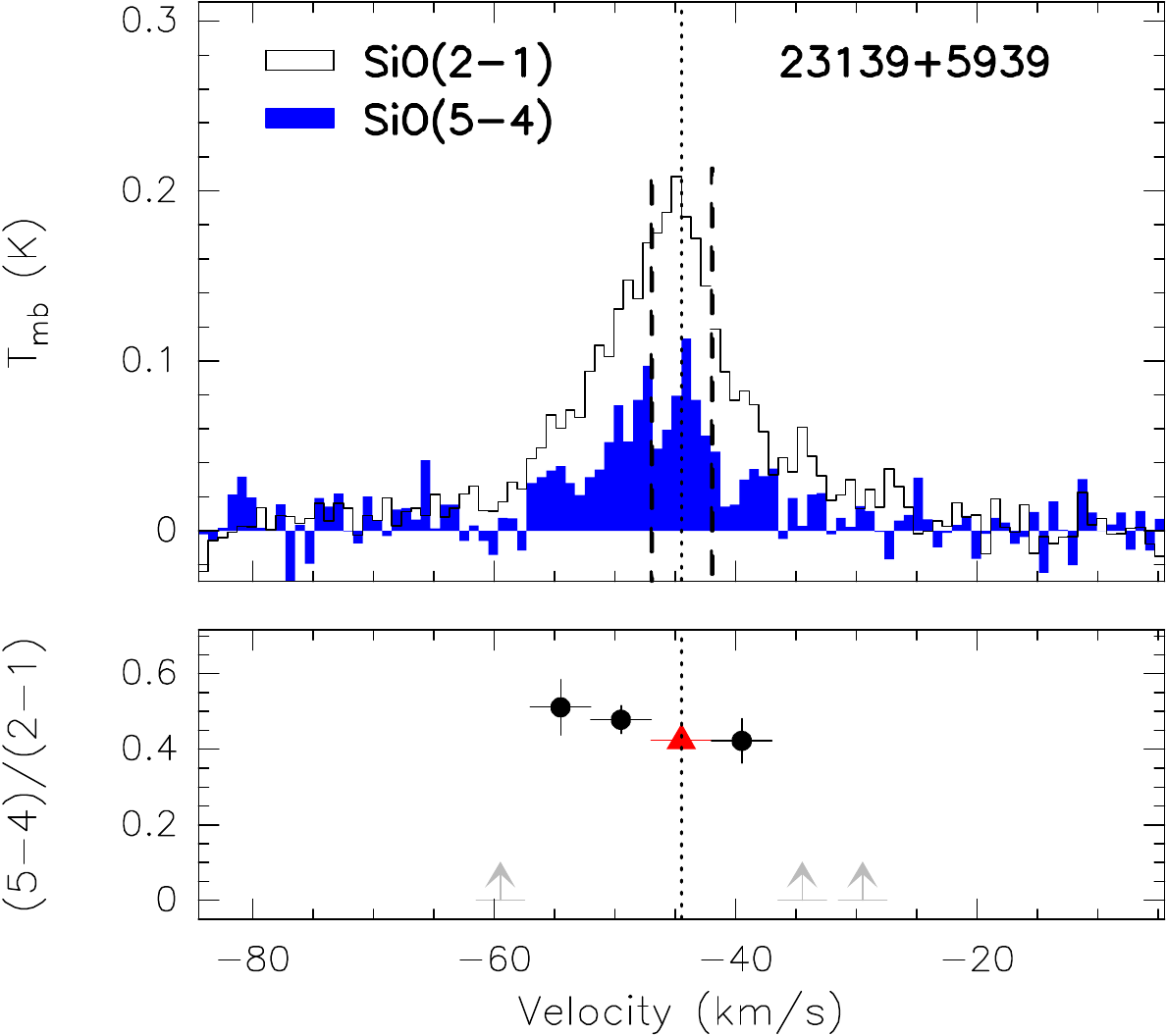, width=0.3\textwidth, angle=0} \\
\end{tabular}
\end{center}
\caption{\label{f:outratio} SiO\,(2--1) and (5--4) spectra obtained from the convolved maps. Each spectrum consists of three different parts: for blue/red-shifted velocities we have extracted the spectrum averaged over the area defined by the 50\% contour level of the blue/red-shifted integrated emission (considering the velocity intervals listed in Table~\ref{t:outflowsio21}), and for systemic velocities the area is defined by the 50\% contour level of the whole integrated emission. The vertical dashed lines indicate the three distinct velocity regimes, and the dotted vertical line corresponds to the systemic velocity. In the bottom panels (below each spectrum) we plot the (5--4)/(2--1) line ratio for velocity intervals of 5~\kms\ width. Red triangles correspond to the line ratio at the systemic velocity.}
\end{figure*}
%----------------------------------------------------------------------

\subsection{SiO excitation\label{s:disoutSiO}}

We can use the SiO\,(2--1) and (5--4) maps of the nine sources with clear detections in both lines, to constrain the physical conditions of the outflows and search for differences in the excitation conditions as a function of velocity. To do this, we first convolved the SiO\,(5--4) and SiO\,(2--1) maps to the same angular resolution, 30\arcsec, and considered two distinct velocity regimes: \emph{systemic velocities}, corresponding to the range $\pm2.5$~\kms\ around the bulk velocity (listed in Table~\ref{t:sample}), and \emph{high velocities}, corresponding to the red- and blue-shifted ($|V-V_\mathrm{LSR}|>2.5$~\kms) outflow velocities.

\begin{figure}[t!]
\begin{center}
\begin{tabular}{c}
 \epsfig{file=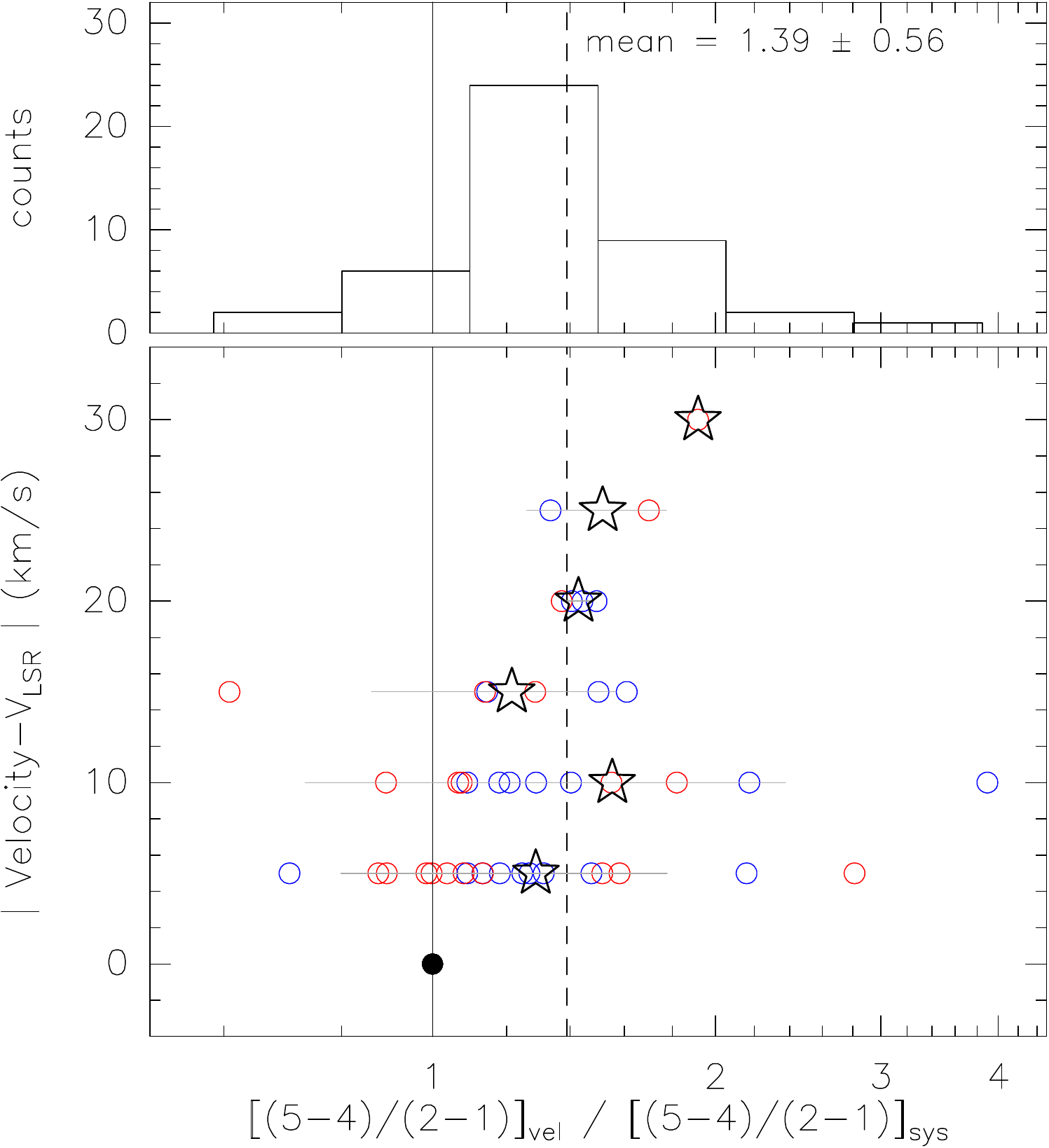, width=0.9\columnwidth, angle=0} \\
\end{tabular}
\end{center}
\caption{\label{f:ratiovelocity} \emph{Bottom panel}: Velocity, with respect to the systemic velocity, against the SiO (5--4)/(2--1) line ratio, normalized with the line ratio measured at systemic velocities. The black dot corresponds to the normalized line ratio equal to unity that corresponds to the systemic velocity. Red and blue circles correspond to the red and blue-shifted SiO emission. Stars (and horizontal gray lines) indicate the mean (and dispersion) value at each velocity. The vertical dashed line indicates the mean value for all the line ratios at velocities higher than the systemic velocity. \emph{Top panel}: Distribution of normalized SiO (5--4)/(2--1) line ratios. The mean (and dispersion) value, corresponding to 1.39 (0.56) is indicated with the vertical dashed line.}
\end{figure}

In Fig.~\ref{f:outratio}, we show the SiO\,(2--1) and (5--4) spectra for the nine outflow sources. These spectra have been obtained after degrading the resolution of all maps to 30\arcsec. In order to take into account that the emission at different velocities comes from different spatial structures: bulk velocity emission and outflow lobes, we extracted the spectra averaged over the area enclosed by the 50\% contour level of the corresponding outflow lobe. At systemic velocities, the average was made over the area enclosed by the 50\% contour level of the whole integrated emission. The different velocity intervals are indicated in the figure with vertical dashed lines. In the bottom panel associated with each spectrum of Fig.~\ref{f:outratio}, we show the (5--4)/(2--1) line ratio as a function of the velocity, considering velocity intervals of 5~\kms. We calculated the line ratio for those velocity intervals in which the line intensity of both transitions is at least 3 times above the noise level, determined by the expression $\sigma=\Delta V\sigma_\mathrm{rms}(N)^{1/2}$, with $\Delta V = 5$~\kms, $\sigma_\mathrm{rms}$ the rms of the spectrum, and $N$ the number of channels \citep[\eg][]{caselli2002a}. We find that for most of the sources, the (5--4)/(2--1) line ratio is typically low at the systemic velocity and increases as we move to red/blue-shifted velocities. This effect is clear in Fig.~\ref{f:ratiovelocity}, in which the x-axis shows the (5--4)/(2--1) line ratio normalized with respect to the line ratio measured at systemic velocities. This normalization permits us to remove calibration errors. For increasing velocities (with respect to the systemic) we find a ratio slightly greater than unity (see star symbols in Fig.~\ref{f:ratiovelocity}). The distribution of the normalized line ratios for red/blue-shifted velocities (see top panel of Fig.~\ref{f:ratiovelocity}) has an average value of 1.39. This result indicates that the excitation conditions for SiO depend on the velocity of the emitting gas. The same trend has been found in some well-studied low-mass protostellar outflows (\eg\ L1157 and L1448: \citealt{nisini2007}).

%----------------------------------------------------------------------
%----------------------------------------------------------------------
%----------------------------------------------------------------------
\begin{figure}[t]
\begin{center}
\begin{tabular}{c}
 \epsfig{file=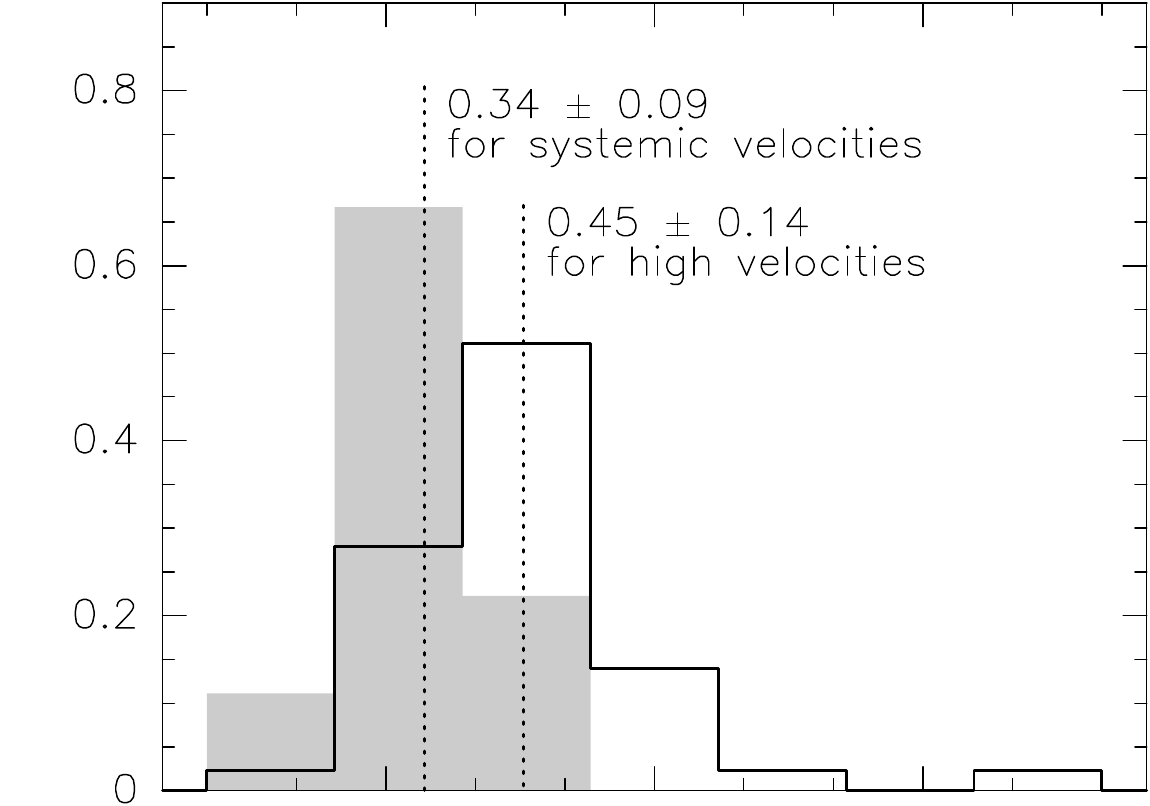, width=0.9\columnwidth, angle=0} \\
 \epsfig{file=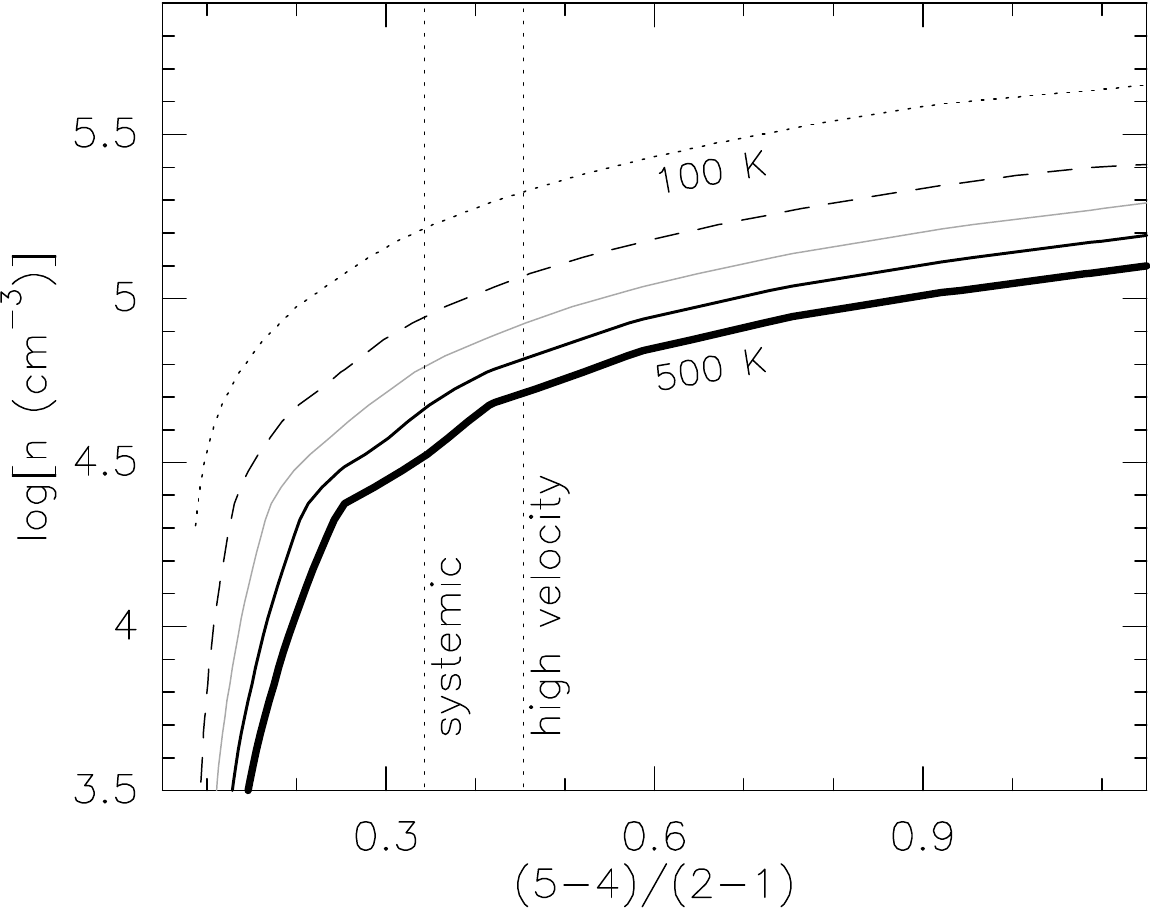, width=0.9\columnwidth, angle=0} \\
\end{tabular}
\end{center}
\caption{\label{f:5421histo} \emph{Top panel}: Distributions of the SiO (5--4)/(2--1) ratio for two different velocity regimes: systemic velocities (gray filled histogram) and high velocities (solid line histogram). The dotted line and numbers indicate the mean ($\pm$standard deviation) value. \emph{Bottom panel}: LVG results produced by the RADEX code. We plot the density versus the SiO\,(5--4)/(2--1) ratio. The curves correspond to kinetic temperatures ranging from 100~K (upper curve) to 500~K (lower curve) in steps of 100~K. The two vertical lines correspond to the mean (5--4)/(2--1) ratios for systemic and high velocities.}
\end{figure}
%----------------------------------------------------------------------

\subsubsection{Collisional excitation calculations}

%and between SiO and He, incorporating levels up to $J=26$ for kinetic temperatures up to 300~K (Dayou \& Balan\c{c}a 2006)
 
The results derived from Figs.~\ref{f:outratio} and \ref{f:ratiovelocity}, indicate that the excitation conditions differ between the gas components at systemic and high velocities. To constrain the corresponding physical conditions, we performed radiative transfer calculations with the RADEX code \citep{vanderTak2007} based on the large velocity gradient (LVG) approximation and assuming a plane-parallel ``slab'' geometry. The molecular data were retrieved from the LAMDA data base\footnote{http://www.strw.leidenuniv.nl/$\sim$moldata/}. The collisional rate coefficients were calculated between SiO and H$_2$, incorporating levels up to $J=40$ for kinetic temperatures up to 2000~K \citep{dayoubalanca2006}. In Fig.~\ref{f:5421histo}-top, we show the distribution of the (5--4)/(2--1) ratio values obtained for the two different velocity regimes: systemic velocities (gray filled histogram) and high velocities (solid line histogram). The mean (and standard deviation) values of the line ratios are $0.34\pm0.09$ and $0.45\pm0.14$ for systemic and high velocities, respectively. In Fig.~\ref{f:5421histo}-bottom, we compare these mean line ratios with the results obtained from the RADEX analysis. RADEX calculations were obtained with a velocity dispersion of 5~\kms, and a column density of $10^{12}$~cm$^{-1}$, \ie\ optically thin emission (the modeled line ratios are insensitive to the column density up to values of $10^{13}$~cm$^{-2}$). The different curves shown in the figure correspond to kinetic temperatures ranging from 100~K to 500~K, which is a range of temperatures typically found toward low- and high-mass outflows \citep[\eg][]{nisini2007, codella2013}. From this analysis we can infer that in general the SiO emission of our outflow sources is generated from shocks in regions with densities ranging from $10^{4.0}$~cm$^{-3}$ to $10^{5.5}$~cm$^{-3}$, for kinetic temperatures in the range 100--500~K. From the mean values of the SiO\,(5--4)/(2--1) ratio (see Fig.~\ref{f:5421histo}-bottom), the range of densities is $10^{4.7}$--$10^{5.3}$~cm$^{-3}$ for high velocities, and $10^{4.4}$--$10^{5.2}$~cm$^{-3}$ for low velocities, which are hints of the gas at high velocities to be produced in regions with larger densities (and/or temperatures) than the gas at systemic velocities.

%In Table~\ref{t:radex}, we list the density ($n$) and thermal pressure ($n\timesT$) at 100~K and 500~K.

\subsection{SiO production\label{s:disprocSiO}}

SiO is known to have an extremely low abundance in dark clouds \citep[\eg][]{ziurys1989} or in photon-dominated regions (PDRs; \eg\ \citealt{schilke2001}) whereas it is easily detected in outflows. Partly as a consequence, most theoretical work on the SiO abundance in star-forming regions has centered on the abundance in outflows and jets where it seems likely that shock chemistry plays a major role \citep[\eg][]{schilke1997, gusdorf2008a, gusdorf2008b, gusdorf2011}. Three general scenarios have been discussed: (1) Si production due to sputtering of grain cores followed by gas phase reactions with O$_2$ or OH, (2) sputtering of SiO mixed into grain mantles, and (3) grain-grain collisions liberating Si from cores. The first and third of these require relatively high shock velocities in order to produce an observable quantity of SiO (at least 25~\kms\ in the case of option 1, though perhaps less in the case of grain-grain collisions). Such high shock velocities seem unlikely for our observed outflows which in general (see $\Delta V_\mathrm{red}$ and $\Delta V_\mathrm{blue}$ in Tables~\ref{t:outflowsio21} and \ref{t:outflowsio54}) show SiO emission at velocities (relative to ambient) less than 30~\kms\ with notable exceptions such as 18316$-$0602 and G23.60$+$0.0M2. Also we note that our observed profiles (see Figs.~\ref{f:outspec} and \ref{f:outratio}) show mainly emission at velocities only a few \kms\ different from the systemic velocity with a high velocity tail suggesting again that low 10--15~\kms\ shocks predominate, similar to what is found in the low-mass protostellar outflow L1157 \citep{nisini2007}. In this sense, it is interesting to note that \citet{gusdorf2008b} conclude that SiO mixed into grain mantles is necessary to explain the observations towards the shock region L1157~B1. This seems plausible also in the regions we have studied though we note that the absence of detections of a solid state SiO feature gives pause for thought. Moreover, the very low SiO abundances found by \citet{schilke2001} towards nearby PDRs are slightly surprising if several percent of silicon are in the form of mantle material. However, present estimates of photodesorption yields from ice mantles are extremely crude and we conclude that the most likely scenario explaining our results is a gradual erosion of ice mantles as the clump evolves resulting in less SiO being ejected into the outflow shocked material at late times.

%%%%%%%%%%%%%%%%%%%%%%%%%%%%%%%%%%%%%%%%%%%%%%%%%%%%%%%%%%%%%%%%%%%%%%%
\section{Summary\label{s:con}}

We have mapped 14 high-mass star-forming regions with the IRAM~30-m telescope in different molecular lines at 1 and 3~mm, with the aim to characterize the properties of molecular outflows associated with high-mass YSOs. The sample was selected from the work of \citet{lopezsepulcre2011}, and contains objects with previous single-pointing SiO molecular outflow emission. Our main conclusions can be summarized as follows:

\begin{itemize}

\item The wide ($\sim$15~GHz) frequency range surveyed has allowed us to simultaneously map several molecular transitions, typically found tracing dense gas (\eg\ N$_2$H$^+$, C$_2$H, NH$_2$D) and outflow (\eg\ SiO, HCO$^+$) emission. Twelve of the fourteen sources, are also detected in the CH$_3$CN line (eight of them showing emission of high excitation $K$ transitions), which is a tracer typically found in association with hot cores.

\item The detection of high-velocity wings in the SiO\,(2--1) and HCO$^+$\,(1--0) lines for all the sources, confirms the presence of outflows in these regions as previously reported by \citet{lopezsepulcre2010, lopezsepulcre2011}. The outflow maps reveal clear bipolar structures in six of the regions.

\item We improved the SEDs of these sources by complementing previous continuum data with Hi-GAL data \citep{molinari2010a, molinari2010b}. From the SEDs we derived dust envelope masses and luminosities in the range 100-3000~\mo\ and 100--50000~\lo, respectively. We calculated the luminosity-to-mass ratio, $L/M$, which is believed a good indicator of the evolutionary stage.

%\item We have improved the SEDs of these sources by complementing previous continuum data with the Hi-GAL data at 500, 350, 250, 160, and 70~$\mu$m \citep{molinari2010a, molinari2010b}. We have fitted a single-temperature, modified black body function to the data at wavelengths $\le$60~$\mu$m. From the fits we have derived the mass and temperature of the dust envelope, and by integrating the observed fluxes at all frequencies, we have derived the bolometric luminosity. The resulting dust envelope masses and luminosities are in the range 100--3000~\mo\ and 100--50000~\lo, respectively. We have also calculated the luminosity-to-mass ratio, $L/M$, which is believed a good indicator of the evolutionary stage of the YSO, with lower values associated with early stages.

%\item The SiO abundance has been determined using two approaches: from the gas mass of the clump, and from the HCO$^+$ emission. On average, the SiO abundance in our sample of outflows is in the range $2$--$8\times10^{-9}$. We have derived the outflow parameters from the SiO and HCO$^+$ lines, obtaining values similar to those typically found in outflows associated with objects of luminosities in the range $10^3$--$10^{4}$~\lo. 

\item We studied the variation of the outflow properties as the object evolves. The SiO outflow energetics seem to remain constant with time (\ie\ $L/M$), while there are some hints of a possible trend suggesting that the HCO$^+$ outflow energetics might increase as the object evolves. Interestingly, we find a decrease of the SiO abundance with time, from $10^{-8}$ to $10^{-9}$. There results are consistent with the variation reported by \citet{lopezsepulcre2011}, and suggest a scenario in which SiO is largely enhanced in the first evolutionary stages, probably due to strong shocks originated by the protostellar jet. As the object evolves, the power of the jet would decrease and so would the SiO abundance. In the meanwhile, the material surrounding the protostar would have been swept up by the jet, and thus the outflow activity, as traced by entrained molecular material (HCO$^+$), would increase with time.

%\item We have studied the variation of the outflow properties with $L/M$, as an indicator of the evolutionary stage, to test and confirm the decrease of SiO luminosity with time reported by \citet{lopezsepulcre2011}. The SiO outflow energetics seem to remain constant with time, while there is a trend that suggests that the HCO$^+$ outflow parameters increase as the object evolves. Interestingly, we find a decrease of the SiO abundance with time. There results, consistent with the variation reported by \citet{lopezsepulcre2011}, suggest a scenario in which SiO is largely enhanced in the first evolutionary stages, probably due to strong shocks originated by the protostellar jet. As the object evolves, the power of the jet decreases and so does the SiO abundance. In the meanwhile, the material surrounding the protostar has been swept up by the jet, and thus the outflow activity, as traced by entrained molecular material (HCO$^+$), increases with time.

\item We find that the SiO\,(5--4)/(2--1) line ratio is typically low at systemic velocities, and increases as we move to red/blue-shifted velocities, as similarly found toward low-mass protostellar outflows \citep[\eg][]{nisini2007}. From radiative transfer calculations done with the RADEX code, we find that, in general, the SiO emission of our outflows is generated in regions with densities $10^{4.0}$--$10^{5.5}$~cm$^{-3}$ and kinetic temperatures 100--500~K, with larger densities and/or temperatures for the high-velocity gas component.

\end{itemize}

%%%%%%%%%%%%%%%%%%%%%%%%%%%%%%%%%%%%%%%%%%%%%%%%%%%%%%%%%%%%%%%%%%%%%%%
\begin{acknowledgements}
A.\ S.-M.\ is grateful to Arturo I.\ G\'omez-Ruiz for fruitful discussions on the analysis of the SiO outflow emission. We thank the anonymous referee for his/her constructive criticisms. The figures of this paper were made with the software package Greg of GILDAS (http://www.iram.fr/IRAMFR/GILDAS).
\end{acknowledgements}

%%%%%%%%%%%%%%%%%%%%%%%%%%%%%%%%%%%%%%%%%%%%%%%%%%%%%%%%%%%%%%%%%%%%%%%
\bibliographystyle{aa}

%%%%%%%%%%%%%%%%%%%%%%%%%%%%%%%%%%%%%%%%%%%%%%%%%%%%%%%%%%%%%%%%%%%%%%%
\begin{appendix}
%%%%%%%%%%%%%%%%%%%%%%%%%%%%%%%%%%%%%%%%%%%%%%%%%%%%%%%%%%%%%%%%%%%%%%%
\section{Figures, molecular line results and spectral energy distributions\label{a:extra}}

%----------------------------------------------------------------------------
\begin{table*}
\caption{\label{t:molecules}Molecular lines detected at 3~mm toward the 14 high-mass young stellar objects\supa}
\centering
\begin{tabular}{l c c c c c c c c c c c c c c c}
\hline\hline
Molecular
&Freq.
\\
transition
&(MHz)
&\#01
&\#02
&\#03
&\#04
&\#05
&\#06
&\#07
&\#08
&\#09
&\#10
&\#11
&\#12
&\#13
&\#14
\\
\hline
NH$_2$D\,(1$_{1,1}$--1$_{0,1}$)\supb		&85926.2780	&Y	&Y	&Y	&Y	&Y	&n	&Y	&Y	&Y	&Y	&Y	&Y	&n	&n	\\
HC$^{15}$N\,(1--0)						&86054.9664	&Y	&n	&n	&Y	&Y	&n	&Y	&Y	&Y	&Y	&n	&Y	&n	&n	\\
SO\,(2$_2$--1$_1$)						&86093.9500	&Y	&n	&Y	&Y	&Y	&n	&Y	&Y	&n	&Y	&n	&Y	&Y	&Y	\\
H$^{13}$CN\,(1$_{2}$--0$_{1}$)\supb		&86340.1630	&Y	&Y	&Y	&Y	&Y	&Y	&Y	&Y	&Y	&Y	&Y	&Y	&Y	&Y	\\
HCO\,(1$_{0,1}$--0$_{0,0}$)\supb			&86670.7600	&Y	&n	&Y	&n	&Y	&n	&n	&n	&n	&n	&n	&n	&n	&Y	\\
H$^{13}$CO$^+$\,(1--0)					&86754.2884	&Y	&Y	&Y	&Y	&Y	&Y	&Y	&Y	&Y	&Y	&Y	&Y	&Y	&Y	\\
SiO\,(2--1)								&86846.9600	&Y	&Y	&Y	&Y	&Y	&Y	&Y	&Y	&Y	&Y	&Y	&Y	&Y	&Y	\\
HN$^{13}$C\,(1--0)						&87090.8252	&Y	&Y	&Y	&Y	&Y	&Y	&Y	&Y	&Y	&Y	&Y	&Y	&Y	&Y	\\
C$_2$H\,(1--0)\supb						&87316.8980	&Y	&Y	&Y	&Y	&Y	&Y	&Y	&Y	&Y	&Y	&Y	&Y	&Y	&Y	\\
HC$_5$N\,(33--32)						&87863.6300	&Y	&n	&n	&n	&n	&n	&n	&n	&Y	&n	&n	&n	&n	&n	\\
HNCO\,(4$_{0,4}$--3$_{0,3}$)				&87925.2370	&Y	&Y	&Y	&Y	&Y	&Y	&Y	&Y	&Y	&Y	&Y	&Y	&n	&Y	\\
HCN\,(1$_{2}$--0$_{0}$)\supb				&88631.8475	&Y	&Y	&Y	&Y	&Y	&Y	&Y	&Y	&Y	&Y	&Y	&Y	&Y	&Y	\\
H$^{15}$NC\,(1--0)						&88865.7150	&Y	&n	&Y	&n	&Y	&n	&Y	&Y	&Y	&n	&n	&Y	&n	&n	\\
HCO$^+$\,(1--0)							&89188.5247	&Y	&Y	&Y	&Y	&Y	&Y	&Y	&Y	&Y	&Y	&Y	&Y	&Y	&Y	\\
HNC\,(1--0)								&90663.5680	&Y	&Y	&Y	&Y	&Y	&Y	&Y	&Y	&Y	&Y	&Y	&Y	&Y	&Y	\\
HC$_3$N\,(10--9)							&90979.0230	&Y	&Y	&Y	&Y	&Y	&Y	&Y	&Y	&Y	&Y	&Y	&Y	&Y	&Y	\\
CH$_3$CN\,(5$_0$--4$_0$)\supc			&91987.0876	&Y	&Y	&Y	&Y	&Y	&n	&Y	&Y	&Y	&Y	&Y	&Y	&Y	&n	\\
$^{13}$CS\,(2--1)						&92494.3080	&Y	&Y	&Y	&Y	&Y	&n	&Y	&Y	&Y	&Y	&Y	&Y	&Y	&Y	\\
N$_2$H$^+$\,(1$_{2,3}$--0$_{1,2}$)\supb	&93173.7642	&Y	&Y	&Y	&Y	&Y	&Y	&Y	&Y	&Y	&Y	&Y	&Y	&Y	&Y	\\
\hline
\end{tabular}
\begin{list}{}{}
\item[\supa] Y: line detected. n: line non detected. The frequencies are obtained from the CDMS (Cologne Database for Molecular Spectroscopy; \citealt{muller2001}) and JPL (Jet Propulsion Laboratory; \citealt{pickett1998}) catalogues.
\item[\supb] Transition and frequency of the main hyperfine component. For a few species, the different hyperfine components are well separated in frequency: for HCO, we detect four hyperfine components at 86670.76~MHz, 86708.36~MHz, 86777.46~MHz and 86777.46~MHz; and for C$_2$H, we detect six hyperfine components at 87284.105~MHz, 87316.898~MHz, 87328.585~MHz, 87401.989~MHz, 87407.165~MHz and 87446.470~MHz.
\item[\supc] For some sources we detect several transits of the $k$-ladder spectrum of the CH$_3$CN\,(5--4) transition, corresponding to the frequencies 91987.0876~MHz ($K=0$), 91985.3141~MHz ($K=1$), 91979.9943~MHz ($K=2$), 91971.1304~MHz ($K=3$), and 91958.7260~MHz ($K=4$).
\end{list}
\end{table*}
%----------------------------------------------------------------------------

\begin{figure*}
\begin{center}
\begin{tabular}[b]{c}
 \epsfig{file=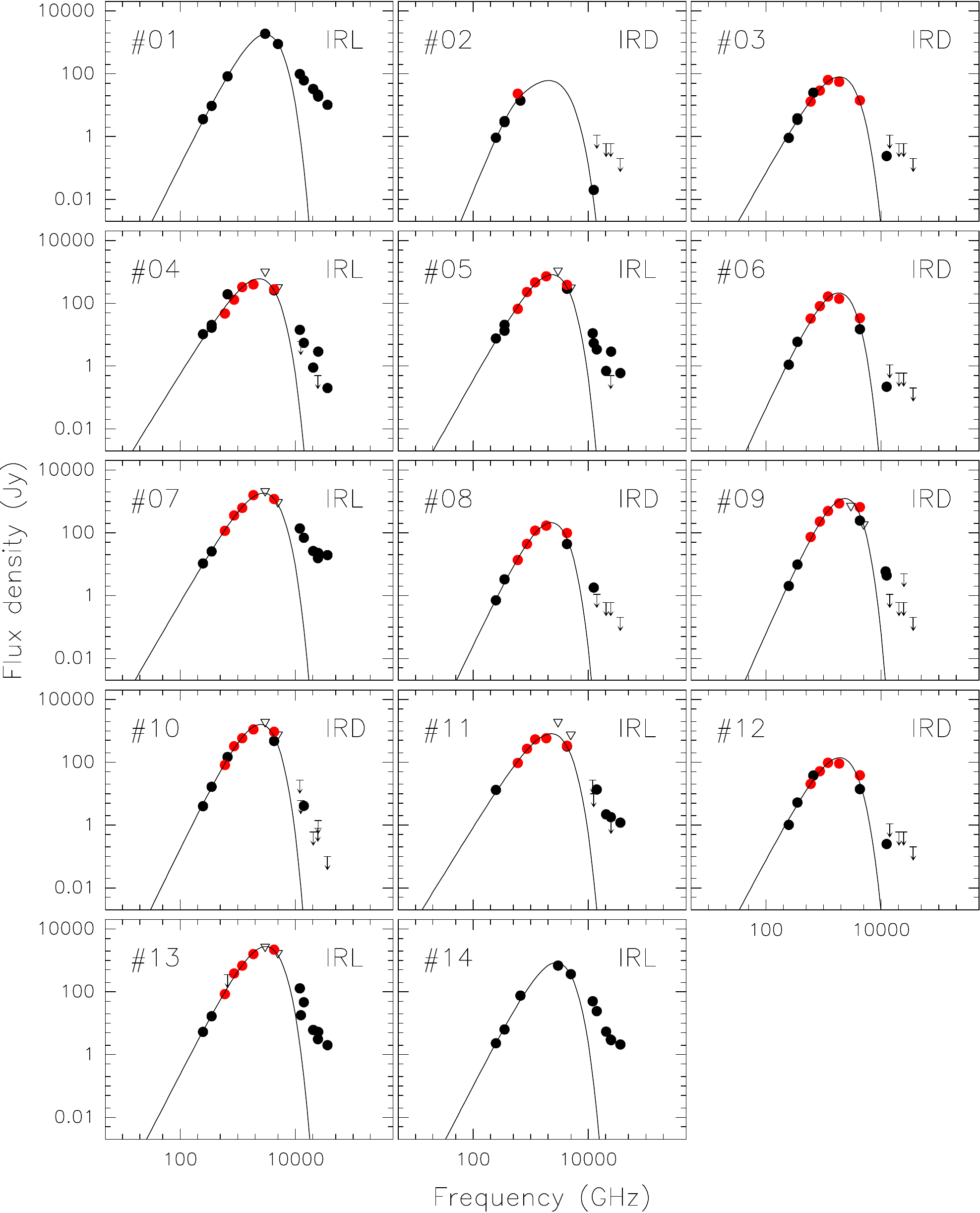, width=0.9\textwidth, angle=0} \\
\end{tabular}
\end{center}
\caption{\label{f:seds}.Spectral energy distributions for the 14 sources listed in Table~\ref{t:sample}. Black dots correspond to measured continuum fluxes in the range 8~$\mu$m to 1.2~mm (see Table~\ref{t:IRfluxes}), while the crosses correspond to upper limits. Red dots correspond to the new Hi-GAL data \citep{molinari2010a, molinari2010b}. White triangles correspond to IRAS fluxes at 60~$\mu$m and 100~$\mu$m not considered in the SED fitting when there is available Hi-GAL data. The solid curve represents the best fit to the data points with wavelengths $\ge 60$~$\mu$m (see Sect.~\ref{s:resSEDs}).}
\end{figure*}

In Figure~\ref{f:summary}, we present different panels for each observed source, including the spectra at 3~mm, zero-order moment (integrated intensity) maps for different molecular lines, and a comparison between the emission at 8~$\mu$m and at sub-millimeter (500~$\mu$m or 850~$\mu$m) wavelengths. In Table~\ref{t:molecules}, we list the detection of molecules toward the 14 high-mass YSOs, while in Table~\ref{t:densegas}, we list the hyperfine structure fit results and physical parameters of the dense cores derived from the N$_2$H$^+$\,(1--0) and C$_2$H\,(1--0) lines. In Table~\ref{t:IRfluxes}, we present continuum fluxes measured for the 14 sources, in the wavelength range from 8.0~$\mu$m to 1.2~mm, including data from different facilities: MSX, \emph{Spitzer}/MIPS, IRAS, \emph{Herschel}/PACS, \emph{Herschel}/SPIRE, JCMT, APEX and IRAM\,30m. Finally, in Fig.~\ref{f:seds} we show the SEDs for the 14 objects, with the observational parameters listed in Table~\ref{t:IRfluxes} and the best fit as explained in Sect.~\ref{s:resSEDs}.

%----------------------------------------------------------------------------
\begin{table*}[ht!]
\caption{\label{t:densegas}Hyperfine structure fit results and physical parameters of dense cores derived from N$_2$H$^+$\,(1--0) and C$_2$H\,(1--0) lines}
\centering
\begin{tabular}{c c c c c c c c c c c c c c c}
\hline\hline

&
&$V_\mathrm{LSR}$\supb
&$\Delta V$\supb
&
&$T_\mathrm{ex}$\supc
&$N_\mathrm{mol}$\supc
&size
&$M_\mathrm{gas}$\supd
\\
\texttt{ID}
&$A\times\tau$\supa
&(\kms)
&(\kms)
&$\tau$\supa
&(K)
&(cm$^{-2}$)
&(\arcsec)
&(\mo)
\\
\hline
\multicolumn{7}{l}{N$_2$H$^+$\,(1--0)} \\
\hline
01 &$1.28\pm0.01$ & \phn$+33.01\pm0.01$ &$2.12\pm0.01$ &$0.17\pm0.01$ &10.68    & \phn$3.25\times10^{12}$ &38 &\phn320	\\%&\phn$0.97\times10^{-7}$	\\
02 &$1.21\pm0.02$ & \phn$+26.88\pm0.01$ &$2.64\pm0.02$ &$0.31\pm0.02$ &\phn6.94 & \phn$3.45\times10^{12}$ &40 &\phn250	\\%&\phn$0.74\times10^{-7}$	\\
03 &$1.72\pm0.01$ & \phn$+26.56\pm0.01$ &$2.79\pm0.01$ &$0.29\pm0.01$ &\phn9.13 & \phn$5.42\times10^{12}$ &33 &	\phn270	\\%&	\phn$0.80\times10^{-7}$	\\
04 &$1.83\pm0.03$ & \phn$+26.50\pm0.01$ &$2.54\pm0.02$ &$0.19\pm0.02$ &13.00    & \phn$6.08\times10^{12}$ &42 &\phn520	\\%&\phn$1.57\times10^{-7}$	\\
05 &$2.00\pm0.01$ & \phn$+43.56\pm0.01$ &$2.80\pm0.01$ &$0.10\pm0.01$ &23.19    & $10.70\times10^{12}$    &31 &\phn990	\\%&\phn$2.98\times10^{-7}$	\\
06 &$1.35\pm0.05$ & $+106.58\pm0.02$    &$3.54\pm0.04$ &$0.31\pm0.04$ &\phn7.43 & \phn$5.20\times10^{12}$ &21 &2010		\\%&\phn$2.01\times10^{-7}$	\\
07 &$1.88\pm0.01$ & \phn$+42.60\pm0.01$ &$3.28\pm0.01$ &$0.23\pm0.01$ &11.45    & \phn$7.59\times10^{12}$ &58 &\phn670	\\%&\phn$5.79\times10^{-7}$	\\
08 &$1.13\pm0.01$ & \phn$+53.48\pm0.01$ &$2.94\pm0.03$ &$0.29\pm0.01$ &\phn6.99 & \phn$3.60\times10^{12}$ &45 &\phn870	\\%&\phn$2.61\times10^{-7}$	\\
09 &$2.44\pm0.02$ & $+113.64\pm0.01$    &$3.33\pm0.01$ &$0.64\pm0.01$ &\phn6.83 & \phn$8.80\times10^{12}$ &39 &4810		\\%&$14.44\times10^{-7}$		\\
10 &$3.89\pm0.01$ & \phn$+57.85\pm0.01$ &$3.03\pm0.01$ &$0.12\pm0.01$ &36.28    & $32.22\times10^{12}$    &40 &5460		\\%&$16.37\times10^{-7}$		\\
11 &$3.45\pm0.02$ & \phn$+57.50\pm0.01$ &$3.36\pm0.01$ &$0.18\pm0.01$ &22.50    & $21.68\times10^{12}$    &43 &4330		\\%&$12.98\times10^{-7}$		\\
12 &$1.72\pm0.01$ & \phn$+59.20\pm0.01$ &$2.71\pm0.01$ &$0.10\pm0.01$ &20.39 	  & \phn$8.10\times10^{12}$ &44 &1700		\\%&\phn$5.11\times10^{-7}$	\\
13 &$0.30\pm0.01$ & \phn$+43.94\pm0.05$ &$4.64\pm0.09$ &$0.10\pm0.03$ &\phn5.98 & \phn$1.52\times10^{12}$ &29 &\phn110	\\%&\phn$0.33\times10^{-7}$	\\
14 &$0.54\pm0.01$ & \phn$-44.51\pm0.01$ &$2.36\pm0.01$ &$0.10\pm0.01$ &\phn8.48 & \phn$1.41\times10^{12}$ &24 &\phn150	\\%&\phn$0.45\times10^{-7}$	\\
\hline
\multicolumn{7}{l}{C$_2$H\,(1--0)} \\
\hline
01 &$1.62\pm0.01$ & \phn$+32.97\pm0.80$ &$2.69\pm2.66$ &$0.10\pm0.10$ &19.37    & $17.20\times10^{13}$    &58 &3650		\\%&$10.95\times10^{-6}$		\\
02 &$0.39\pm0.05$ & \phn$+26.79\pm0.04$ &$2.56\pm0.14$ &$0.46\pm0.30$ &\phn3.68 & \phn$4.09\times10^{13}$ &43 &\phn290	\\%&\phn$0.88\times10^{-6}$	\\
03 &$0.55\pm0.03$ & \phn$+26.35\pm0.02$ &$2.97\pm0.07$ &$1.05\pm0.15$ &\phn3.32 & \phn$9.31\times10^{13}$ &41 &\phn630	\\%&\phn$1.89\times10^{-6}$	\\
04 &$1.21\pm0.10$ & \phn$+26.55\pm0.04$ &$3.14\pm0.10$ &$2.23\pm0.30$ &\phn3.34 & $21.10\times10^{13}$    &55 &2790		\\%&\phn$8.36\times10^{-6}$	\\
05 &$1.12\pm0.01$ & \phn$+43.83\pm0.01$ &$3.10\pm0.01$ &$0.10\pm0.01$ &14.30    & $11.30\times10^{13}$    &37 &1320		\\%&\phn$3.95\times10^{-6}$	\\
06 &$0.60\pm0.07$ & $+106.35\pm0.06$	&$3.85\pm0.18$ &$0.14\pm0.38$ &\phn7.34 & \phn$5.83\times10^{13}$    &48 &3620		\\%&$10.86\times10^{-6}$		\\
07 &$0.81\pm0.01$ & \phn$+42.41\pm0.01$ &$4.35\pm0.02$ &$0.10\pm0.01$ &11.24    & $10.10\times10^{13}$    &51 &1740		\\%&\phn$5.22\times10^{-6}$	\\
08 &$0.57\pm0.07$ & \phn$+53.57\pm0.04$ &$2.68\pm0.15$ &$0.61\pm0.31$ &\phn3.78 & \phn$5.94\times10^{13}$ &44 &1210		\\%&\phn$3.64\times10^{-6}$	\\
09 &$0.51\pm0.01$ & $+113.51\pm0.01$	&$3.69\pm0.06$ &$0.10\pm0.01$ &\phn8.03 & \phn$4.77\times10^{13}$    &35 &1770		\\%&\phn$5.31\times10^{-6}$	\\
10 &$1.25\pm0.05$ & \phn$+58.00\pm0.02$ &$3.34\pm0.06$ &$0.72\pm0.11$ &\phn4.65 & $12.10\times10^{13}$    &30 &1040		\\%&\phn$3.11\times10^{-6}$	\\
11 &$1.49\pm0.03$ & \phn$+57.62\pm0.01$ &$4.04\pm0.04$ &$1.36\pm0.07$ &\phn3.95 & $21.40\times10^{13}$    &42 &3620		\\%&$10.86\times10^{-6}$		\\
12 &$0.71\pm0.03$ & \phn$+59.23\pm0.02$ &$3.02\pm0.06$ &$0.32\pm0.10$ &\phn5.17 & \phn$5.75\times10^{13}$ &43 &1000		\\%&\phn$3.00\times10^{-6}$	\\
13 &$0.64\pm0.02$ & \phn$+43.87\pm0.05$ &$5.58\pm0.12$ &$0.10\pm0.02$ &\phn9.42 & \phn$9.45\times10^{13}$ &31 &\phn660	\\%&\phn$1.98\times10^{-6}$	\\
14 &$0.76\pm0.01$ & \phn$-44.45\pm0.01$ &$2.96\pm0.03$ &$0.10\pm0.02$ &10.65    & \phn$6.25\times10^{13}$ &32 &1010		\\%&\phn$3.02\times10^{-6}$	\\
\hline
\end{tabular}
\begin{list}{}{}
\item[\supa] Parameters obtained from the CLASS fit with $A=f[J_\nu(T_\mathrm{ex})-J_\nu(T_\mathrm{bg})]$, and $\tau$ the optical depth of the main hyperfine, adopted to be the ($1_{2,3}$--$0_{1,2}$) line for the N$_2$H$^+$ and the ($1_{2,2}$--$0_{1,1}$) line for the C$_2$H.
\item[\supb] Systemic velocity, $V_\mathrm{lsr}$, and linewidth, $\Delta V$, both in \kms.
\item[\supc] Excitation temperature, $T_\mathrm{ex}$, in K, and molecular column density, $N_\mathrm{mol}$, in cm$^{-2}$.
\item[\supd] Mass of the gas calculated as $M_\mathrm{gas}=(N_\mathrm{mol}/X)\,\mu\,m_\mathrm{H}$\,Area, where Area$=\pi\,\mathrm{size}/2$, $m_\mathrm{H}$ is the mass of the hydrogen atom, $\mu$ is the molecular weight ($=16$ for N$_2$H$^+$ and $=14$ for C$_2$H), and $X$ is the abundance with respect to H$_2$ (fixed to $3\times10^{-10}$ for N$_2$H$^+$ and $3\times10^{-9}$ for C$_2$H; \eg\ \citealt{huggins1984, caselli2002b, beuther2008, padovani2009, busquet2011, frau2012}).
\end{list}
\end{table*}
%----------------------------------------------------------------------------

%\end{appendix}
%\end{document}
\clearpage
%----------------------------------------------------------------------
\begin{figure*}
\begin{center}
\begin{tabular}[b]{c c}
 \vspace{0.5cm}
 \epsfig{file=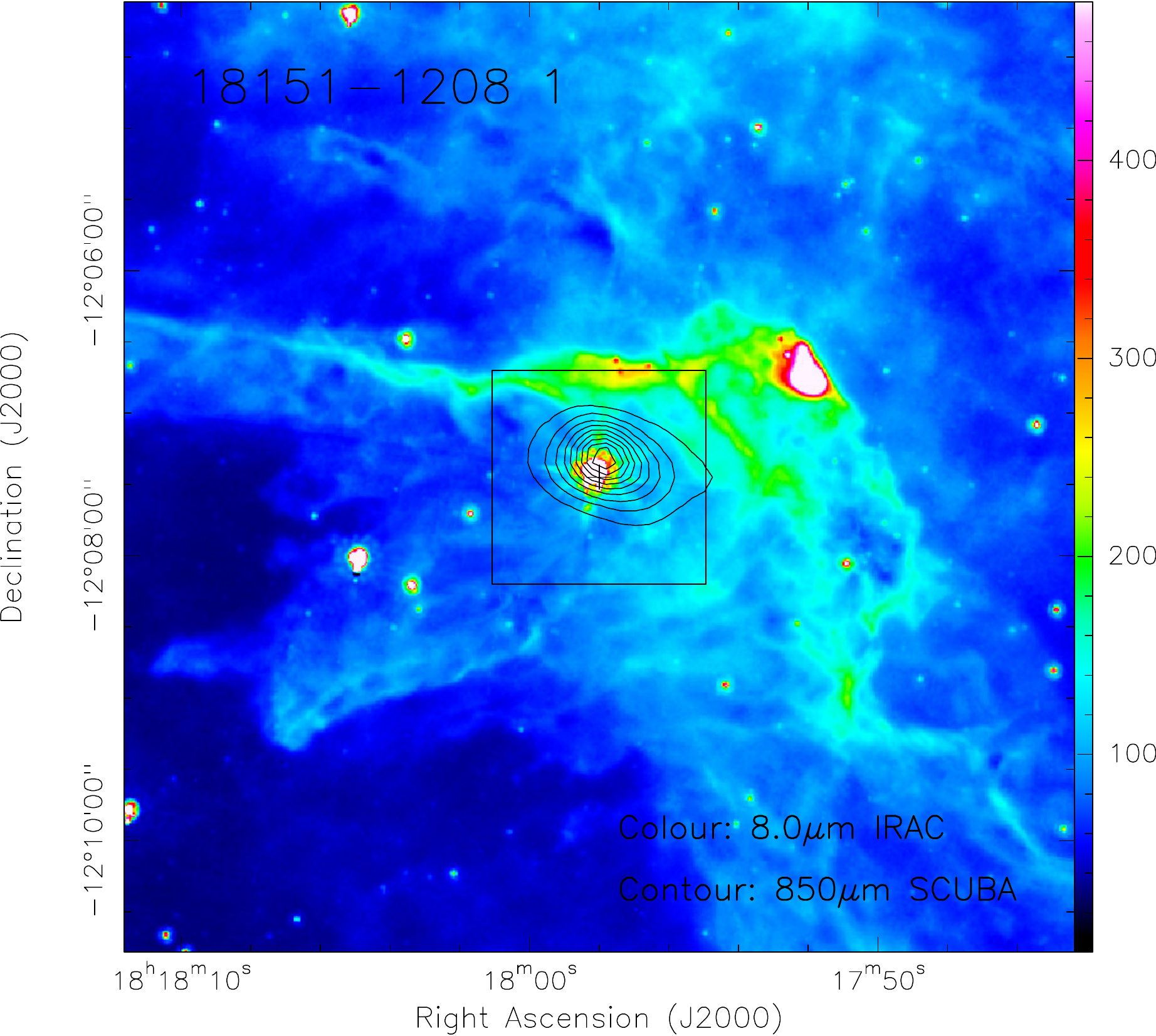, width=1.0\columnwidth, angle=0} &
 \epsfig{file=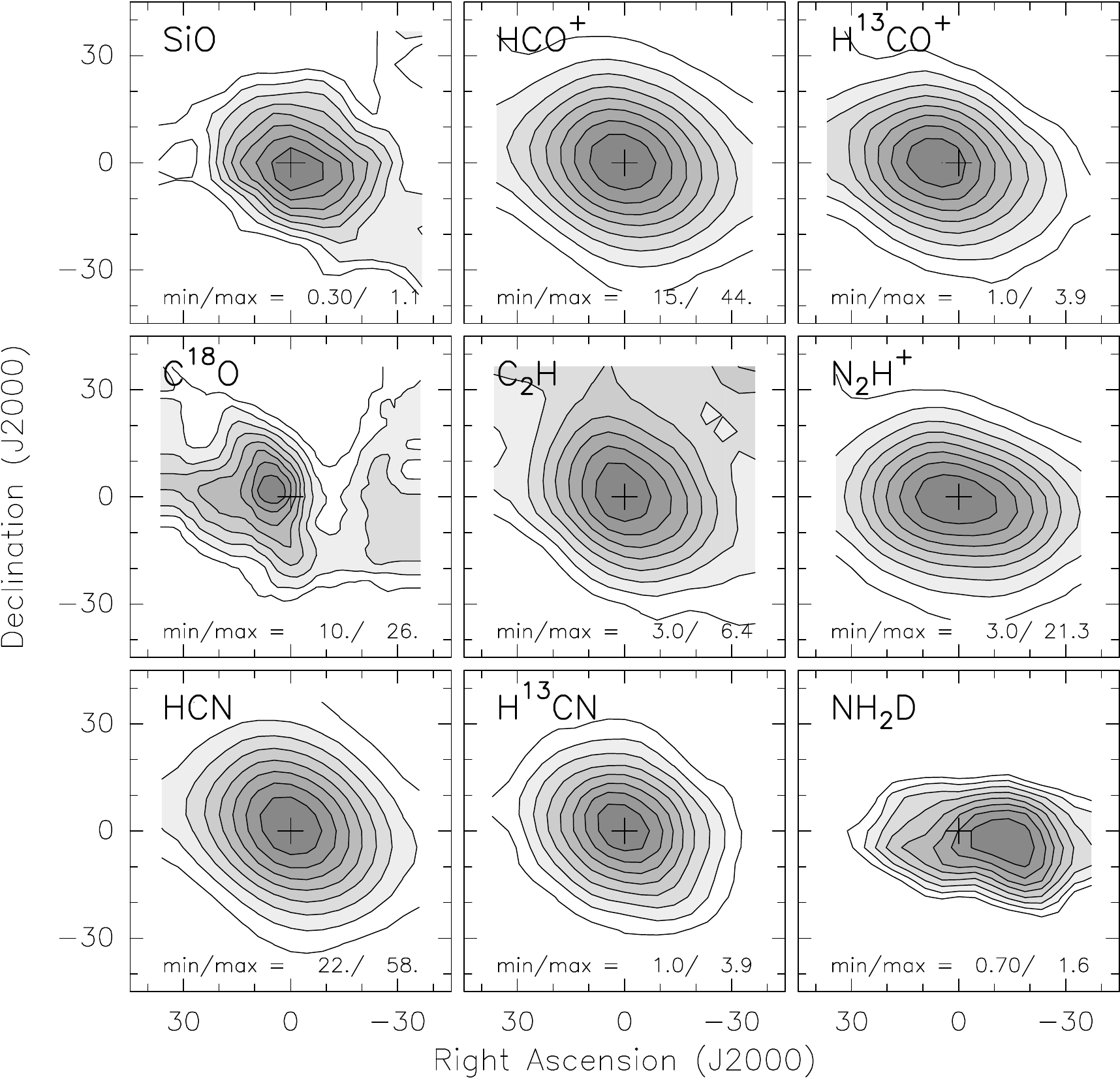, width=0.9\columnwidth, angle=0} \\
% \vspace{1cm}
 \multicolumn{2}{c}{\epsfig{file=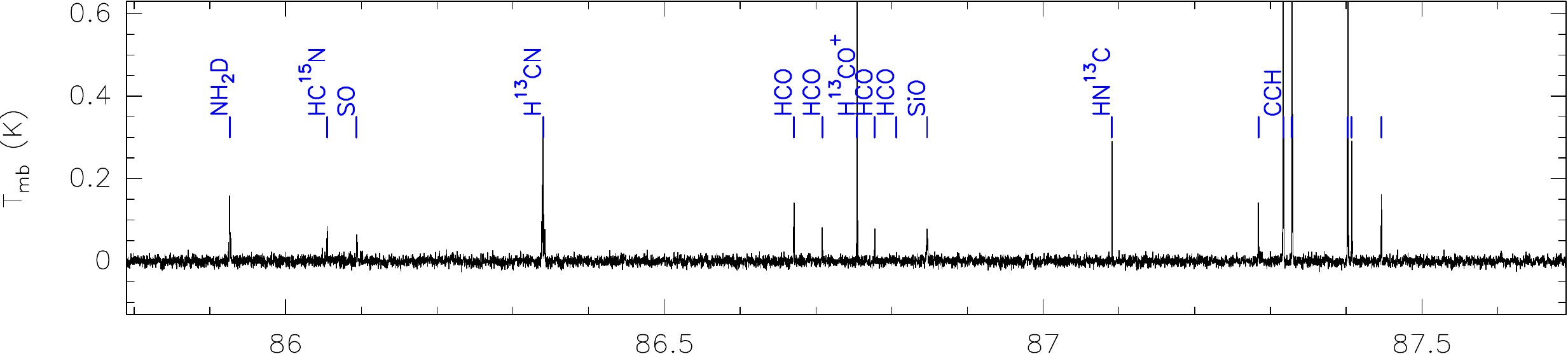, width=0.8\textwidth, angle=0}} \\
 \multicolumn{2}{c}{\epsfig{file=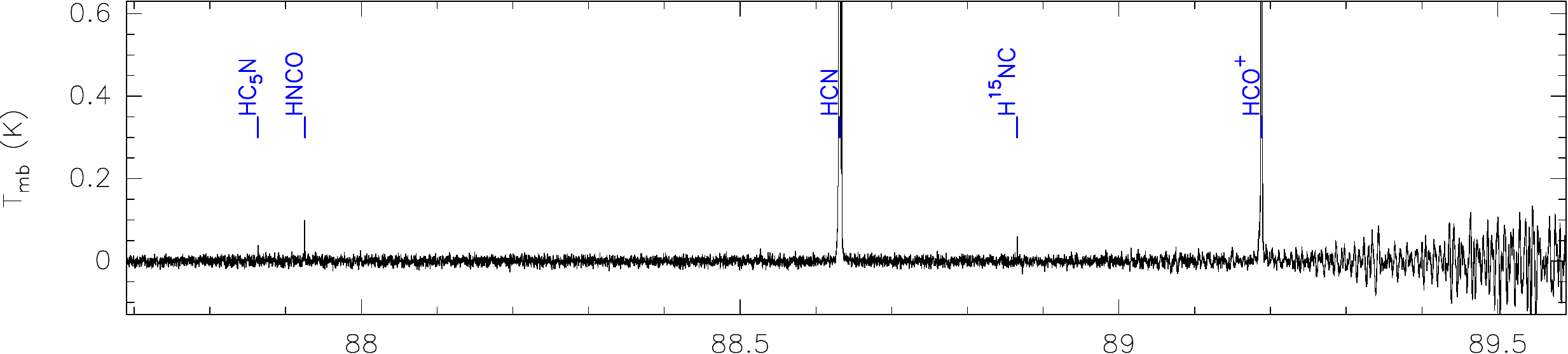, width=0.8\textwidth, angle=0}} \\
 \multicolumn{2}{c}{\epsfig{file=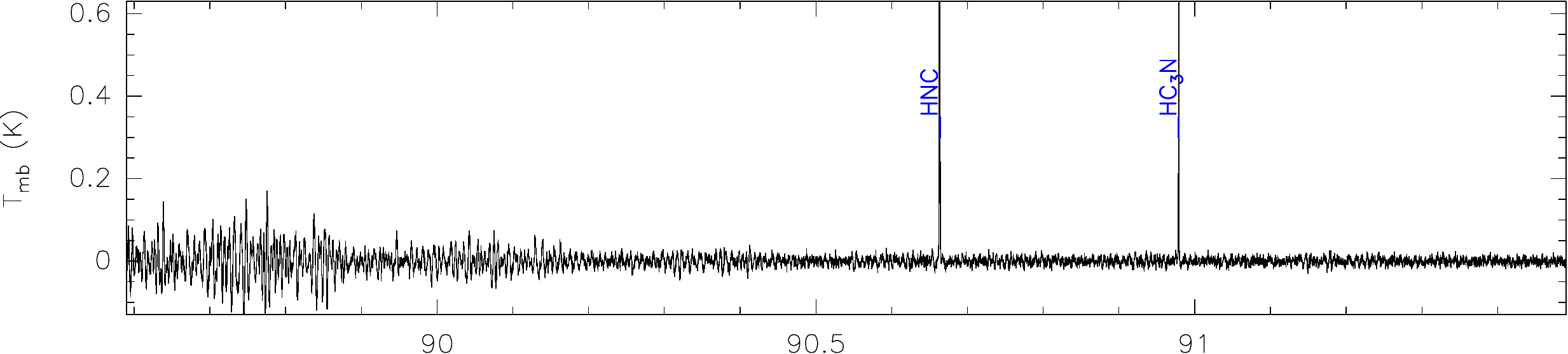, width=0.8\textwidth, angle=0}} \\
 \multicolumn{2}{c}{\epsfig{file=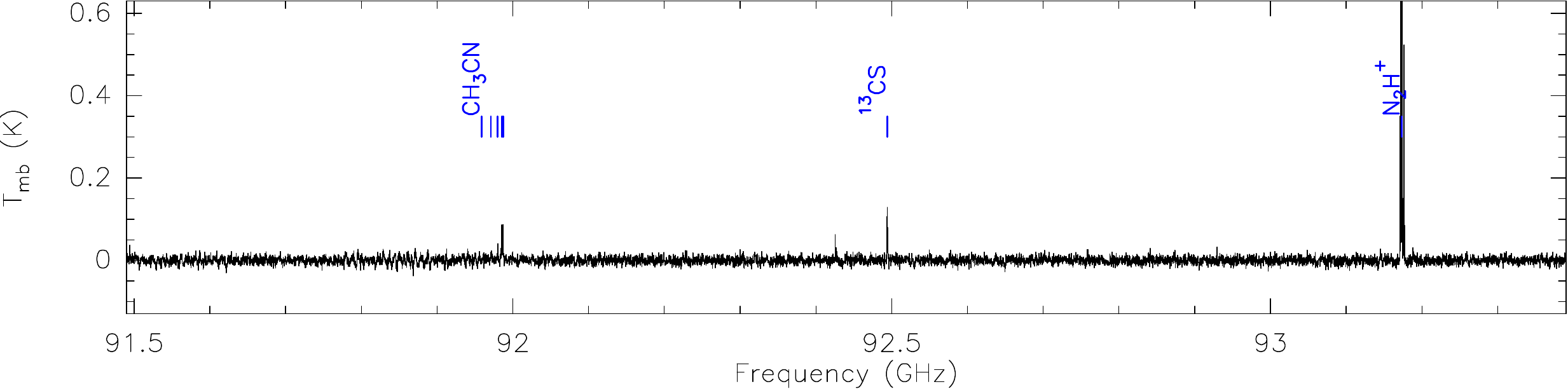, width=0.8\textwidth, angle=0}} \\
\end{tabular}
\end{center}
\caption{\label{f:summary}\emph{Top-left}: Hi-GAL 500~$\mu$m (or JCMT 850~$\mu$m) continuum image in contours overlaid on the \emph{Spitzer}/IRAC 8.0~$\mu$ (or MSX 8.3~$\mu$m) image in color scale. The box show the region mapped with the IRAM~30-m telescope. \emph{Top-right}: nine panels with the zero-order moment (velocity-integrated intensity) maps of different molecular species (labeled in the top-left corner of each panel). The minimum and maximum contour level (in units K~\kms) is indicated in the bottom part of each panel. \emph{Bottom}: four panels with the IRAM~30-m spectra at 3~mm, covering a frequency range from 87.6~GHz to 93.4~GHz.}
\end{figure*}
\begin{figure*}
\ContinuedFloat
\begin{center}
\begin{tabular}[b]{c c}
 \vspace{0.5cm}
 \epsfig{file=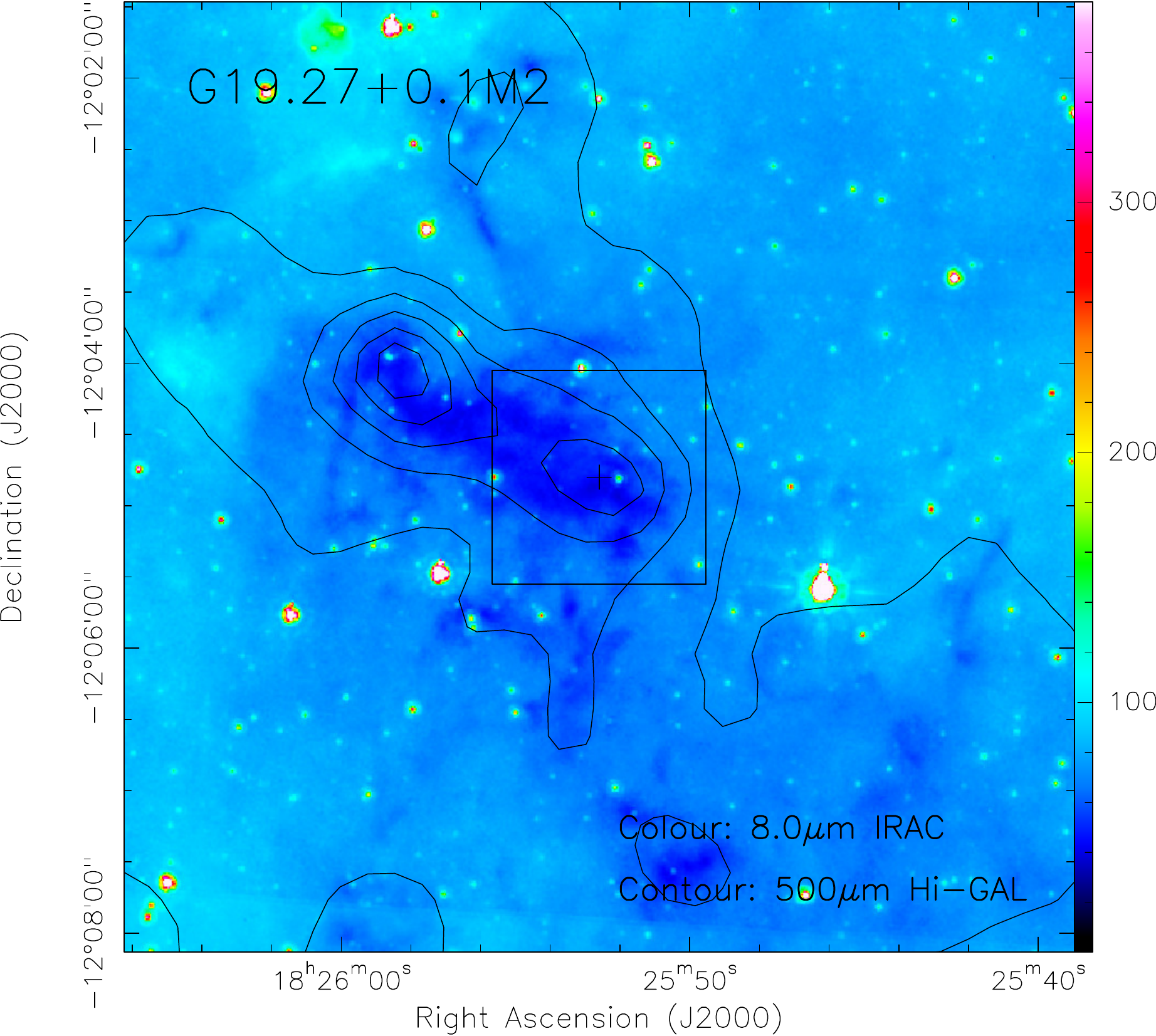, width=1.0\columnwidth, angle=0} &
 \epsfig{file=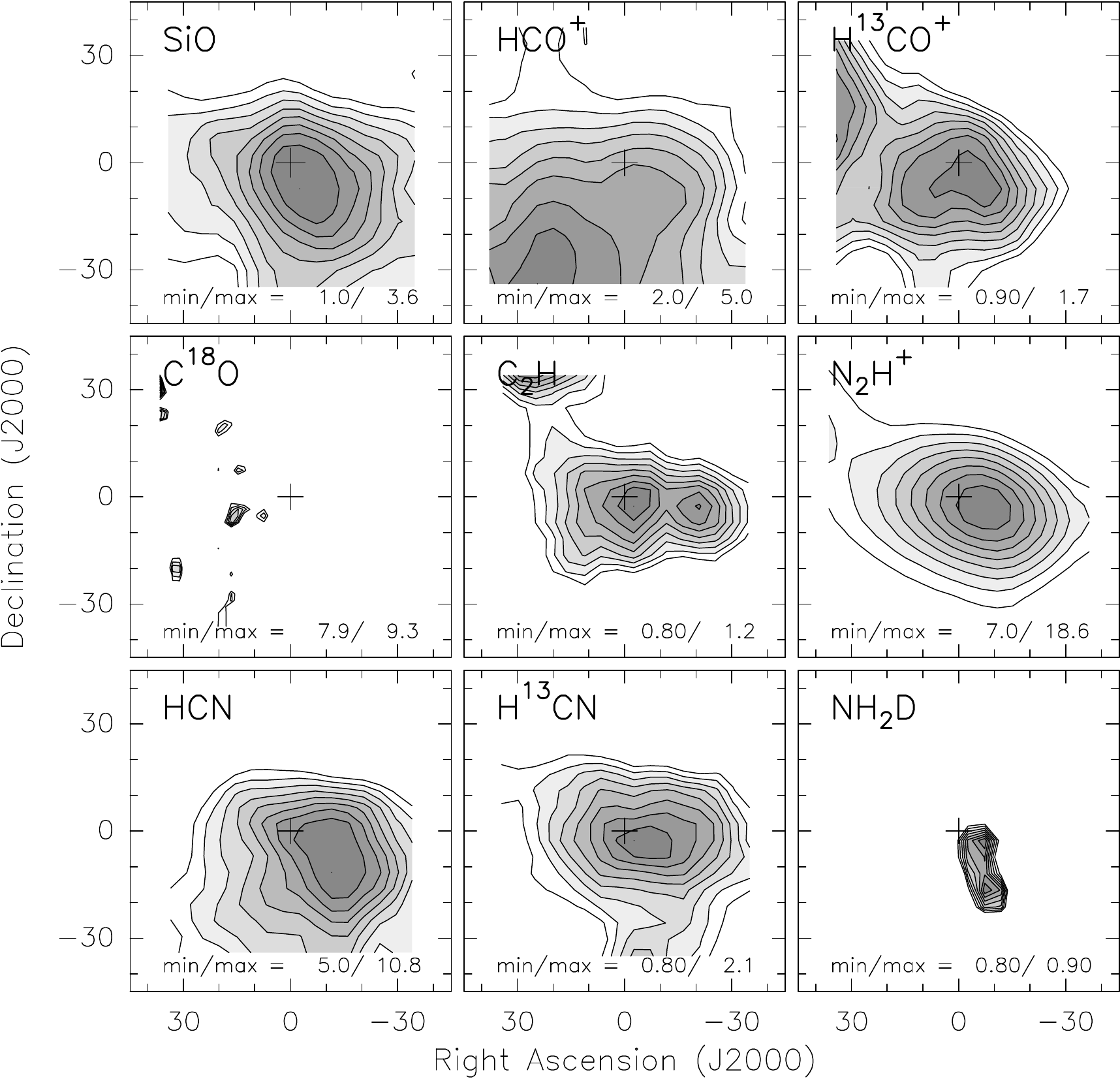, width=0.9\columnwidth, angle=0} \\
% \vspace{1cm}
 \multicolumn{2}{c}{\epsfig{file=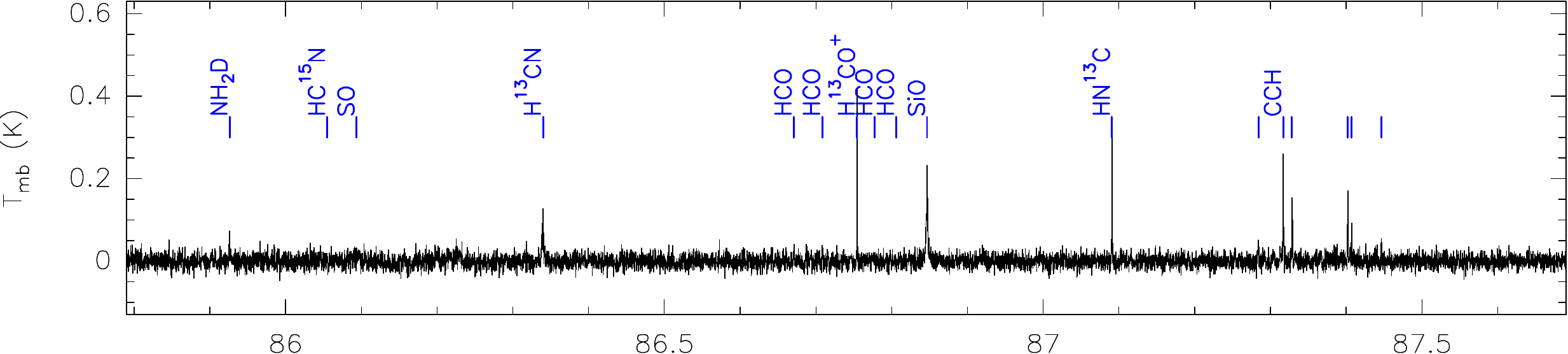, width=0.8\textwidth, angle=0}} \\
 \multicolumn{2}{c}{\epsfig{file=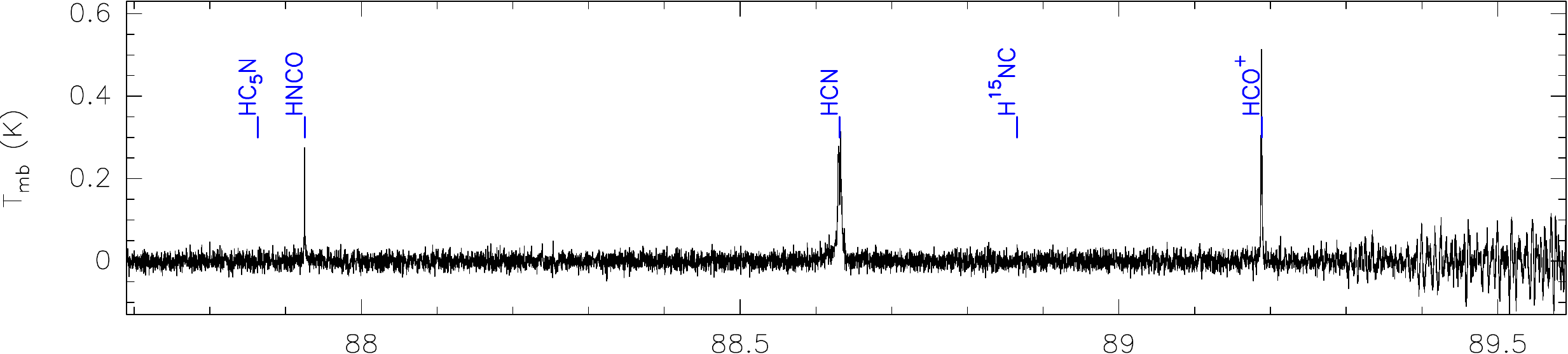, width=0.8\textwidth, angle=0}} \\
 \multicolumn{2}{c}{\epsfig{file=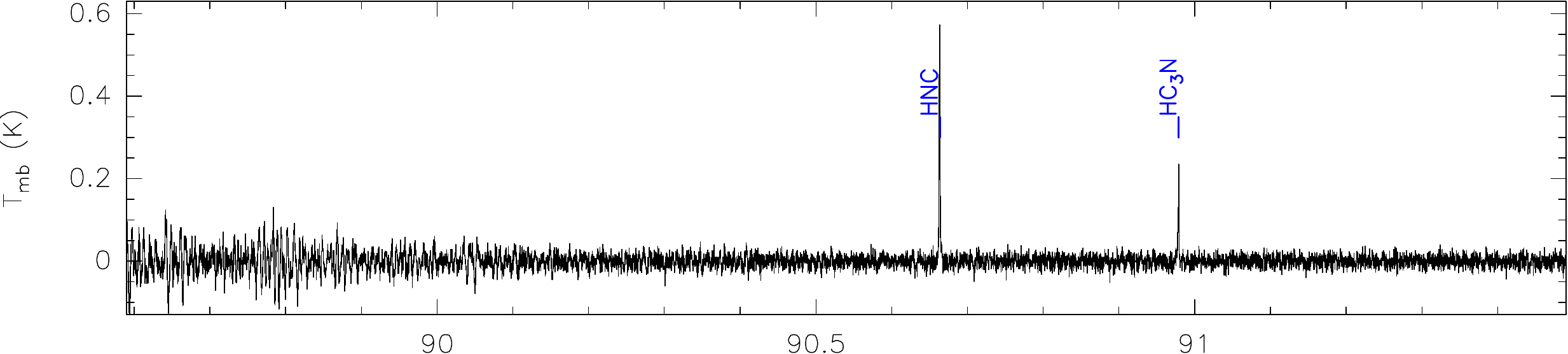, width=0.8\textwidth, angle=0}} \\
 \multicolumn{2}{c}{\epsfig{file=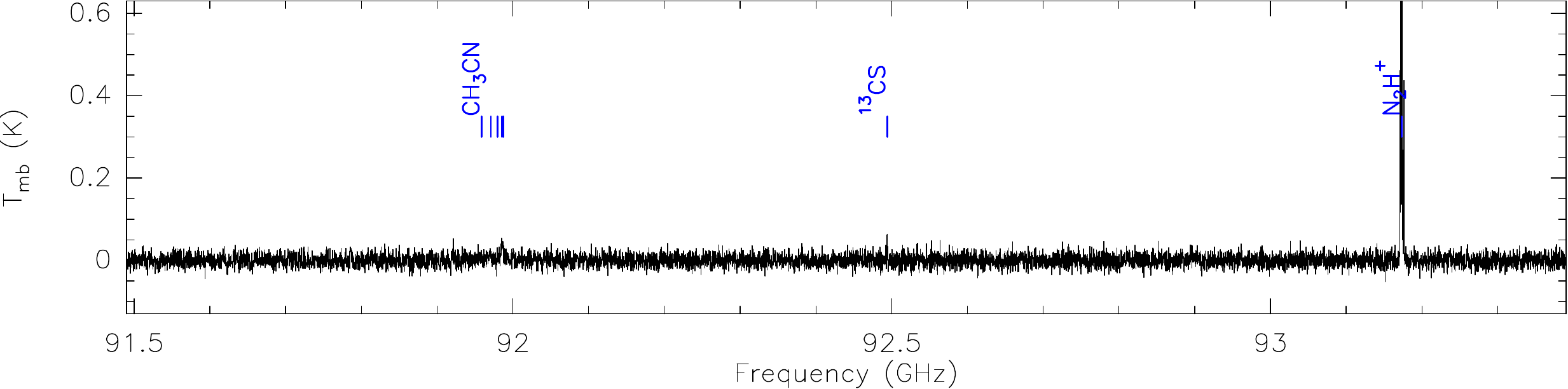, width=0.8\textwidth, angle=0}} \\
\end{tabular}
\end{center}
\caption{continued, for G19.27$+$0.1\,M2.}
\end{figure*}
\begin{figure*}
\ContinuedFloat
\begin{center}
\begin{tabular}[b]{c c}
 \vspace{0.5cm}
 \epsfig{file=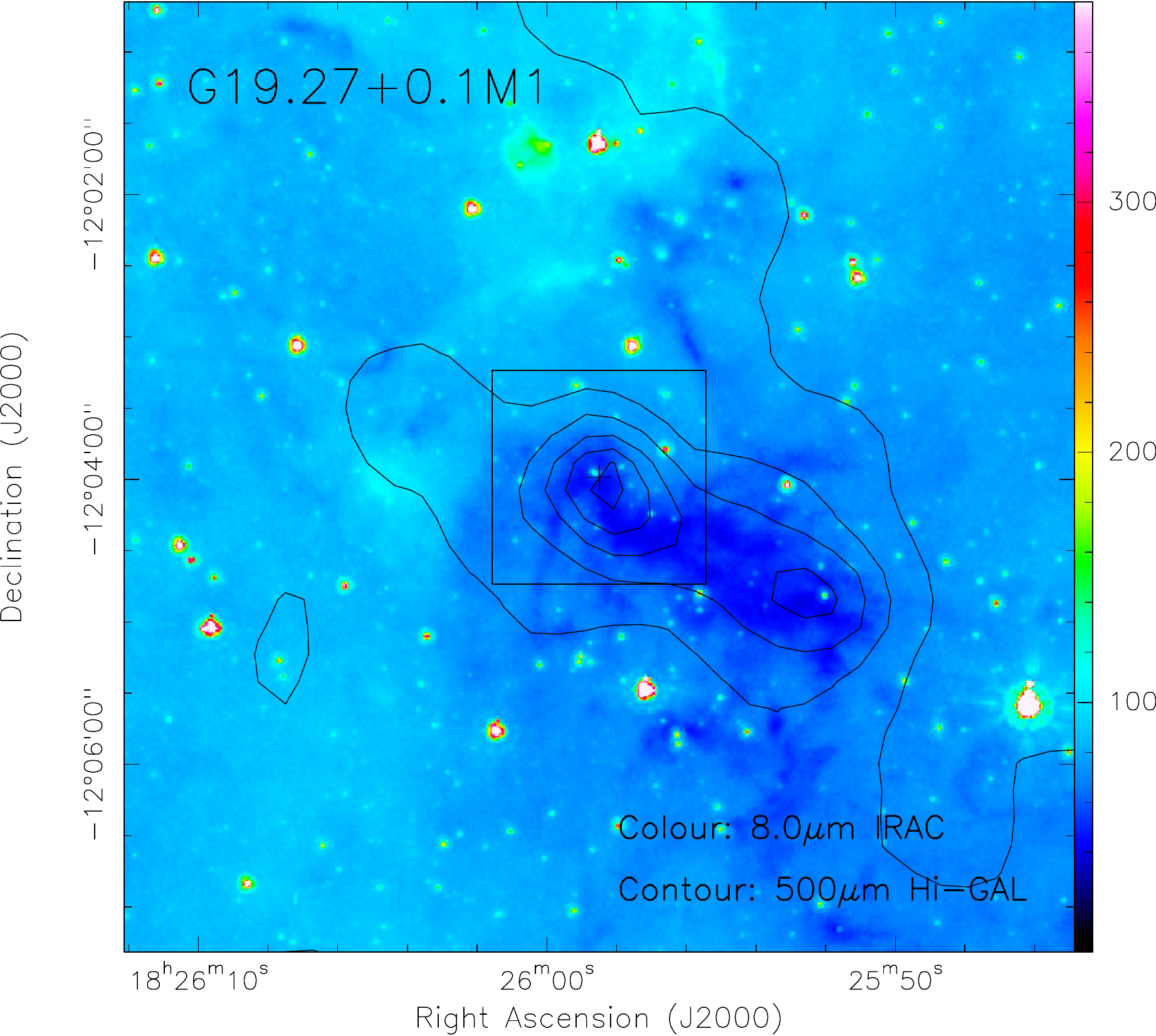, width=1.0\columnwidth, angle=0} &
 \epsfig{file=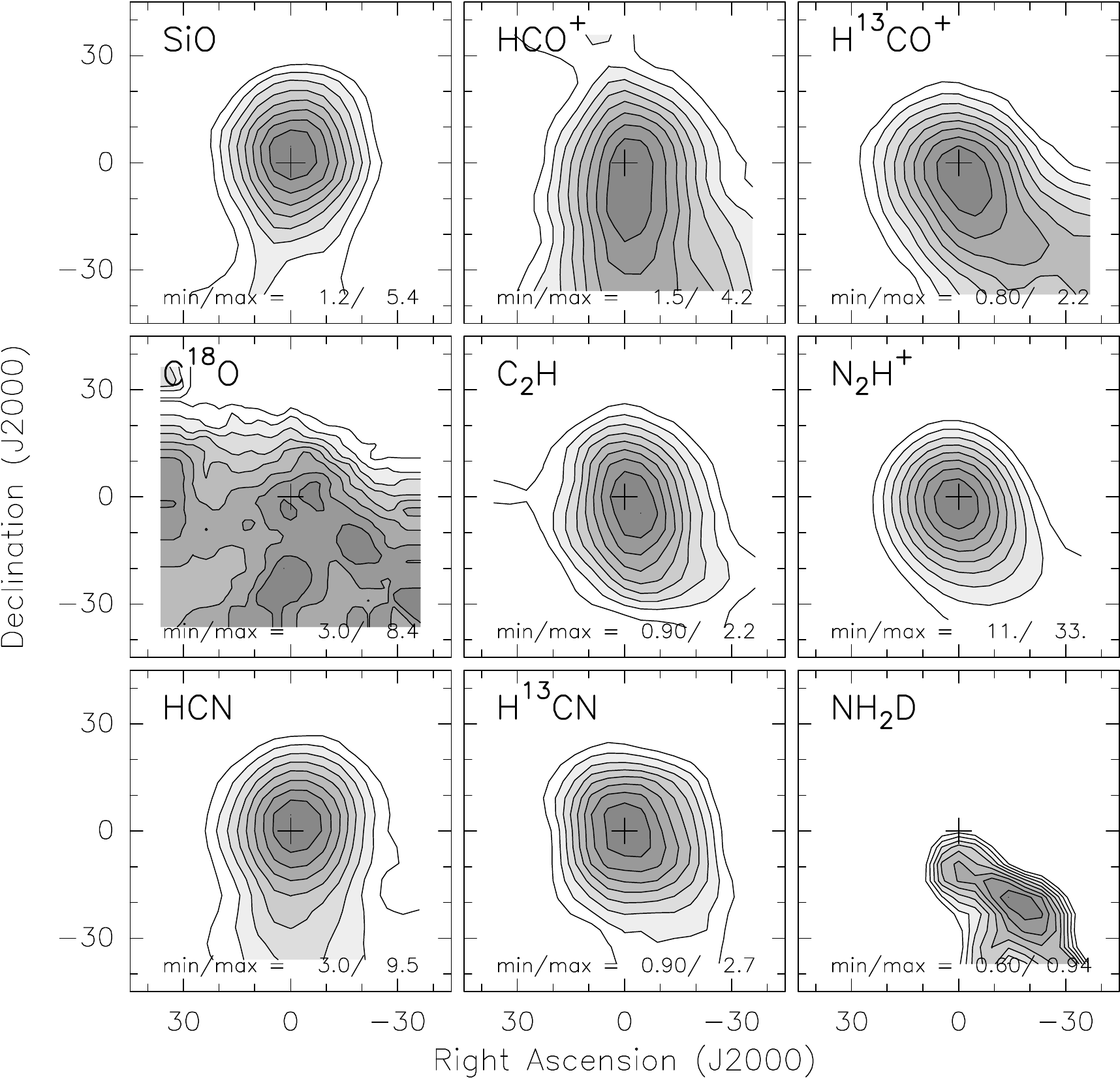, width=0.9\columnwidth, angle=0} \\
% \vspace{1cm}
 \multicolumn{2}{c}{\epsfig{file=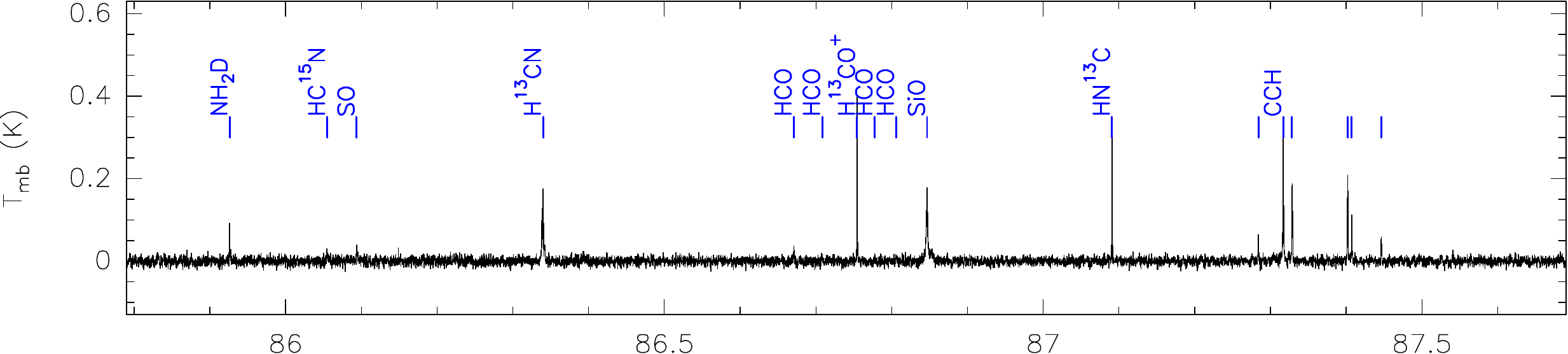, width=0.8\textwidth, angle=0}} \\
 \multicolumn{2}{c}{\epsfig{file=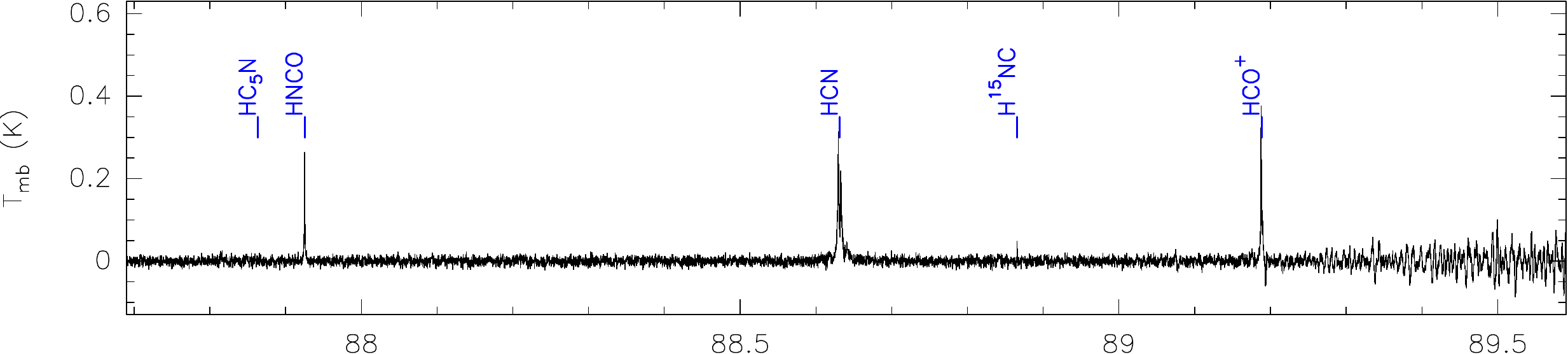, width=0.8\textwidth, angle=0}} \\
 \multicolumn{2}{c}{\epsfig{file=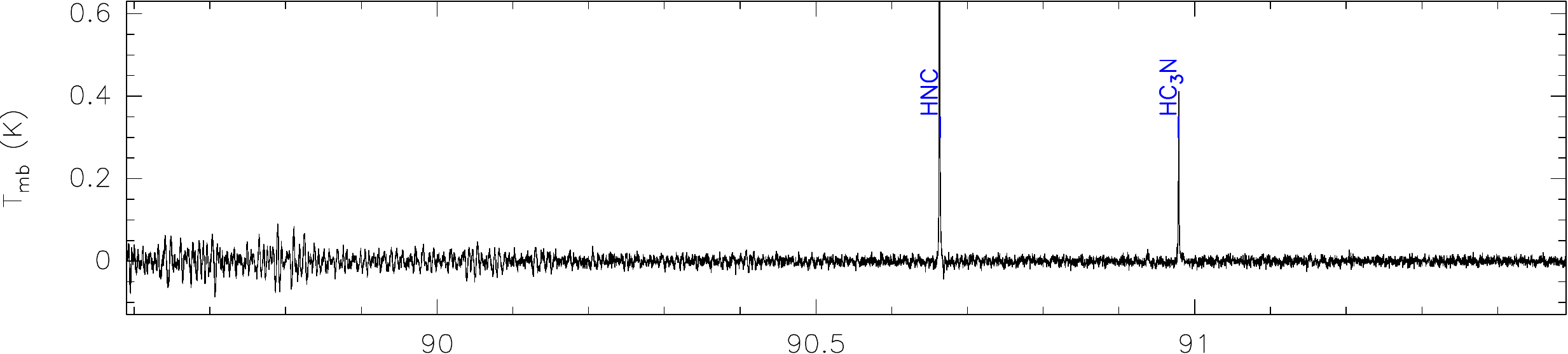, width=0.8\textwidth, angle=0}} \\
 \multicolumn{2}{c}{\epsfig{file=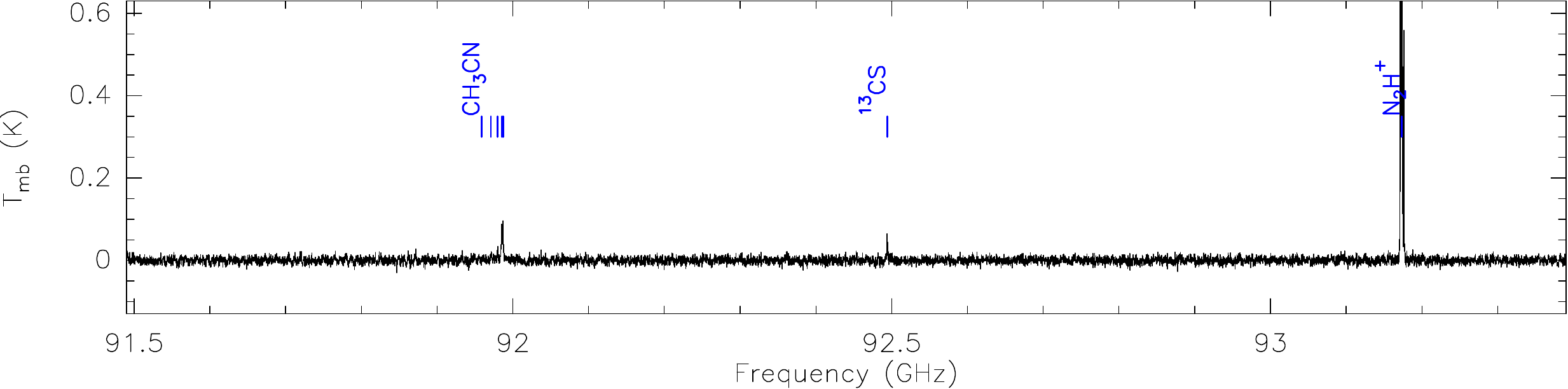, width=0.8\textwidth, angle=0}} \\
\end{tabular}
\end{center}
\caption{continued, for G19.27$+$0.1\,M1.}
\end{figure*}
\begin{figure*}
\ContinuedFloat
\begin{center}
\begin{tabular}[b]{c c}
 \vspace{0.5cm}
 \epsfig{file=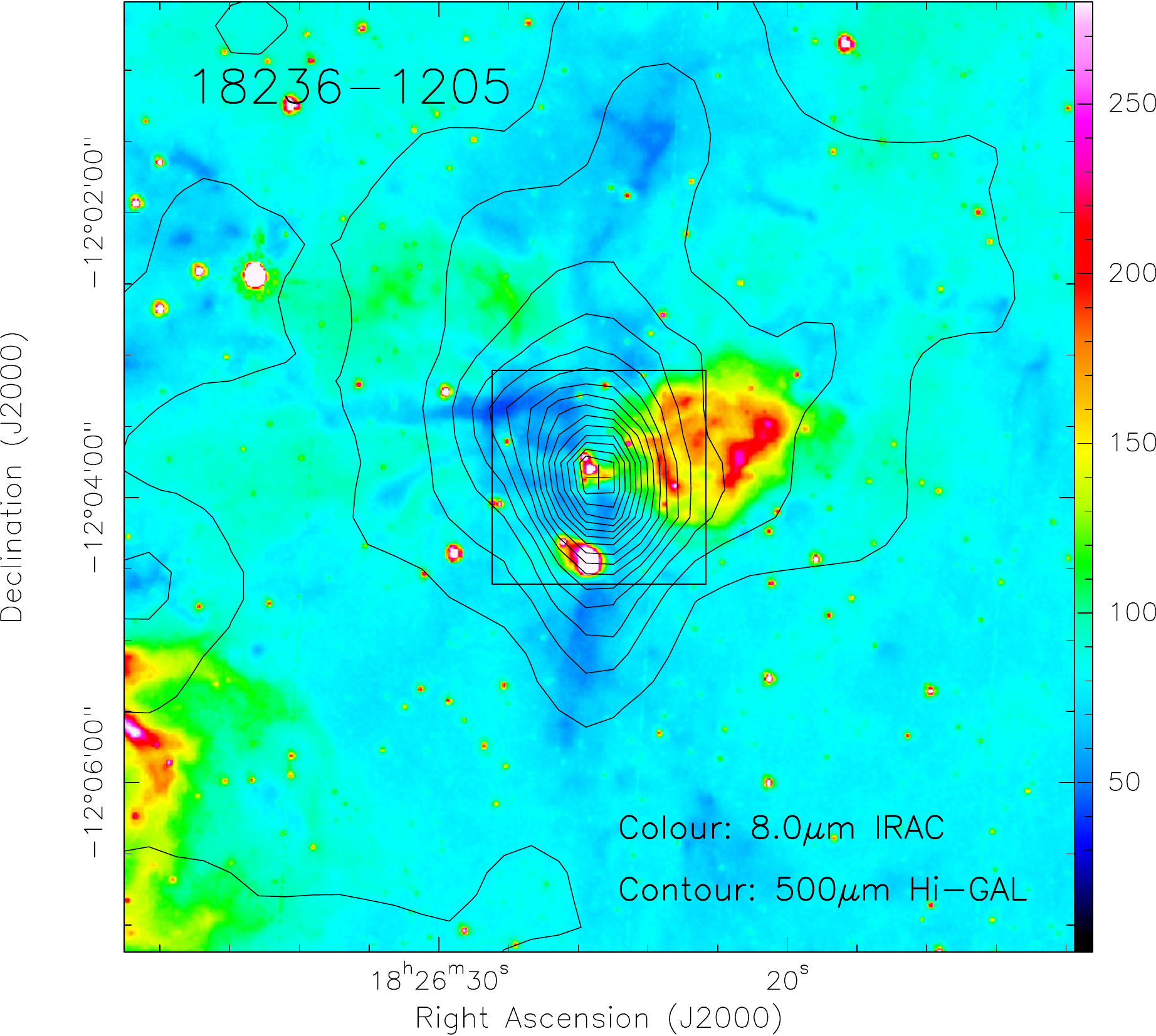, width=1.0\columnwidth, angle=0} &
 \epsfig{file=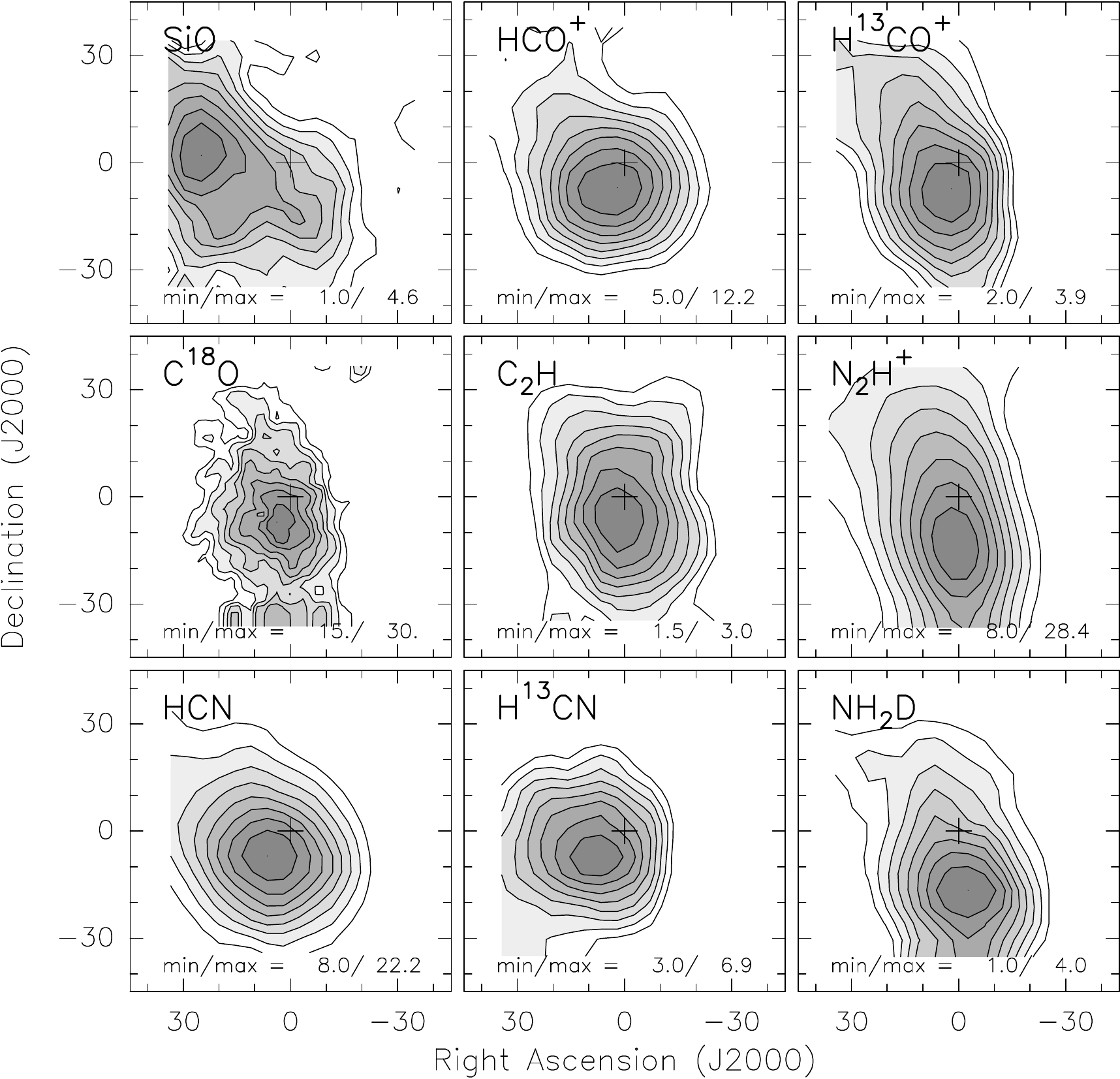, width=0.9\columnwidth, angle=0} \\
% \vspace{1cm}
 \multicolumn{2}{c}{\epsfig{file=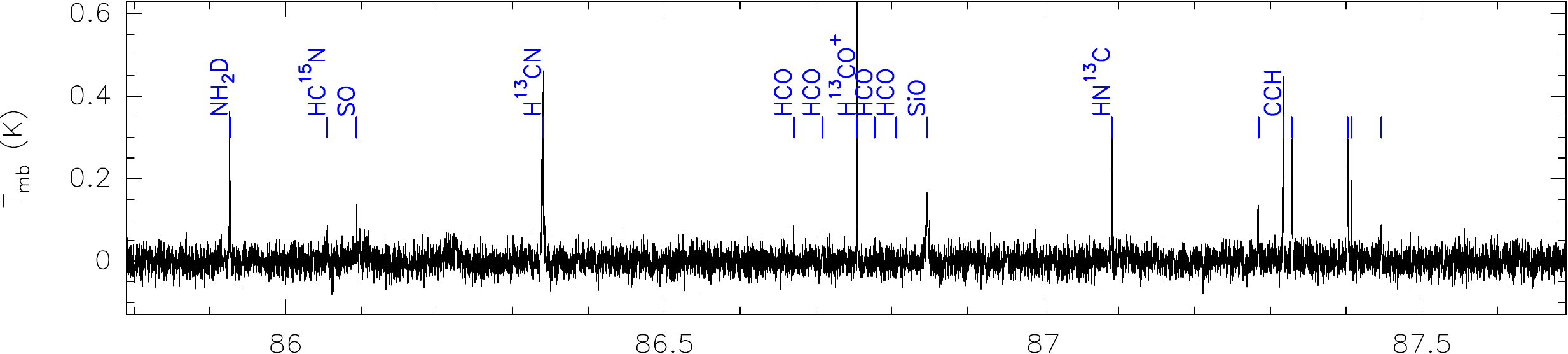, width=0.8\textwidth, angle=0}} \\
 \multicolumn{2}{c}{\epsfig{file=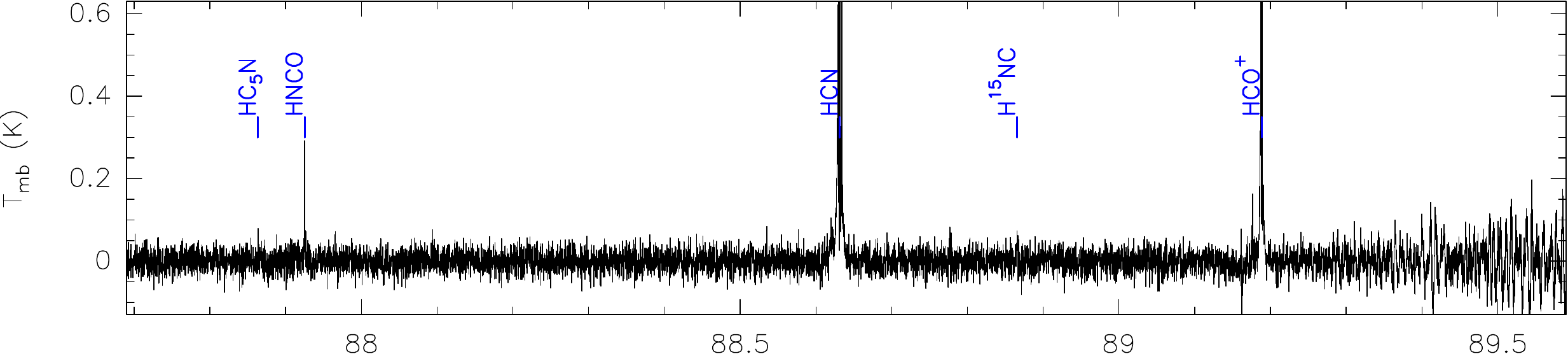, width=0.8\textwidth, angle=0}} \\
 \multicolumn{2}{c}{\epsfig{file=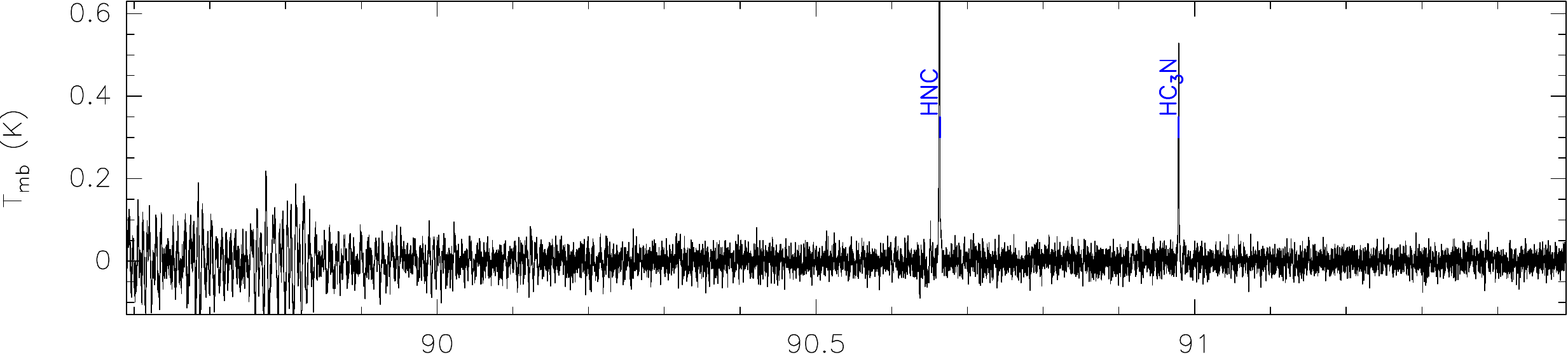, width=0.8\textwidth, angle=0}} \\
 \multicolumn{2}{c}{\epsfig{file=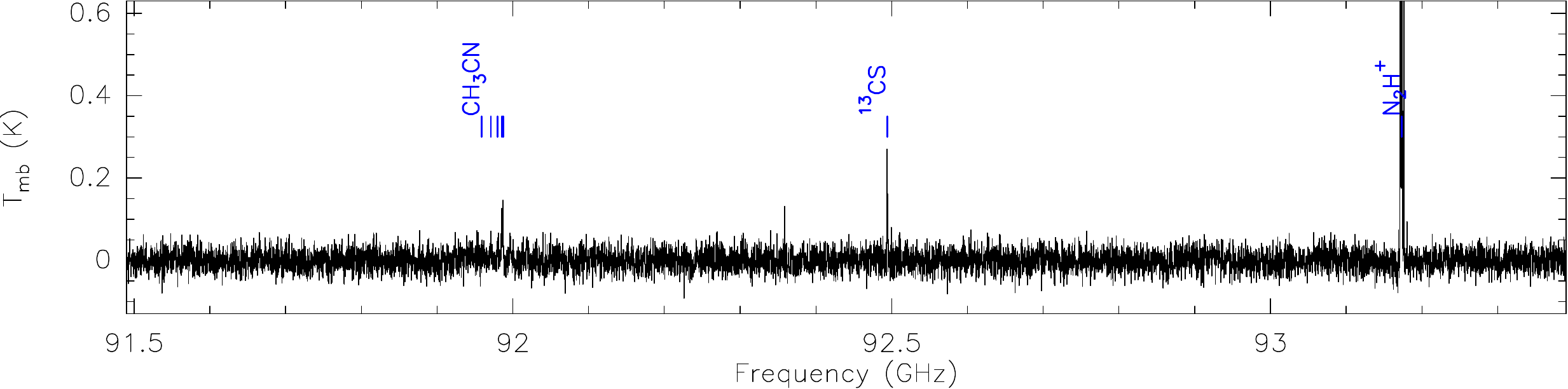, width=0.8\textwidth, angle=0}} \\
\end{tabular}
\end{center}
\caption{continued, for 18236$-$1205.}
\end{figure*}
\begin{figure*}
\ContinuedFloat
\begin{center}
\begin{tabular}[b]{c c}
 \vspace{0.5cm}
 \epsfig{file=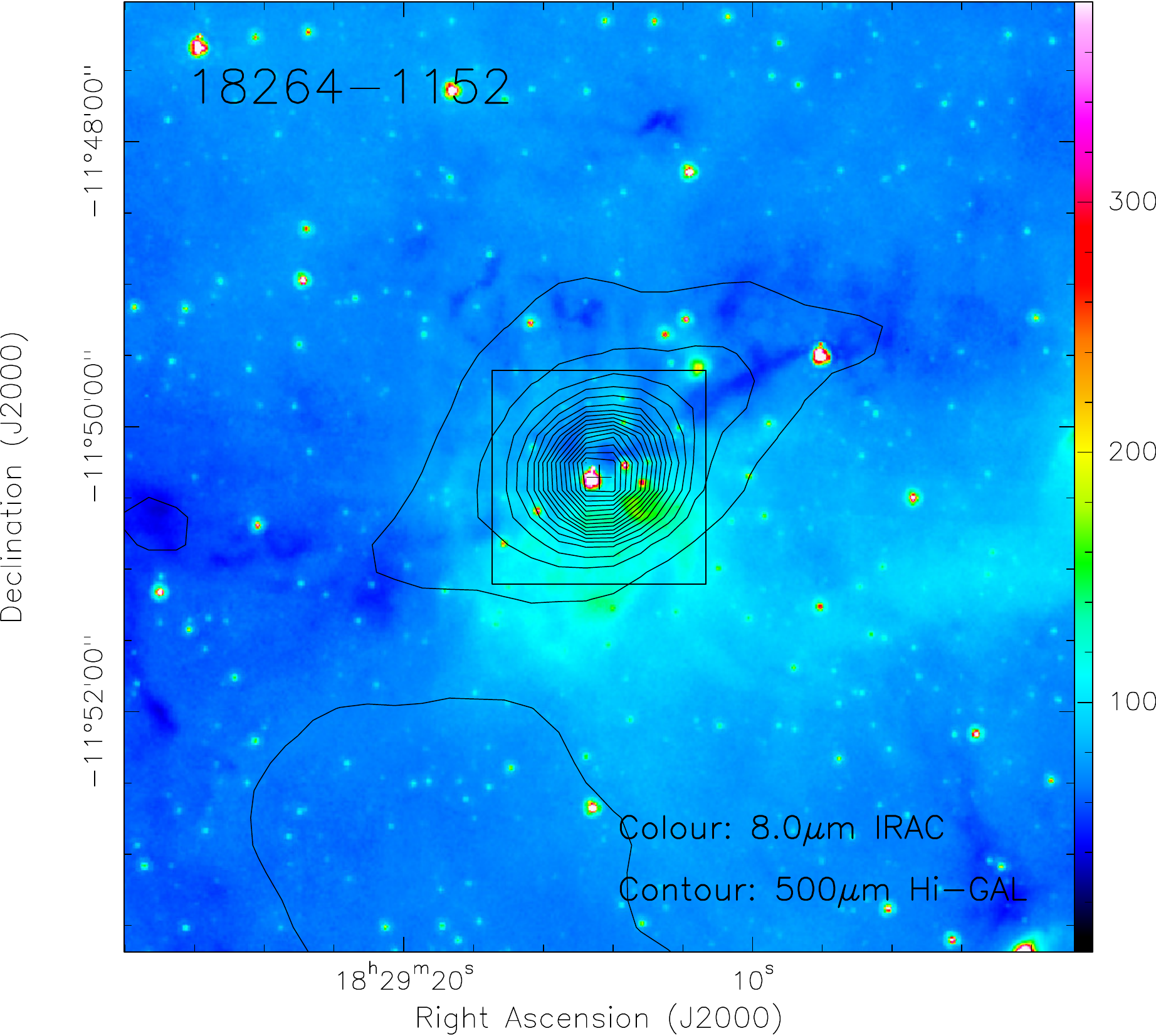, width=1.0\columnwidth, angle=0} &
 \epsfig{file=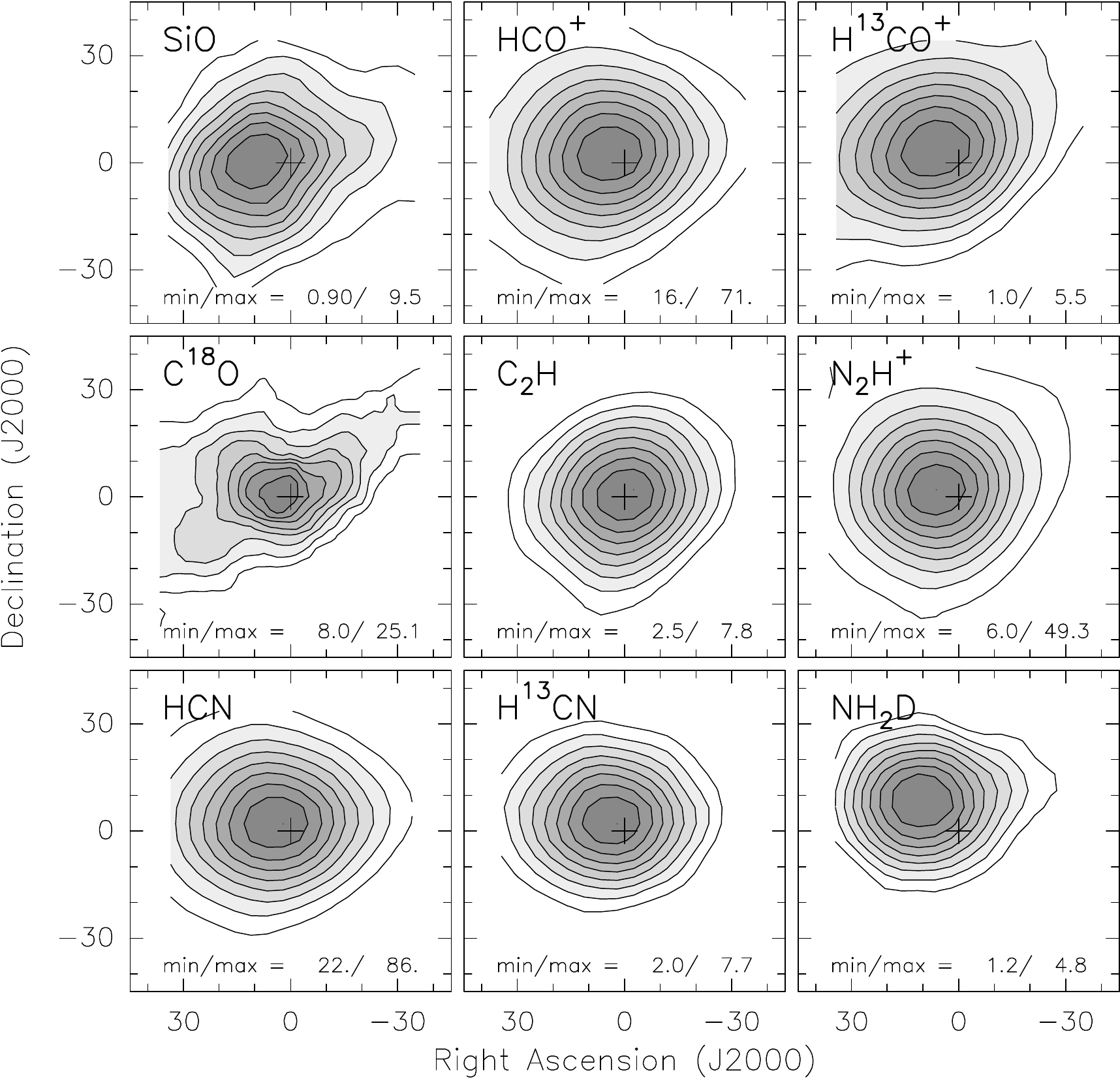, width=0.9\columnwidth, angle=0} \\
% \vspace{1cm}
 \multicolumn{2}{c}{\epsfig{file=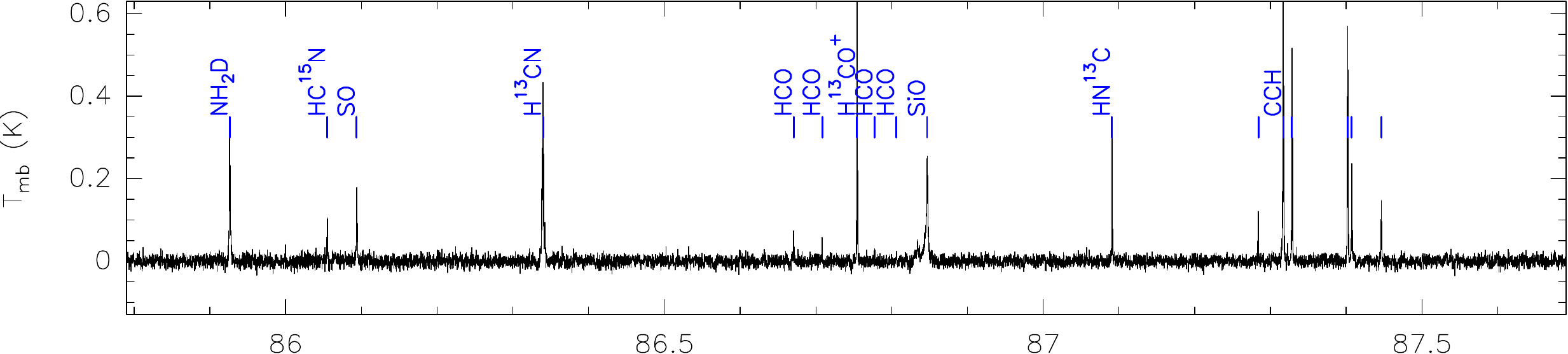, width=0.8\textwidth, angle=0}} \\
 \multicolumn{2}{c}{\epsfig{file=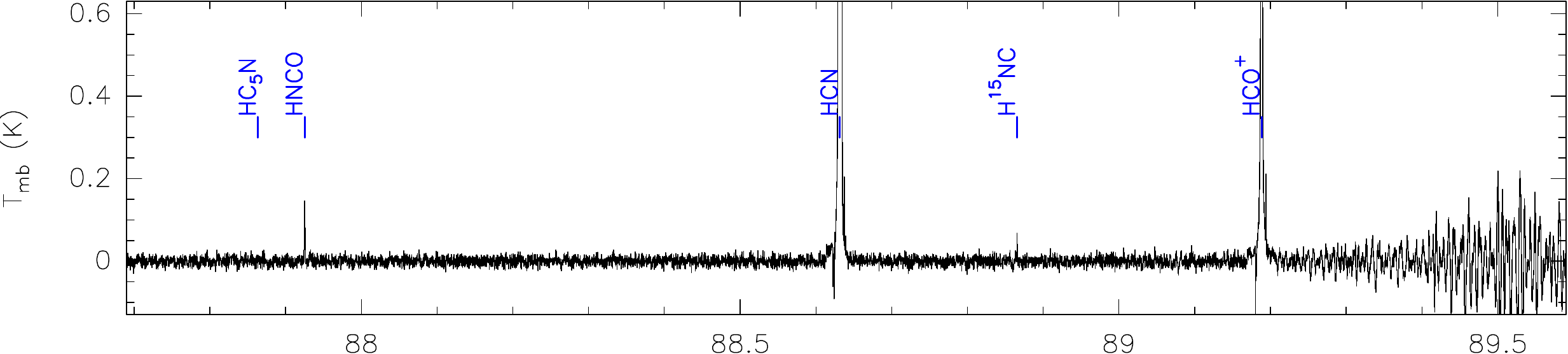, width=0.8\textwidth, angle=0}} \\
 \multicolumn{2}{c}{\epsfig{file=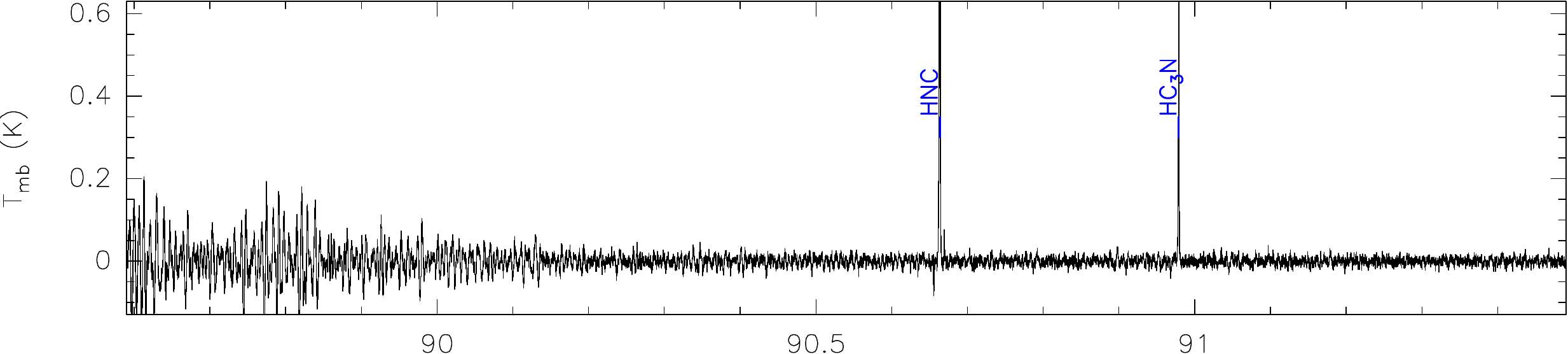, width=0.8\textwidth, angle=0}} \\
 \multicolumn{2}{c}{\epsfig{file=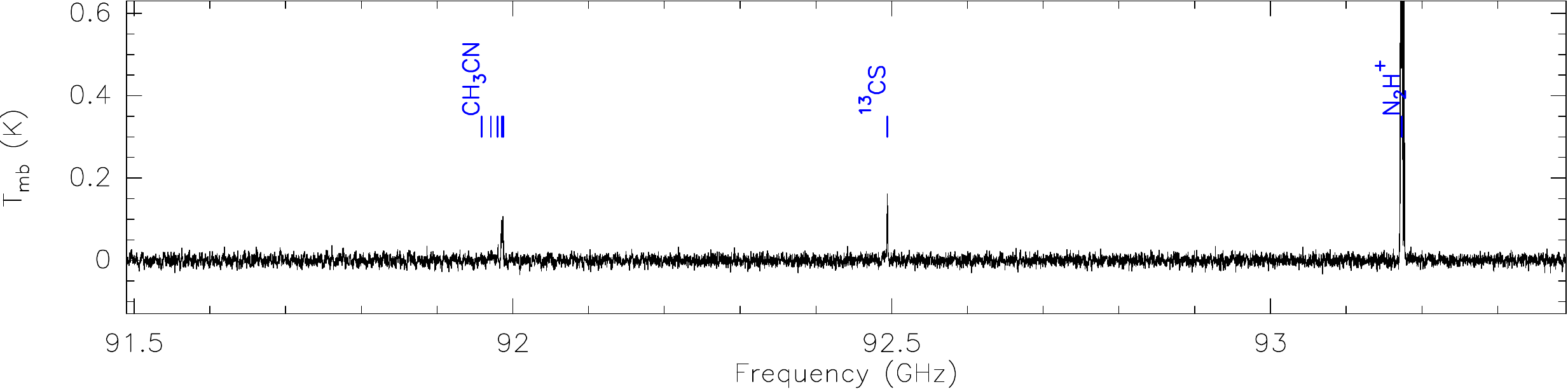, width=0.8\textwidth, angle=0}} \\
\end{tabular}
\end{center}
\caption{continued, for 18264$-$1152.}
\end{figure*}
\begin{figure*}
\ContinuedFloat
\begin{center}
\begin{tabular}[b]{c c}
 \vspace{0.5cm}
 \epsfig{file=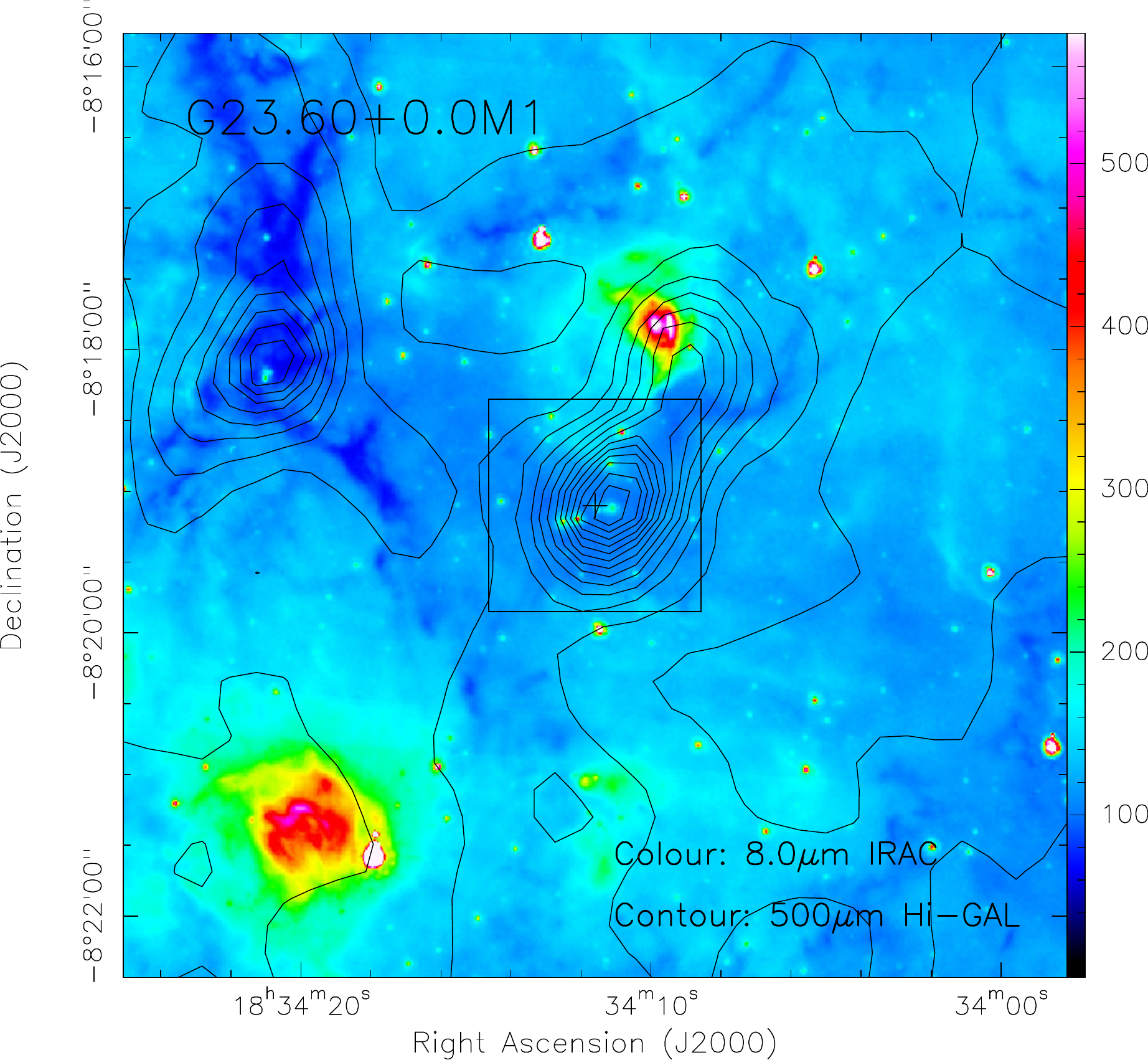, width=1.0\columnwidth, angle=0} &
 \epsfig{file=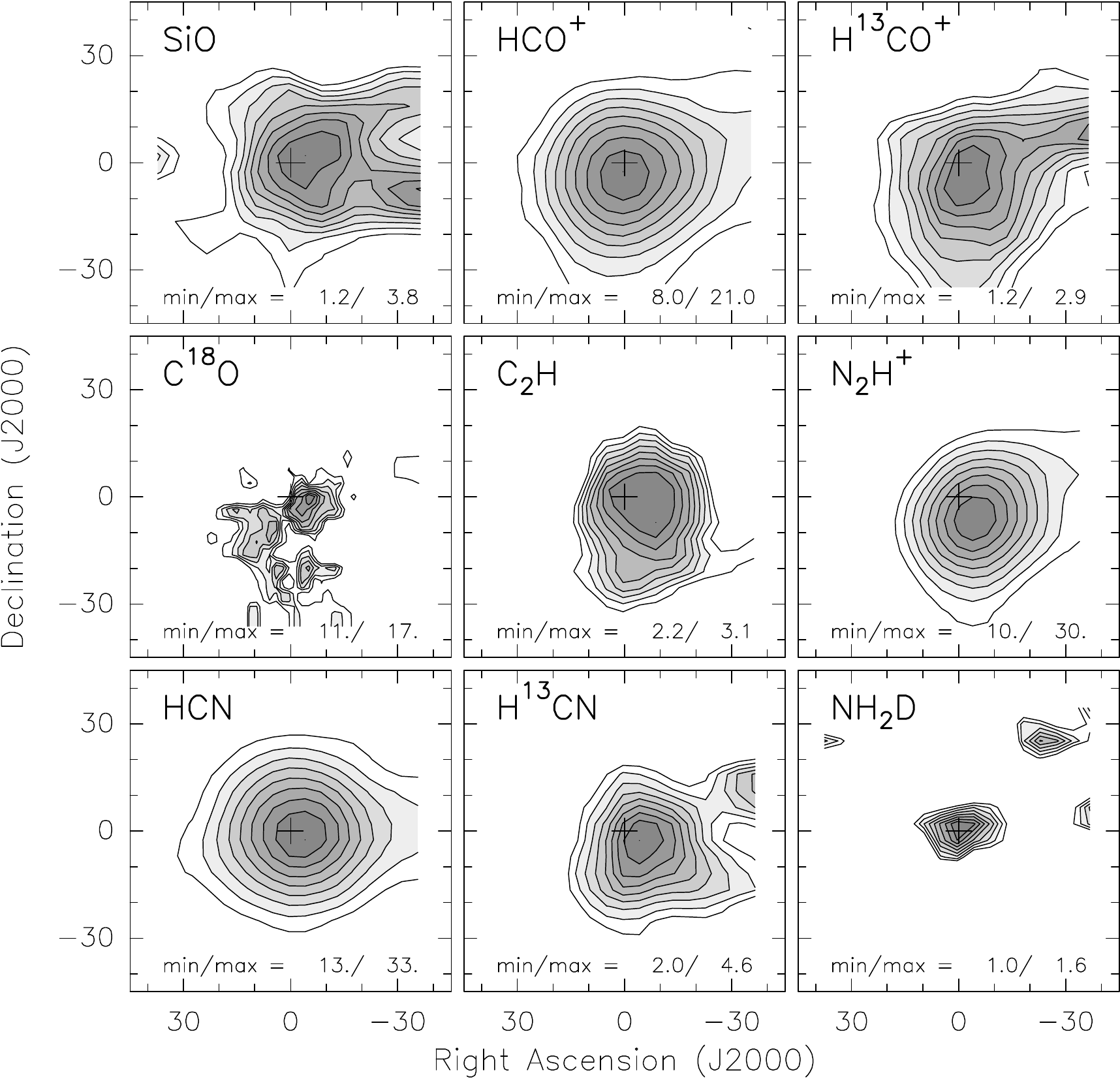, width=0.9\columnwidth, angle=0} \\
% \vspace{1cm}
 \multicolumn{2}{c}{\epsfig{file=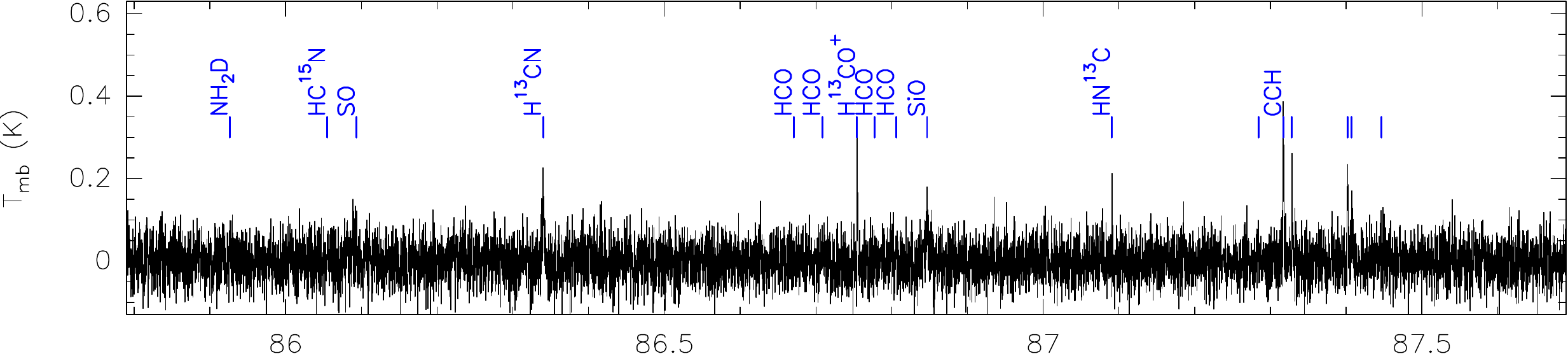, width=0.8\textwidth, angle=0}} \\
 \multicolumn{2}{c}{\epsfig{file=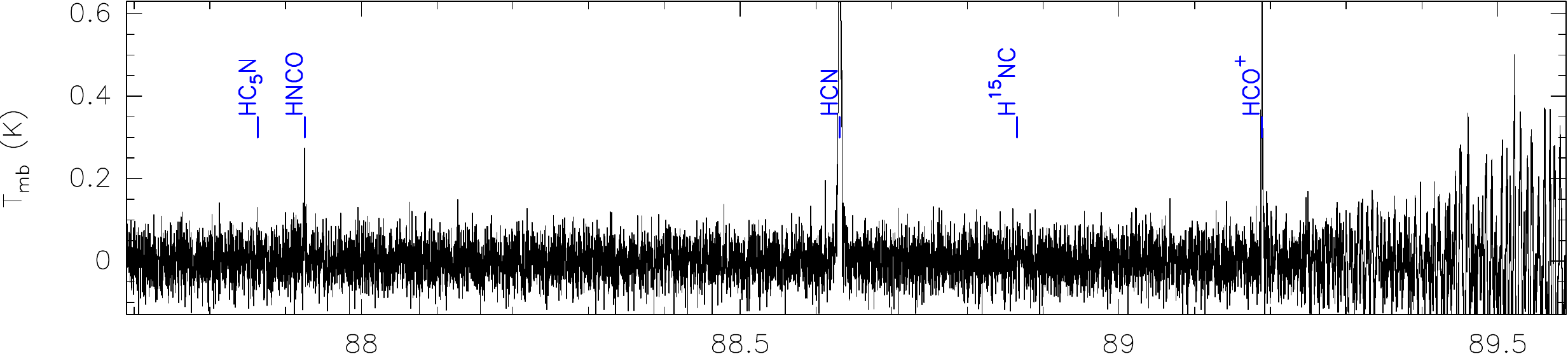, width=0.8\textwidth, angle=0}} \\
 \multicolumn{2}{c}{\epsfig{file=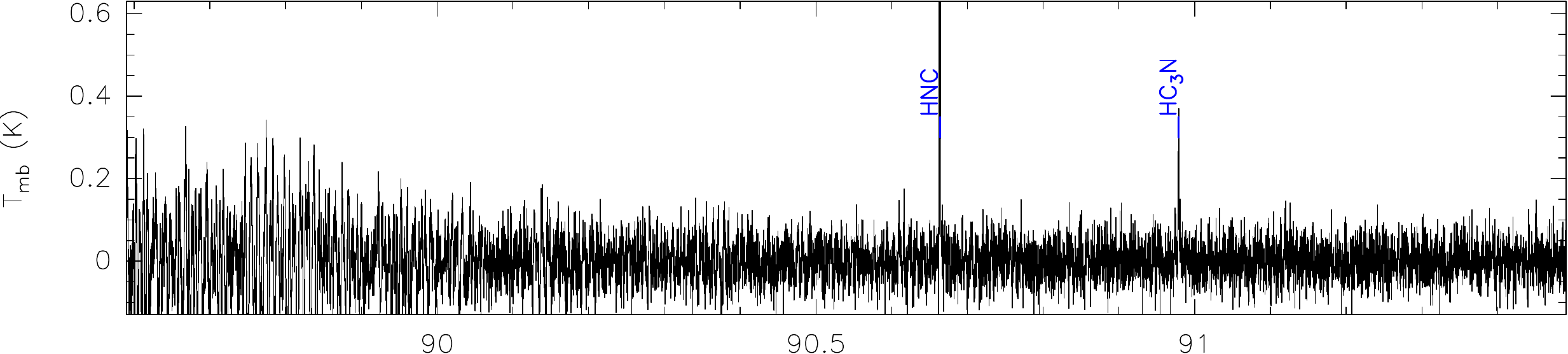, width=0.8\textwidth, angle=0}} \\
 \multicolumn{2}{c}{\epsfig{file=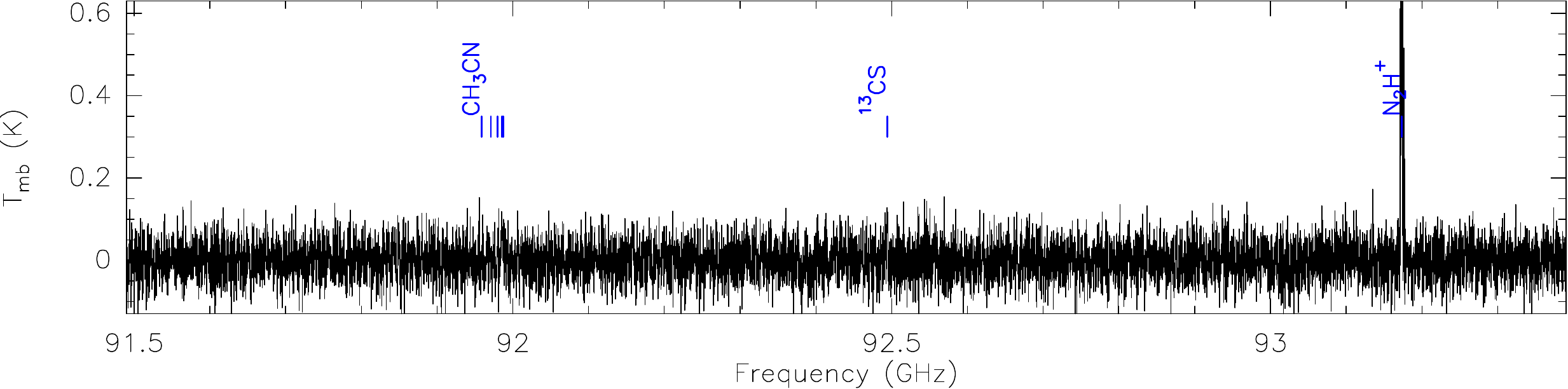, width=0.8\textwidth, angle=0}} \\
\end{tabular}
\end{center}
\caption{continued, for G23.60$+$0.0\,M1.}
\end{figure*}
\begin{figure*}
\ContinuedFloat
\begin{center}
\begin{tabular}[b]{c c}
 \vspace{0.5cm}
 \epsfig{file=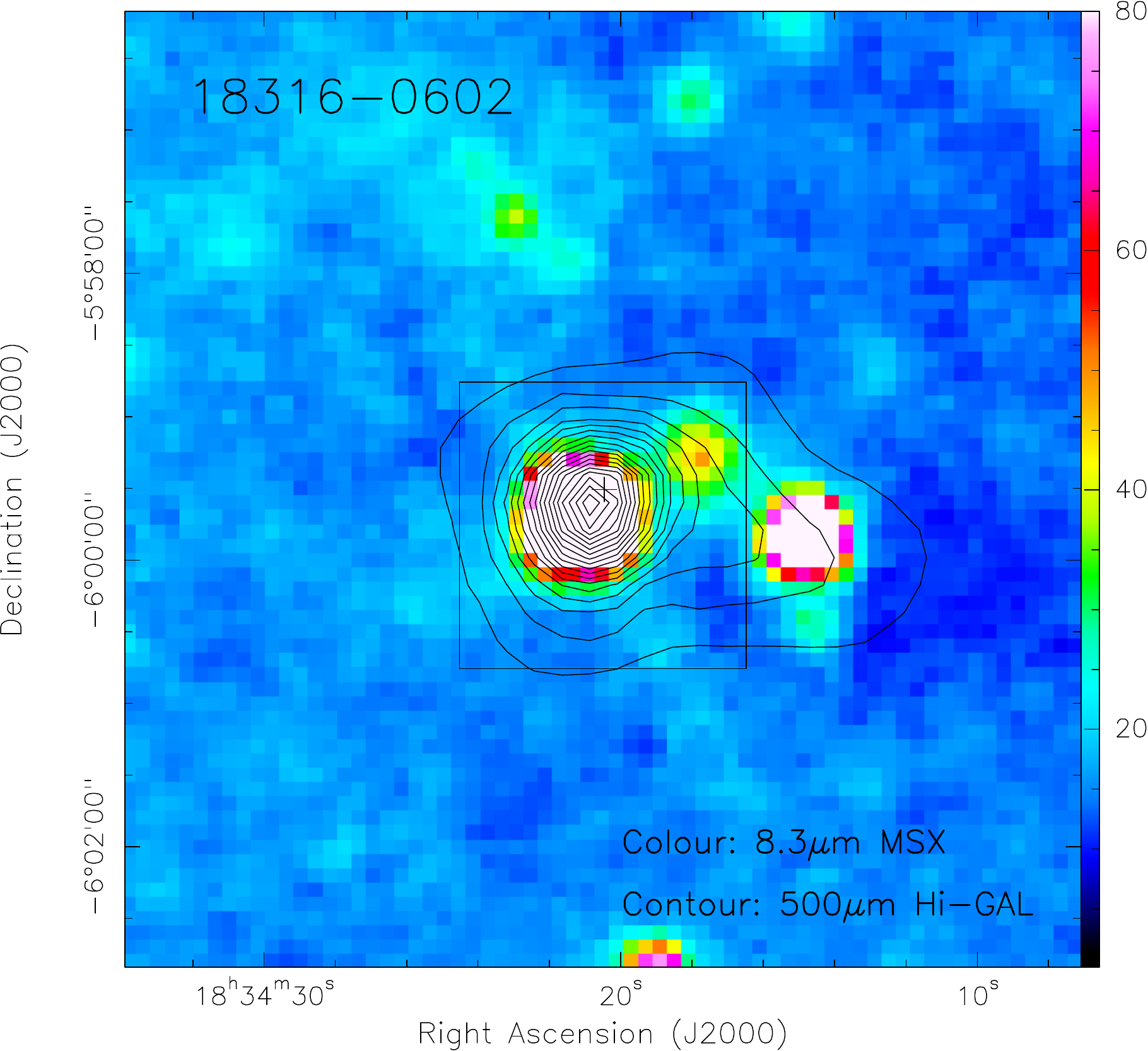, width=1.0\columnwidth, angle=0} &
 \epsfig{file=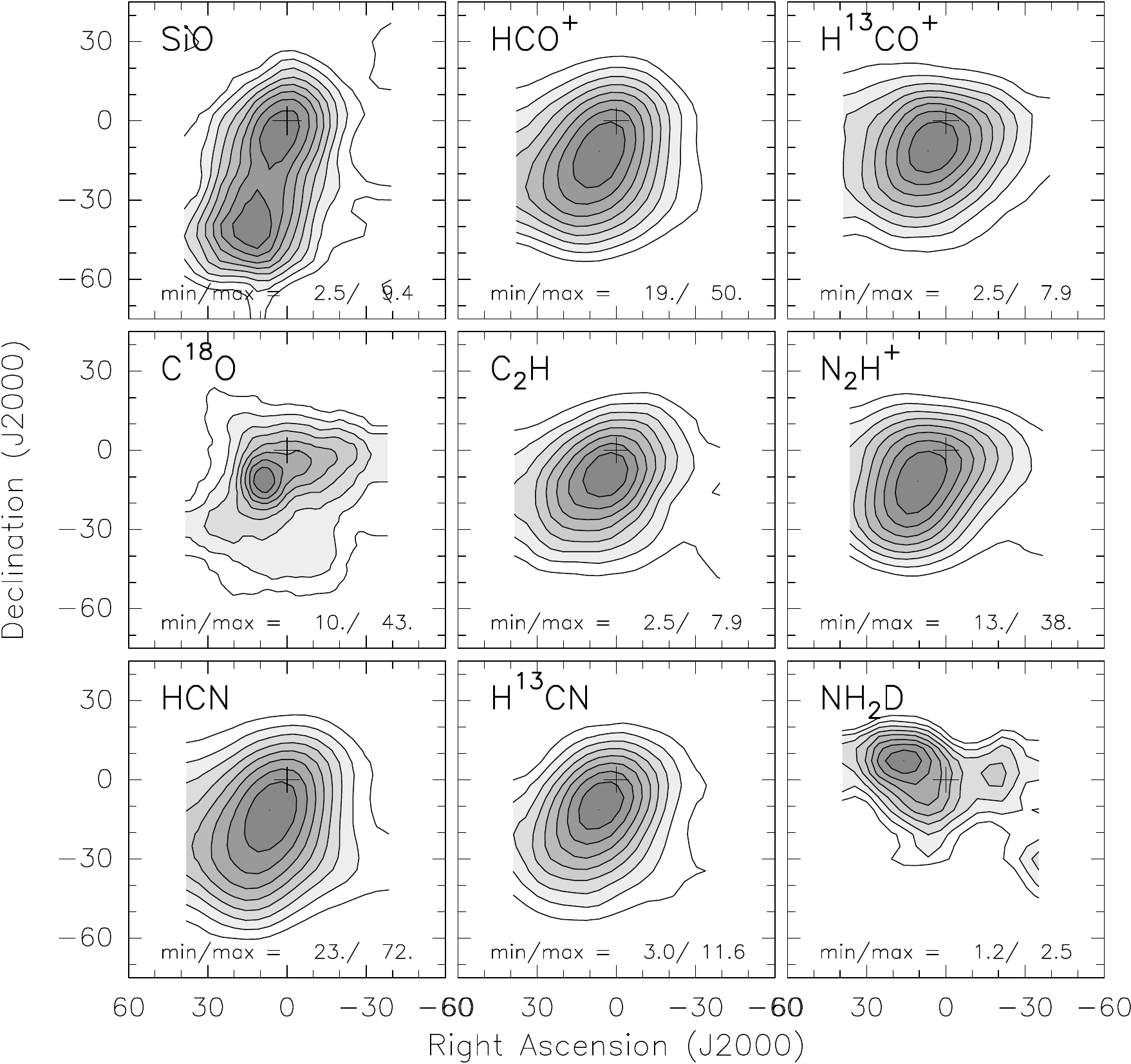, width=0.9\columnwidth, angle=0} \\
% \vspace{1cm}
 \multicolumn{2}{c}{\epsfig{file=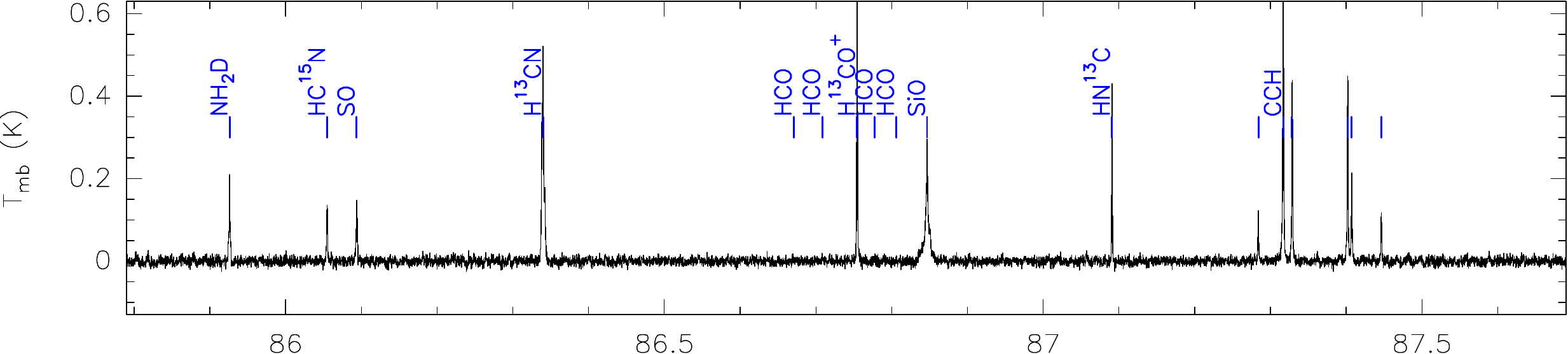, width=0.8\textwidth, angle=0}} \\
 \multicolumn{2}{c}{\epsfig{file=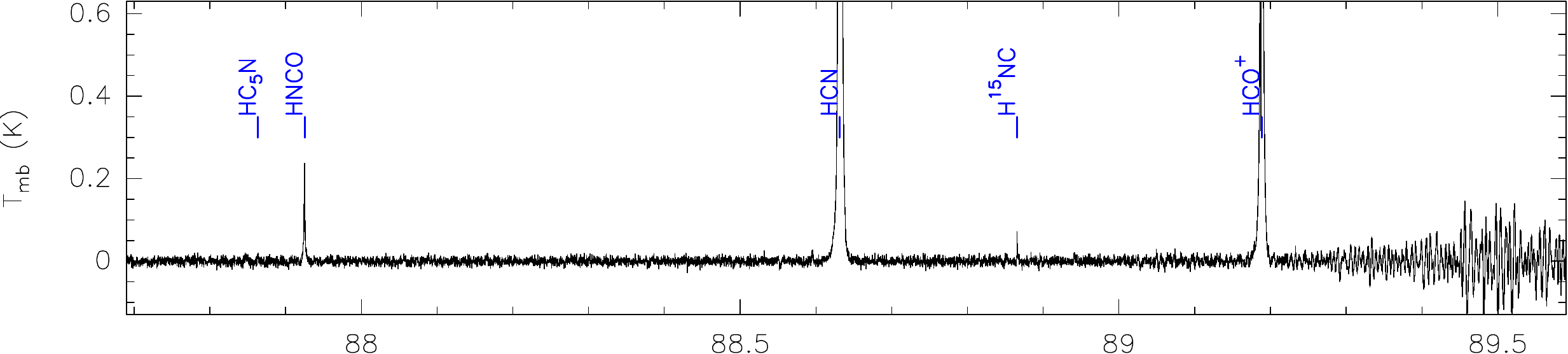, width=0.8\textwidth, angle=0}} \\
 \multicolumn{2}{c}{\epsfig{file=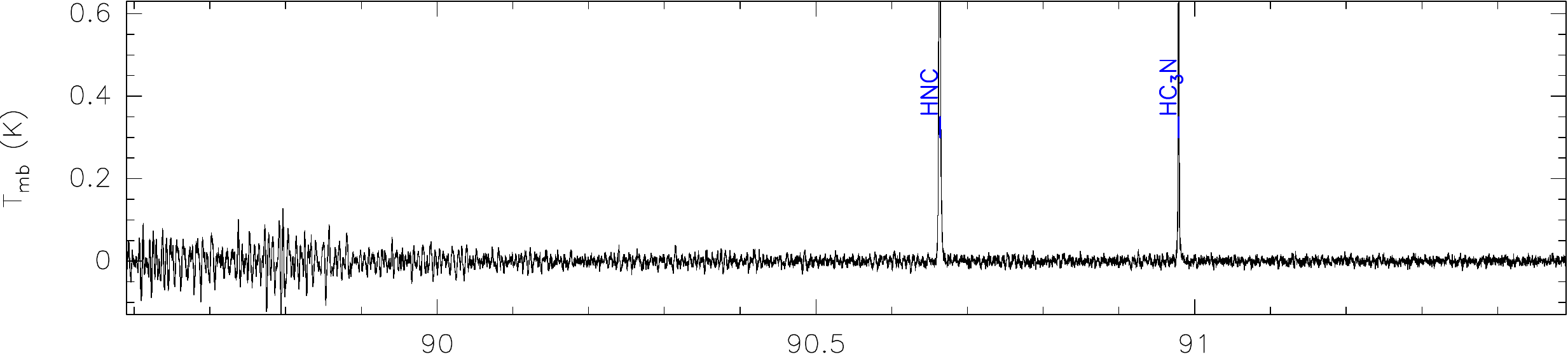, width=0.8\textwidth, angle=0}} \\
 \multicolumn{2}{c}{\epsfig{file=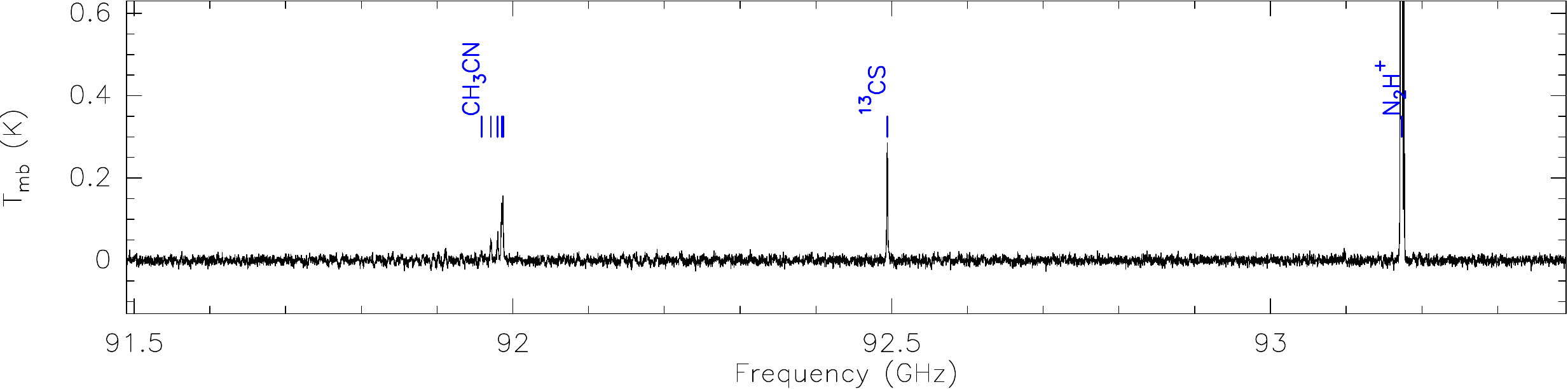, width=0.8\textwidth, angle=0}} \\
\end{tabular}
\end{center}
\caption{continued, for 18316$-$0602.}
\end{figure*}
\begin{figure*}
\ContinuedFloat
\begin{center}
\begin{tabular}[b]{c c}
 \vspace{0.5cm}
 \epsfig{file=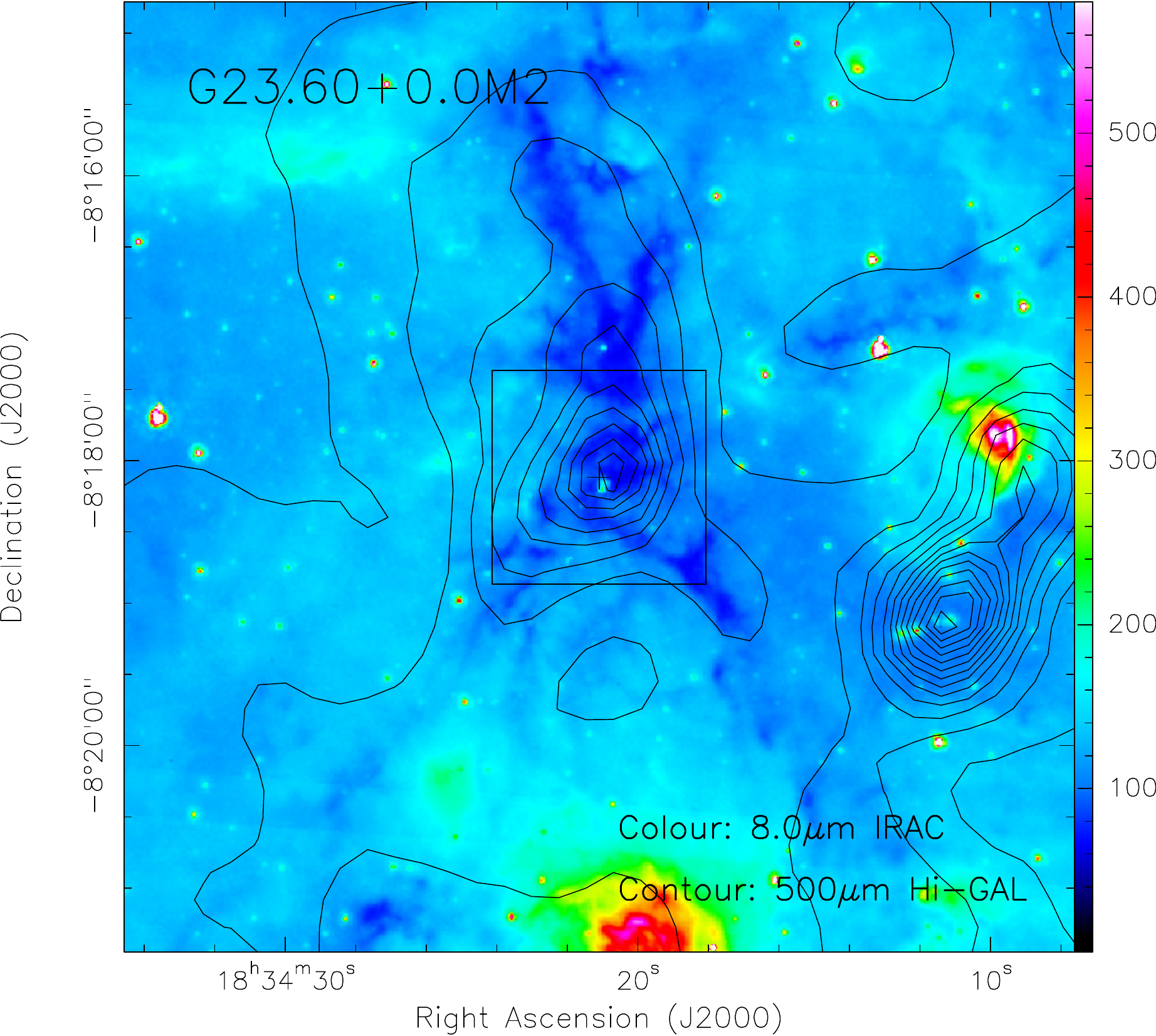, width=1.0\columnwidth, angle=0} &
 \epsfig{file=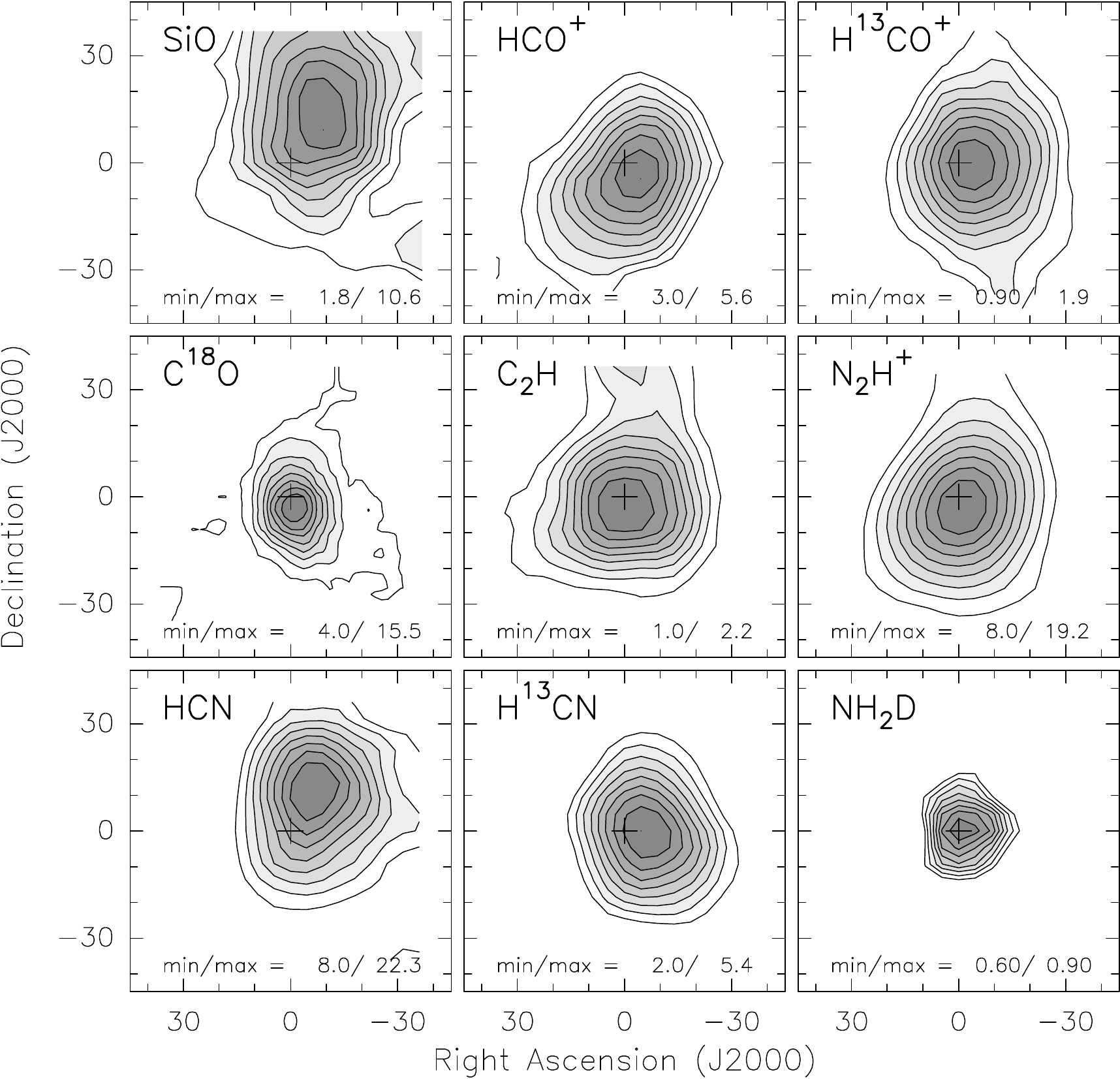, width=0.9\columnwidth, angle=0} \\
% \vspace{1cm}
 \multicolumn{2}{c}{\epsfig{file=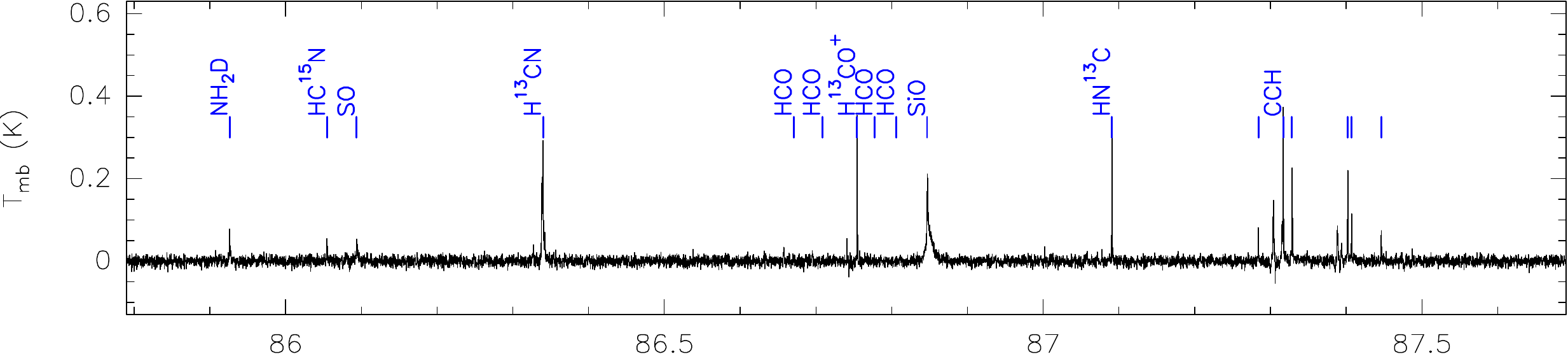, width=0.8\textwidth, angle=0}} \\
 \multicolumn{2}{c}{\epsfig{file=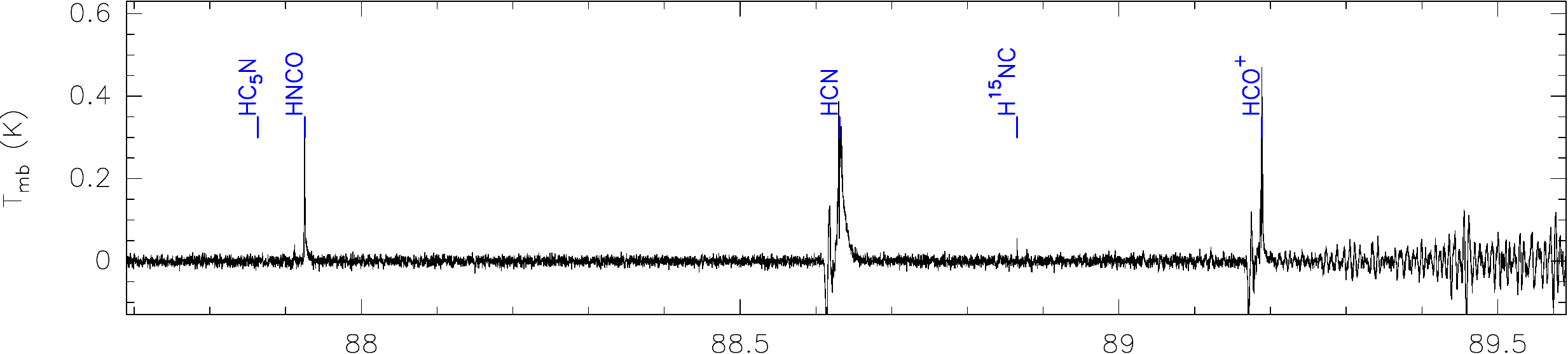, width=0.8\textwidth, angle=0}} \\
 \multicolumn{2}{c}{\epsfig{file=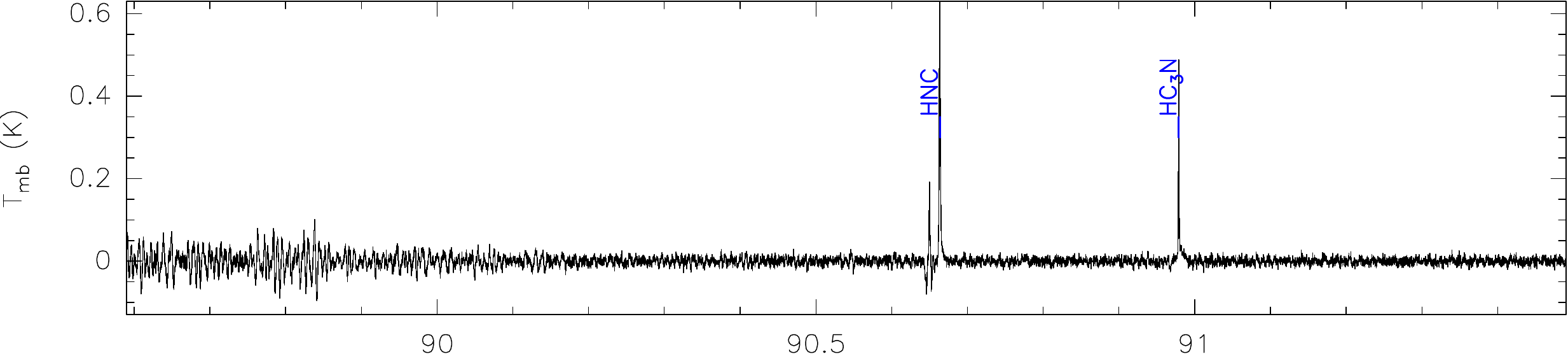, width=0.8\textwidth, angle=0}} \\
 \multicolumn{2}{c}{\epsfig{file=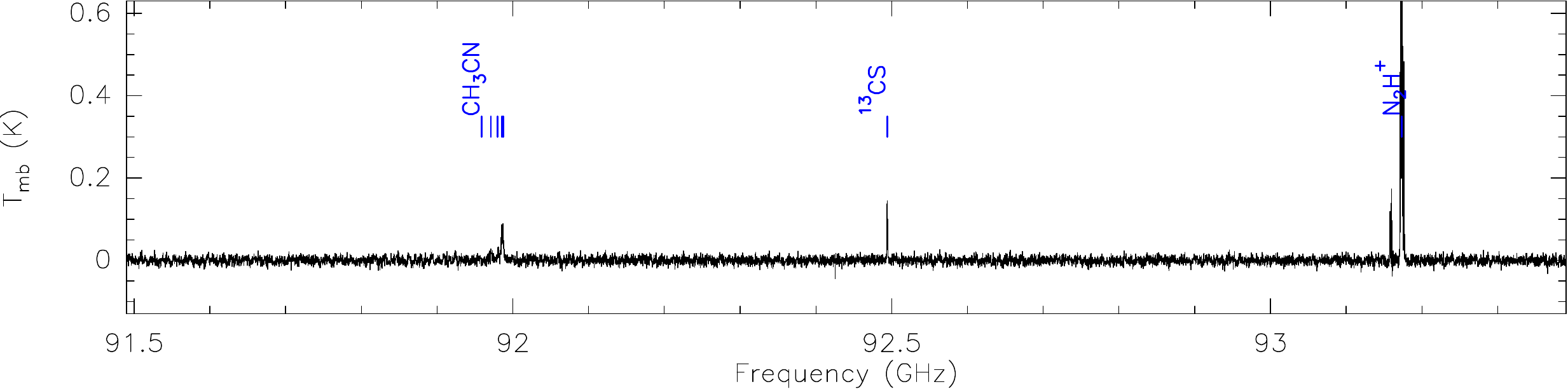, width=0.8\textwidth, angle=0}} \\
\end{tabular}
\end{center}
\caption{continued, for G23.60$+$0.0\,M2.}
\end{figure*}
\begin{figure*}
\ContinuedFloat
\begin{center}
\begin{tabular}[b]{c c}
 \vspace{0.5cm}
 \epsfig{file=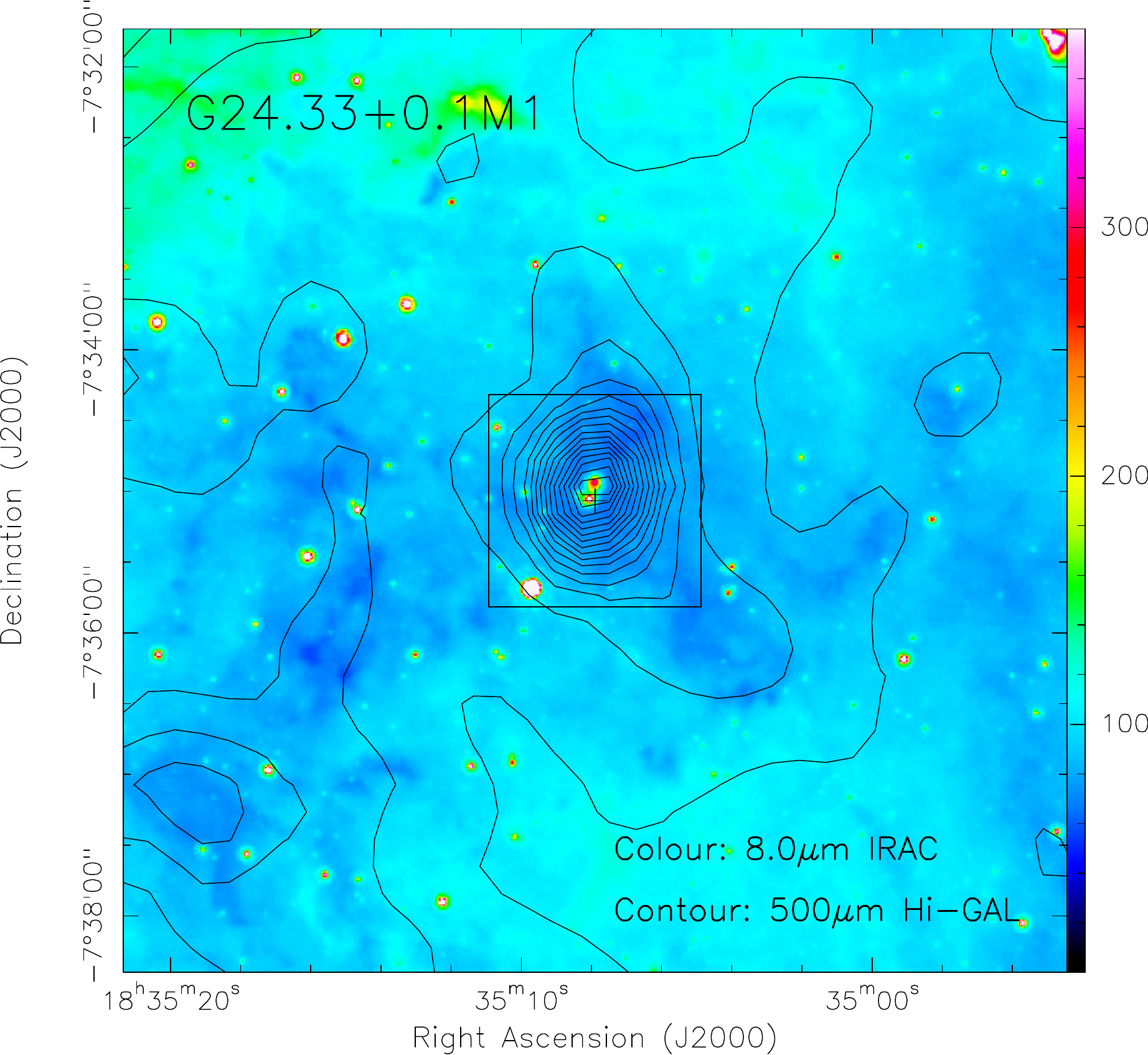, width=1.0\columnwidth, angle=0} &
 \epsfig{file=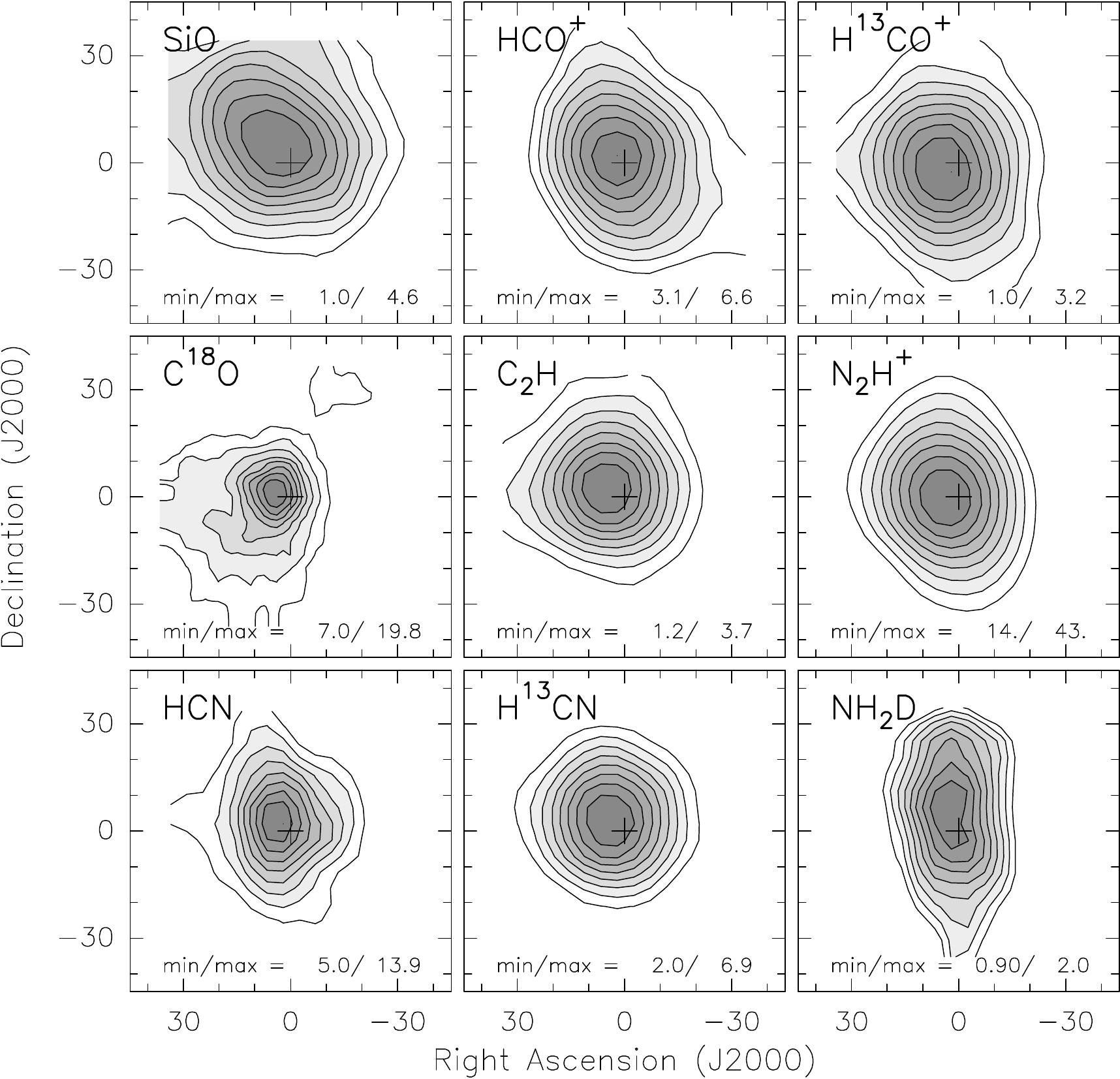, width=0.9\columnwidth, angle=0} \\
% \vspace{1cm}
 \multicolumn{2}{c}{\epsfig{file=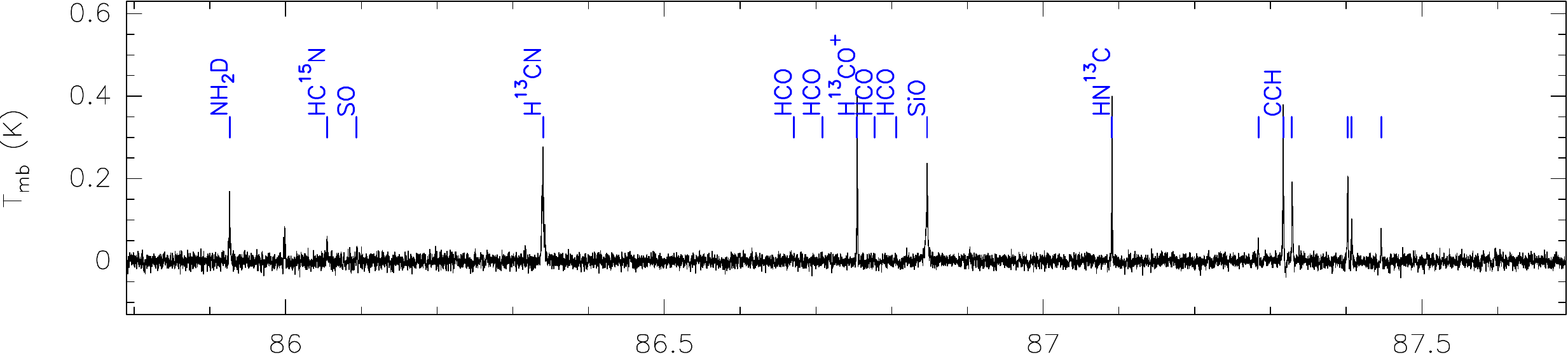, width=0.8\textwidth, angle=0}} \\
 \multicolumn{2}{c}{\epsfig{file=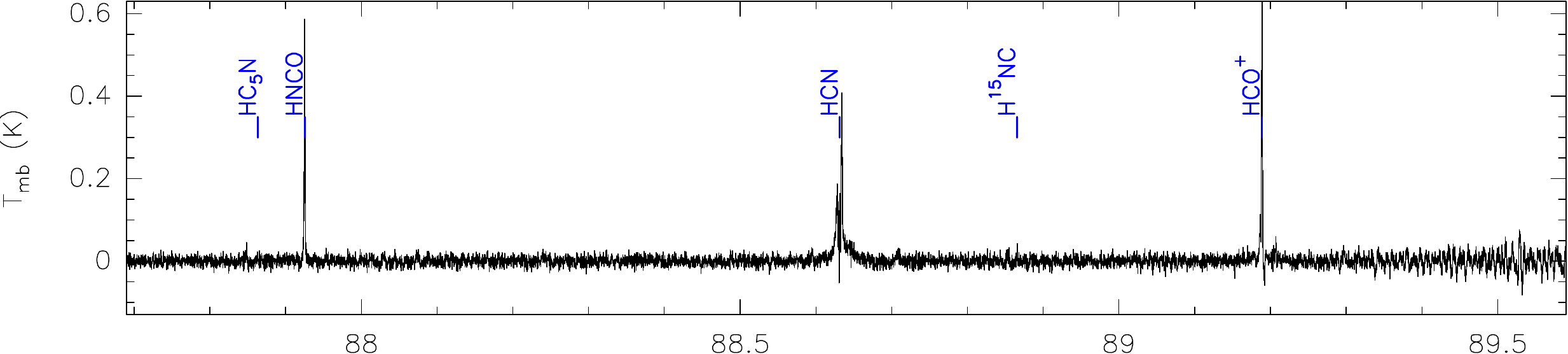, width=0.8\textwidth, angle=0}} \\
 \multicolumn{2}{c}{\epsfig{file=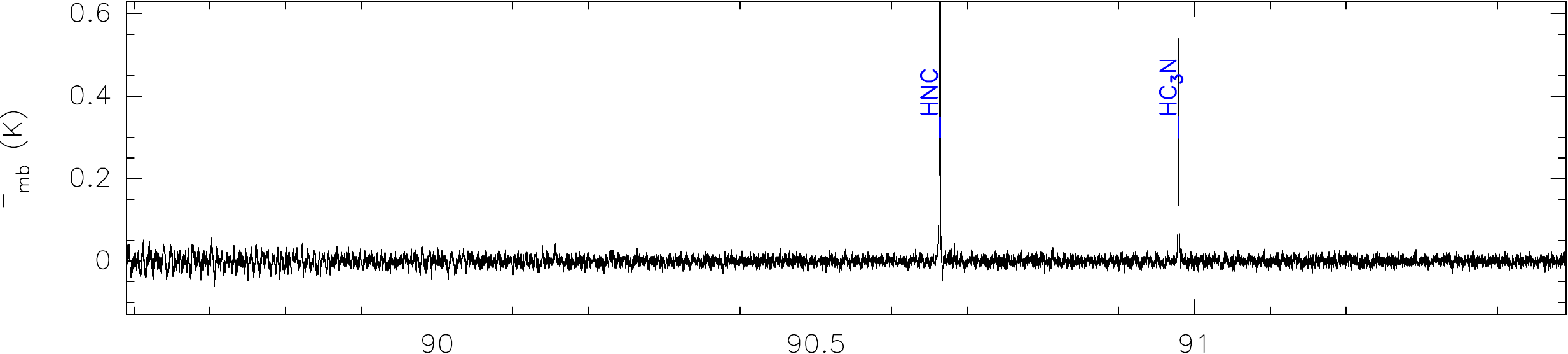, width=0.8\textwidth, angle=0}} \\
 \multicolumn{2}{c}{\epsfig{file=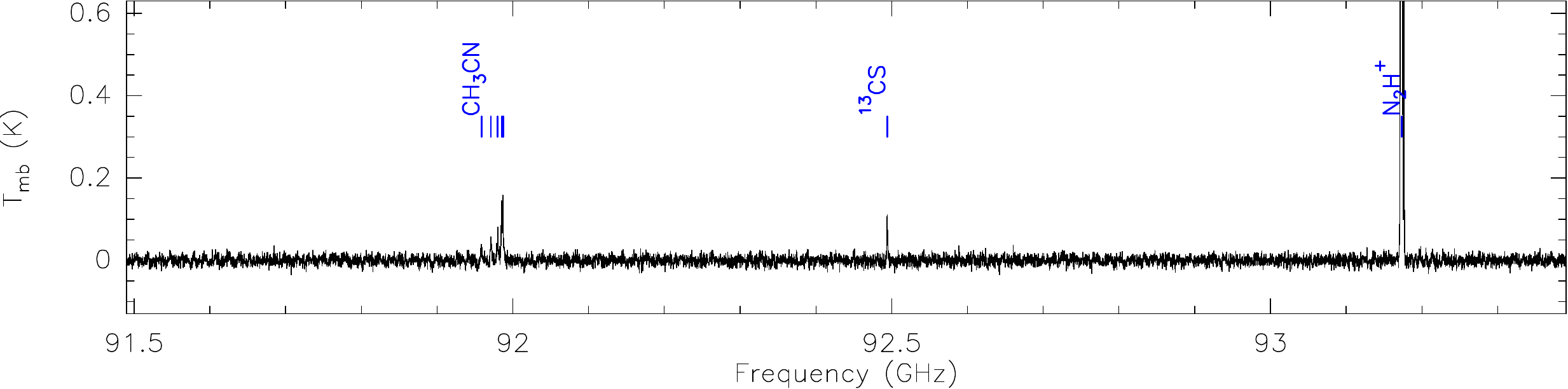, width=0.8\textwidth, angle=0}} \\
\end{tabular}
\end{center}
\caption{continued, for G24.33$+$0.1\,M1.}
\end{figure*}
\begin{figure*}
\ContinuedFloat
\begin{center}
\begin{tabular}[b]{c c}
 \vspace{0.5cm}
 \epsfig{file=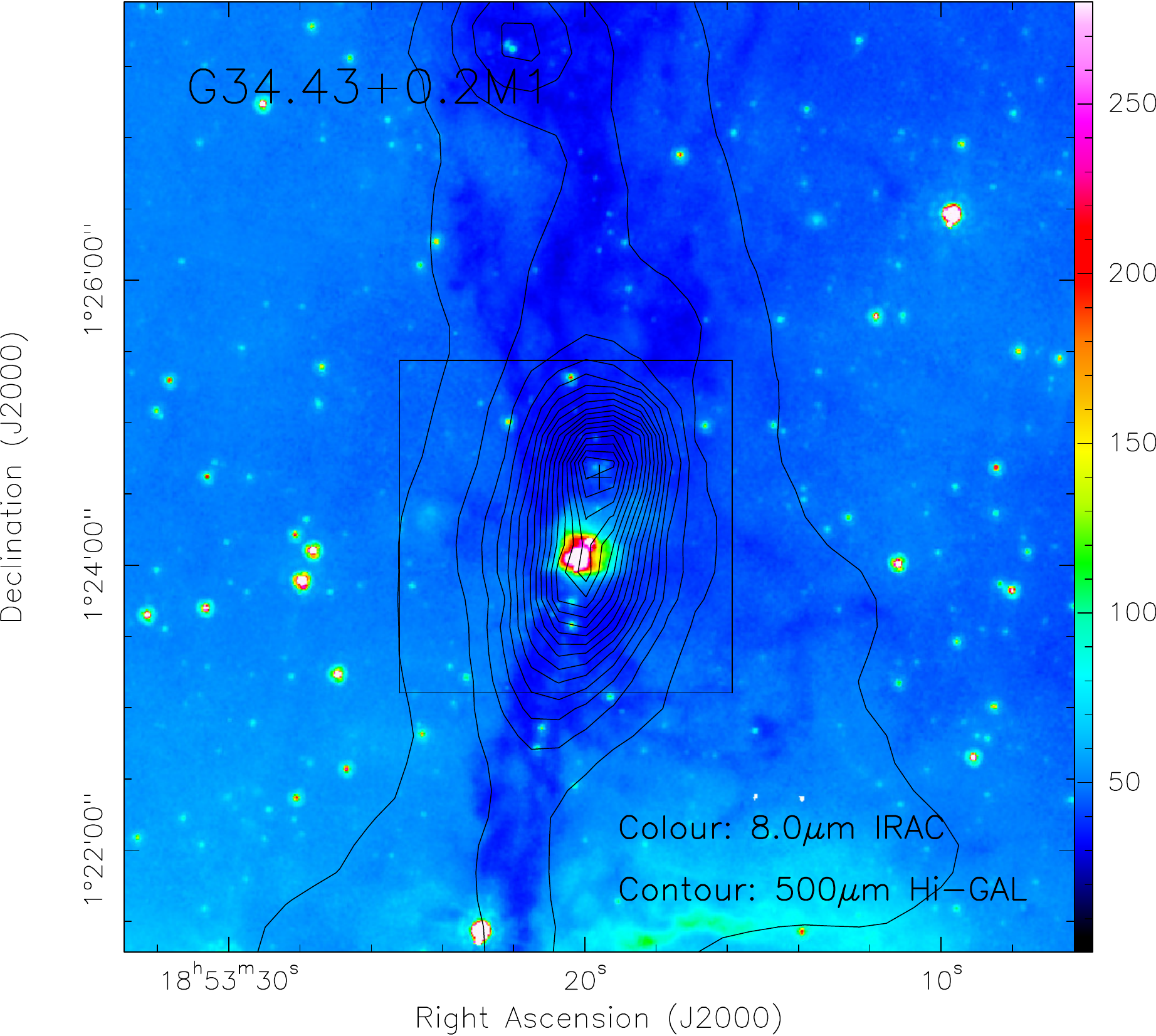, width=1.0\columnwidth, angle=0} &
 \epsfig{file=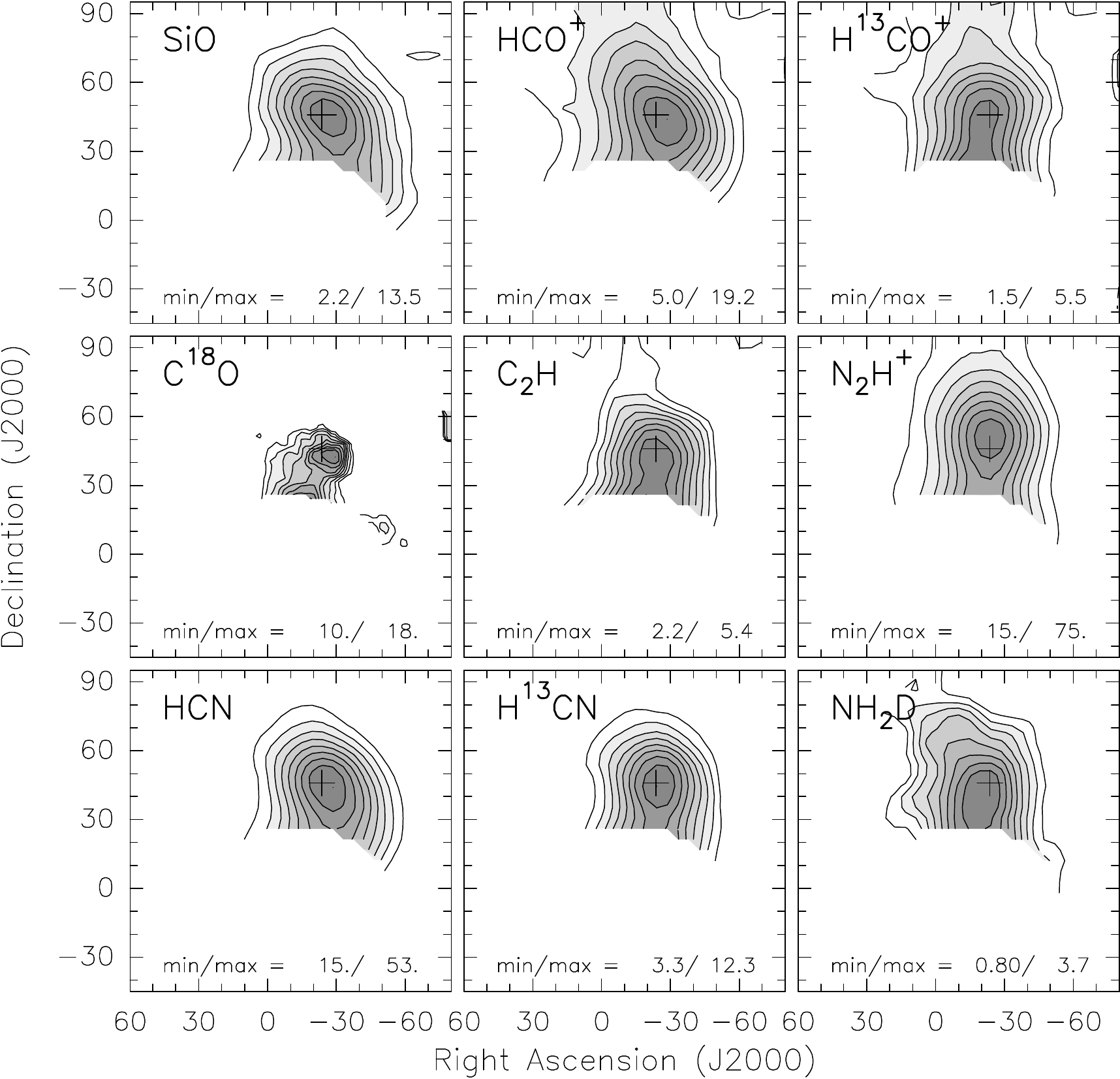, width=0.9\columnwidth, angle=0} \\
% \vspace{1cm}
 \multicolumn{2}{c}{\epsfig{file=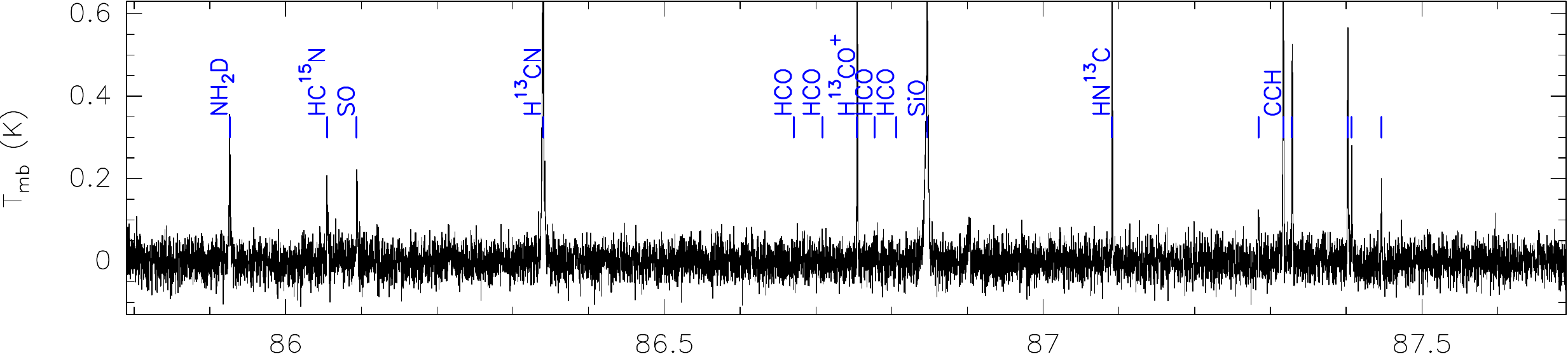, width=0.8\textwidth, angle=0}} \\
 \multicolumn{2}{c}{\epsfig{file=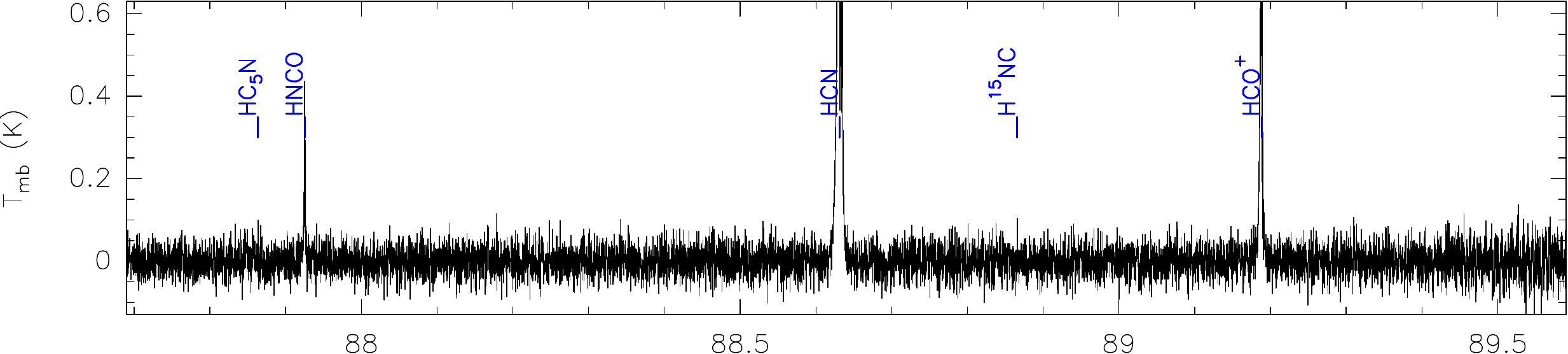, width=0.8\textwidth, angle=0}} \\
 \multicolumn{2}{c}{\epsfig{file=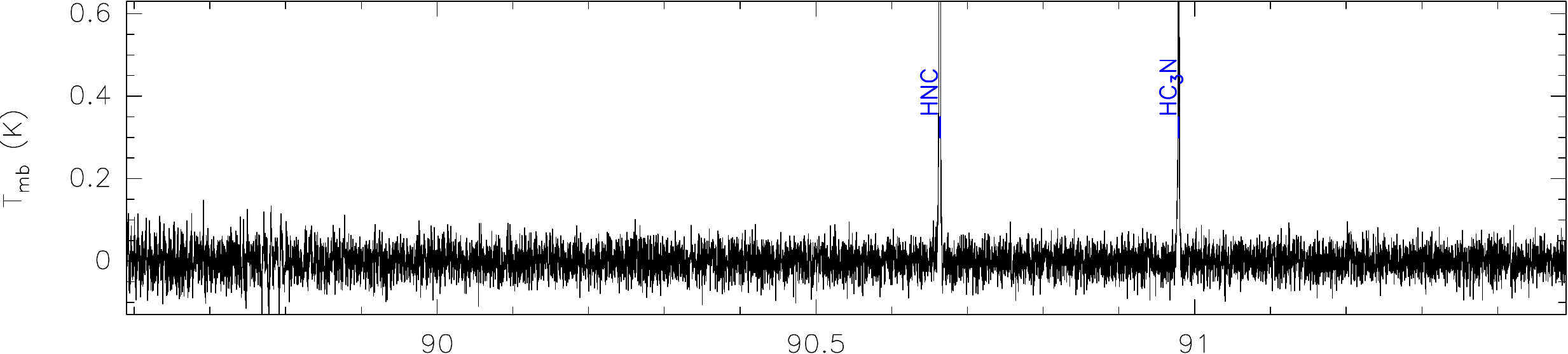, width=0.8\textwidth, angle=0}} \\
 \multicolumn{2}{c}{\epsfig{file=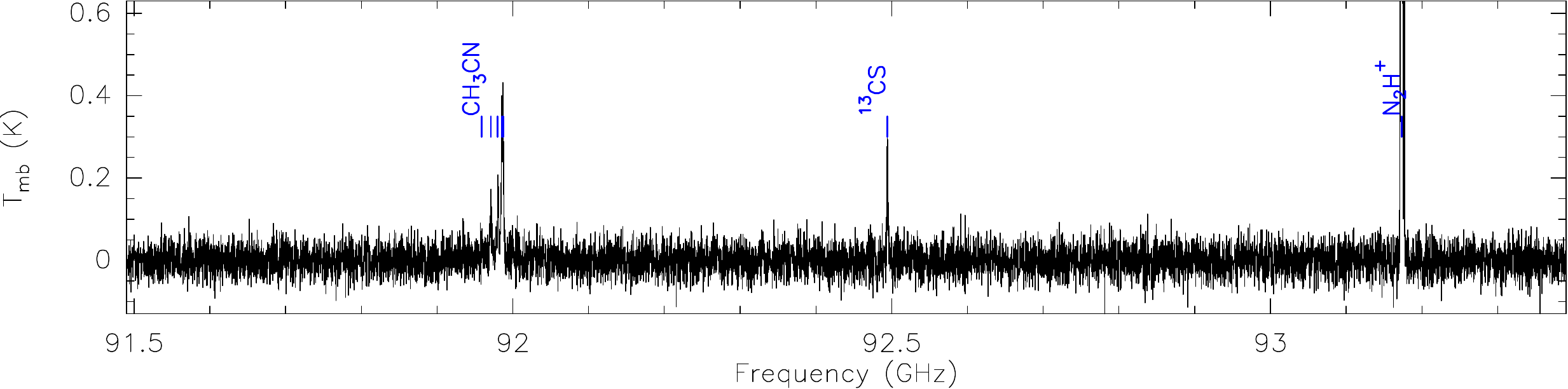, width=0.8\textwidth, angle=0}} \\
\end{tabular}
\end{center}
\caption{continued, for G34.43$+$0.2\,M1.}
\end{figure*}
\begin{figure*}
\ContinuedFloat
\begin{center}
\begin{tabular}[b]{c c}
 \vspace{0.5cm}
 \epsfig{file=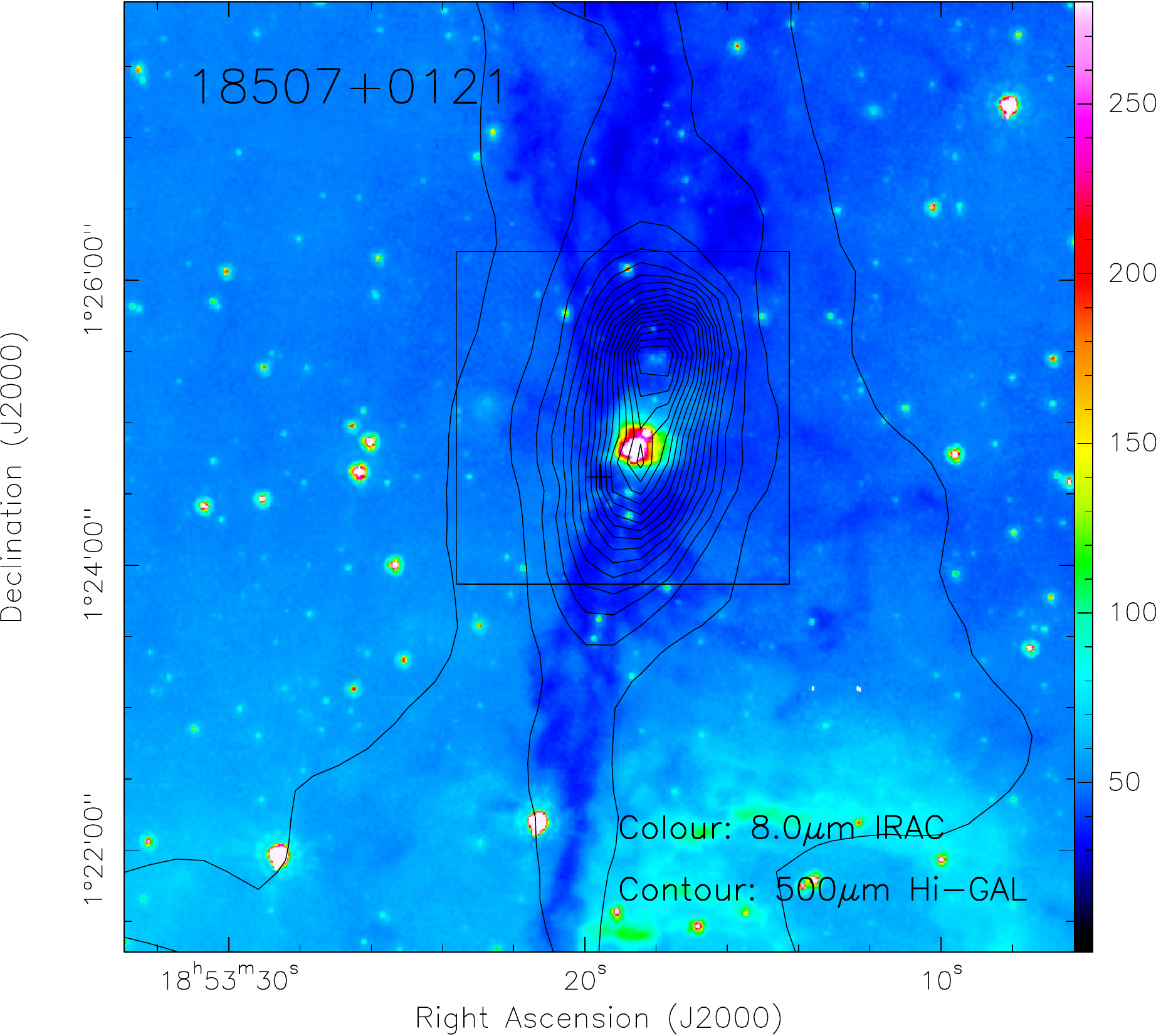, width=1.0\columnwidth, angle=0} &
 \epsfig{file=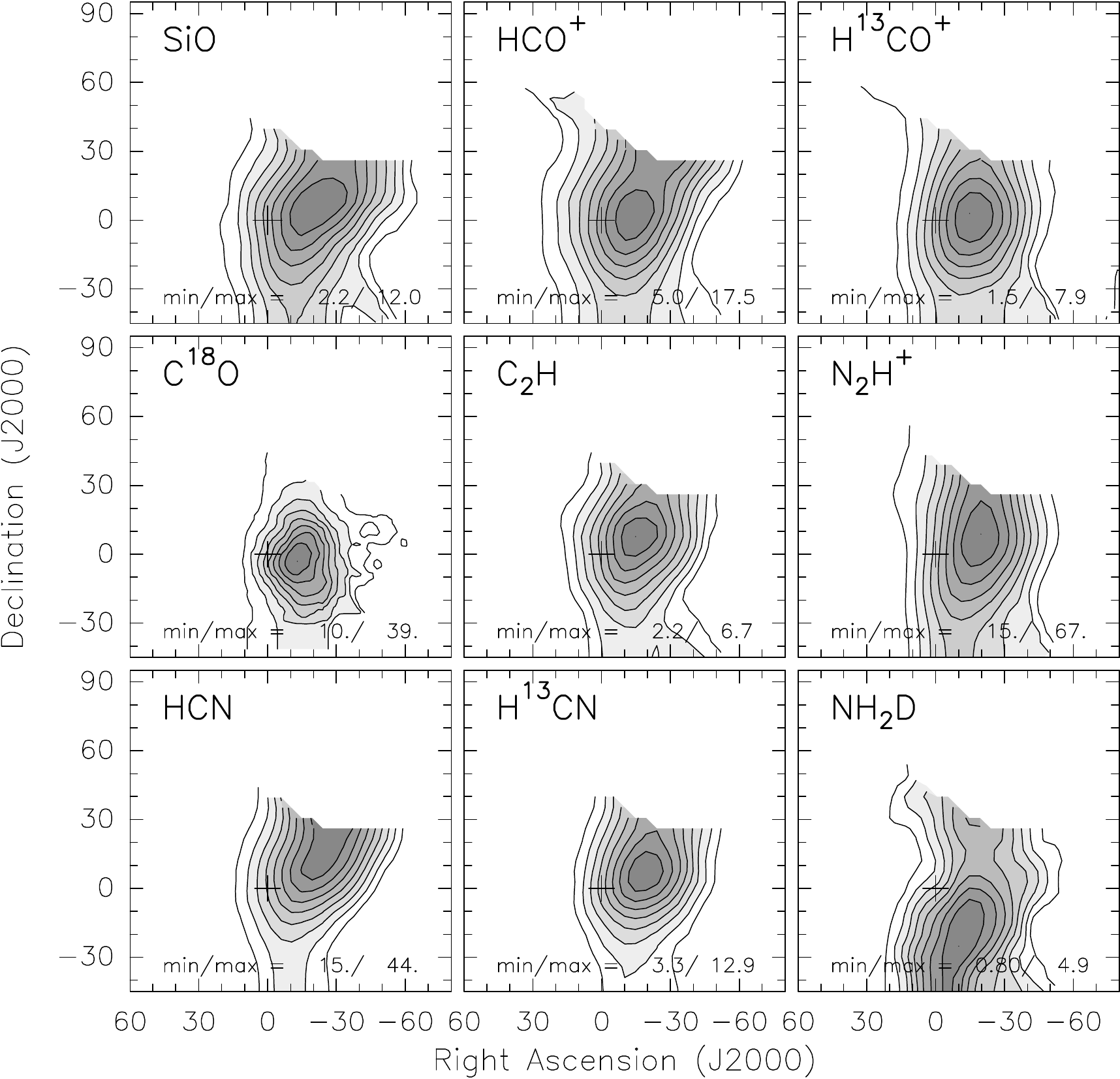, width=0.9\columnwidth, angle=0} \\
% \vspace{1cm}
 \multicolumn{2}{c}{\epsfig{file=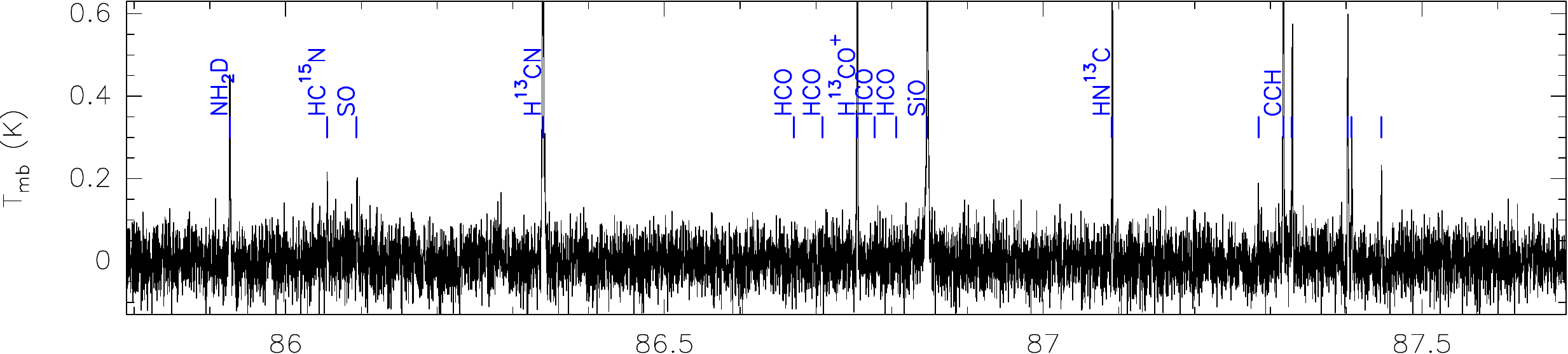, width=0.8\textwidth, angle=0}} \\
 \multicolumn{2}{c}{\epsfig{file=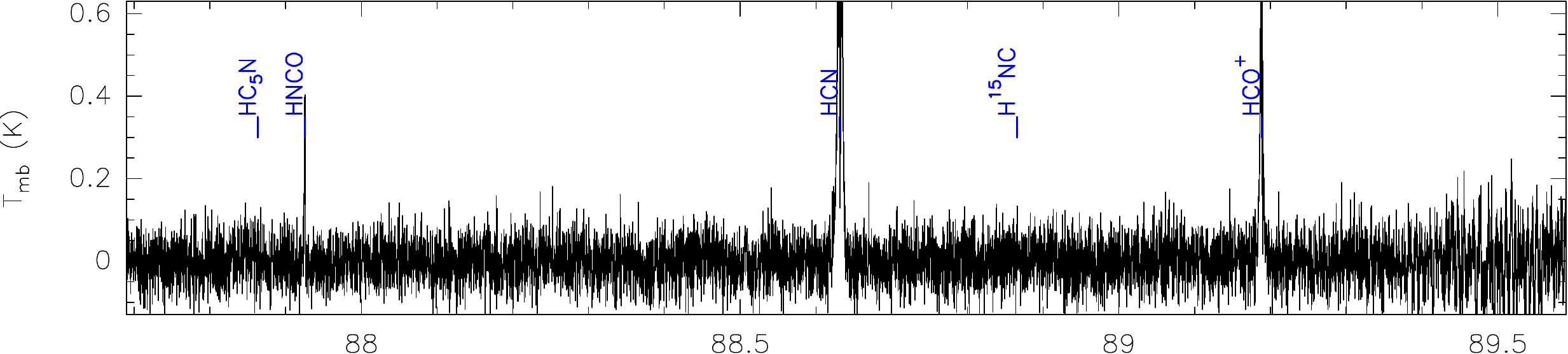, width=0.8\textwidth, angle=0}} \\
 \multicolumn{2}{c}{\epsfig{file=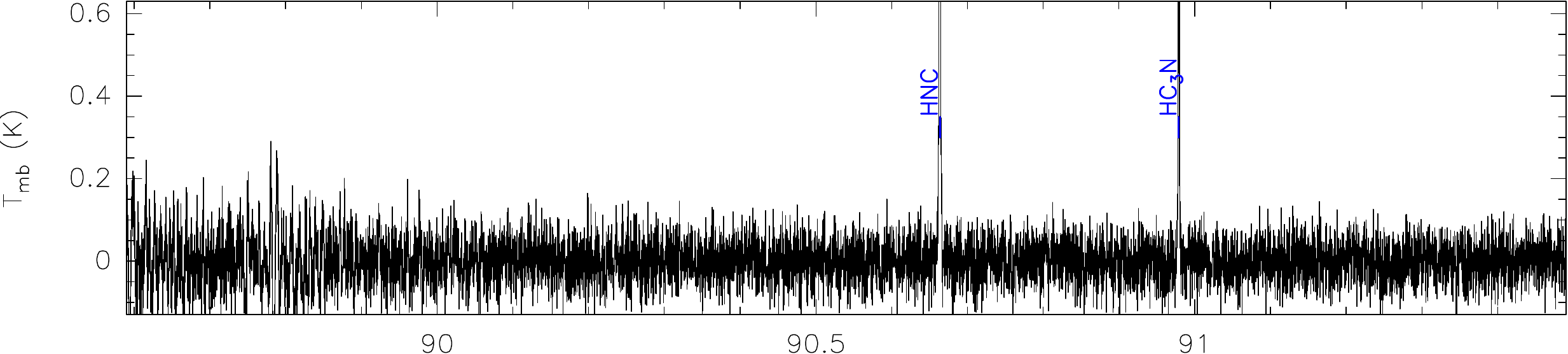, width=0.8\textwidth, angle=0}} \\
 \multicolumn{2}{c}{\epsfig{file=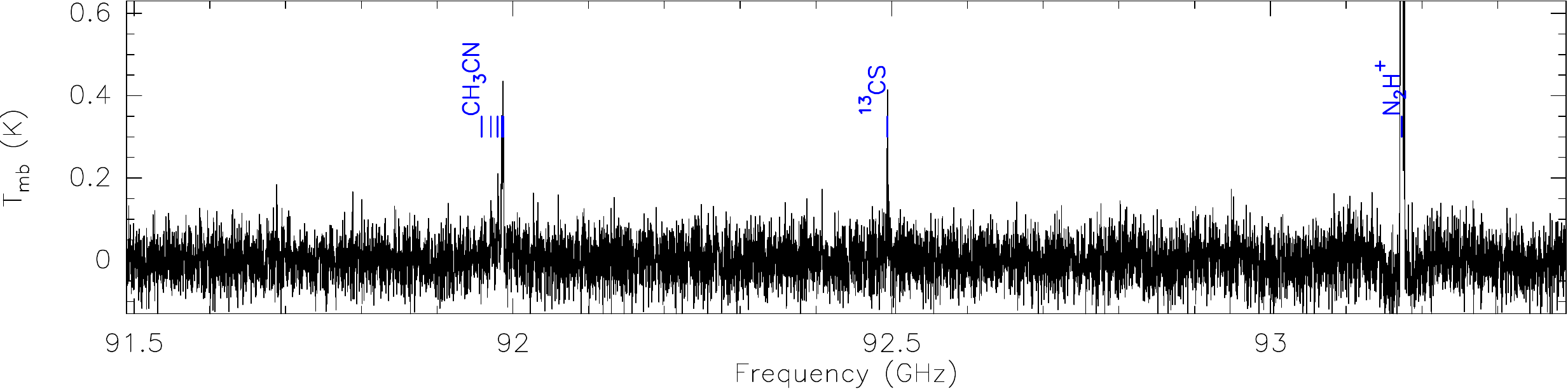, width=0.8\textwidth, angle=0}} \\
\end{tabular}
\end{center}
\caption{continued, for 18507$+$0121.}
\end{figure*}
\begin{figure*}
\ContinuedFloat
\begin{center}
\begin{tabular}[b]{c c}
 \vspace{0.5cm}
 \epsfig{file=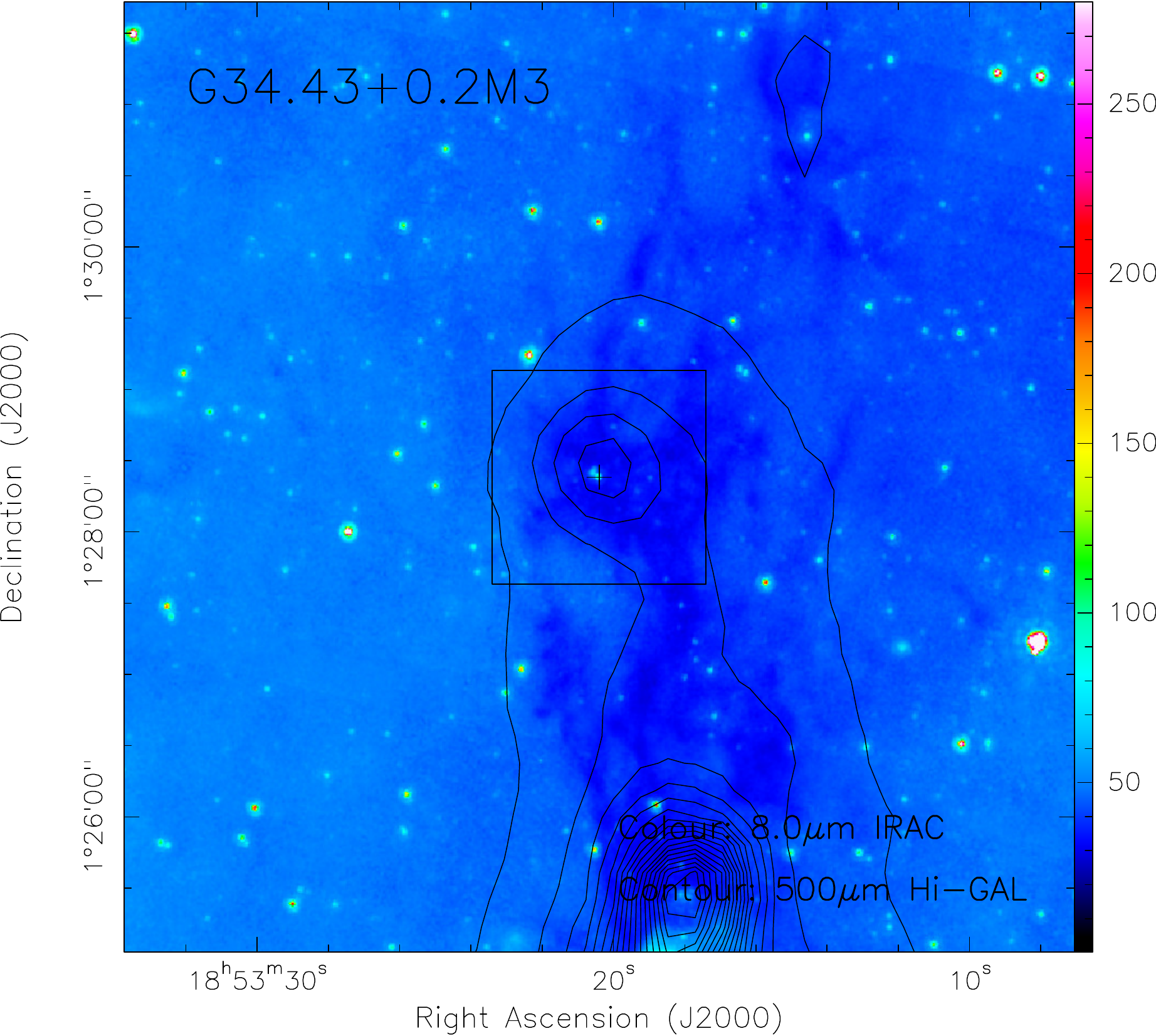, width=1.0\columnwidth, angle=0} &
 \epsfig{file=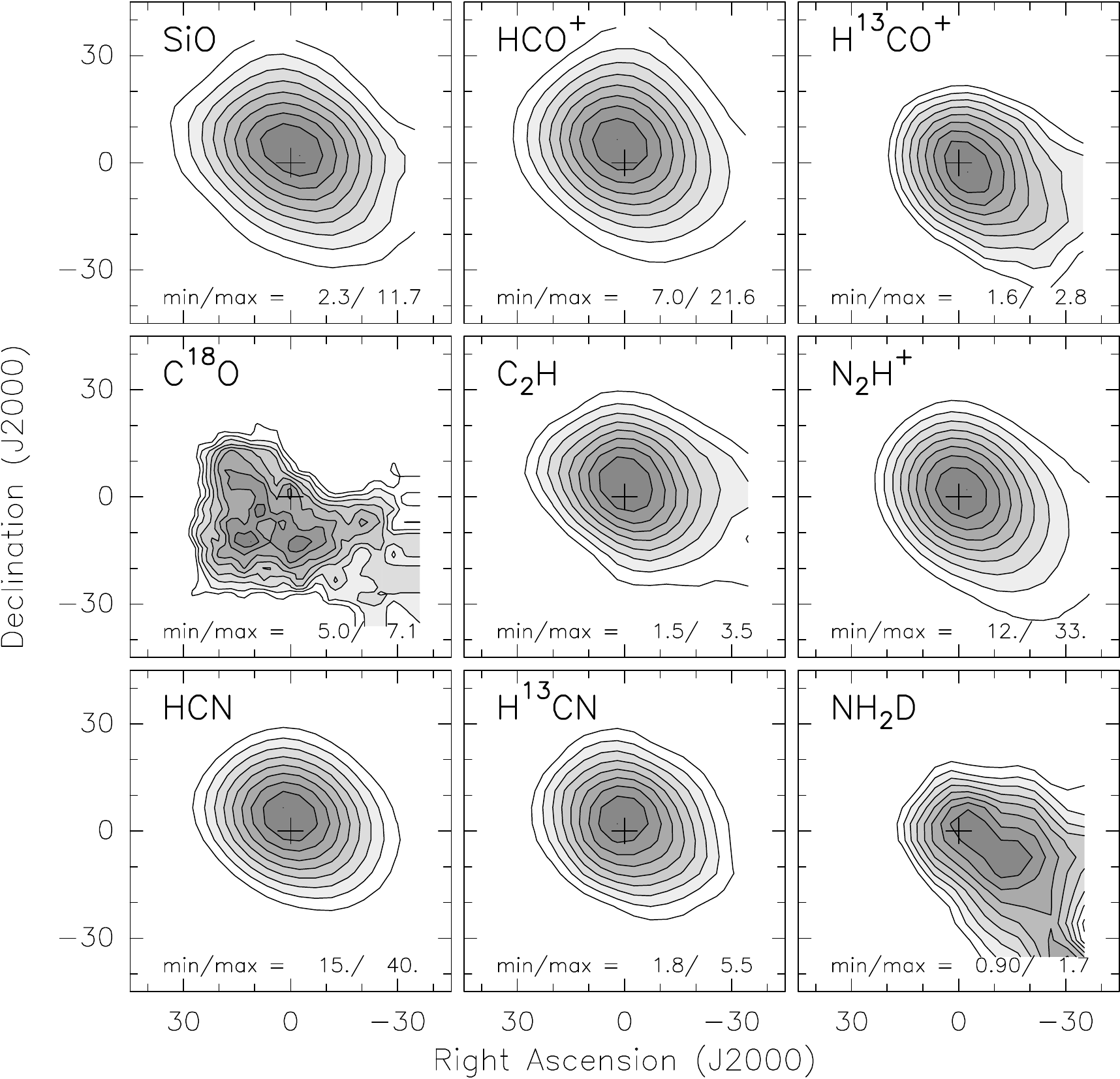, width=0.9\columnwidth, angle=0} \\
% \vspace{1cm}
 \multicolumn{2}{c}{\epsfig{file=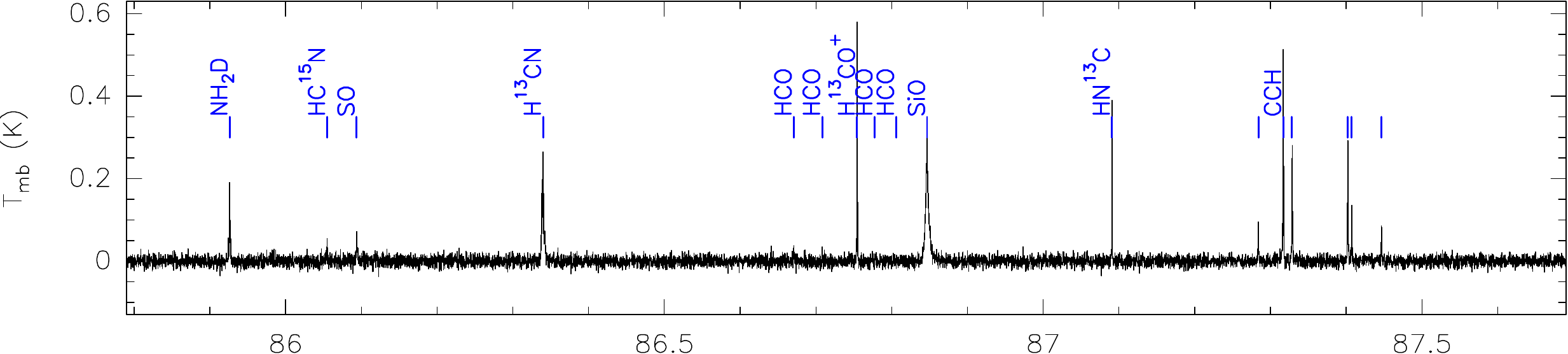, width=0.8\textwidth, angle=0}} \\
 \multicolumn{2}{c}{\epsfig{file=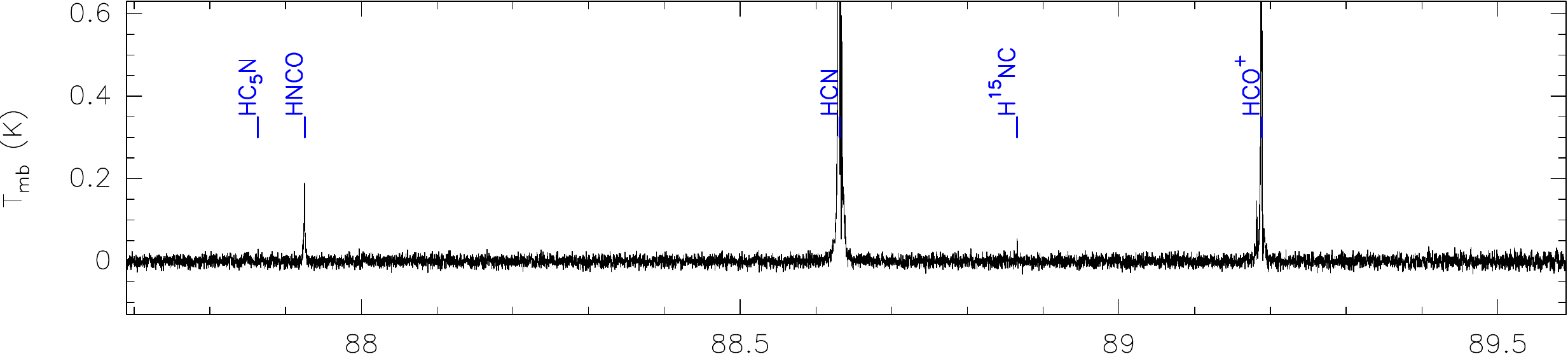, width=0.8\textwidth, angle=0}} \\
 \multicolumn{2}{c}{\epsfig{file=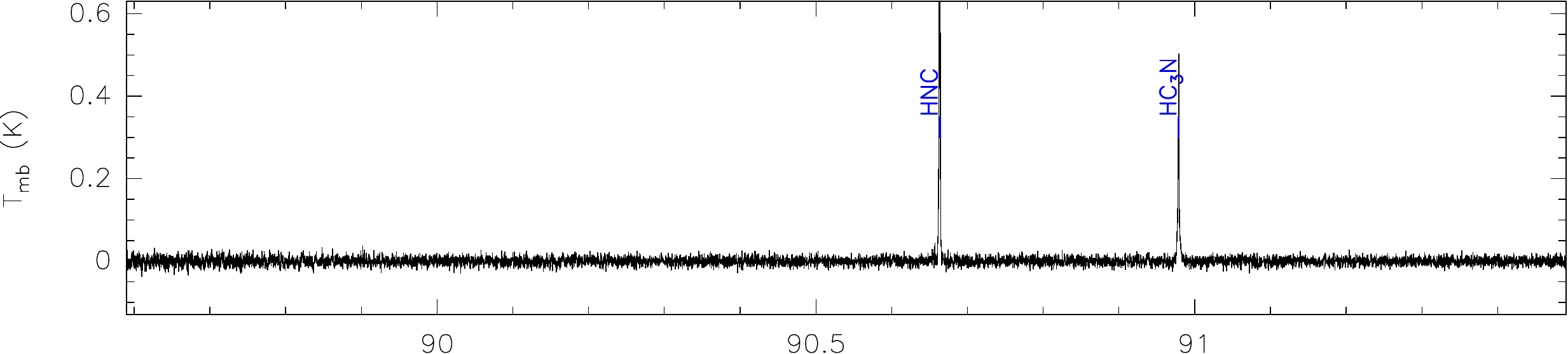, width=0.8\textwidth, angle=0}} \\
 \multicolumn{2}{c}{\epsfig{file=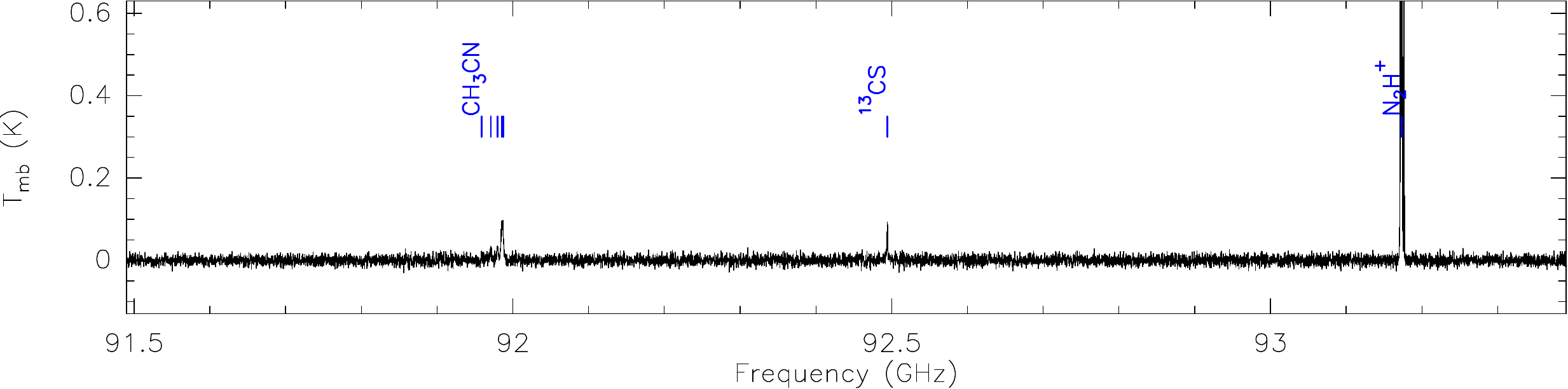, width=0.8\textwidth, angle=0}} \\
\end{tabular}
\end{center}
\caption{continued, for G34.43$+$0.2\,M3.}
\end{figure*}
\begin{figure*}
\ContinuedFloat
\begin{center}
\begin{tabular}[b]{c c}
 \vspace{0.5cm}
 \epsfig{file=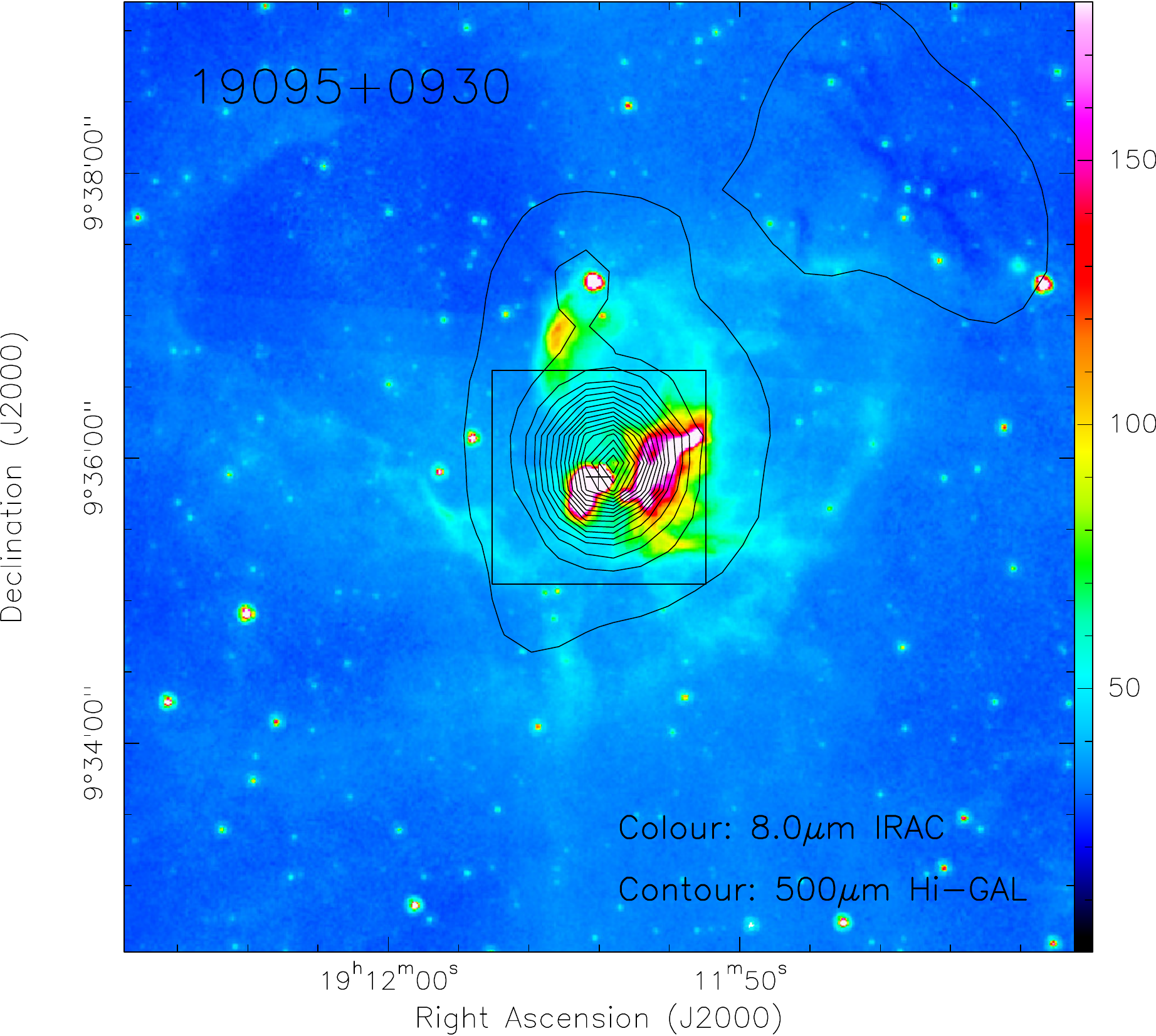, width=1.0\columnwidth, angle=0} &
 \epsfig{file=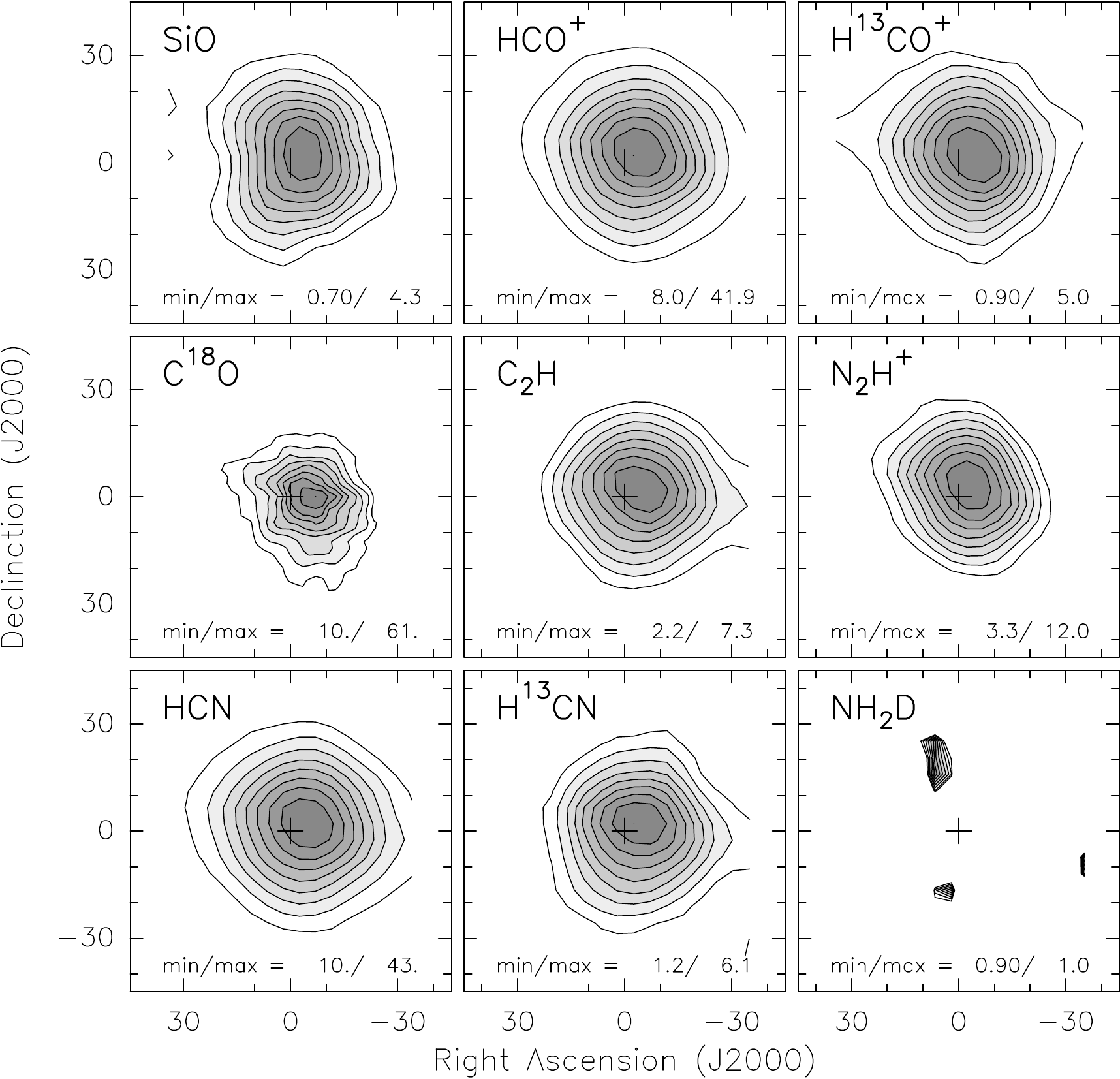, width=0.9\columnwidth, angle=0} \\
% \vspace{1cm}
 \multicolumn{2}{c}{\epsfig{file=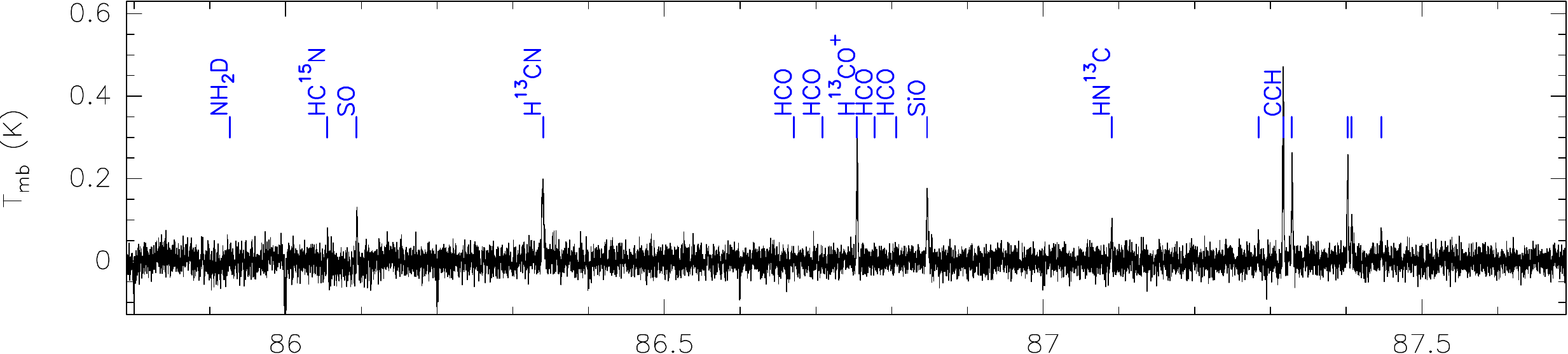, width=0.8\textwidth, angle=0}} \\
 \multicolumn{2}{c}{\epsfig{file=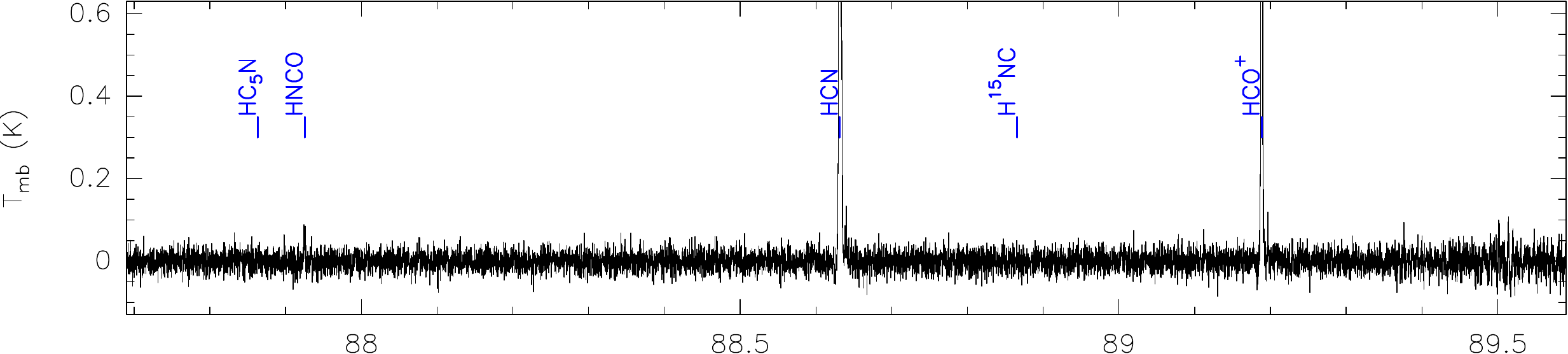, width=0.8\textwidth, angle=0}} \\
 \multicolumn{2}{c}{\epsfig{file=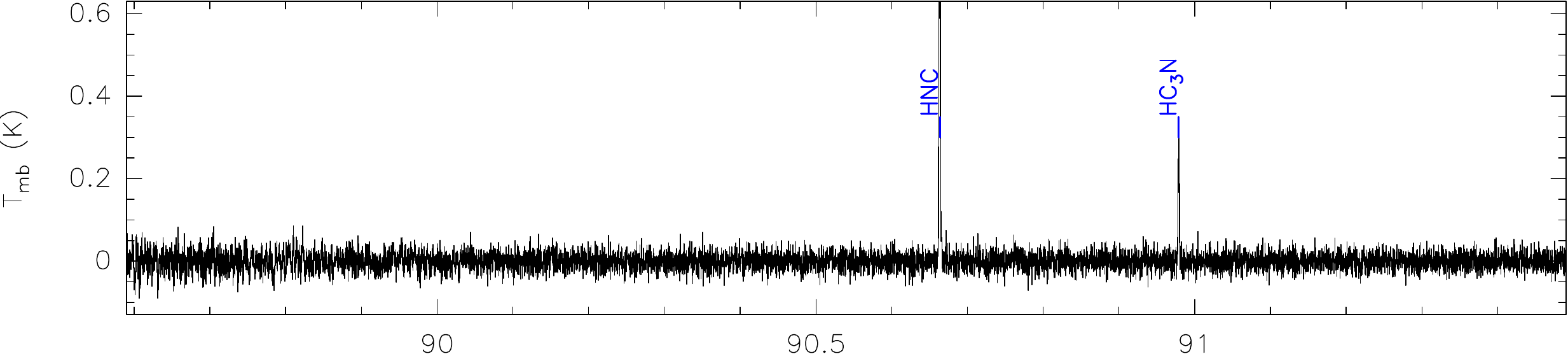, width=0.8\textwidth, angle=0}} \\
 \multicolumn{2}{c}{\epsfig{file=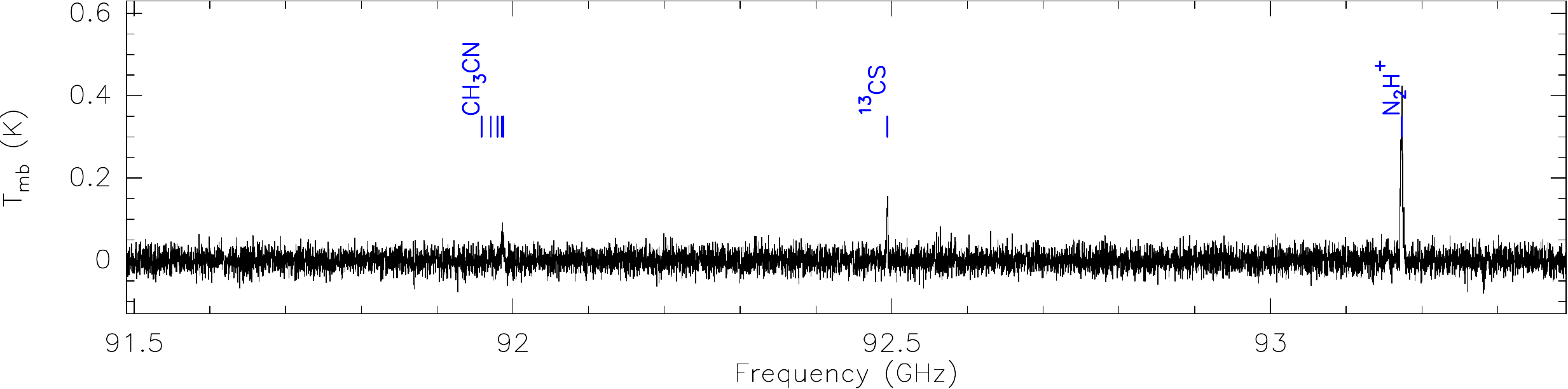, width=0.8\textwidth, angle=0}} \\
\end{tabular}
\end{center}
\caption{continued, for 19095$+$0930.}
\end{figure*}
\begin{figure*}
\ContinuedFloat
\begin{center}
\begin{tabular}[b]{c c}
 \vspace{0.5cm}
 \epsfig{file=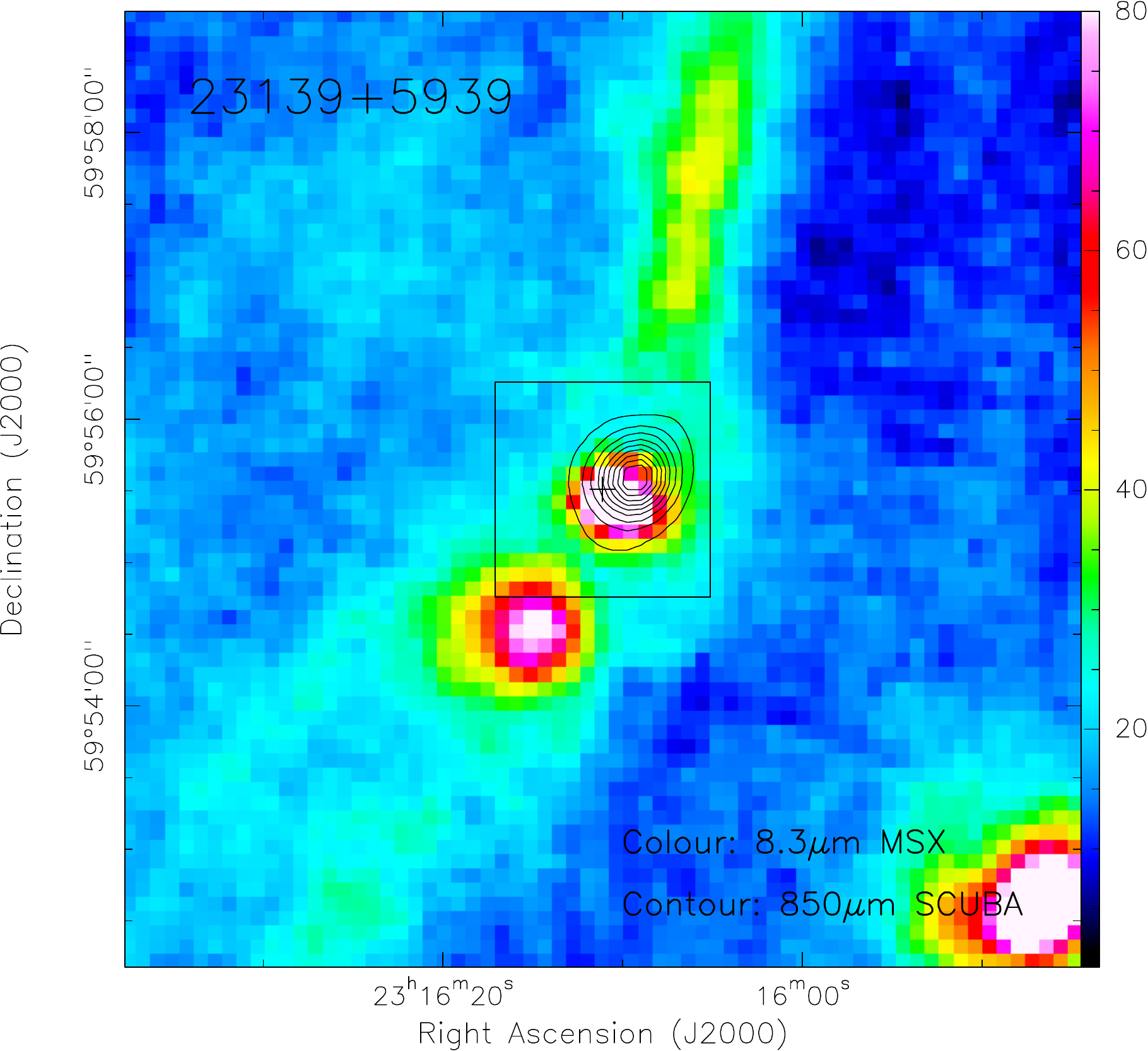, width=1.0\columnwidth, angle=0} &
 \epsfig{file=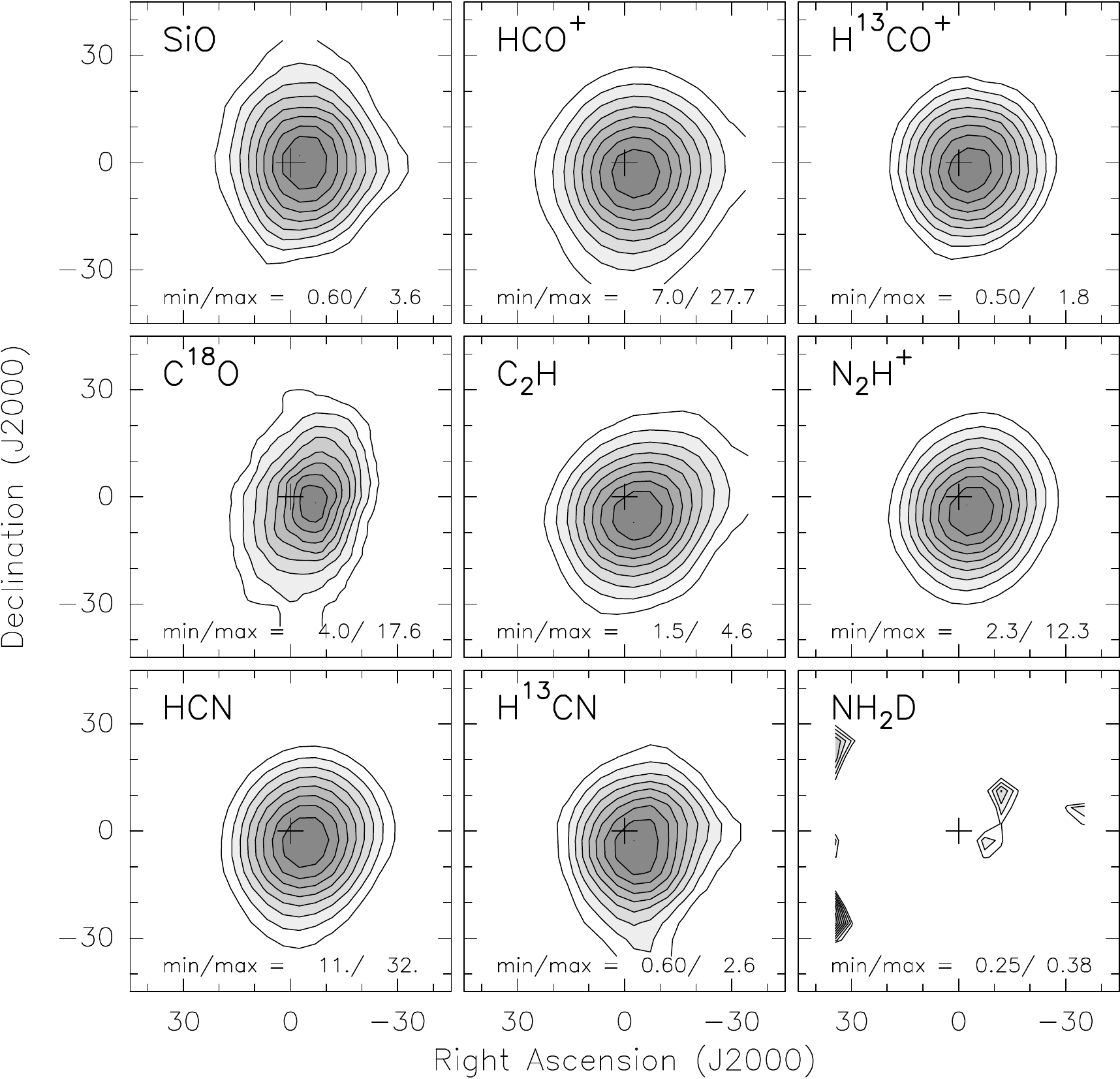, width=0.9\columnwidth, angle=0} \\
% \vspace{1cm}
 \multicolumn{2}{c}{\epsfig{file=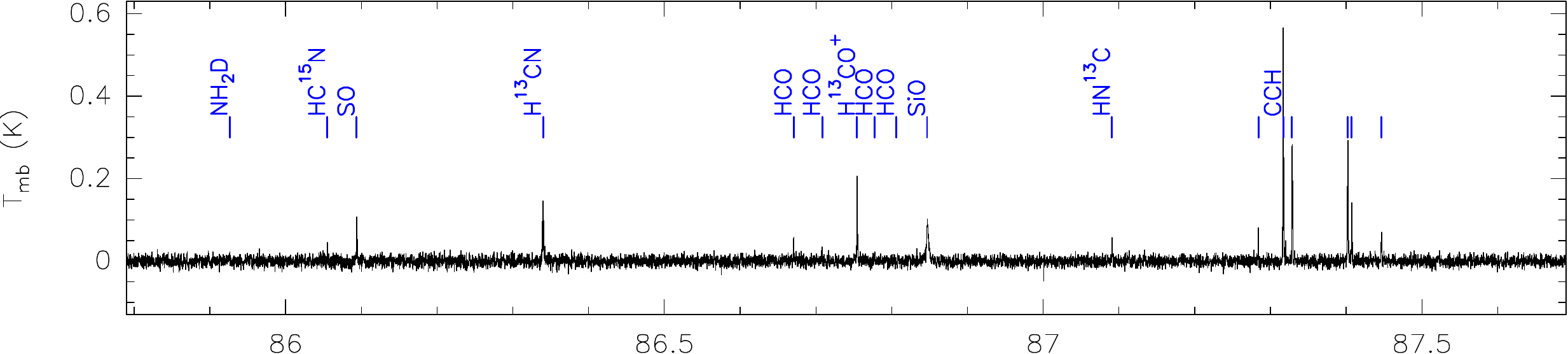, width=0.8\textwidth, angle=0}} \\
 \multicolumn{2}{c}{\epsfig{file=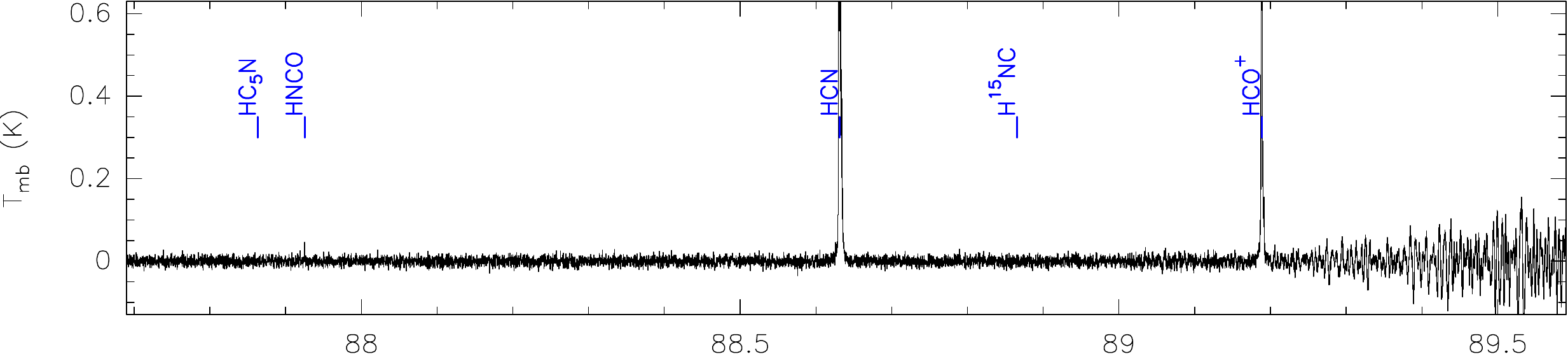, width=0.8\textwidth, angle=0}} \\
 \multicolumn{2}{c}{\epsfig{file=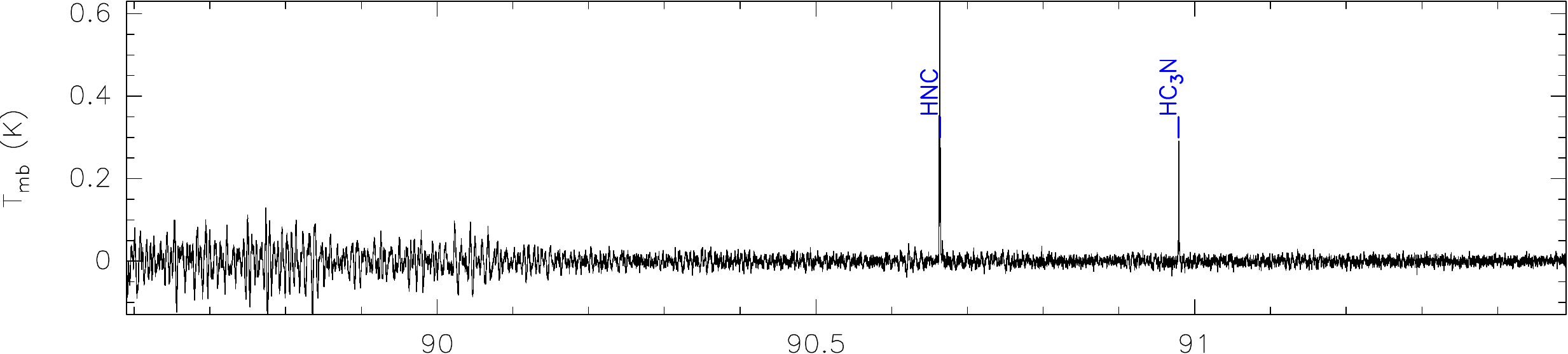, width=0.8\textwidth, angle=0}} \\
 \multicolumn{2}{c}{\epsfig{file=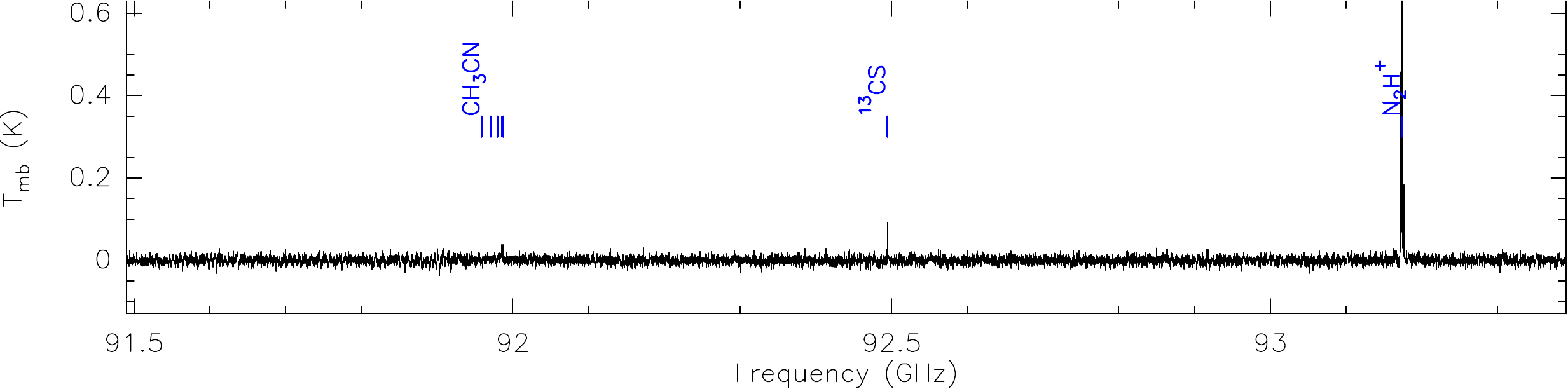, width=0.8\textwidth, angle=0}} \\
\end{tabular}
\end{center}
\caption{continued, for 23139$+$5939.}
\end{figure*}
%
%----------------------------------------------------------------------------

%----------------------------------------------------------------------------
\begin{table*}[ph!]
\caption{\label{t:IRfluxes}Measured continuum fluxes, in Jy, from mid-IR to millimeter wavelengths}
\centering
\begin{tabular}{l r c c c c c c c c c c c c c c c c c c}
\hline\hline
&
&\multicolumn{14}{c}{Spectral Energy Distributions for our sources}
\\
\cline{3-16}
&
&\#01 &\#02 &\#03 &\#04 &\#05 &\#06 &\#07 &\#08 &\#09 &\#10 &\#11 &\#12 &\#13 &\#14
\\
\hline
%						&01		&02		&03		&04		&05		&06		&07		&08		&09		&10		&11		&12		&13		&14		\\
MSX\supa		&8.3~$\mu$m	&10.3	&$<$0.2	&$<$0.2	&0.2		&0.6		&$<$0.2	&19.5	&$<$0.2	&$<$0.2	&$<$0.1	&1.2		&$<$0.2	&2.0		&2.1		\\
IRAS\supb	&12~$\mu$m	&19.0	&---		&---		&2.9		&2.9		&---		&22.8	&---		&$<$5	&$<$1.4	&$<$1.4	&---		&5.3		&2.9		\\
MSX\supa		&12.1~$\mu$m	&21.8	&$<$0.6	&$<$0.6	&$<$0.5	&$<$0.5	&$<$0.6	&15.6	&$<$0.6	&$<$0.6	&$<$0.8	&1.8		&$<$0.6	&3.1		&3.0		\\
MSX\supa		&14.7~$\mu$m	&33.0	&$<$0.6	&$<$0.6	&0.9		&0.7		&$<$0.6	&26.3	&$<$0.6	&$<$0.6	&$<$0.6	&2.2		&$<$0.6	&6.1		&5.4		\\
MSX\supa		&21.3~$\mu$m	&61.8	&$<$1.1	&$<$1.1	&5.5		&3.4		&$<$1.1	&69.7	&$<$1.1	&$<$1.1	&4.1		&13.6	&$<$1.1	&46.7	&24.3	\\
MIPS\supc	&24~$\mu$m	&---		&0.02	&0.24	&$>$6	&5.4		&0.22	&---		&1.8		&4.4		&$>$6	&$>$10	&0.25	&$>$18	&---		\\
IRAS\supb	&25~$\mu$m	&98.6	&---		&---		&14.3	&11.1	&---		&137.6	&---		&5.9		&$<$27.2	&$<$27.2	&---		&129.1	&50.1	\\
IRAS\supb	&60~$\mu$m	&891		&---		&---		&324		&317		&---		&958		&---		&187		&$<$765	&$<$765	&---		&1725	&367		\\
MIPS\supc	&70~$\mu$m	&---		&---		&---		&260		&291		&15		&---		&44		&245		&479		&321		&14		&---		&---		\\
HiGAL\supd	&70~$\mu$m	&---		&---		&14.3	&280.8	&386.3	&33.7	&1200	&98.4	&659.8	&942.0	&338.4	&38.7	&2212	&---		\\
IRAS\supb	&100~$\mu$m	&1891	&---		&---		&1028	&1104	&---		&2136	&---		&727		&$<$1948	&$<$1948	&---		&2744	&$<$686	\\
HiGAL\supd	&160~$\mu$m	&---		&---		&55.6	&400.2	&725.5	&139.5	&1590	&169.2	&866.0	&1120	&587.0	&91.7	&1598	&---		\\
HiGAL\supd	&250~$\mu$m	&---		&---		&63.8	&328.8	&464.5	&165.0	&623		&117.6	&496.3	&585.5	&538.6	&97.0	&682.2	&---		\\
HiGAL\supd	&350~$\mu$m	&---		&---		&29.3	&130.4	&229.2	&81.7	&359		&44.2	&229.1	&326.3	&270.1	&52.1	&385.8	&---		\\
JCMT\supe	&450~$\mu$m	&83		&14		&25		&194		&---		&---		&---		&---		&---		&148		&---		&38		&$<$351	&76		\\
HiGAL\supd	&500~$\mu$m	&---		&23.3	&13.1	&47.3	&66.4	&32.8	&116		&13.6	&73.7	&82.3	&95.5	&20.5	&84.9	&---		\\
JCMT\supe	&850~$\mu$m	&9.53	&2.88	&3.83	&20.4	&13.4	&---		&---		&---		&---		&16.7	&---		&5.26	&16.7	&6.32	\\
APEX\supf	&850~$\mu$m	&---		&3.13	&3.43	&16.8	&20.6	&5.95	&25.6	&3.29	&9.75	&---		&---		&---		&---		&---		\\
---\supg		&1.2~mm		&3.60	&0.92	&0.91	&10.4	&7.6		&1.11	&10.6	&0.71	&2.03	&4.01	&13.2	&1.02	&5.3		&2.30	\\
% HiGAL fluxes from a circular area
%HiGAL	&70~$\mu$m	&---		&28		&45		&287		&355		&123		&941		&175		&573		&808		&254		&47		&1840	&---		\\
%HiGAL	&160~$\mu$m	&---		&157		&208		&609		&778		&411		&1230	&421		&997		&1230	&628		&188		&1450	&---		\\
%HiGAL	&250~$\mu$m	&---		&139		&178		&439		&524		&324		&609		&292		&465		&609		&544		&172		&485		&---		\\
%HiGAL	&350~$\mu$m	&---		&71		&87		&196		&237		&152		&375		&130		&280		&375		&254		&94		&250		&---		\\
%HiGAL	&500~$\mu$m	&---		&34		&40		&78		&90		&64		&153		&55		&104		&153		&106		&44		&74		&---		\\
% HiGAL fluxes from a hand-defined polygon
%HiGAL	&70~$\mu$m	&---		&0.2		&7.2		&268		&352		&17.3	&1200	&64.4	&537		&784		&220		&22.3	&2020	&---		\\
%HiGAL	&160~$\mu$m	&---		&9.2		&22.1	&622		&780		&170		&1590	&108		&844		&1150	&526		&52.7	&1680	&---		\\
%HiGAL	&250~$\mu$m	&---		&10.7	&58.7	&510		&485		&207		&623		&121		&434		&517		&646		&75.3	&543		&---		\\
%HiGAL	&350~$\mu$m	&---		&28.1	&30.7	&197		&175		&98.7	&359		&77.8	&216		&315		&328		&46.7	&257		&---		\\
%HiGAL	&500~$\mu$m	&---		&12.8	&9.6		&65.7	&80.9	&35.8	&116		&19.1	&72.9	&122		&84.2	&28.1	&77.7	&---		\\
\hline
\end{tabular}
\begin{list}{}{}
\item[\supa] Fluxes obtained from the catalog of the Midcourse Space Experiment \citep[MSX;][]{price1999}.
\item[\supb] Fluxes obtained from the point source catalog v2.1 of the Infrared Astronomical Satellite \citep[IRAS;][]{neugebauer1984}.
\item[\supc] Fluxes obtained from the Spitzer satellite, and the instrument Multiband Imaging Photometer for Spitzer \citep[MIPS;][]{rieke2004}.
\item[\supd] Fluxes obtained from the Hi-GAL/Herschel project \citep{molinari2010a, molinari2010b}.
\item[\supe] Fluxes obtained from the SCUBA Legacy survey \citep{difrancesco2008}.
\item[\supf] Fluxes obtained from the ATLASGAL project \citep{schuller2009}.
\item[\supg] At 1.2~mm we used the single-dish surveys of \citet{beuther2002a, faundez2004, hill2005, rathborne2006}, carried out with the IRAM~30\,m telescope or with the SEST (Swedish-ESO 15m Submillimeter Telescope).
\end{list}
\end{table*}
%----------------------------------------------------------------------------

%----------------------------------------------------------------------
%
%----------------------------------------------------------------------------

\end{appendix}
\end{document}